\documentclass[preprint]{aastex6}
\usepackage{natbib}
\usepackage{amssymb}
\newcommand{\Bl}{B_{\rm los}}
\newcommand{\Bz}{B_{\rm z}}

\newcommand{\x}{\mbox{\boldmath$x$}}
\newcommand{\grad}{\mbox{\boldmath$\nabla$}}
\newcommand{\TP}{{\rm TP}}
\newcommand{\TN}{{\rm TN}}
\newcommand{\FP}{{\rm FP}}
\newcommand{\FN}{{\rm FN}}
\newcommand{\Heidke}{{\rm HSS}}
\newcommand{\True}{{\rm H\&KSS}}
\newcommand{\Appleman}{{\rm ApSS}}
\newcommand{\Brier}{{\rm BSS}}
\newcommand{\RC}{{\rm RC}} 
\newcommand{\CC}{C1.0+, 24\,hr}
\newcommand{\MS}{M1.0+, 12\,hr}
\newcommand{\ML}{M5.0+, 12\,hr}
\newcommand{\MCD}[1]{MCD\##1}

\usepackage{longtable}

\begin{document}

\title{A Comparison of Flare Forecasting Methods, I: Results from the ``All-Clear'' Workshop}

\author{G.~Barnes and K.D.~Leka}
\affil{NWRA, 3380 Mitchell Ln., Boulder, CO 80301, USA}
\email{graham@nwra.com, leka@nwra.com}

\author{C.J.~Schrijver}
\affil{Lockheed Martin Solar and Astrophysics Laboratory, 3251 Hanover St., Palo Alto, CA 94304, USA}
\email{schrijver@lmsal.com}

\author{T.~Colak, R.~Qahwaji, O.W.~\replaced{Ahmed}{Ashamari}}
\affil{School of Computing Informatics and Media, University of Bradford, Bradford, UK}
\email{Tufancolak@hotmail.com, r.s.r.qahwaji@bradford.ac.uk, omarashamari@gmail.com}

\author{Y.~Yuan}
\affil{Space Weather Research Laboratory, New Jersey Institute of Technology, Newark, NJ 07102, USA}
\email{yy46@njit.edu}

\author{J.~Zhang}
\affil{Department of Physics and Astronomy, George Mason University, 4400 University Dr., Fairfax, VA 22030, USA}
\email{jzhang7@gmu.edu}

\author{R.T.J.~McAteer}
\affil{Department of Astronomy, New Mexico State University, P.O. Box 30001, MSC 4500, Las Cruces, NM 88003-8001, USA}
\email{mcateer@nmsu.edu}

\author{D.S.~Bloomfield\altaffilmark{1}, P.A.~Higgins, P.T.~Gallagher}
\affil{Astrophysics Research Group, School of Physics, Trinity College Dublin, College Green, Dublin 2, Ireland}
\email{shaun.bloomfield@tcd.ie, pohuigin@gmail.com, peter.gallagher@tcd.ie}

\author{D.A.~Falconer}
\affil{Heliophysics and Planetary Science Office, ZP13, Marshall Space Flight Center, Huntsville, AL 35812, USA ; Physics Department, University of Alabama in Huntsville, AL 35899, USA ; Center for Space Plasma and Aeronomic Research, University of Alabama in Huntsville, AL 35899, USA}
\email{david.a.falconer@nasa.gov}

\author{M.K.~Georgoulis}
\affil{Research Center for Astronomy and Applied Mathematics (RCAAM), Academy of Athens,  4 Soranou
Efesiou Street, Athens 11527, Greece}
\email{Manolis.Georgoulis@academyofathens.gr}

\author{M.S.~Wheatland}
\affil{Sydney Institute for Astronomy, School of Physics, The University of Sydney, NSW 2006, Australia}
\email{michael.wheatland@sydney.edu.au}

\author{C.~Balch}
\affil{NOAA Space Weather Prediction Center, 325 Broadway Ave, Boulder, CO 80305, USA}
\email{Christopher.Balch@noaa.gov}

\author{T.~Dunn and E.L.~Wagner}
\affil{NWRA, 3380 Mitchell Ln., Boulder, CO 80301, USA}
\email{tdunn@nwra.com, wagneric@nwra.com}

\altaffiltext{1}{\added{Department of Physics and Electrical Engineering, Northumbria
University, Newcastle Upon Tyne, NE1 8ST, UK}}

\begin{abstract} 
Solar flares produce radiation which can have an almost immediate effect on the
near-Earth environment, making it crucial to forecast flares in order to
mitigate their negative effects.  The number of published approaches to flare
forecasting using photospheric magnetic field observations has proliferated,
with varying claims about how well each works.  Because of the different
analysis techniques and data sets used, it is essentially impossible to compare
the results from the literature.  This problem is exacerbated by the low event
rates of large solar flares.  The challenges of forecasting rare events have
long been recognized in the meteorology community, but have yet to be fully
acknowledged by the space weather community.  During the interagency workshop
on ``all clear'' forecasts held in Boulder, CO in 2009, the performance of a
number of existing algorithms was compared on common data sets, specifically
line-of-sight magnetic field and continuum intensity images from MDI, with
consistent definitions of what constitutes an event.  We demonstrate the
importance of making such systematic comparisons, and of using standard
verification statistics to determine what constitutes a good prediction scheme.
When a comparison was made in this fashion, no one method clearly outperformed
all others, which may in part be due to the strong correlations among the
parameters used by different methods to characterize an active region.  For
M-class flares and above, the set of methods tends towards a weakly positive
skill score (as measured with several distinct metrics), with no participating
method proving substantially better than climatological forecasts. 
\end{abstract}


\section{Introduction}
\label{sec:intro}

Solar flares produce X-rays which can have an almost immediate effect on the
near-Earth environment, especially the terrestrial ionosphere.  With only eight
minutes delay between the event occurring and its effects at Earth, it is
crucial to be able to forecast solar flare events in order to mitigate negative
socio-economic effects.  As such, it is desirable to be able to predict when
a solar flare event will \added{occur} and how large it will be prior to observing the flare
itself.  In the last decade, the number of published approaches to flare
forecasting using photospheric magnetic field observations has proliferated,
with widely varying evaluations about how well each works \citep[e.g.,][in
addition to references for each method described,
below]{Abramenko2005b,Jing_etal_2006,dfa3,McAteer_etal_2005,Schrijver2007,BarnesLeka2008,MasonHoeksema2010,Yu_etal_2010b,YangX_etal_2013,Boucheronetal2015,AlGhraibahetal2015}.

Some of the discrepancy in reporting success arises from how success is
evaluated, a problem exacerbated by the low event rates typical of large solar
flares.  The challenges of forecasting when event rates are low have long been
recognized in the meteorology community \citep[e.g.,][]{Murphy1996}, but have
yet to be fully acknowledged by the space weather community.  The use of
climatological skill scores \citep{Woodcock1976,JolliffeStephenson2003}, which
account for event climatology and in some cases for underlying sample
discrepancies, enables a more informative assessment of forecast performance
\citep{BarnesLeka2008,Balch2008,Bloomfield_etal_2012,Crown2012}.

Comparisons of different studies are also difficult because of differences in
data sets, and in the definition of an event used.  The requirements and
limitations of the data required for any two techniques may differ (e.g., in
the field-of-view required, imposed limits on viewing angle, and the data
required for a training set).  Event definitions vary in the temporal window
(how long a forecast is applicable), the latency (time between the observation
and the start of the forecast window), and more fundamentally in what
phenomenon constitutes an event, specifically the flare magnitude.

A workshop was held in Boulder, CO in 2009 to develop a framework to compare
the performance of different flare forecasting methods.  The workshop was
sponsored jointly by the NASA/Space Radiation Analysis Group and the National
Weather Service/Space Weather Prediction Center, hosted at the National Center
for Atmospheric Research/High Altitude Observatory, with data preparation and
analysis for workshop participants performed by NorthWest Research Associates
under funding from NASA/Targeted Research and Technology program.   In addition
to presentations by representatives of interested commercial entities and
federal agencies, researchers presented numerous flare forecasting methods. 

The focus of the workshop was on ``all-clear'' forecasts, namely predicting
time intervals during which {\it no} flares occur that are over a given
intensity (as measured using the peak GOES 1--8\,\AA\ flux). 
\replaced{For users of the forecasts, it can be more useful to know when no
event will occur, however, most forecasting methods focus on simply predicting
the probability that a flare will occur.  The results presented here are thus
not specific to all-clear forecasts.}
{For users of these forecasts, it can be useful to know when no event will
occur because the cost of a missed event is much higher than the cost of a
false alarm.  However, most forecasting methods focus on simply predicting the
probability that a flare will occur.  Therefore, the results presented here
focus on comparing flare predictions and are not specific to all-clear
forecasts.}

The workshop made a first attempt at direct comparisons between methods.  Data
from the Solar and Heliospheric Observatory/Michelson Doppler Imager
\citep[{\it SoHO}/MDI;][]{mdi} were prepared and distributed, and it was
requested that participants with flare-prediction algorithms use their own
methods to make predictions from the data.  The data provided were for a
particular time and a particular active region (or group in close proximity).
That is, the predictions are made using single snapshots and do not include the
evolution of the active regions.  Thus, the data were not ideal for many of the
methods.  Using time-series data likely increases the information available and
hence the potential for better forecasts, but at the time of the workshop and
as a starting point for building the infrastructure required for comparisons,
only daily observations were provided \citep[however, see][where time series
data were presented]{ffc2}. 

The resulting predictions were collected, and standard verification statistics
were calculated for each method.  For the data and event definitions
considered, no method achieves values of the verification statistics that are
significantly larger than all the other methods, and there is considerable room
for improvement for all the methods.  There were some trends common to the
majority of the methods, most notably that higher values of the verification
statistics are achieved for smaller event magnitudes. 

The data preparation is described in Section~\ref{sec:data}, and
Section~\ref{sec:sss} provides an overview of how to evaluate the performance
of forecasts.  The methods are summarized in Section~\ref{sec:methods}, with
more details given in Appendix~\ref{app:methods}, and sample results are
presented in Section~\ref{sec:results}.  The results and important trends are
discussed in Section~\ref{sec:discussion}.  Finally, Appendix~\ref{app:website}
describes how to access the data used during the workshop, along with many of
the results. 

\section{Workshop Data}
\label{sec:data}

The data prepared and made available for the workshop participants constitutes
the basic level of data that was usable for the majority of methods compared.
Some methods could make use of more sophisticated data or time series or a
different wavelength, but the goal for this particular comparison is to provide
all methods with the same data, so the only differences are in the methods, not
in the input data.

The database prepared for the workshop is comprised of line-of-sight magnetic
field data from the newest MDI calibration (Level 1.8\footnote{See
\url{http://soi.stanford.edu/magnetic/Lev1.8/}.}) for the years 2000--2005
inclusive.  The algorithms for region selection and for extracting sub-areas
are described in detail below (\S~\ref{sec:mdidata}).  The event data are solar
flares with peak GOES flux magnitude C1.0 and greater, associated with an
active region (see \S~\ref{sec:eventdata}).  All these data are available for
the community to view and test new methods\footnote{See
\url{http://www.cora.nwra.com/AllClear/}.} (see
Appendix~\ref{app:website} for details).

\subsection{Selection and Extraction of MDI data}
\label{sec:mdidata}

The data set provided for analysis contains sub-areas extracted from the
full-disk magnetogram and continuum images from the {\it SoHO}/MDI.  These
extracted magnetogram and intensity image files, presented in FITS format, are
taken close to noon each day, specifically daily image \#0008 from the {\tt
M\_96m} magnetic field data series, which was generally obtained between 12:45
and 12:55 UT, and image \#0002 from the {\tt Ic\_6h} continuum intensity
series, generally obtained before 13:00 UT. 

To extract regions for a given day, the list of daily active-region coordinates
was used, as provided by the National Oceanic and Atmospheric Administration
(NOAA), and available through the National Center for Environmental Information
(NCEI)\footnote{See \url{http://www.ngdc.noaa.gov/stp/stp.html}.}.  The
coordinates were rotated to the continuum image/magnetic field time using
differential rotation and the synodic apparent solar rotation rate.  A box was
centered on the active region coordinates whose size reflects the NOAA listed
size of the active region in micro-hemispheres (but not adjusted for any
evolution between the issuance and time of the magnetogram or continuum image).
\replaced{A minimum active region size of 100\,$\mu$H was chosen to impose a
minimum box size of $125\arcsec \times 125\arcsec$.  The extracted box size was
also scaled with the location on the solar disk to reflect only the intrinsic
reported area and roughly preserve the area on the Sun contained within each
box regardless of observing angle.} {A minimum active region size of
100\,$\mu$H was chosen, corresponding to a minimum box size of $125\arcsec
\times 125\arcsec$ at disk center.  The extracted box size was scaled according
to the location on the solar disk to reflect the intrinsic reported area and to
roughly preserve the area on the Sun contained within each box regardless of
observing angle. This procedure most noticeably decreases the horizontal size
near the solar east/west limbs, although the vertical size is impacted
according to the region's latitude.} The specifics of the scalings and minimum
(and maximum) sizes were chosen empirically for ease of processing, and do not
necessarily reflect any solar physics beyond the reported distribution of
active region sizes reported by NOAA.

\begin{figure}
\epsscale{1.0}
\plottwo{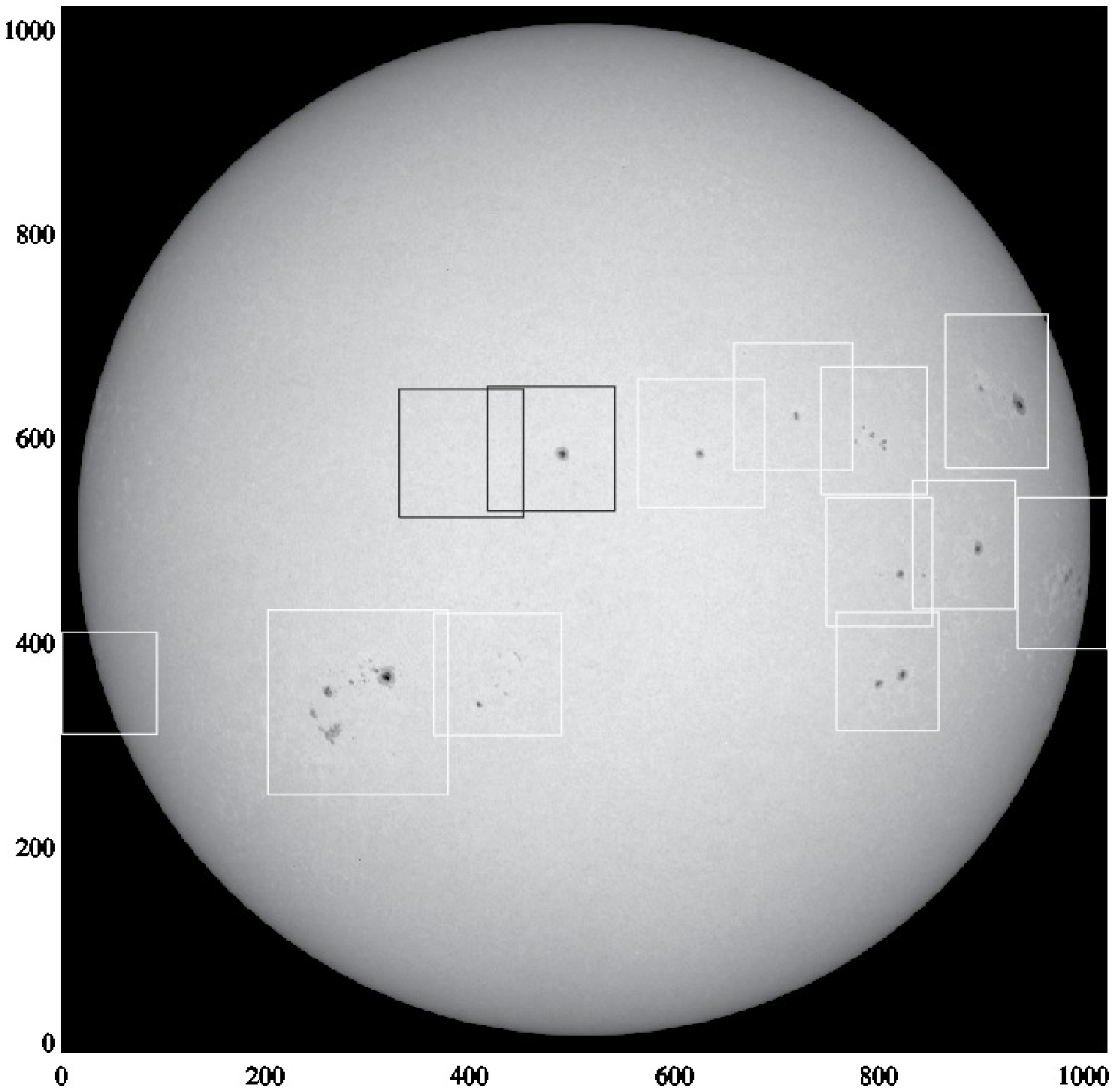}{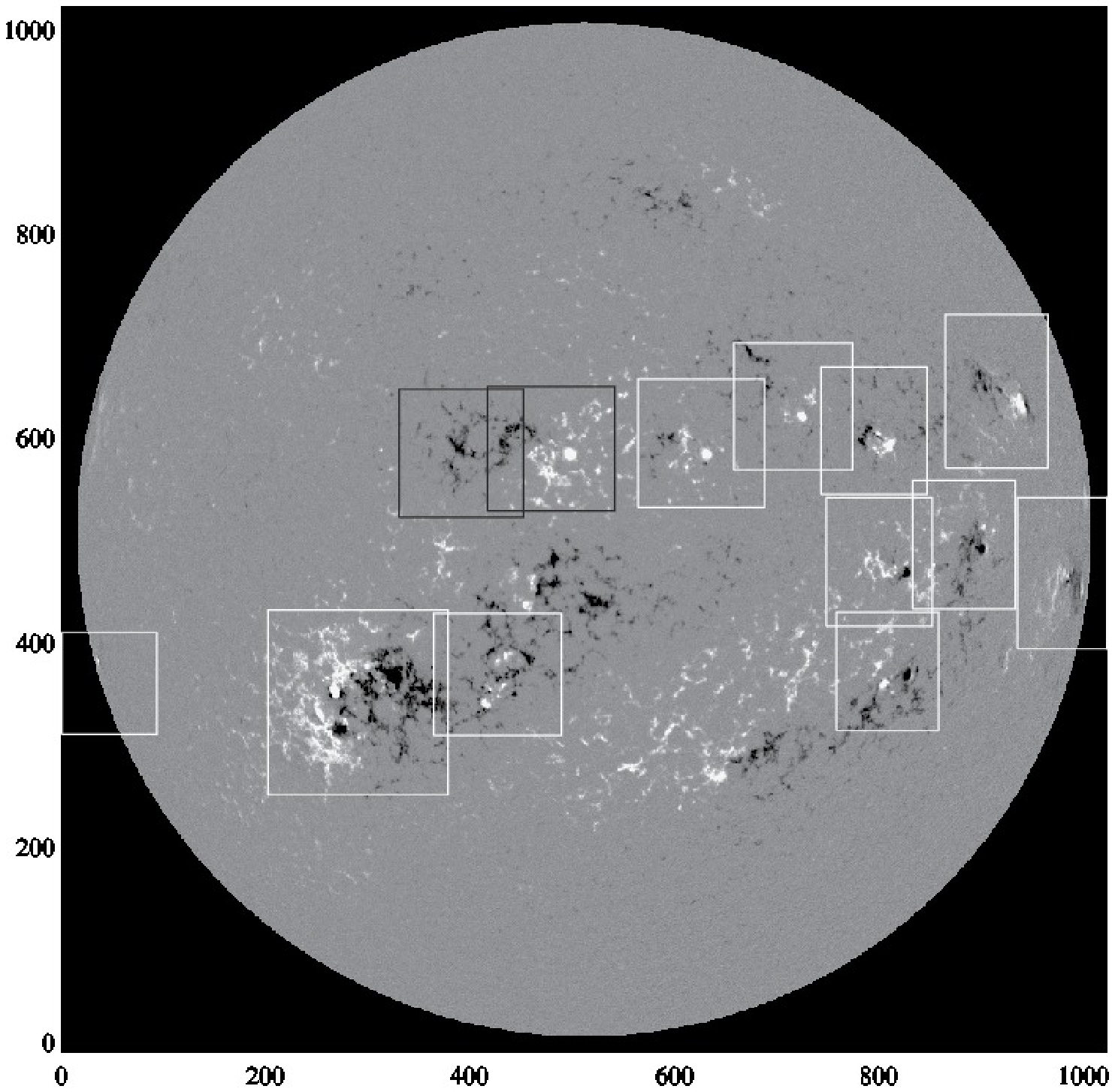}
\plottwo{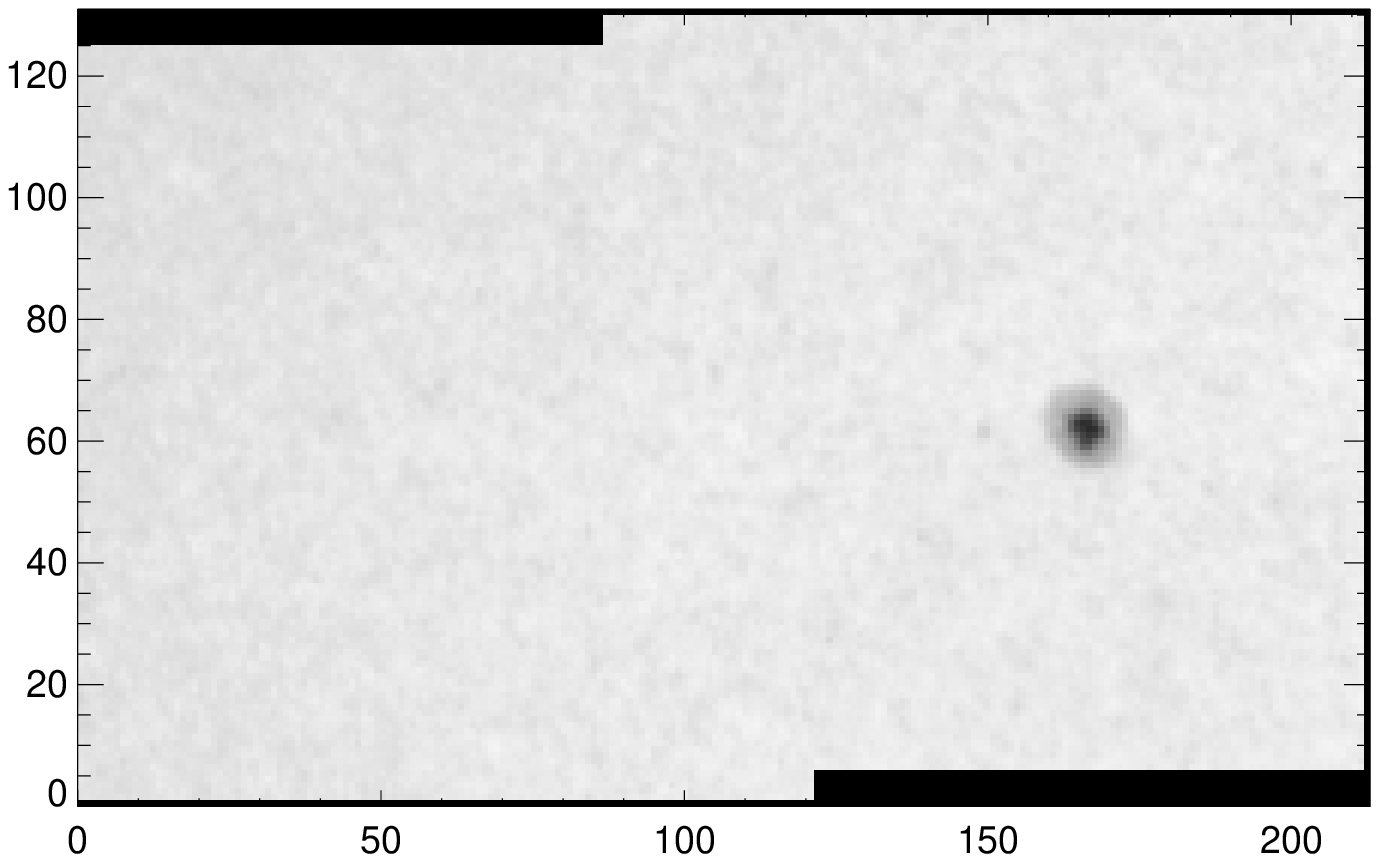}{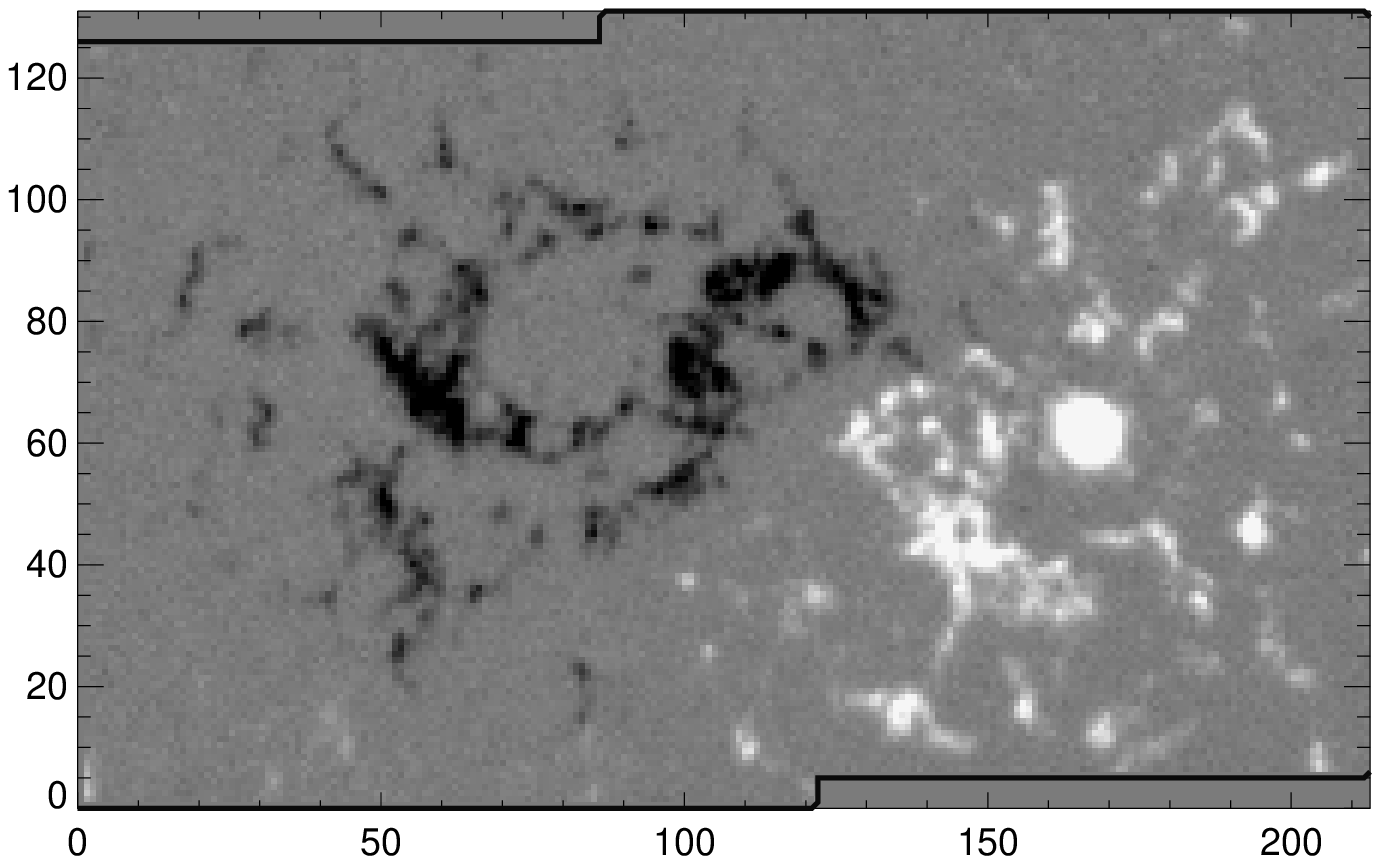}
\caption{Examples of the active region patches extracted from full-disk images
for 2002 January 3. {\bf Top:} each white rectangle includes a single NOAA
numbered active region; each black rectangle includes a single NOAA numbered
active region which is judged to be part of a cluster of active regions
\replaced{which is}{and} subsequently treated together.  The clustering
criterion is a function of the relative box sizes of the regions in question,
hence while other boxes overlap, the overlap and respective areas do not meet
the criterion.  {\bf Left:} MDI full-disk continuum intensity image with
extracted areas indicated.  {\bf Right:} MDI full-disk line of sight magnetic
field image, scaled to $\pm 500$\,G with the same extracted areas indicated.
{\bf Bottom:} example of the extraction of a cluster. The black area in the
figures above (in this case NOAA ARs 09766, 09765 on the left, right
respectively) shows the cluster with the areas on the periphery of the cluster
zeroed out. {\bf Left:} the continuum intensity, and {\bf Right:} the line of
sight magnetic field image, scaled to $\pm 500$\,G.  In the latter case, a
contour indicates the non-overlap zeroed-out area.
}
\label{fig:MDI}
\end{figure}

During times of high activity, this simple approach to isolate active regions
using rectangular arrays often creates two boxes which overlap by a significant
amount.  Such an overlap, especially when it includes strong-field areas from
another active region, introduces a double-counting bias into the
flare-prediction statistics.  To avoid this, regions were merged when an
overlapping criterion based on the geometric mean of the regions' respective
areas and the total was met.  If two or more of these boxes overlap such that
${\mathcal{A}}/\sqrt{({\rm area\ box}_1 \times {\rm area\ box}_2)} > 0.95$,
where ${\mathcal{A}}$ is the area of overlap (in image-grid pixels) of the two
boxes, then the two boxes were combined into one region or ``merged cluster''.
Clustering was restricted to occur between regions in the same hemisphere.  No
restrictions were imposed to limit the clustering, and in some cases more than
two (at most six) regions are clustered together.  In practice, in fewer than 10
cases, the clustering algorithm was over-ridden by hand in order to prevent
full-Sun clusters, in which case the manual clustering was done in such a way
as to separate clusters along areas of minimum overlap.  The clustering is
performed in the image-plane, although as mentioned above the box sizes took
account of the location on the observed disk.  When boxes merged, a new
rectangle was drawn around the merged boxes.  The area not originally included
in a single component active region's box was zeroed out.  An example of the
boxes for active regions and a 2-region cluster for 2002 January 3 is shown in
Figure~\ref{fig:MDI}.  {\tt JPEG} images of all regions and clusters similar to
Figure~\ref{fig:MDI} (top) are available at the workshop website.
Note that a morphological
analysis method based on morphological erosion and dilation has been
used as a robust way to group or reject neighboring active regions
\citep{ZhangWangLiu2010} although it is not implemented here.

The final bounds of each extracted magnetogram file are the starting
point for extracting an accompanying continuum file. The box was shifted
to adjust for the time difference between the acquisition time of the
magnetogram and that of the continuum file. If the time difference was
greater than four hours, continuum extracted files were not generated
for that day.  Also, if the MDI magnetogram {\tt 0008} file was unavailable
and the magnetogram nearest noon was obtained more than 96
min from noon, then neither the continuum nor the magnetogram extracted
files were made for that day.

Data were provided as FITS files, with headers derived from the
original but modified to include all relevant pointing information for
the extracted area and updated ephemeris information.  Additionally, the
NOAA number, the NOAA-reported area (in $\mu$H) and number of spots, and
the Hale and McIntosh classifications (from the most recent NOAA report)
are also included.  The number of regions is included in the header,
which is $>1$ only for clusters. In the case of clusters, all of the
classification data listed above are included separately for each NOAA
region in the cluster.

There was no additional stretching or re-projection performed on the data;
the images were presented in plane-of-the-sky image coordinates.
No pre-selection was made for an observing-angle limit, so many of the
boxes are close to the solar limb. Similarly, no selection conditions
were imposed for active-region size, morphology or flaring history,
beyond the fact that a NOAA active region number was required. The result
is 12,965 data points (daily extracted magnetograms) between 2000--2005.

\subsection{Event Lists}
\label{sec:eventdata}

Event lists were constructed from flares recorded in the NCEI archives. Three
definitions of event were considered:
\begin{itemize}
\item at least one C1.0 or greater flare within 24\,hr after the observation,
\item at least one M1.0 or greater flare within 12\,hr after the observation, and 
\item at least one M5.0 or greater flare within 12\,hr after the observation.
\end{itemize}
Table~\ref{tbl:flare_rate} shows the flaring sample size for each event
definition.  No distinction is made between one and multiple flares above the
specified threshold: a region was considered a member of the flaring sample
whether one or ten flares occurred that satisfied the event definition.  This
flaring sample size defines the climatological rate of one or more flares
occurring for each definition which in turn forms the baseline forecast against
which results are compared.

\begin{center}
\begin{deluxetable}{lcc}
\tablecolumns{3}
\tablewidth{0pc}
\tablecaption{Flare Event Rates}
\tablehead{
\colhead{Event Definition} & \colhead{Number in Event Sample} & 
\colhead{Fraction In Event Sample}}
\startdata
\CC &   2609 & 0.201 \\
\MS &  \ 400 & 0.031 \\
\ML & \ \ 93 & 0.007 \\
\enddata
\label{tbl:flare_rate}
\end{deluxetable}
\end{center}

A subtlety arises because not all of the observations occur at exactly the same
time of day, whereas the event definitions use fixed time intervals ({\it i.e.}
12 or 24\,hr).  As such it is possible for a particular flare to be
double-counted. For example, if magnetograms were obtained at 12:51\,UT on day
\#1 and then 12:48\,UT on day\#2, and a flare occurred at 12:50\,UT on day \#2
then both days would be part of the \CC\ flaring sample. This situation only
arises for the 24\,hr event interval and is likely to be extremely rare.
However, it means that not all the events are strictly independent.

The NOAA active region number associated with each event from NCEI is used to
determine the source of the flare. When no NOAA active region number is
assigned, a flare is assumed to not have come from any visible active region,
although it is possible that the flare came from an active region but no
observations were available to determine the source of the flare.  For
large-magnitude flares, the vast majority of events ($\approx 85$\% for M-class
and larger flares, $\approx 93$\% for M5.0 and larger, and $\approx 93$\% for
X-class flares) are associated with an active region, but a substantial
fraction of small flares are not ($\approx$ 38\% for all C-class flares).

\section{Overview of Evaluation Methods}
\label{sec:sss}

To quantify the performance of binary, categorical forecasts, contingency
tables and a variety of skill scores are used
\citep[e.g.,][]{Woodcock1976,JolliffeStephenson2003}.  A contingency table
(illustrated in Table~\ref{tbl:contingency_eg}),\footnote{The reader should
note that many published contingency tables flip axes relative to each other.
We follow the convention of \citet{Woodcock1976} and
\citet{Bloomfield_etal_2012}, but this is opposite, for example, to that of
\citet{JolliffeStephenson2003,Crown2012}.} summarizes the performance of
categorical forecasts in terms of the number of true positives, $\TP$ (hits),
true negatives, $\TN$ (correct rejections), false positives, $\FP$ (false
alarms), and false negatives, $\FN$ (misses).  The elements of the contingency
table can be combined in a variety of ways to obtain a single number
quantifying the performance of a given method. 

\begin{deluxetable}{lcc}
\tablecolumns{3}
\tablewidth{0pc}
\tablecaption{Example Contingency Table}
\tablehead{\colhead{} & \multicolumn{2}{c}{Predicted}}
\startdata 
Observed & {\it Event} & {\it No Event}  \\ \hline
{\it Event}    & True Positive  ($\TP$, hit)         & False Negative ($\FN$, miss)  \\
{\it No Event} & False Positive ($\FP$, false alarm) & True Negative  ($\TN$, correct negative) \\
\enddata
\tablecomments{The number of events is $\TP + \FN$, the number of non-events is
$\FP + \TN$, and the sum of all entries, $N=\TP + \FP + \TN + \FN$, is the
sample size.}
\label{tbl:contingency_eg}
\end{deluxetable}

One quantity that at least superficially seems to measure forecast performance
is the {\it Rate Correct}. This is simply the fraction correctly predicted, for
both event and no-event categories, 
\begin{equation}
\RC = (\TP + \TN)/N, 
\end{equation}
where $N=\TP+\FP+\FN+\TN$ is the total number of forecasts.  A perfect forecast
has $\RC=1$, while a set of completely incorrect forecasts has $\RC=0$.  The
accuracy is an intuitive score, but can be misleading for very unbalanced
event/no-event ratios (e.g., $\TP + \FN << \FP + \TN$) such as larger flares
because it is possible to get a very high accuracy by always forecasting no
event \citep[see, for example][for an extensive discussion of this issue in the
context of the famous ``Finley Affair'' in tornado forecasting]{Murphy1996}. A
forecast system that always predicts no event has $\RC = \TN/(\TN + \FN)$,
which approaches one as the event/no-event ratio goes to zero (i.e., $\FN \ll
\TN$).  A widely used approach to avoid this issue is to normalize the
performance of a method to a reference forecast by using a skill score
\citep{Woodcock1976,JolliffeStephenson2003,SWJ,Bloomfield_etal_2012}\footnote{See
also
\url{http://www.cawcr.gov.au/projects/verification/\#What\_makes\_a\_forecast\_good}.}.

A generalized skill score takes the form: 
\begin{equation}
{\rm Skill} = \frac{A_{\rm forecast} - A_{\rm reference}}
{A_{\rm perfect} - A_{\rm reference}},
\end{equation}
%
where $A_{\rm forecast}$ is the accuracy of the method under consideration,
which can be any measure of how well the forecasts correspond to the observed
outcome.  $A_{\rm perfect}$ is the accuracy of a perfect forecast (i.e., the
entire sample is forecast correctly), and $A_{\rm reference}$ is the expected
accuracy of the reference method.  Skill scores are referred to by multiple
names, having been rediscovered by different authors over spans of decades; we
follow the naming convention used in \citet{Woodcock1976}.  Each skill score
has advantages and disadvantages, and quantifies the performance with slightly
different emphasis, but in general, ${\rm Skill}=1$ is perfect performance,
${\rm Skill}=0$ is no improvement over a reference forecast, and ${\rm
Skill}<0$ indicates worse performance than the reference.  

For binary, categorical forecasts, a measure of the forecast accuracy is the
rate correct,
\begin{equation}
A_{\rm forecast} = (\TP + \TN)/N, 
\end{equation}
and the corresponding accuracy of perfect forecasts is $A_{\rm perfect}=1$.
Three standard skill scores based on different reference forecasts are
described below.  

\noindent{\it Appleman's Skill Score} (\Appleman) uses the unskilled predictor
({\it i.e.}, the climatological event rate) as a reference:
\begin{equation}
A_{\rm reference} = {\TN + \FP \over N},
\end{equation}
for the case that the number of events is less than the number of non-events
($\TP + \FN<\TN + \FP$), as is typically the case for large flares.  When the
converse is true ($\TP + \FN>\TN + \FP$),
\begin{equation}
A_{\rm reference} = {\TP + \FN \over N}.
\end{equation}
\Appleman\ treats the cost of each type of error (miss and false alarm) as
equal.

\noindent{\it The Heidke Skill Score} (\Heidke) uses a random forecast as a
reference.  Assuming that the event occurrences and the forecasts for events
are statistically independent, the probability of a hit ($\TP$) is the product
of the event rate with the forecast rate, and the probability of a correct
rejection ($\TN$) is the product of the rate of non events with the rate of
forecasting non events.  Thus the reference accuracy is 
\begin{equation}
A_{\rm reference} 
= {(\TP + \FN) \over N} {(\TP + \FP) \over N} + {(\TN + \FN) \over N} 
{(\TN + \FP) \over N}.
\end{equation}
The Heidke skill score is very commonly used, but the random reference forecast
has to be used carefully since the quality scale has a dependence on the event 
rate (climatology).

\noindent{\it Hanssen \& Kuipers' Discriminant} (\True) uses a reference
accuracy
\begin{equation}
A_{\rm reference} = {\FN (\TN + \FP)^2 + \FP (\TP + \FN)^2 \over 
N \big [\FN (\TN + \FP) + \FP (\TP + \FN) \big ]}
\end{equation}
constructed such that both the random and unskilled predictors score zero.
The Hanssen \& Kuipers' Discriminant, also known as the True Skill Statistic,
can be written as the sum of two ratio tests, one for events (the probability
of detection) and one for non-events (the false alarm rate), 
\begin{equation}\label{eqn:true}
\True = {\TP \over \TP + \FN} - {\FP \over \FP + \TN}.
\end{equation}
As such it is not sensitive to differences in the size of the event and
no-event samples, provided the samples are drawn from the same populations.
This can be particularly helpful when comparing studies performed on different
data sets \citep{HanssenKuipers1965,Bloomfield_etal_2012}. 

\begin{figure}
\plottwo{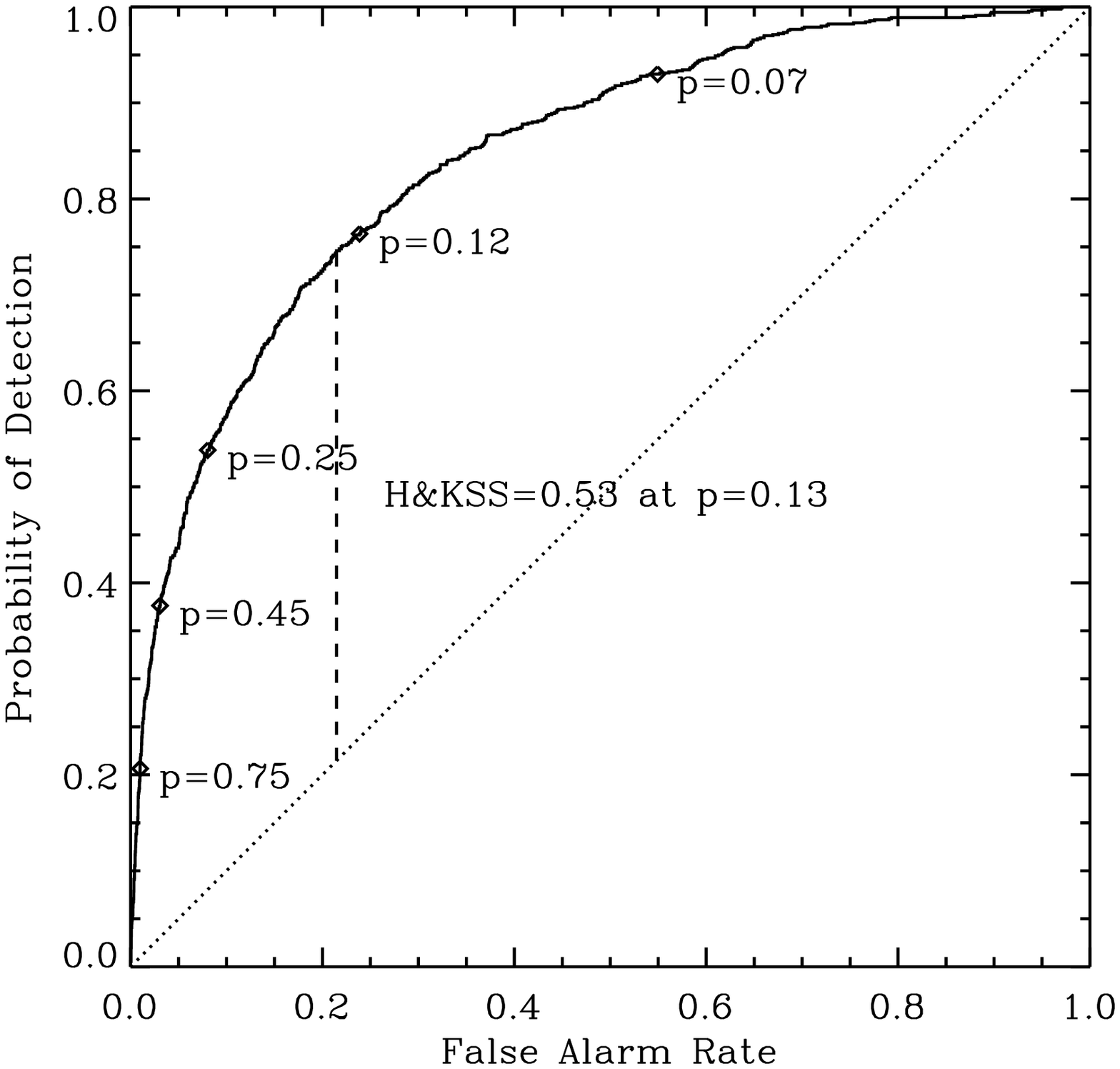}{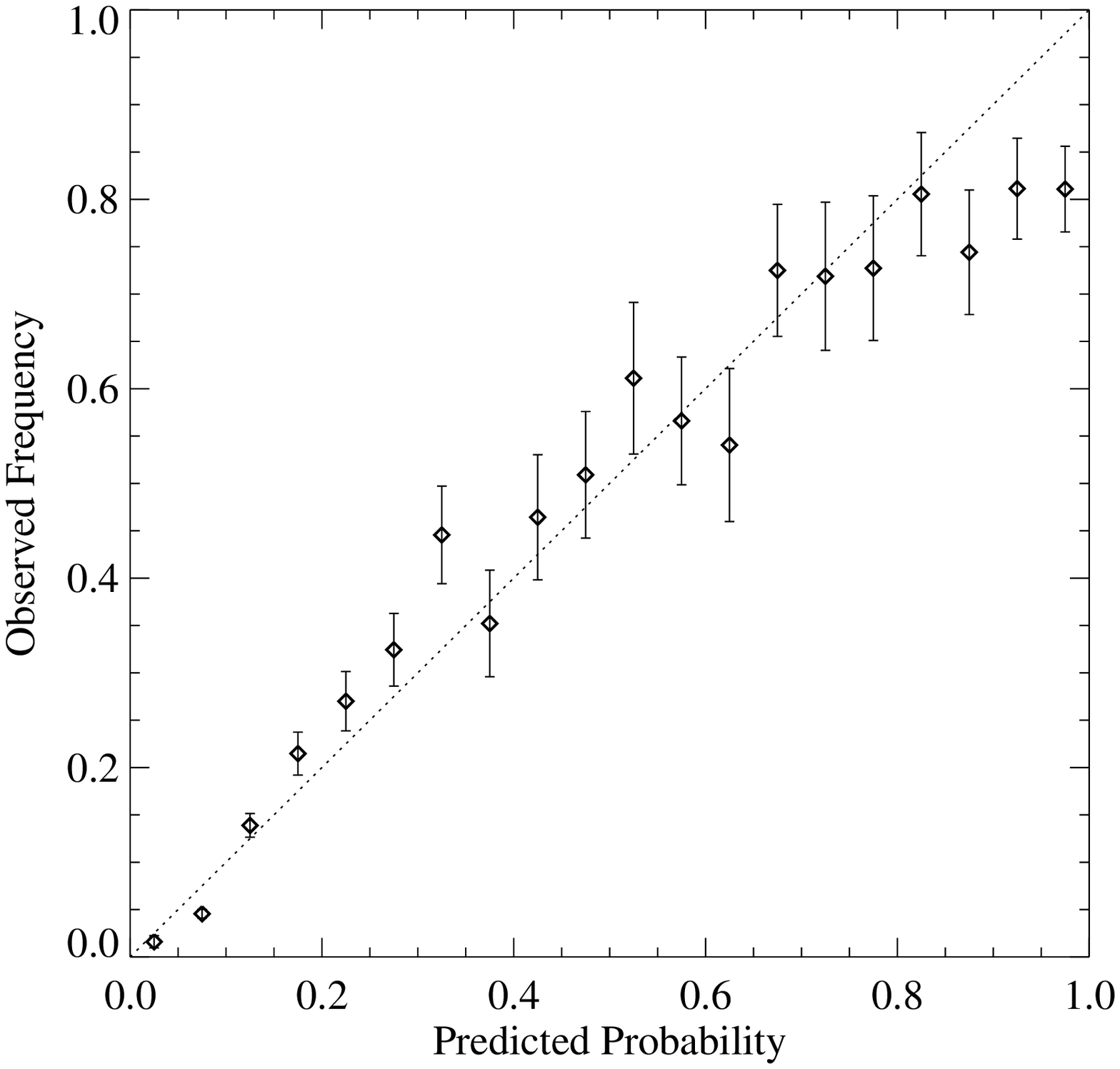}
\caption{Example forecast verification plots for results using a sub-set of
data (\MCD{1}, discussed in Section~\ref{sec:results}) from one method (the
machine-learning BBSO method, see \S\ref{sec:asap}), for the \CC\ event
definition.  
{\it Left}: A Receiver Operating Characteristic (ROC) plot shows the
probability of detection as a function of the false alarm rate by varying the
threshold above which a region is predicted to produce a flare.  Example
thresholds with $p \in [0.07, 0.12, 0.25, 0.45, 0.75]$ are labeled. In this
case the maximum \True\ occurs for $p=0.13$, and is indicated by a dashed 
vertical line. 
{\it Right}: A reliability plot in which the observed frequency of flaring is
plotted as a function of the forecast probability.  Perfect reliability occurs
when all points lie on the $x=y$ line. For this case, there is a slight
tendency to overprediction (i.e., points lying below and to the right of the
$x=y$) in the three largest probability bins.  The error bars are based on the
sample sizes in each relevant bin.
}
\label{fig:rocrel}
\end{figure}

One way to visualize the \True\ is to use a probability forecast to generate a
Receiver Operating Characteristic (ROC) curve (Figure~\ref{fig:rocrel}, left),
in which the probability of detection (POD, first term on the right hand side
of equation~(\ref{eqn:true})) is plotted as a function of the false alarm rate
(FAR, second term on the right hand side of equation~(\ref{eqn:true})) by
varying the probability threshold above which a region is predicted to flare.
When the threshold is set to one, all regions are forecast to remain flare
quiet, hence $\TP=\FP=0$, which corresponds to the point (0,0) on the ROC
diagram; when the threshold is set to zero, all regions are forecast to flare,
hence $\FN=\TN=0$, which corresponds to the point (1,1) on the ROC diagram.
For perfect forecasts, the curve consists of line segments from (0,0) to (0,1)
then from (0,1) to (1,1).  A method that has an ROC curve that stays close to
POD=1 while the FAR drops will be good at issuing all-clear forecasts; a method
that has an ROC curve that stays close to FAR=0 while the POD rises will be
good at forecasting events.

The flare forecasting methods discussed here generally predict the probability
of a flare of a given class occurring, rather than a binary, categorical
forecast.  A measure of accuracy for probabilistic forecasts is the mean square
error \added{(MSE)}, 
\begin{equation}
A_{\rm forecast} = {\rm MSE}(p_f,o) = \langle (p_f -o)^2 \rangle 
\end{equation}
where $p_f$ is the forecast probability, and $o$ is the observed outcome ($o=0$
for no event, $o=1$ for an event).  Perfect accuracy corresponds to a mean
square error of zero, $A_{\rm perfect}=0$.

\noindent The {\it Brier Skill Score} (\Brier) uses the climatological event
rate as a reference forecast with a corresponding accuracy
\begin{equation}
A_{\rm reference} = {\rm MSE}(\langle o\rangle,o) 
\end{equation}
thus  
\begin{equation}
\Brier = {{\rm MSE}(p_f,o) - {\rm MSE}(\langle o\rangle,o) \over 
0 - {\rm MSE}(\langle o\rangle,o)}. 
\end{equation}

The Brier Skill Score can be complemented by a reliability plot, which compares
the predicted probabilities with observed event rates, as demonstrated in
Figure~\ref{fig:rocrel}, right.  To construct a reliability plot, predicted
probability intervals are selected, and the frequency of observed events within
each interval is determined.  This observed frequency is then plotted {\it
vs.}~the predicted probability, with error bars estimated based on the number
of points which lie in each bin \citep[e.g.,][]{Wheatland2005}. Predictions
with perfect reliability lie along the $x=y$ line, with observed frequency
equal to predicted probability. Points lying above the line indicate
underprediction while points lying below the line indicate overprediction.
Perfect reliability is {\em not} enough to guarantee perfect forecasts. For
example, climatology has perfect reliability with a single point lying on the
$x=y$ line, but does not resolve events and non-events.

In order to make a meaningful comparison of the performance of methods, it is
important not just to use an appropriate metric such as a skill score, but also
to estimate the uncertainty in the metric.  For the present study, no
systematic attempt was made to estimate the uncertainties.  However, this has
been done for several individual methods  using either bootstrap or jackknife
approaches \citep[e.g.,][]{EfronGong1983}.  For the nonparametric discriminant
analysis described in \S\ref{sec:nwra_mag}, a bootstrap method estimate of
the one sigma uncertainties gives values of order $\pm0.01, ~\pm0.02, ~\pm0.03$
for the \CC, \MS, and \ML\ sets, respectively.  Given these values, all skill
scores are quoted to two decimal places.  A ``+0.00'' or ``-0.00'' result
signals that there was an extremely small value on that side of zero.  

All the methods produce probabilistic forecasts, so it was necessary to pick a
probability threshold to convert to the categorical forecasts needed for the
binary, categorical skill scores presented here.  That is, a threshold
probability was selected such that any forecast probability over the threshold
was considered to be a forecast for an event, and anything less was considered
to be a forecast for a non-event.  \cite{Bloomfield_etal_2012} found that for
the Poisson method, the best \True\ and \Heidke\ are typically produced by
picking a threshold that depends on the ratio $\FN/\FP$, with $\FN/\FP \approx
1$ giving the best \Heidke, and $\FN/\FP \approx N_{\rm event}/N_{\rm no\,
event}$ giving the best \True.  A similar approach to
\cite{Bloomfield_etal_2012} of stepping through the probability threshold was
followed for each combination of method, skill score and event definition using
the optimal data set for that method to determine the value that produced the
maximum skill score.  The Tables in Appendix~\ref{app:methods} include the
probability thresholds used, and ROC plots are presented for each method, with
the best \True\ shown by selecting the appropriate threshold.

\section{Overview of Prediction Methods}
\label{sec:methods}

Each participant in the workshop was invited to make predictions based on the
data set provided (\S~\ref{sec:data}), and the event definitions described
(\S~\ref{sec:eventdata}).  Generally, forecasting methods consist of two parts:
(1) parameterization of the observational solar data to characterize the target
active region, such as calculating the total magnetic flux, the length of
strong neutral lines, {\it etc.}, and (2) a statistical method by which prior
parameters or flaring activity is used to evaluate a particular target's
parameters and predict whether or not it is going to flare.  

The data analysis used by the methods broadly falls into two categories: those
which characterize the photosphere (magnetic field and/or continuum intensity)
directly (\S~\ref{sec:asap}, \ref{sec:bbso}, \ref{sec:WLSG2},
\ref{sec:nwra_mag}, \ref{sec:R}, \ref{sec:GCD}, \ref{sec:smart_ccnn},
\ref{sec:poisson}), and those which characterize the coronal magnetic field
based on the photospheric magnetic field (\S~\ref{sec:beff}, \ref{sec:MCT}).
These are supplemented by an event statistics approach
(\S~\ref{sec:wheatland}), which uses only the past flaring history, and thus
serves as a baseline against which to compare the other methods.  

A variety of statistical methods are employed to produce the forecasts from the
parameterizations.  At one end of this spectrum are methods based on a
McIntosh-like classification \citep{mcintosh90} from which a historical flaring
rate is employed as a look-up table.  At the other end are sophisticated
machine-learning techniques that generally do not employ {\it a priori}
classifications.  It may be possible to improve forecasts by combining the
parameterization used by one group with the prediction algorithm of another
group, but for the results presented here, no such attempt is explicitly made. 

For each method, a brief description is provided here, with a more detailed
description and some summary metrics of the performance of each method on its
optimal subset of the data given in Appendix~\ref{app:methods}.  A comparison
of methods on common data sets is presented in Section~\ref{sec:results}. 

\noindent{\bf The Effective Connected Magnetic Field - M.~Georgoulis,
\ref{sec:beff}}

The analysis presented in \citet{GeorgoulisRust2007} describes the coronal
magnetic connectivity using the effective connected magnetic field strength
($B_{\rm eff}$). The $B_{\rm eff}$ parameter is calculated following inference
of a connectivity matrix in the magnetic-flux distribution of the target active
region.  From the distribution of $B_{\rm eff}$, flare forecasts are generated
using Bayesian inference and Laplace's rule of succession \citep{Jaynes2003}. 

\noindent{\bf Automated Solar Activity Prediction (ASAP) - T.~Colak, R.~Qahwaji,
\ref{sec:asap}}

The {\it Automated Solar Activity Prediction}
\citep[ASAP;][]{ColakQahwaji2008,ColakQahwaji2009} system uses a
feature-recognition system to generate McIntosh classifications for active
regions from MDI white-light images.  From the McIntosh classifications, a
machine-learning system is used to make forecasts.

\noindent{\bf Big Bear Solar Observatory/Machine Learning Techniques - Y.~Yuan,
\ref{sec:bbso}}

The method developed at NJIT \citep{Yuan_etal_2010,Yuan_etal_2011} computes
three parameters describing an active region: the total unsigned magnetic flux,
the length of the strong-gradient neutral line, and the total magnetic energy
dissipation, following \citet{Abramenko_etal_2003}.  Ordinal logistic
regression and support vector machines are used to make predictions.

\noindent{\bf Total Nonpotentiality of Active Regions - D.~Falconer, 
\ref{sec:WLSG2}}

Two parameters are calculated in the approach of \citet{Falconer_etal_2008}:  a
measure of the free magnetic energy based on the presence of strong gradient
neutral lines, $WL_{SG2}$, and the total unsigned magnetic flux.  A
least-squares power-law fit to the event rates is constructed as a function of
these parameters, and the predicted event rate is converted through Poisson
statistics to the probability of an event in the forecast interval.  The
rate-fitting algorithm is best for larger flares, and so no forecasts were made
for the \CC\ events.  

\noindent{\bf Magnetic Field Moment Analysis and Discriminant Analysis -
K.D.~Leka, G.~Barnes, \ref{sec:nwra_mag}}

The NWRA moment analysis parameterizes the observed magnetic field, its spatial
derivatives, and the character of magnetic neutral lines using the first four
moments (mean, standard deviation, skew and kurtosis), plus totals and net
values when appropriate \citep{params}.  The neutral line category includes a
variation on the $\mathcal{R}$ parameter described in Section~\ref{sec:R}.
Nonparametric Discriminant Analysis \citep[NPDA; e.g.,][]{ken83,Silverman1986}
is combined with Bayes's theorem to produce a probability forecast using pairs
of variables simultaneously.

\noindent{\bf Magnetic Flux Close to High-Gradient Polarity Inversion Lines - 
C.~Schrijver, \ref{sec:R}}

\citet{Schrijver2007} proposed a parameter $\mathcal{R}$, measuring the flux
close to high gradient polarity inversion lines, as a proxy for the emergence
of current-carrying magnetic flux.  For the results here, the $\mathcal{R}$
parameter was calculated as part of the NWRA magnetic field analysis
(Section~\ref{sec:nwra_mag}), but is also included in the parameterizations by
other groups ({\it e.g.,} SMART, see Section~\ref{sec:smart_ccnn}), with
slightly different implementations (see Section~\ref{sec:implement}).  For the
results presented here, predictions using $\mathcal{R}$ were made using
one-variable NPDA (Section~\ref{sec:nwra_mag}). 

\noindent{\bf Generalized Correlation Dimension - R.T.J.~McAteer, \ref{sec:GCD}}

The Generalized Correlation (akin to a fractal) Dimension $D_{\rm BC}$
describes the morphology of a flux concentration (active region)
\citep{McAteer_etal_2010}.  The generalized correlation dimensions were
calculated for ``$q$-moment'' values from 0.1 to 8.0.  For the results
presented here, predictions with these fractal-related parameterizations were
made using one-variable NPDA (Section~\ref{sec:nwra_mag}). 

\noindent{\bf Magnetic Charge Topology and Discriminant Analysis - G.~Barnes,
K.D.~Leka, \ref{sec:MCT}}

In a Magnetic Charge Topology coronal model \citep[MCT;][and references
therein]{mct}, the photospheric field is partitioned into flux concentrations
with each one represented as a point source. The potential field due to these
point sources is used as a model for the coronal field, and determines the flux
connecting each pair of sources. This model is parameterized by quantities such
as the number, orientation, and flux in the connections
\citep{mct,dfa2,BarnesLeka2008}, including a quantity, $\phi_{2, \rm tot}$,
that is very similar to $B_{\rm eff}$ (Section~\ref{sec:beff}).  Two-variable
NPDA with cross-validation is used to make a prediction
\citep[Section~\ref{sec:nwra_mag},][]{dfa2}. 

\noindent{\bf Solar Monitor Active Region Tracker with Cascade Correlation
Neural Networks - P.A.~Higgins, O.W.~Ahmed, \ref{sec:smart_ccnn}}

The SMART2 code package \citep{Higgins_etal_2011} computes twenty parameters,
including measures of the area and flux of each active region, properties of
the spatial gradients of the field, the length of polarity separation lines,
and the measures of the flux near strong gradient polarity inversion lines
$\mathcal{R}$ \citep[Section~\ref{sec:R},][]{Schrijver2007} and $WL_{SG2}$
\citep[Section~\ref{sec:WLSG2},][]{Falconer_etal_2008}.  These parameters are
used to make forecasts using the Cascade Correlation Neural Networks method
\citep[CCNN;][]{Qahwaji_etal_2008} using the SMART-ASAP algorithm
\citep{Ahmedetal2013}. 

\noindent{\bf Event Statistics - M.S.~Wheatland, \ref{sec:wheatland}}

The event statistics method \citep{whe04a} predicts flaring probability for
different flare sizes using only the flaring history of observed active
regions.  The method assumes that solar flares (the events) obey a power-law
frequency-size distribution and that events occur randomly in time, on short
timescales following a Poisson process with a constant mean rate.  Given a past
history of events above a small size, the method infers the current mean rate
of events subject to the Poisson assumption, and then uses the power-law
distribution to infer probabilities for occurrence of larger events within a
given time.  Three applications of the method were run for these tests: active
region forecasts for which a minimum of five prior events was required for a
prediction, active region forecasts for which ten prior events were required,
and a full-disk prediction.  

\noindent{\bf Active Region McIntosh Class Poisson Probabilities -
D.S.~Bloomfield, P.A.~Higgins, P.T.~Gallagher, \ref{sec:poisson}}

The McIntosh-Poisson method uses the historical flare rates from 
McIntosh active region classifications to make forecasts using Poisson 
probabilities \citep{Gallagheretal2002,Bloomfield_etal_2012}. The 
McIntosh class was obtained for each region on a given day by 
cross-referencing the NOAA region number(s) provided in the NOAA Solar 
Region Summary file for that day. Unfortunately, forecasts for the 
\ML\ event definition could not be produced by this approach 
because the event rates in the historical data \citep{Kildahl1980} were only 
identified by GOES class bands (i.e., C, M, or X) and not the complete 
class and magnitude.

\section{Comparison of Method Performance}
\label{sec:results}

A comparison is presented here of the different methods' ability to forecast a
solar flare for select definitions of an event (as described in
Section~\ref{sec:eventdata}).  The goal is not to identify any method as a
winner or loser.  Rather, the hope is to identify successful trends being used
to identify the flare-productivity state of active regions, as well as failing
characteristics, to assist with future development of prediction methods.  The
focus is on the Brier skill score, since methods generally return probability
forecasts, but the Appleman skill score is used to indicate the performance on
categorical forecasts.  The \Appleman\ effectively treats each type of error
(misses and false alarms) as equally important, and so gives a good overall
indication of the performance of a method.  In practice, which skill score is
chosen does not greatly change the ranking of the methods, or the overall
conclusions, with a few exceptions discussed in more detail below. 

\subsection{Data Subsets}

The request was made for every method to provide a forecast for each and every
dataset provided.  Many methods, as alluded to in the descriptions in
Appendix~\ref{app:methods}, had restrictions on where it was believed they
would perform reliably, and so each method did not provide a forecast in every
case.  The resulting variation in the sample sizes, as shown in the summary
tables, is not fully accounted for in the skill scores reported
(Section~\ref{sec:sss}), meaning that direct comparisons between methods with
different sample sets is not reliable.  Even the Hanssen \& Kuipers' skill
score is not fully comparable among datasets if the samples of events and
non-events are drawn from different populations, for example all regions 
versus only those regions with strong magnetic neutral lines. 

To account for the different samples, three additional datasets are considered
for performance comparison.  The first is all data (AD), with an unskilled
forecast (the climatology, or event rate) used if a method did not provide a
forecast for that particular target.  In this way, forecasts are produced for
all data for all methods.  However, this approach penalizes methods that
produce forecasts for only a limited subset of data.

\replaced{The second approach is to construct maximum common datasets (MCD) for
which all methods provided forecasts.  Two were constructed, one (\MCD{1}) for
all methods except the event-statistics, and a second (\MCD{2}) which included
the additional event-statistics restrictions imposed for 10 prior events.
}{The second approach is to extract the largest subset from AD for which all
methods provided forecasts.  Two such maximum common datasets were constructed,
one (\MCD{1}) for all methods except the event-statistics, while the second
(\MCD{2}) which included the additional event-statistics restrictions imposed
by requiring at least ten prior events.} One method (MSFC/Falconer,
\S~\ref{sec:WLSG2}) did not return results for \CC, so strictly speaking these
should be null sets.  However, for the purposes of this manuscript, that method
was removed for constructing the MCDs for \CC.  A drawback to the MCD approach
arises for the methods that were trained using larger data samples, {\it i.e.}
samples which included regions that were not part of the MCD.  For methods that
trained on AD, for example, many additional regions were used for training,
while for methods with the most restrictive assumptions, almost all the regions
used for training are included in \MCD{1}, hence the impact of using the
maximum common datasets varies from method to method.  In the case of \MCD{1},
where the primary restriction is on the distance of regions from disk center,
this may not be a large handicap since the inherent properties of the regions
are not expected to change based on their location on the disk, although the
noise in the data and the magnitude of projection effects do change.  However,
for \MCD{2}, the requirement of a minimum number of prior events means the
samples are drawn from very different populations.  For {\it any} method,
training on a sample from one population then forecasting on a sample drawn
from a different population adversely affects the performance of the method. 

The magnitude of the changes in the populations from which the samples are
drawn can be roughly seen in the changes in the event rates shown in
Table~\ref{tbl:samplecomps}.  Between AD and \MCD{1}, the event rates change by
no more than about 10\%; between AD and \MCD{2}, the event rates change by up
to an order of magnitude for the smaller event sizes such that \MCD{2} for \CC\
is the only category with more events than non-events.  The similarity in the
event rates of AD and \MCD{1} is consistent with the hypothesis that they are
drawn from similar underlying populations.  However, based on the changes in
the event rates, it is fairly certain that \MCD{2} is drawn from different
populations than AD and \MCD{1}.  Despite this difference, measures of the
performance of the methods are presented for all three subsets to illustrate
the magnitude of the effect and the challenge in making meaningful comparisons
of how well methods perform.

\begin{center}
\begin{deluxetable}{|l|rrr|rrr|rrr|}
\tablecolumns{10}
\tablewidth{0pc}
\tablecaption{Sample Sizes of All Data (AD) {\it vs.} Maximum Common Datasets 
(MCDs)} 
\tablehead{
\colhead{Event} & \colhead{Event} & \colhead{No} & 
\colhead{Event} & \colhead{Event} & \colhead{No} & 
\colhead{Event} & \colhead{Event} & \colhead{No} & \colhead{Event} \\
\colhead{List} & \colhead{} & \colhead{Event} & 
\colhead{Rate} & \colhead{} & \colhead{Event} & 
\colhead{Rate} & \colhead{} & \colhead{Event} & \colhead{Rate}
}
\startdata 
& \multicolumn{3}{c|}{AD} & \multicolumn{3}{c|}{\MCD{1}} & \multicolumn{3}{c|}{\MCD{2}} \\ 
\CC & 2609 & 10356 & 0.201 & 789 & 3751 & 0.174 & 249 & 128 & 0.660 \\
\MS &  400 & 12565 & 0.031 & 102 & 3162 & 0.031 &  70 & 220 & 0.241 \\
\ML &   93 & 12872 & 0.007 &  26 & 3633 & 0.007 &  21 & 270 & 0.072 \\
\enddata
\label{tbl:samplecomps}
\end{deluxetable}
\end{center}

\subsection{Method Performance Comparisons}
\label{sec:best}

The Appleman and Brier skill scores for each method are listed for each event
definition in Tables~\ref{tbl:bestAD}-\ref{tbl:bestMCD2}, separately for each
of the three direct-comparison data sets: All Data, Maximum Comparison Datasets
\#1 (without the event-statistics restrictions) and \#2 (with the further
restrictions from event statistics).  As described in \S\ref{sec:sss}, the
probability thresholds for generating the binary, categorical classifications
used for calculating the \Appleman\ were chosen to maximize the \Appleman\
computed for each event definition using each method's optimal data set (see
Appendix~\ref{app:methods})\footnote{A complete set of the probability
forecasts for all methods is published in the machine-readable format.}
When a method produces more than one forecast (e.g., the Generalized
Correlation Dimension, \S\ref{sec:GCD}, which produces a separate forecast for
each $q$ value), the one with the highest Brier Skill Score is presented.
Using a different skill score to select which forecast is presented generally
results in the same forecast being selected, so the results are not sensitive
to this choice.

Recalling that skill scores are normalized to unity, none of the methods
achieves a particularly high skill score.  No method for any event definition
achieves an \Appleman\ or \Brier\ value greater than 0.4, and for the large
event magnitudes, the highest skill score values are close to 0.2.  Thus there
is considerable room for improvement in flare forecasting.  The skill scores
for some methods are much lower than might be expected from prior published
results.  This is likely a combination of the data set provided here not being
optimal for any of the algorithms, and variations in performance based on the
particular time interval and event definitions being considered.

In each category of event definition and for most data sets, at least three
methods perform comparably given a typical uncertainty in the skill score, so
there is no single method that is clearly better than the others for flare
prediction in general.  For a specific event definition, some methods achieve
significantly higher skill scores.  There is a tendency for the machine
learning algorithms to produce the best categorical forecasts, as evidenced by
some of the highest \Appleman\ values in
Tables~\ref{tbl:bestAD}-\ref{tbl:bestMCD2}, and for nonparametric discriminant
analysis to produce the best probability forecasts, as evidenced by some of the
highest \Brier\ values. 

For rare events, most of the machine learning methods (ASAP, SMART2/CCNN, and 
to a lesser degree BBSO) produce negative \Brier\ values, even when the value
of the \Appleman\ for the method is positive.  This is likely a result of the
training of the machine learning algorithms, which were generally optimized on
one or more of the categorical skill scores with a probability threshold of
0.5.  The maximum \True\ is obtained for a probability threshold that is much
smaller than 0.5 when the event rate is low \citep{Bloomfield_etal_2012}.  When
a machine learning algorithm is trained to maximize the \True\ with a threshold
of 0.5, it compensates for the threshold being higher than optimal by
overpredicting (see the reliability plots in Appendix~\ref{app:methods}).  That
is, by imposing a threshold of 0.5 and systematically overpredicting, a 
similar classification table is produced as when
a lower threshold is chosen and not overpredicting.  The former results in higher categorical skill score values, but
reduces the value of the \Brier\ because the \Brier\ does not use a threshold
but is sensitive to overprediction.  In contrast, discriminant analysis is
designed to produce the best probabilistic forecasts and so it tends to have
high reliability (does not overpredict or underpredict). This results in good
\Brier\ and \Appleman\ values.

\begin{figure}
\plottwo{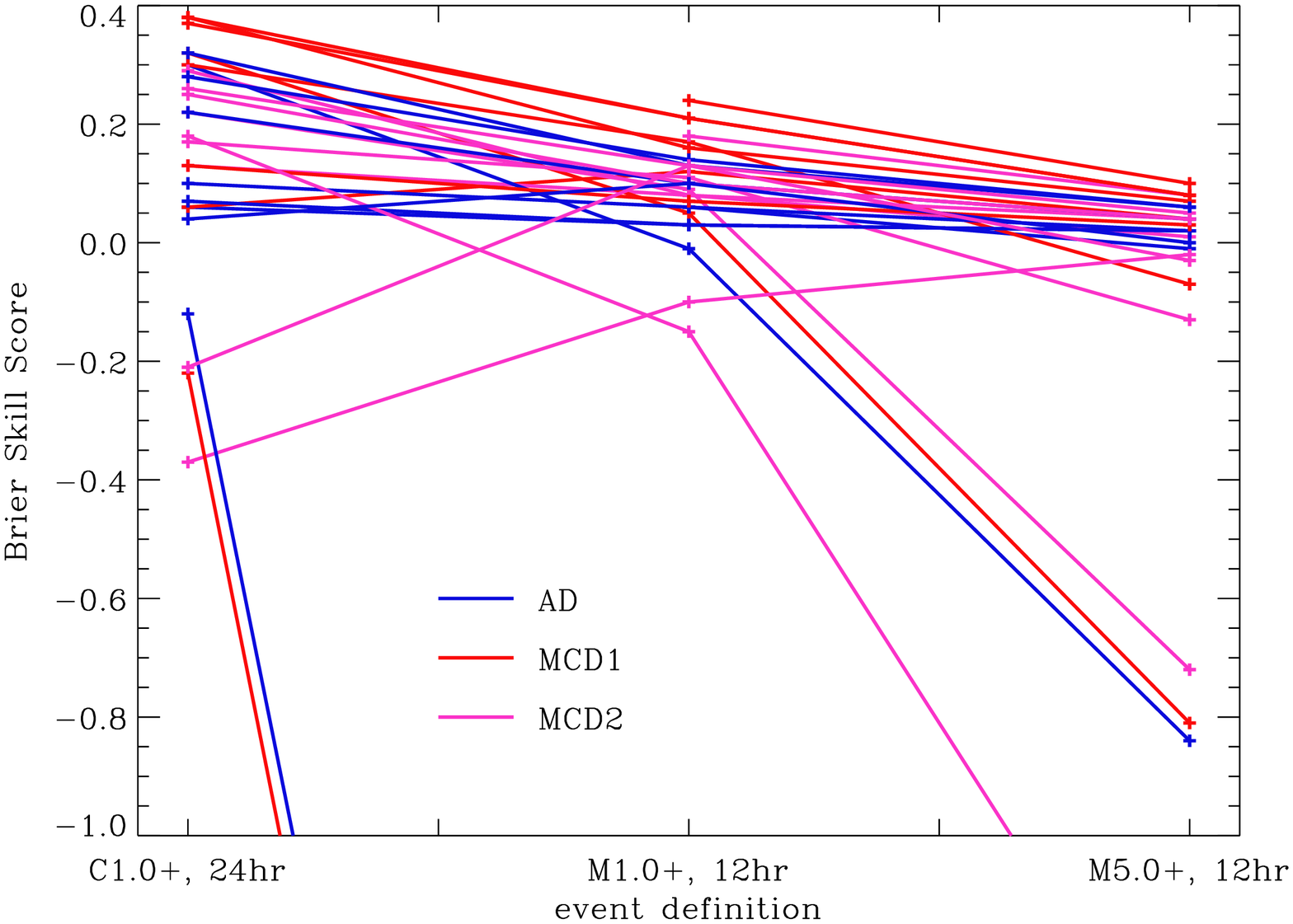}{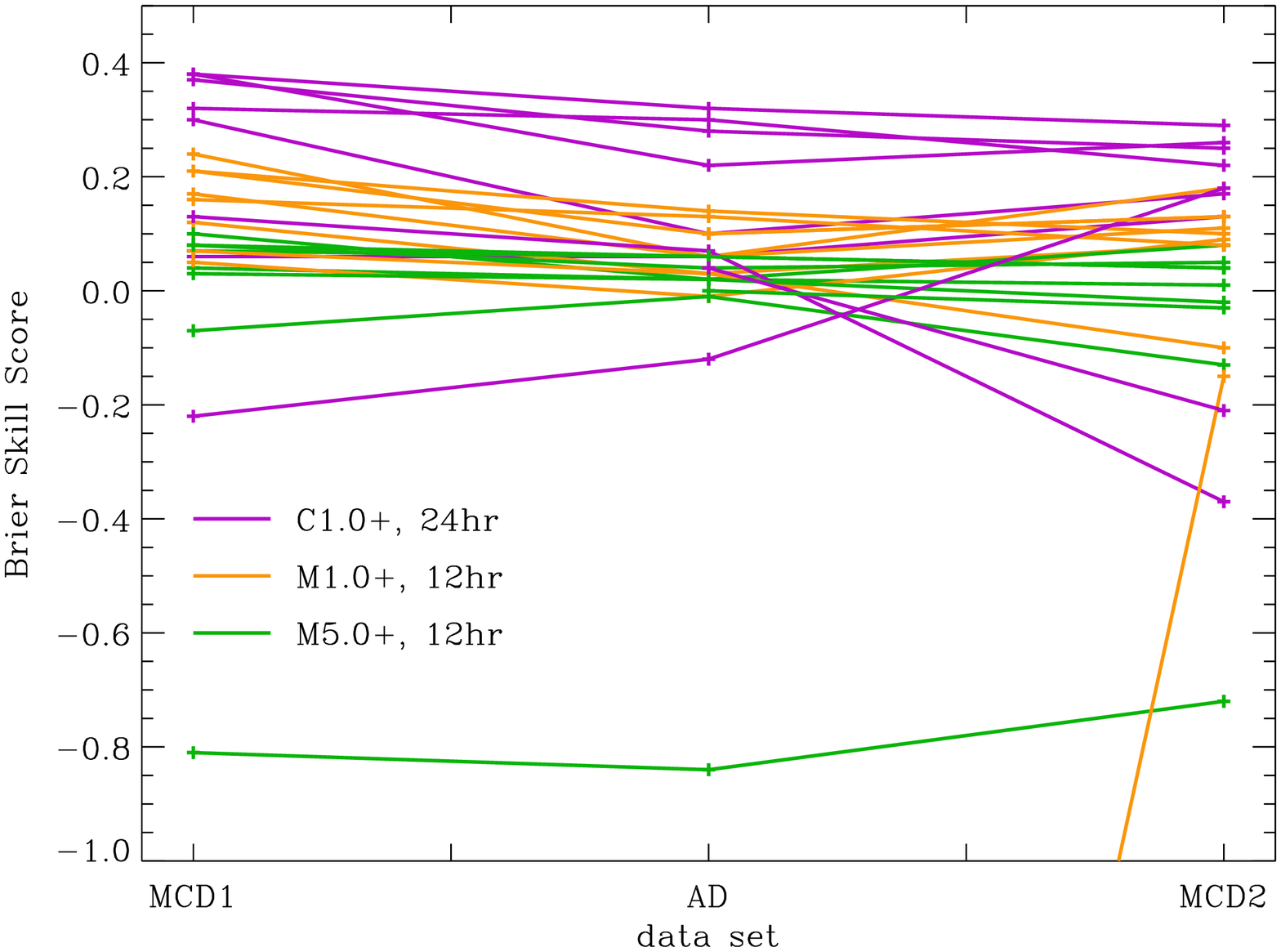}

\plottwo{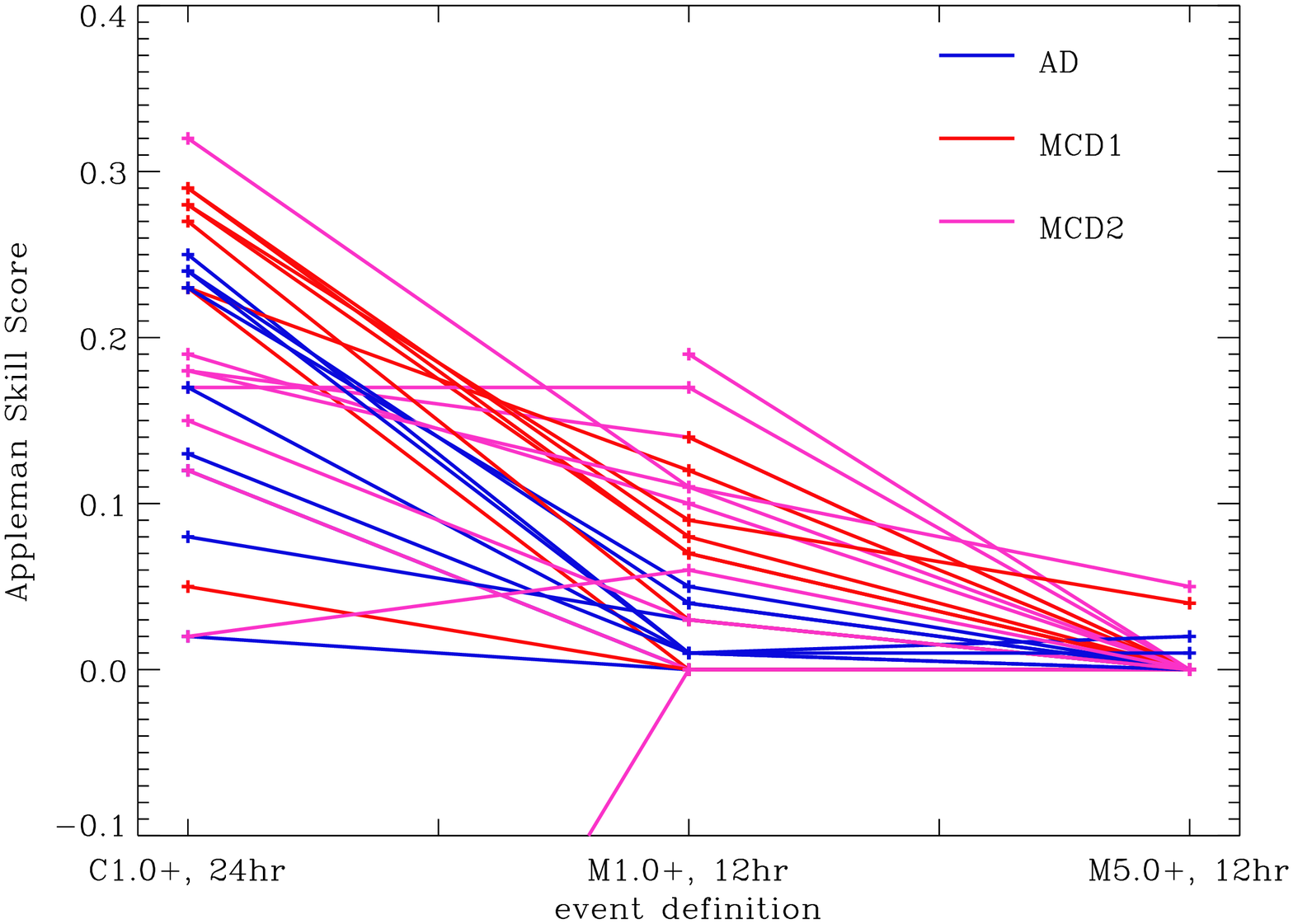}{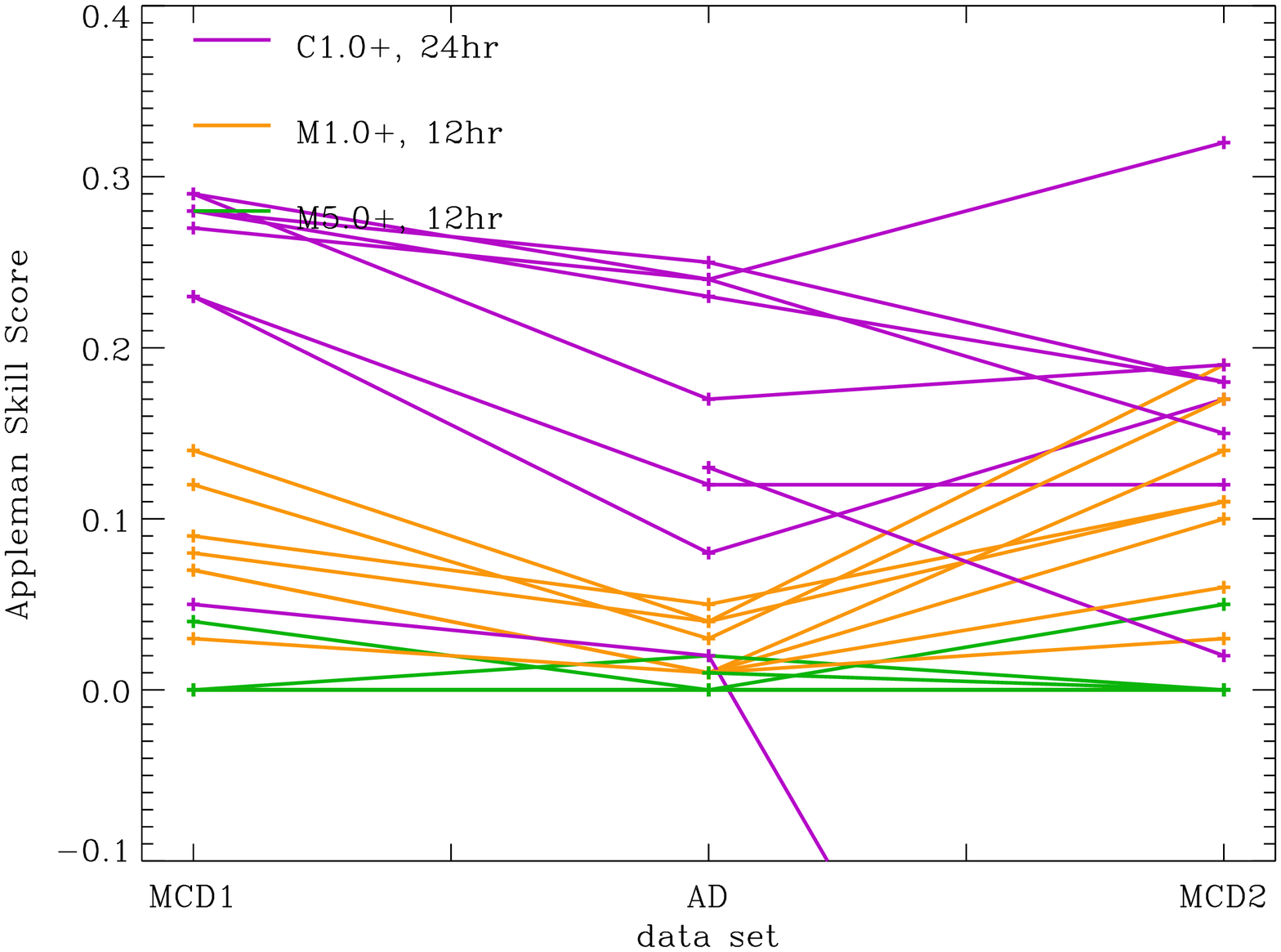}
\caption{Skill scores from different methods for different data sets and
different event definitions. Top: the Brier Skill Score and Bottom:
the Appleman Skill Score as a function of the event definition (left) and
the data set used (right).  Not all methods produced forecasts for all
event definitions and data sets, so a few points are missing from the plots.
Several trends are clearly present in the results: methods generally perform
better on smaller magnitude events, whether evaluated based on the Brier or the
Appleman skill score, and most methods perform better on \MCD{1} than on other
datasets.  The ranking of methods changes between the different data sets,
showing the importance of using consistent data sets when comparing forecasting
algorithms.
}
\label{fig:skillscores}
\end{figure}

Skill scores from the different methods for the three data sets and different
event definitions are shown in Figure~\ref{fig:skillscores}.  Several trends
are seen in the results. From the left panels, it can be seen that forecasting
methods perform better on smaller magnitude events, whether evaluated based on
the Brier or the Appleman skill score, with the most notable exceptions being
ones for which the Brier skill score is negative for \CC, including the event
statistics method.  The event statistics method uses the small events to
forecast the large events, and thus is not as well suited to forecasting
smaller events.  The other exceptions are for the \MCD{2}, and thus are likely
a result of a mismatch between the training and the forecasting data sets.  The
overall trend for most methods is likely due to the smaller sample sizes and
lower event rates for the \MS\ and \ML\ categories, and holds for AD and both
MCD sets.  The smaller sample sizes make it more difficult to train forecasting
algorithms, and the lower event rates result in smaller prior probabilities for
an event to occur.

The right panels of Figure~\ref{fig:skillscores} show that almost all methods
achieve higher skill scores on \MCD{1} than on AD, although the improvement is
modest.  For methods that did not provide a forecast for every region, this is
simply a result of the methods making better predictions than climatology. For
methods providing a forecast for every region, it suggests that restricting the
forecasts to close to disk center improves the quality of the forecasts,
although the effect is not large.  However, methods trained on AD and then
applied to \MCD{1} may show a more substantial improvement if trained on
\MCD{1}.  One of the main restrictions in many methods is the distance from
disk center due to projection effects.  Thus it is likely that improved results
would be achieved by using vector magnetograms.

Most methods also achieve higher skill scores on \MCD{2} than on AD, but a
considerable fraction have lower skill scores, and there is more variability in
the changes between AD and \MCD{2}, as measured by the standard deviation of
the change in skill score, than in the changes between AD and \MCD{1}.  This
supports the hypothesis that AD and \MCD{1} draw from similar populations,
while \MCD{2} draws from significantly different populations.  The ranking of
methods changes somewhat between AD and \MCD{1}, but more significantly between
AD and \MCD{2}.  This highlights the danger of using a subset of data to
compare methods, particularly when the subset is drawn from a different
population than the set used for training.

\begin{center}
\begin{deluxetable}{llrrrrrr}
\tablecolumns{8}
\tablewidth{0pc}
\tabletypesize{\small}
\tablecaption{Performance on All Data with Reference Forecast
\label{tbl:bestAD}}
\tablehead{
\colhead{Parameter/} & \colhead{Statistical} & \multicolumn{2}{c}{\CC} & 
\multicolumn{2}{c}{\MS} & \multicolumn{2}{c}{\ML} \\ 
\colhead{Method}     & \colhead{Method}      & \colhead{\Appleman} & \colhead{\Brier} & 
\colhead{\Appleman} & \colhead{\Brier} & \colhead{\Appleman} & \colhead{\Brier} 
}
\startdata 
${\rm B}_{\rm eff}$        & Bayesian      & 0.12 &\ 0.06 &\ 0.00 &\ 0.03 &\ 0.00 &\ 0.02 \\
ASAP                       & Machine       & 0.25 &\ 0.30 &\ 0.01 & -0.01 &\ 0.00 & -0.84 \\
BBSO                       & Machine       & 0.08 &\ 0.10 &\ 0.03 &\ 0.06 &\ 0.00 & -0.01 \\
$WL_{SG2}$                 & Curve fitting &  N/A &\  N/A &\ 0.04 &\ 0.06 &\ 0.00 &\ 0.02 \\
NWRA MAG 2-VAR             & NPDA          & 0.24 &\ 0.32 &\ 0.04 &\ 0.13 &\ 0.00 &\ 0.06 \\
$\log(\mathcal{R})$        & NPDA          & 0.17 &\ 0.22 &\ 0.01 &\ 0.10 &\ 0.02 &\ 0.04 \\
GCD                        & NPDA          & 0.02 &\ 0.07 &\ 0.00 &\ 0.03 &\ 0.00 &\ 0.02 \\
NWRA MCT 2-VAR             & NPDA          & 0.23 &\ 0.28 &\ 0.05 &\ 0.14 &\ 0.00 &\ 0.06 \\
SMART2	                   & CCNN          & 0.24 & -0.12 &\ 0.01 & -4.31 &\ 0.00 & -11.2 \\
Event Statistics, 10 prior & Bayesian      & 0.13 &\ 0.04 &\ 0.01 &\ 0.10 &\ 0.01 &\ 0.00 \\
McIntosh	 	   & Poisson 	  & 0.15 &\ 0.07 &\ 0.00 & -0.06 & N/A & N/A  \\
\enddata
\tablecomments{An entry of N/A indicates that the method did not provide
forecasts for this event definition.}
\end{deluxetable}
\end{center}

\begin{center}
\begin{deluxetable}{llrrrrrr}
\tablecolumns{8}
\tablewidth{0pc}
\tabletypesize{\small}
\tablecaption{Performance on Maximum Common Dataset \#1
\label{tbl:bestMCD1}}
\tablehead{
\colhead{Parameter/} & \colhead{Statistical} & \multicolumn{2}{c}{\CC} &
\multicolumn{2}{c}{\MS} & \multicolumn{2}{c}{\ML} \\
\colhead{Method}     & \colhead{Method}      & \colhead{\Appleman} & \colhead{\Brier} &
\colhead{\Appleman} & \colhead{\Brier} & \colhead{\Appleman} & \colhead{\Brier}
}
\startdata 
${\rm B}_{\rm eff}$        & Bayesian      & 0.23 & 0.06 & 0.00 & 0.12 & 0.00 & 0.04 \\
ASAP                       & Machine       & 0.29 & 0.32 & 0.07 & 0.05 & 0.00 &-0.81 \\
BBSO                       & Machine       & 0.24 & 0.30 & 0.12 & 0.17 & 0.00 &-0.07 \\
$WL_{SG2}$                 & Curve fitting &  N/A &  N/A & 0.14 & 0.24 & 0.00 & 0.10 \\
NWRA MAG 2-VAR             & NPDA          & 0.30 & 0.38 & 0.08 & 0.16 & 0.00 & 0.07 \\
$\log(\mathcal{R})$        & NPDA          & 0.29 & 0.38 & 0.07 & 0.21 & 0.00 & 0.08 \\
GCD                        & NPDA          & 0.05 & 0.13 & 0.00 & 0.07 & 0.00 & 0.03 \\
NWRA MCT 2-VAR             & NPDA          & 0.29 & 0.37 & 0.09 & 0.21 & 0.04 & 0.08 \\
SMART2                     & CCNN          & 0.27 &-0.22 & 0.03 &-4.46 & 0.00 &-12.49\\
Event Statistics, 10 prior & Bayesian      & N/A &  N/A &  N/A &  N/A &  N/A &  N/A \\
McIntosh		   & Poisson	  & 0.12 &-0.03 & 0.00 &-0.05 &  N/A &  N/A \\
\enddata
\tablecomments{An entry of N/A indicates that the method did not provide
forecasts for this event definition.}
\end{deluxetable}
\end{center}

\begin{center}
\begin{deluxetable}{llrrrrrr}
\tablecolumns{8}
\tablewidth{0pc}
\tabletypesize{\small}
\tablecaption{Performance on Maximum Common Dataset \#2
\label{tbl:bestMCD2}}
\tablehead{
\colhead{Parameter/} & \colhead{Statistical} & \multicolumn{2}{c}{\CC} &
\multicolumn{2}{c}{\MS} & \multicolumn{2}{c}{\ML} \\
\colhead{Method}     & \colhead{Method}      & \colhead{\Appleman} & \colhead{\Brier} &
\colhead{\Appleman} & \colhead{\Brier} & \colhead{\Appleman} & \colhead{\Brier}
} 
\startdata
${\rm B}_{\rm eff}$        & Bayesian      & 0.12 & 0.13 & 0.00 & 0.08 & 0.00 & 0.01 \\
ASAP                       & Machine       & 0.22 & 0.22 & 0.14 & 0.09 & 0.00 &-0.72 \\
BBSO                       & Machine       & 0.23 & 0.17 & 0.17 & 0.11 & 0.00 &-0.13 \\
$WL_{SG2}$                 & Curve fitting &  N/A &  N/A & 0.19 & 0.18 & 0.00 & 0.08 \\
NWRA MAG 2-VAR             & NPDA          & 0.38 & 0.29 & 0.11 & 0.08 & 0.00 & 0.04 \\
$\log(\mathcal{R})$        & NPDA          & 0.23 & 0.26 & 0.10 & 0.13 & 0.00 & 0.05 \\
GCD                        & NPDA          & -0.47 &-0.37 & 0.00 & -0.10 & 0.00 &-0.02 \\
NWRA MCT 2-VAR             & NPDA          & 0.23 & 0.25 & 0.11 & 0.10 & 0.05 & 0.04 \\
SMART2                     & CCNN          & 0.15 & 0.18 & 0.03 &-0.15 & 0.00 &-1.47 \\
Event Statistics, 10 prior & Bayesian      & 0.05 &-0.21 & 0.06 & 0.13 & 0.00 & -0.03 \\
McIntosh                   & Poisson       & 0.02 &-0.09 & 0.00 & 0.01 &  N/A &  N/A \\
\enddata
\tablecomments{An entry of N/A indicates that the method did not provide
forecasts for this event definition.}
\end{deluxetable}
\end{center}

\begin{figure}
\epsscale{1.0}
\plotone{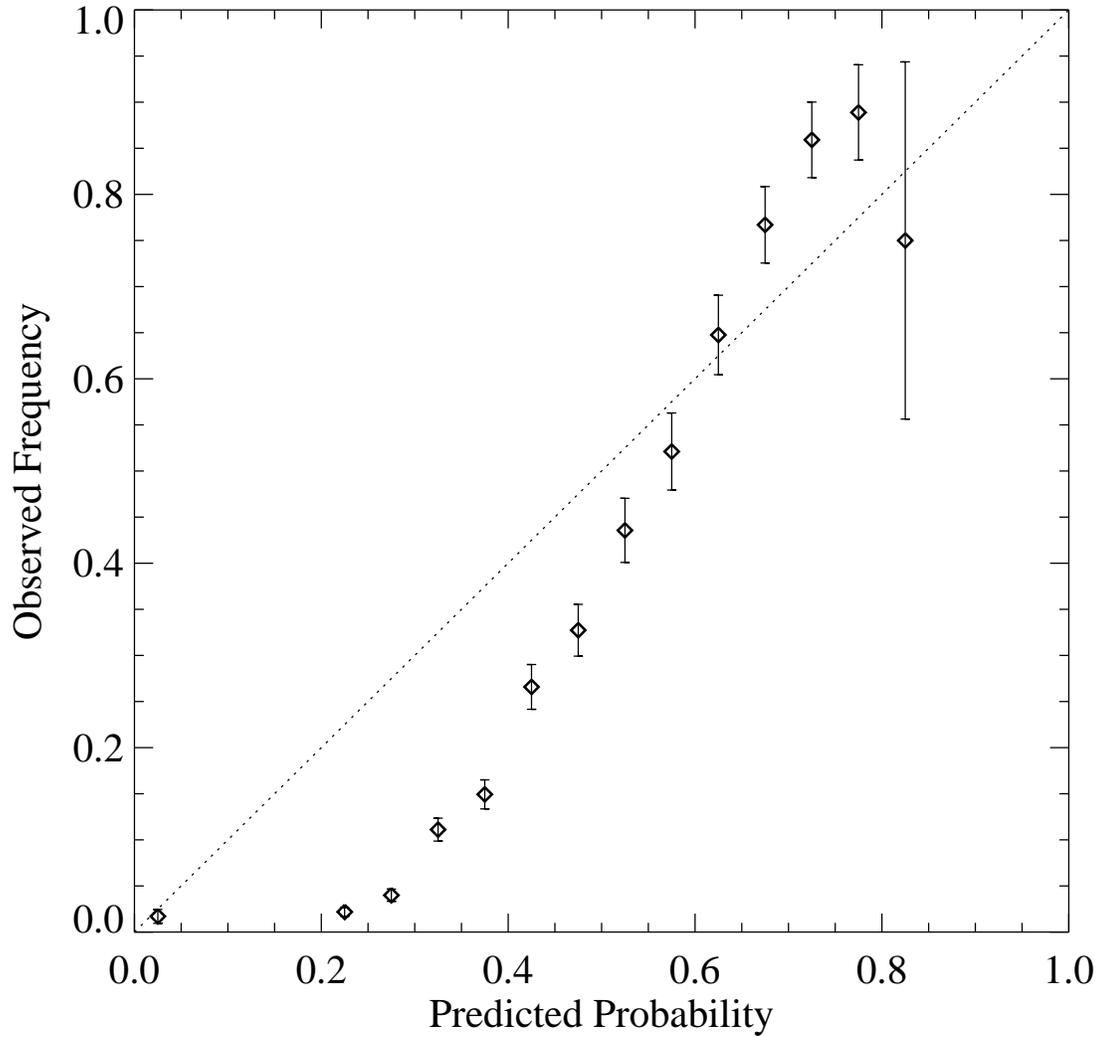}
\caption{Reliability plot for the \MCD{1}, \CC\ forecasts based on Bayesian
statistics and the ${\rm B}_{\rm eff}$ method, for comparison with
Figure~\ref{fig:rocrel}, right.  The two forecasts have essentially identical
Appleman skill scores, but very different Brier Skill scores.}
\label{fig:reliability2}
\end{figure}

As discussed, different skill scores emphasize different aspects of
performance.  This is demonstrated by the results for the $B_{\rm eff}$ and the
BBSO methods shown in Table~\ref{tbl:bestMCD1}, for \CC\ using \MCD{1}.  The
forecasts using these methods result in essentially the same values of the
\Appleman\ of $0.23$ and $0.24$, respectively.  However, the corresponding
\Brier\ values of $0.06$ and $0.30$ are quite different.  This is graphically
illustrated in comparing Figure~\ref{fig:reliability2} with
Figure~\ref{fig:rocrel}, right.  The $B_{\rm eff}$ method
(Fig.~\ref{fig:reliability2}) systematically overpredicts for forecast
probabilities less than about 0.6, but slightly underpredicts for larger
forecast probabilities.  Thus the probabilistic forecasts result in a small
\Brier, but by making a categorical forecast of an event for any region with a
probability greater than 0.55, the method produces a much higher \Appleman.
The SMART/CCNN method produces a similar systematic under- and overprediction
(see \S\ref{sec:smart_ccnn}) while most methods (e.g., the BBSO method shown in
Fig.~\ref{fig:rocrel}) have little systematic over- or underprediction, so the
Brier and Appleman skill scores are similar.

\begin{figure}
\epsscale{0.3}
\plotone{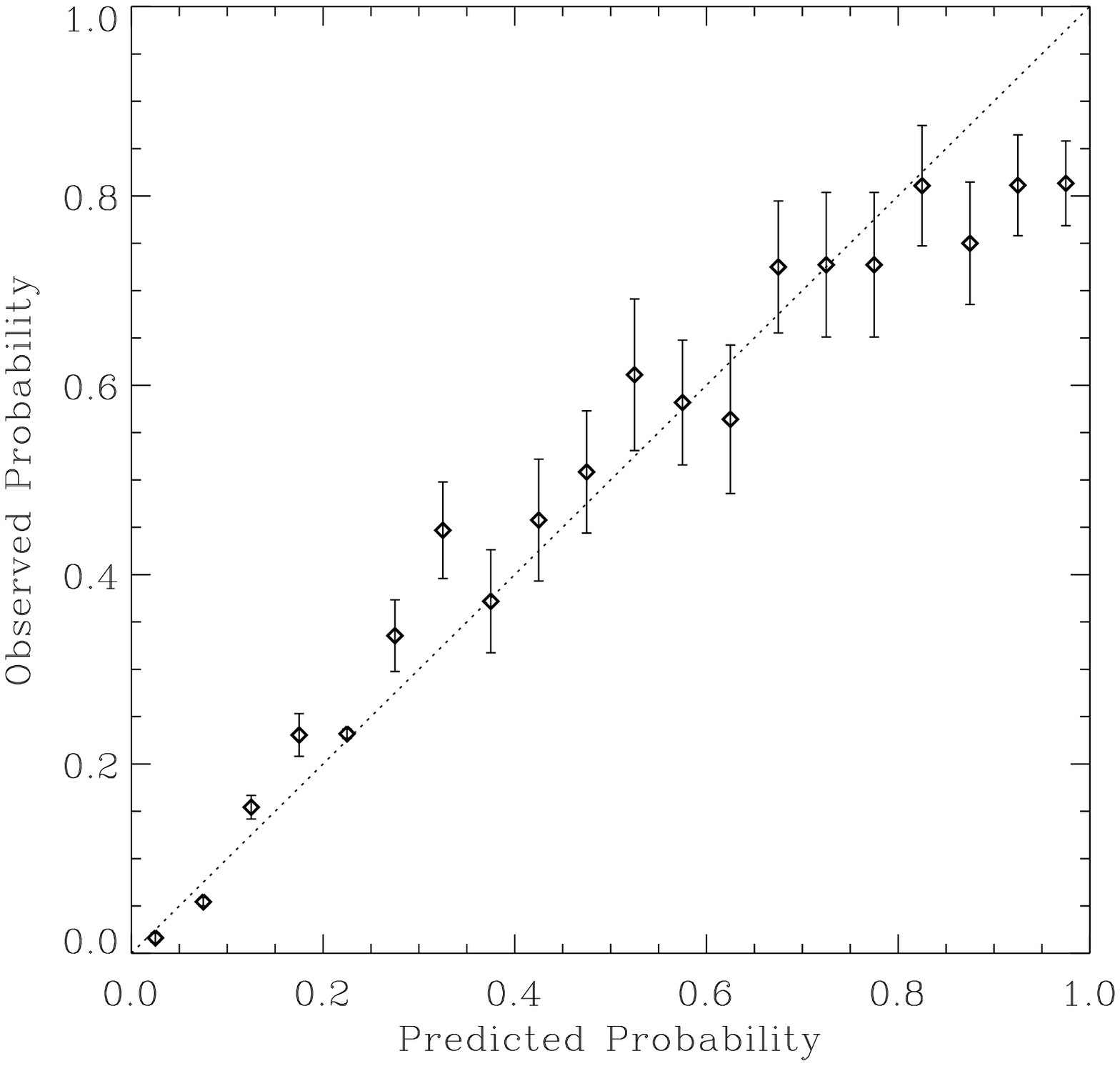}
\plotone{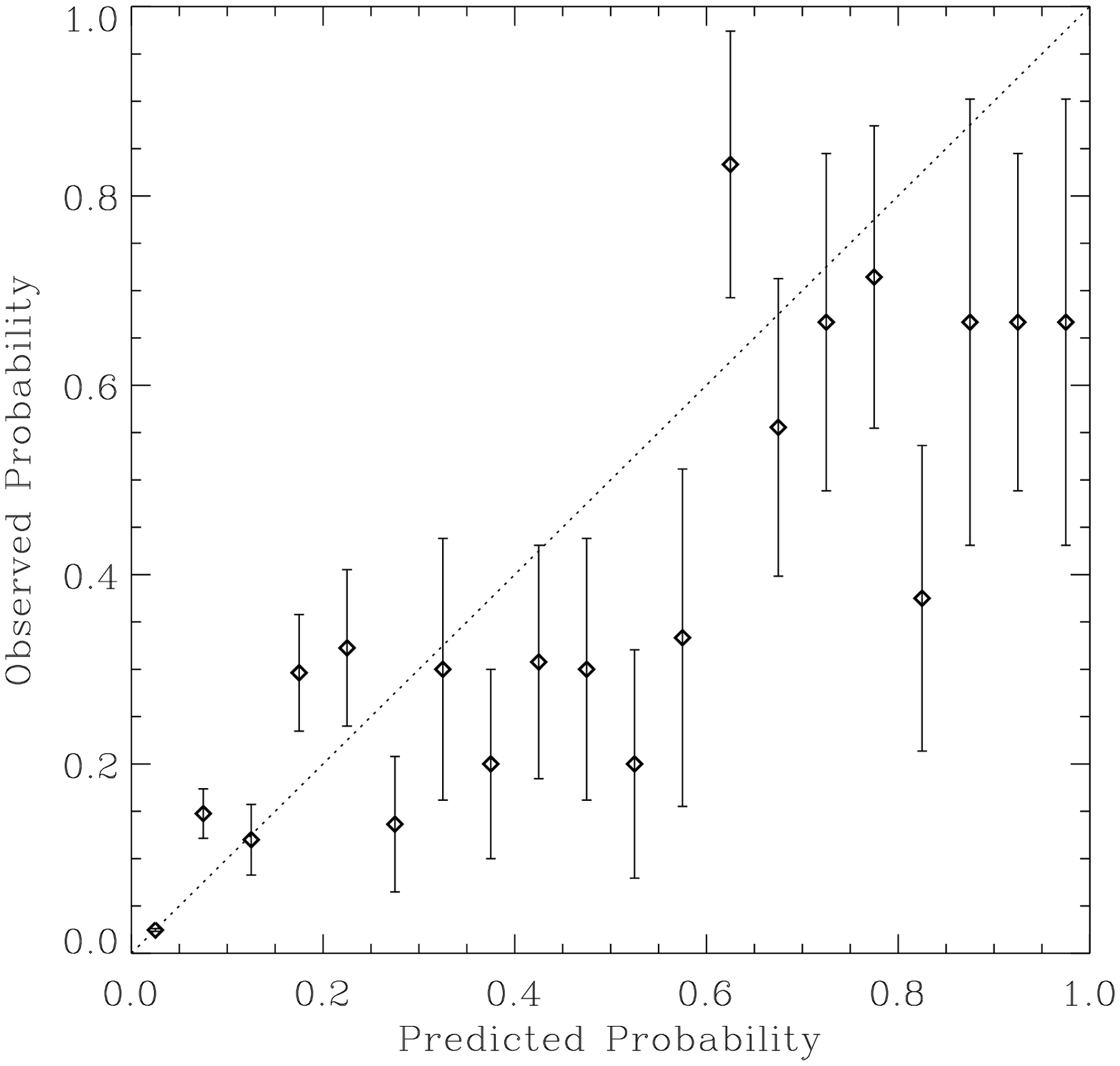}
\plotone{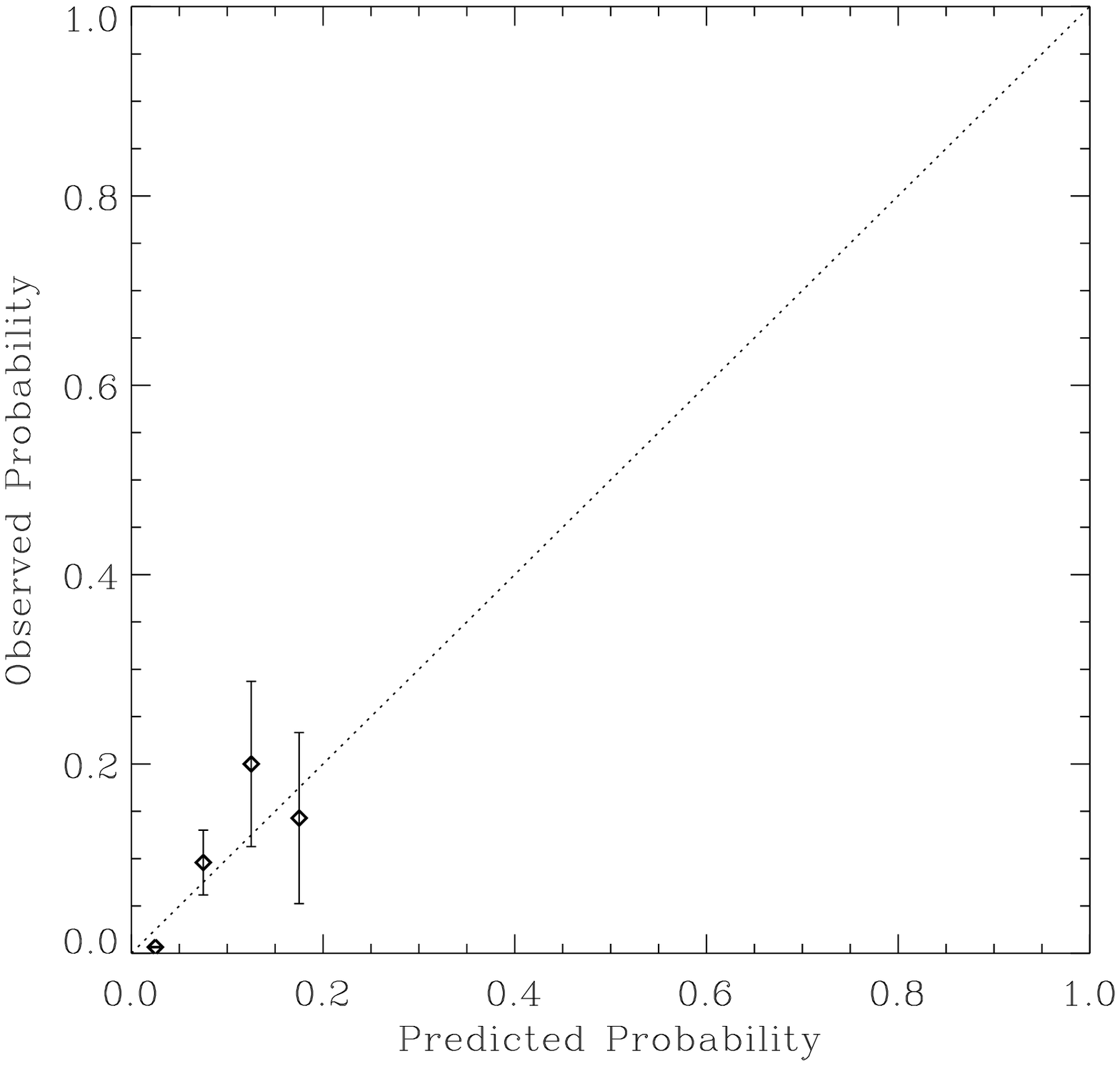}
\caption{Reliability plots for the BBSO predictions for (from left to right)
the \CC, \MS, \ML\ event definitions and All Data treatment (reference forecast
is used when no forecast was otherwise returned).  Note the increasingly poor
performance with event size and the increasing size of the error bars. This
reflects the decreasing sample size for the larger events.}
\label{fig:reliability_with_size}
\end{figure}

Inspection of reliability plots for a single method
(Figure~\ref{fig:reliability_with_size}) shows a phenomenon common to most
methods: the maximum forecast probability typically decreases with increasing
event size, so most methods only produce low-probability forecasts for, say,
\ML\ event definitions.  This explains the small values of the Appleman skill
score for larger event definitions as very few or no regions have high enough
forecast probabilities to be considered a predicted event in a categorical
forecast.  It also suggests that all-clear forecasts have more promise than
general forecasts. However, attention to the possibility of missed events would
be critical from an operations point of view.

\subsection{Differences and Similarities in Approach}
\label{sec:implement}

All groups were given the same data, and many computed the same parameters.
However, implementations differ significantly, so values for the same parameter
are substantially different in some instances.  In other cases, two parameters
computed using completely different algorithms lead to parameter values that
are extremely well correlated.

The total unsigned magnetic flux of a region, $\sum \vert B_z \vert$, is often
considered a standard candle for forecasting. Larger regions have long being
associated with greater propensity for greater-sized events, and the total flux
is a direct measure of region size, hence it provides a standard for
flare-forecast performance.  Four groups calculated the total magnetic flux for
this exercise, and provided the value for each target region.  By necessity,
since the MDI data provide only the line-of-sight component of the field
vector, approximations were made, which varied between groups, and one group
(NWRA) calculated the flux in two ways, using different approximations.  There
are also different thresholds used to mitigate the influence of noise, and
different observing-angle limits beyond which some groups do not calculate this
parameter.  How large are the effects of these different assumptions and
approximations on the inferred value of the total unsigned flux, and what is
the impact on flare forecasting?

\begin{figure}
\epsscale{1.0}
\plotone{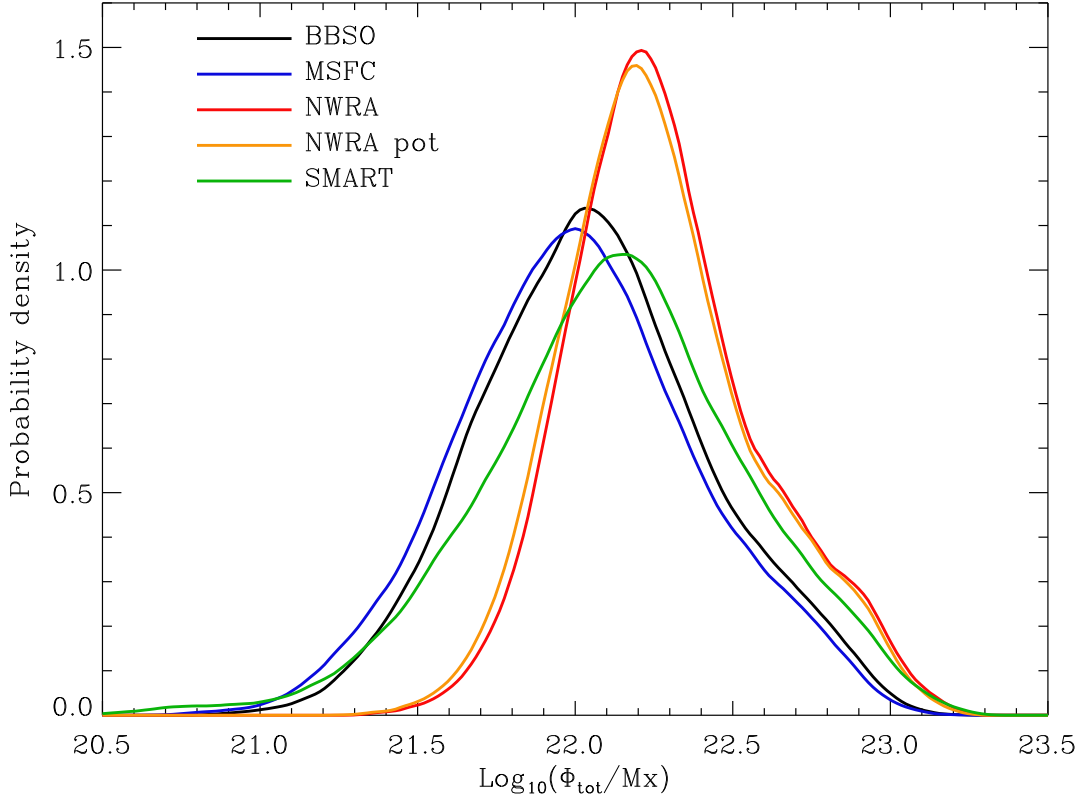}
\caption{The distribution of the total unsigned flux, $\Phi_{\rm tot}$, as computed
by different groups for the \MCD{1}.  There are considerable differences in both 
the width of the distribution and the location of its peak among the different 
implementations.
}
\label{fig:fluxtotdist}
\end{figure}

The distributions of the total flux parameter for the four groups are shown in
Figure~\ref{fig:fluxtotdist}, estimated from the \MCD{1}, which includes
the same regions.  There are considerable differences among the distributions.
In general, the NWRA values of the flux are larger than the other groups,
although the SMART values have similar peak values, but with a tail to lower
values than seen in the NWRA flux distribution.  The NWRA distributions are
also more sharply peaked than the other groups' distributions.

\begin{figure}
\plottwo{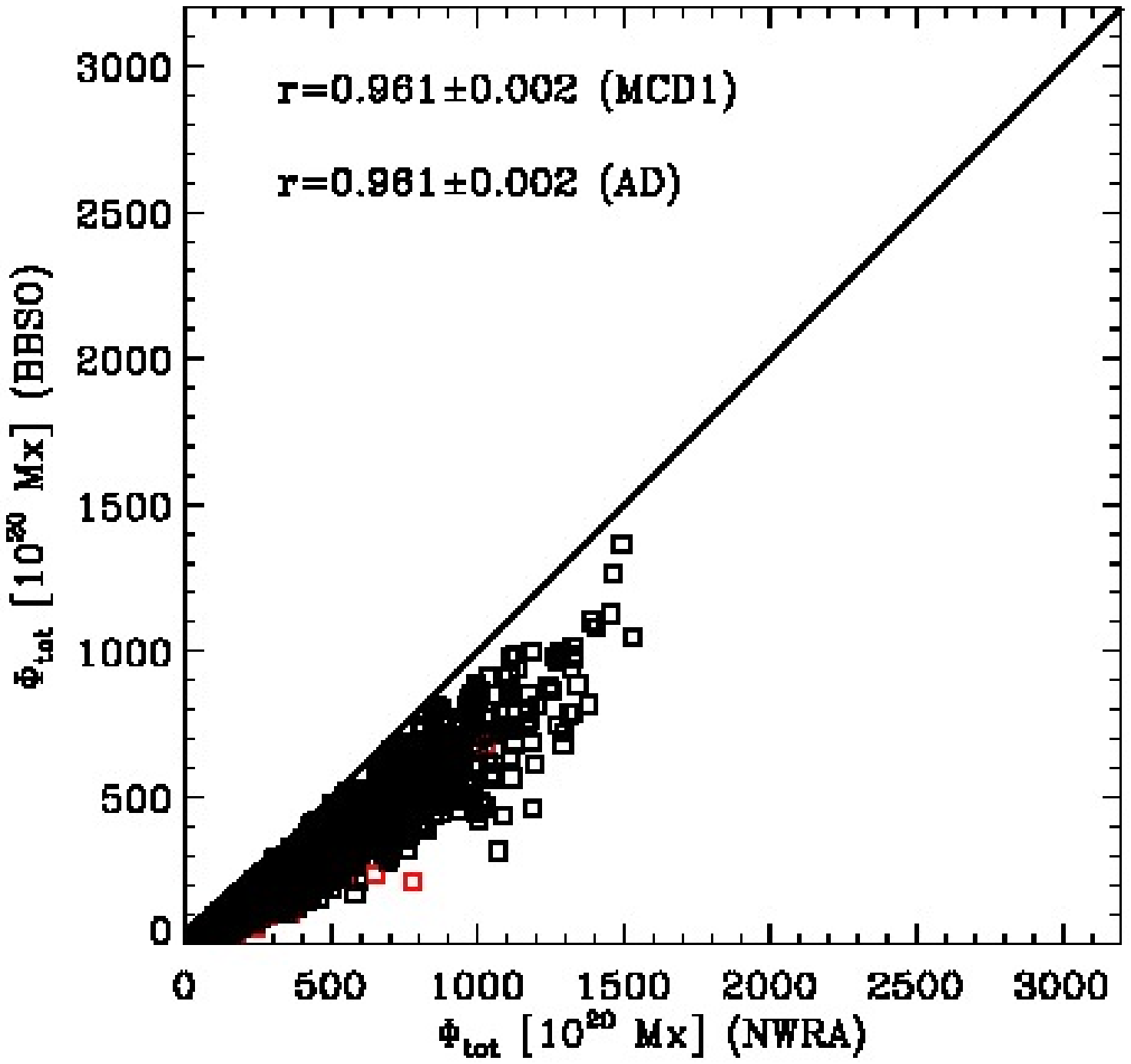}
{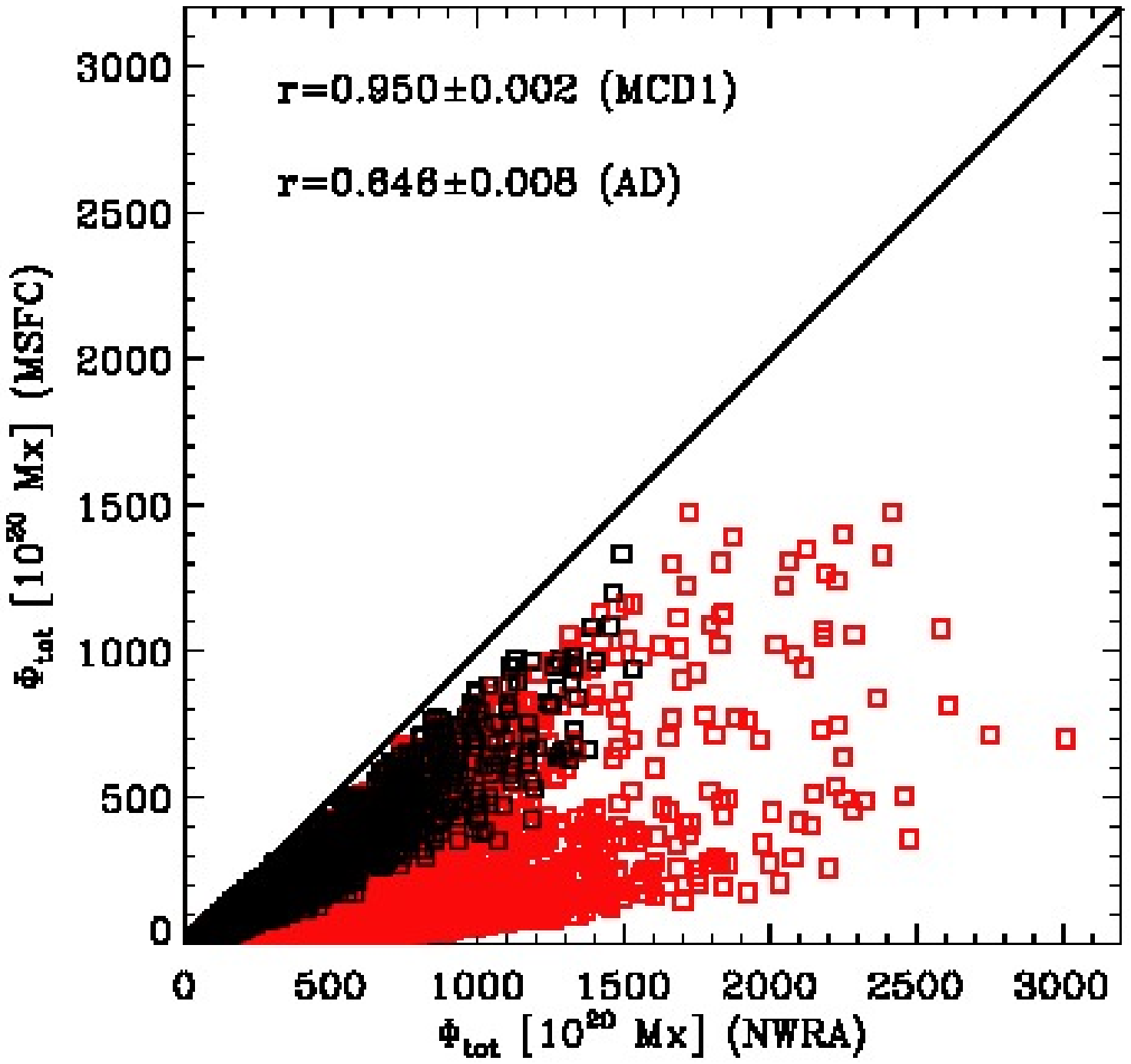}

\plottwo{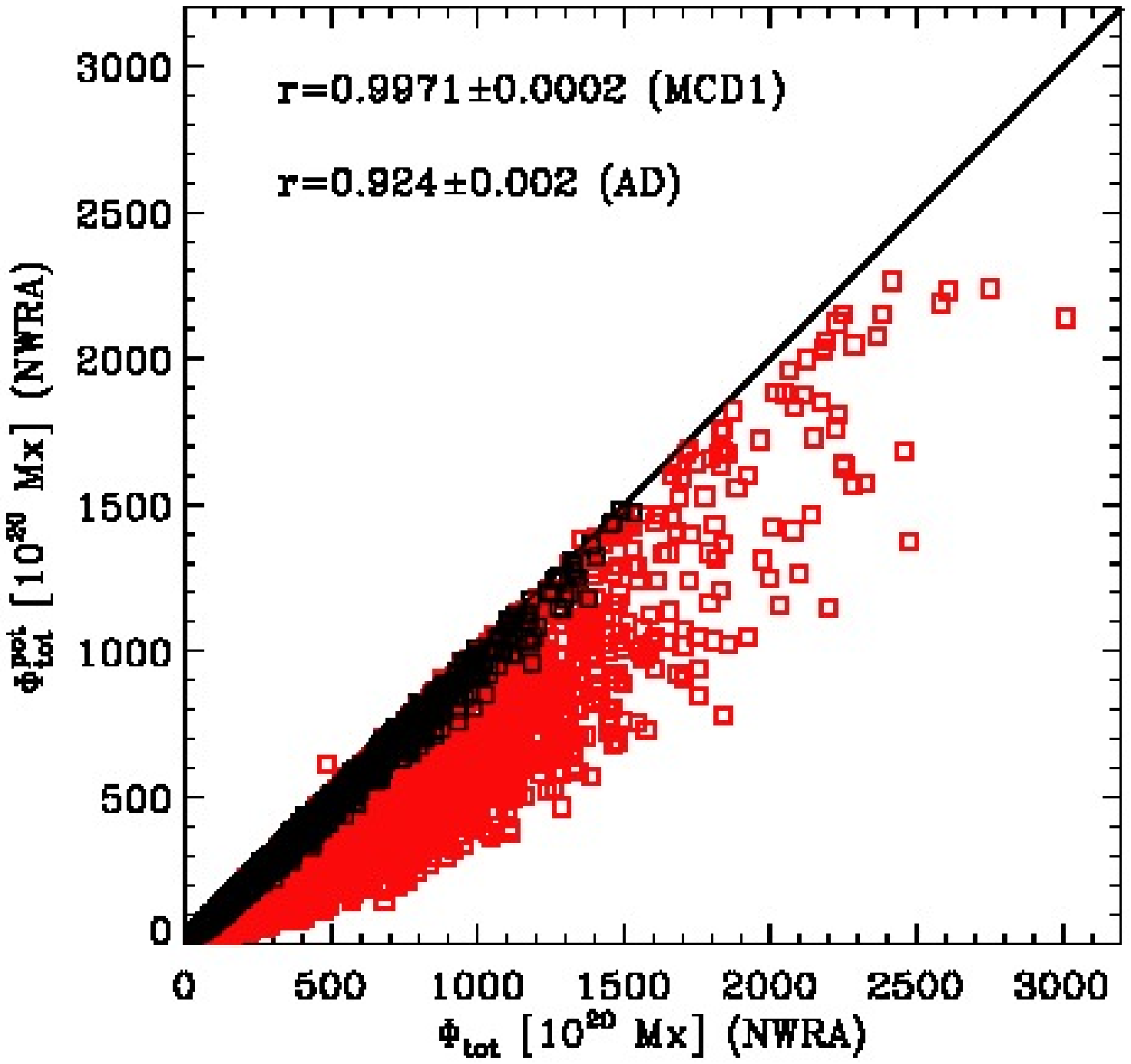}
{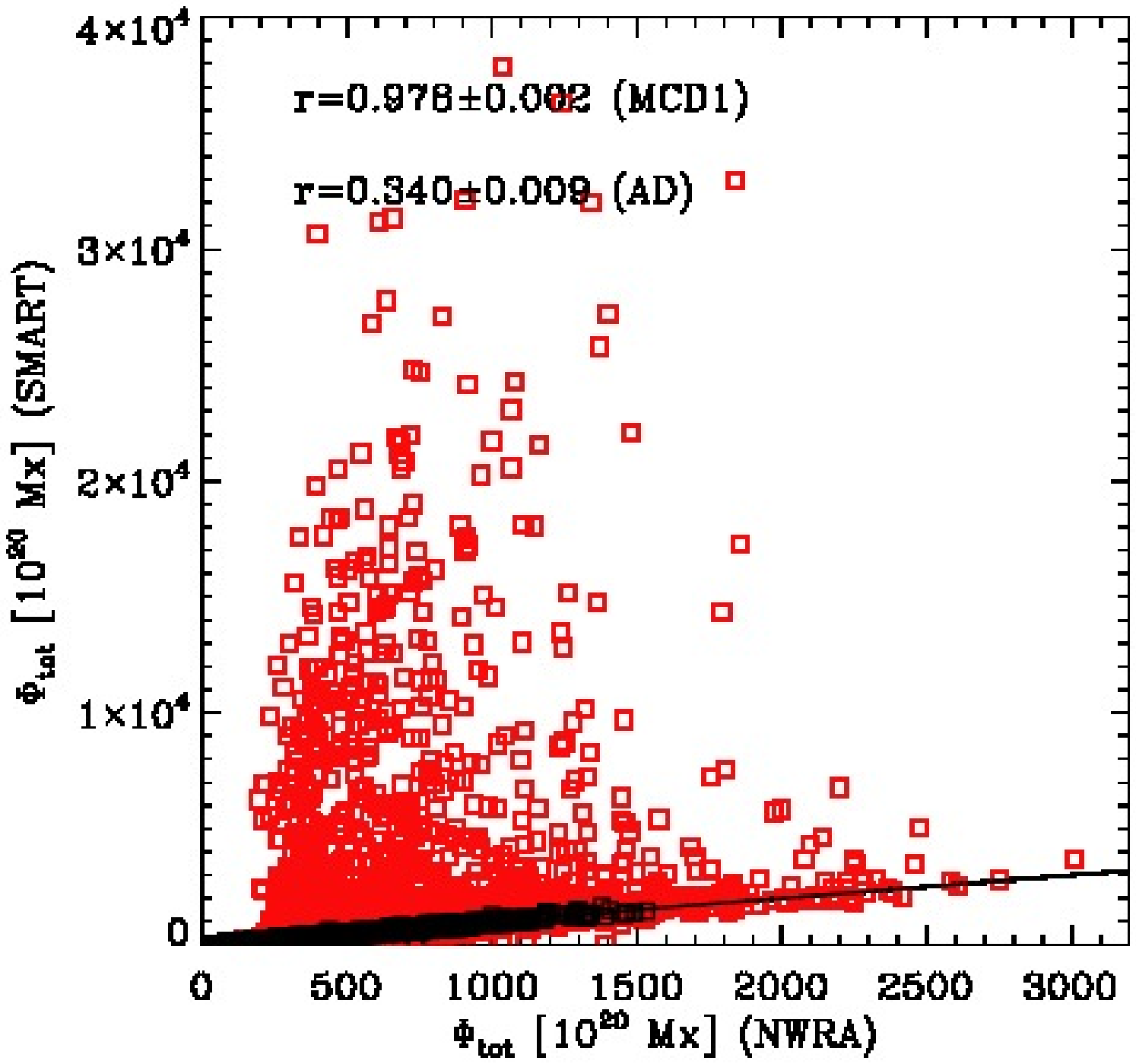}
\caption{The total unsigned flux, $\Phi_{\rm tot}$, as computed by different
groups, plotted as a function of $\Phi_{\rm tot}$ computed by NWRA, with the
Pearson correlation coefficient shown in each plot.  Black points are part of
the \MCD{1} used to compute the distributions in Figure~\ref{fig:fluxtotdist},
while red points are not.  Although the values computed by some groups show
strong correlation (e.g., BBSO and NWRA), others show only moderate (e.g.,
MSFC and NWRA) or weak correlations (e.g., SMART and NWRA).  The correlations
are stronger when considering only the \MCD{1} points, indicating that most of
the scatter is a result of the treatment of projection effects.  In addition to
scatter, there are systematic difference among the values from the different
groups, even for the \MCD{1} points.
}
\label{fig:fluxtotcomp}
\end{figure}

To understand these differences, Figure~\ref{fig:fluxtotcomp} shows the values
for each group plotted versus the NWRA values, for all regions for which the
flux was computed by that group. The NWRA method is used as the reference
because it has a value for every region in the data set.  The values for
regions in the \MCD{1} (black) generally show much less scatter than when all
available regions are considered.  However, there are systematic offsets in
values for NWRA versus BBSO and MSFC even for the \MCD{1}.  The \MCD{1} only
includes regions relatively close to disk center, so much of the scatter may be
a result of how projection effects are accounted for. This is particularly
apparent for the SMART values. 

How much influence do the variations in the total flux resulting from different
implementations have on forecasting flares?  To separate out the effects of the
statistical method, NPDA was applied to all of the total-flux related
parameters.  The forecast performance solely as a single-variable parameter
with NPDA is summarized in Table~\ref{tbl:fluxtotcomp}.  Because different
approximations may not influence how well the determination of total flux works
toward the limb, entries are included using both all data files (without
restriction), and using only the maximum common dataset.

\begin{center}
\begin{deluxetable}{ccccc}
\tablecolumns{5}
\tablewidth{0pc}
\tablecaption{NPDA Forecasts from Total Flux, \CC}
\tablehead{
\colhead{Group/} & \multicolumn{2}{c}{Appleman Skill Score} & \multicolumn{2}{c}{Brier Skill Score} \\ 
\colhead{I.D.} & \colhead{\MCD{1}} & \colhead{All Data} & \colhead{\MCD{1}} & \colhead{All Data} 
}
\startdata
BBSO                  & $0.19\pm0.02$ & $0.06\pm0.01$ & $0.268\pm0.016$ & $0.103\pm0.006$\\
MSFC                  & $0.19\pm0.02$ & $0.13\pm0.01$ & $0.265\pm0.015$ & $0.182\pm0.008$\\
NWRA $\Phi_{\rm los}$ & $0.18\pm0.02$ & $0.14\pm0.01$ & $0.276\pm0.015$ & $0.224\pm0.008$\\
NWRA $\Phi_{\rm pot}$ & $0.18\pm0.02$ & $0.18\pm0.01$ & $0.276\pm0.014$ & $0.246\pm0.008$\\
SMART                 & $0.17\pm0.02$ & $0.06\pm0.01$ & $0.267\pm0.015$ & $0.204\pm0.008$\\
\enddata
\label{tbl:fluxtotcomp}
\end{deluxetable}
\end{center}

Despite differences in the inferred values of the total unsigned flux, the
resulting skill scores for the \MCD{1} are effectively the same for all the
implementations.  The AD forecasting results use a climatology forecast where a
method did not provide a total flux measurement (because it was beyond their
particular limits, for example).  This difference can be seen in the variation
between the results for AD and \MCD{1}, in particular for the BBSO
implementation, which had the most restrictive condition on the distance from
disk center for computing the total unsigned flux.  Overall, the different
implementations result in significantly different skill scores for AD.  

There is also evidence that different approximations applied to the $\Bl$ to
retrieve an estimate of $\Bz$, the radial field, can make a difference.  The
NWRA potential field implementation performs nearly as well on AD as on the
\MCD{1}, and better than the other implementations on AD.  This suggests that,
for flare forecasting, the potential field approximation is better than the
$\mu$-correction when only measurements of the line of sight component are
available.

\begin{figure}
\epsscale{1.0}
\plottwo{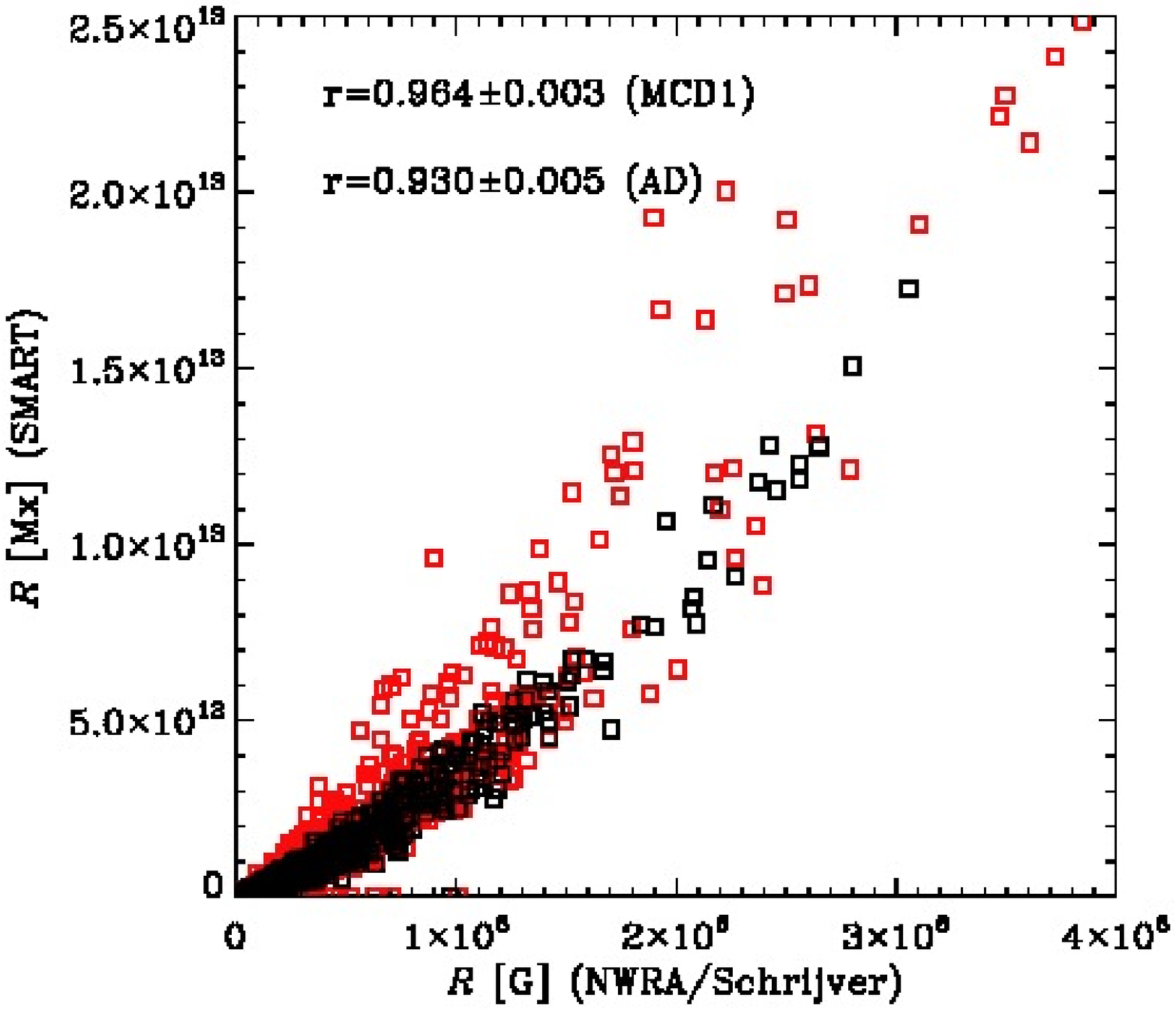}{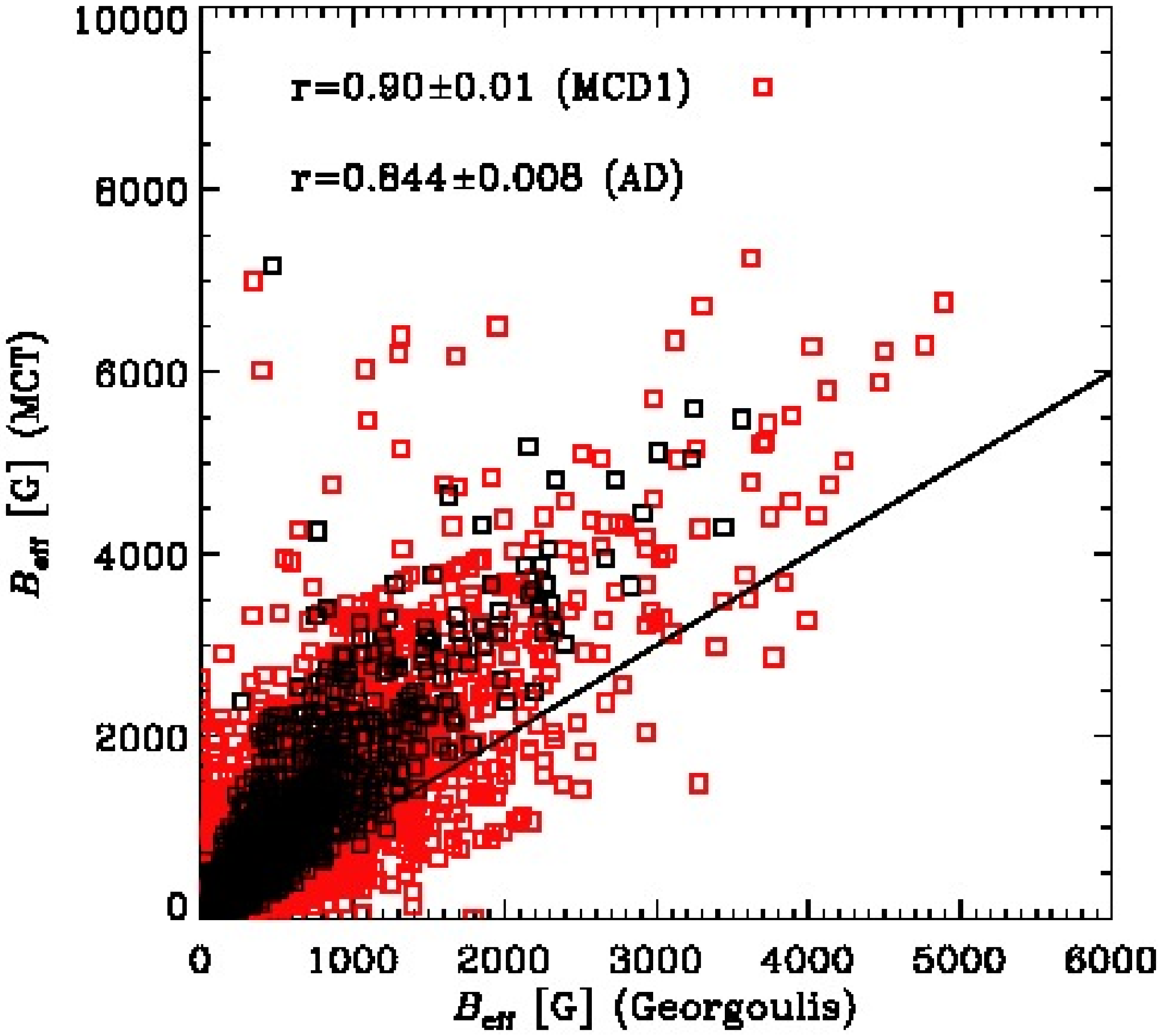}
\caption{Values of the same parameter obtained with different implementations.
Left: a measure of the strong gradient polarity inversion lines,
$\mathcal{R}$ proposed by \cite{Schrijver2007} as computed by two different
groups. 
Right: a measure of the connectivity of the coronal magnetic field,
$B_{\rm eff}$ proposed by \cite{GeorgoulisRust2007} as computed by two
different groups. In both cases, there are noticeable differences in the
values of the parameter depending on implementation.  
}
\label{fig:corr_implement}
\end{figure}

Details of the implementation are also important for other parameters.
Figure~\ref{fig:corr_implement} illustrates the impact of implementation on the
measure $\mathcal{R}$ of the strong gradient polarity inversion lines proposed
by \cite{Schrijver2007}, and on the measure $B_{\rm eff}$ of the connectivity
of the coronal magnetic field proposed by \cite{GeorgoulisRust2007} as computed
by different groups.  The difference is more pronounced in the connectivity
measure.  Although the same mathematical formula for $B_{\rm eff}$ is used, the
differences are due to the distinct approaches followed for partitioning a
magnetogram to determine the point sources, and for inferring the connectivity
matrix for a given set of sources, as noted in \S\ref{sec:MCT}.

\begin{figure}
\plottwo{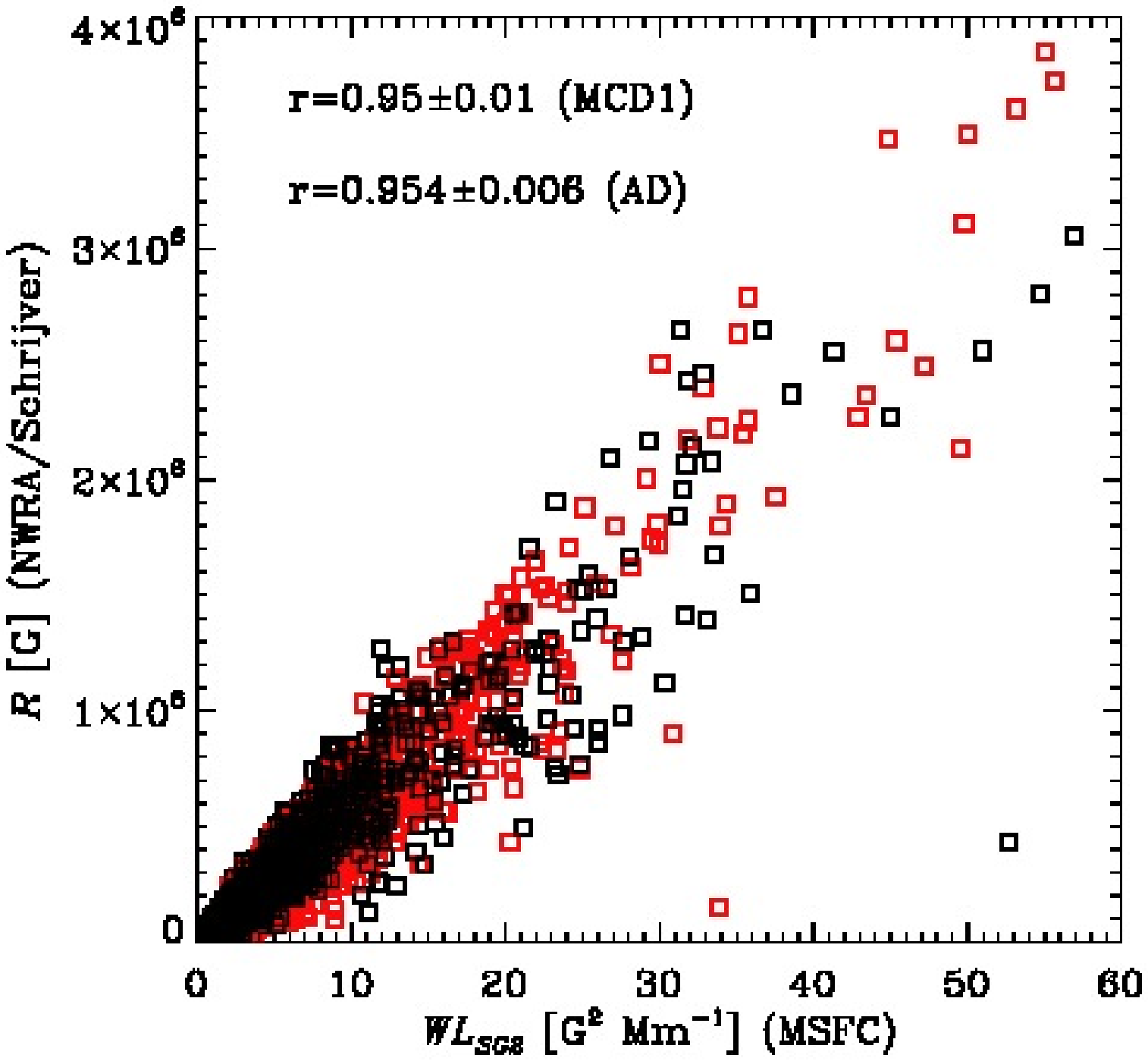}{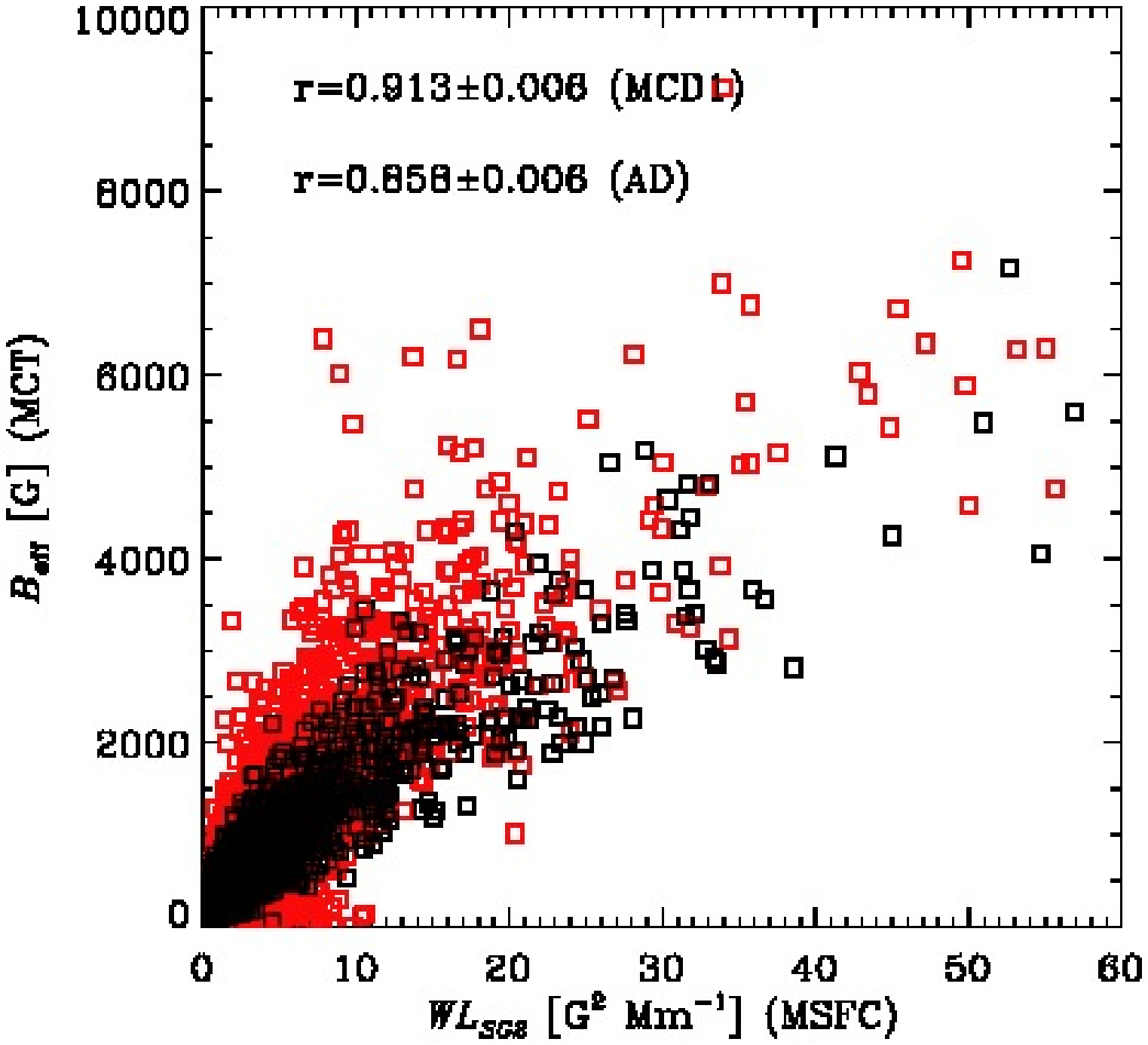}
\caption{Values of parameters characterizing different physical quantities.
Left: two measures of the strong gradient polarity inversion lines,
$\mathcal{R}$ proposed by \cite{Schrijver2007} and $WL_{SG2}$ proposed by
\cite{Falconer_etal_2008}.
Right: the $WL_{SG2}$ measure of the strong gradient polarity inversion
lines calculated by \cite{Falconer_etal_2008} and the $B_{\rm eff}$ measure of
the coronal magnetic connectivity proposed by \cite{GeorgoulisRust2007}.
Although these measures are based on different physical quantities, they are as
strongly correlated as different implementations of $B_{\rm eff}$ (see
Figure~\ref{fig:corr_implement}, right).
}
\label{fig:corr_different}
\end{figure}

In contrast to the differences in a parameter as implemented by different
groups, parameters proposed by different researchers can also be strongly
correlated.  For example, the parameter $\mathcal{R}$ proposed by Schrijver and
the parameter $WL_{SG2}$ proposed by Falconer are both measures of strong
gradient polarity inversion lines. The methods by which they are calculated are
quite different, but the linear correlation coefficient between the two is
$r=0.95$ (Figure~\ref{fig:corr_different}, left).  Perhaps even more surprising
is that $WL_{SG2}$ is also strongly correlated with the parameter $B_{\rm eff}$ 
of \cite{GeorgoulisRust2007}, which is a measure of the connectivity of the 
coronal magnetic field (Figure~\ref{fig:corr_different}, right).


Two conclusions can be drawn from this exercise.  First, implementation details
can greatly influence the resulting parameter values, although this makes
surprisingly little difference in the forecasting ability of the quantities
considered.  Second, there may be a limited amount of information available for
flare forecasting from only the line of sight magnetic field without additional
modeling.  Even if additional modeling is used, such as when the coronal
connectivity is used to determine $B_{\rm eff}$, solar active regions that are
small tend to be simpler, larger regions tend to be more complex, and
differentiating those with imminent flare potential remains difficult.

\section{Discussion and Speculation}
\label{sec:discussion}

During the workshop and subsequent group discussions, a few salient points
regarding flare forecasting methods emerged, and are discussed here.

Using NOAA ARs may be a less than optimal approach for forecasting.  Sometimes
there is no obvious photospheric division in the magnetic field of two NOAA ARs
with clear coronal connectivity. Since the flux systems are physically
interacting, they may be best treated as a single entity for the purpose of
prediction.  Yet identifying them thus can result in extremely large fractions
of the Sun becoming a single forecasting target.  Growing sunspot groups can
occasionally be flare productive prior to acquiring a number by NOAA, and this
will bias the results.  Many groups are working on better active-region
identification methods, but testing each is beyond the scope of this paper.
Ideally, we would like to know where on the Sun a flare will occur, independent
of the assignment of an active region number.

The data used here were not ideal for any method. Each had different specific
requirements on the data needed for a forecast, and it was difficult to
accommodate these needs.  Only by inserting a reference forecast when methods
did not provide one, or by restricting the comparisons to the maximum common
datasets, which relied upon only data for which all methods could provide
forecasts, could a systematic comparison be made.  For most combinations of
event definition and data set, the best three methods resulted in comparable
skill scores, so that no one method was clearly superior to the rest, and even
which three methods resulted in the highest skill scores varied with event
definition and data set.  This result emphasizes that comparison of
reported skill scores is impossible unless the underlying data, limits,
treatment of missing forecasts, and event rates, are all standardized.
Ideally, a common data set should be established in advance, such that all
methods can train and forecast on exactly the same data, thus circumventing
many of the difficulties in making a comparison.

There is a limit to how much information is available in a single line-of-sight
magnetogram.  Many of the parameters used by different methods and groups are
correlated with each other, meaning that they bring no independent information
to the forecasts.  More concerning is that parameter calculation and
implementation differences, for even such a simple quantity as the total
magnetic flux in a region, can significantly change the value of the parameter.
Surprisingly, this had only a minor effect on forecasting ability for the cases
considered.  The use of vector field data and prior flare history may improve
forecast performance by providing independent information.  

Defining an event based on the peak output in a particular wavelength band
(GOES 1-8\,\AA) is not based on the physics of flares.  The soft X-ray
signature (and hence magnitude) is not a direct measure of the total energy
released in a reconnection event \citep{Emslieetal2012}, yet the parameters
presently being computed are generally measures of the total energy of an
active region.  Considering how the energy release is partitioned between
particles, thermal heating and bulk flow may lead to improved forecasts for the
peak soft X-ray flux.  Likewise, defining the validity period for a forecast
based on a time interval related to the Earth's rotational period (12\,hr or
24\,hr for the results presented here) has no basis in the physics of flares.
Studies looking at either longer \citep{Falconeretal2014} or shorter
\citep{AlGhraibahetal2015} validity periods do not show a substantial increase
in performance.  Nevertheless, considering the evolution of an active region on
an appropriate time scale may better indicate when energy will be released and
thus lead to improved forecasts.

The forecast of an all-clear is an easier problem for large events than
providing accurate event forecasts because there must be sufficient energy
present for a large event to occur.  When no active region has a large amount
of free energy, and particularly during times when no active region faces the
Earth, geo-effective solar activity is low in general and the possibility of a
large flare is lower still.  The low event rate for large events does mean that
it is relatively easy to achieve a low false alarm rate.  Still, during epochs
when regions are on the disk, the inability of the prediction methods
(admittedly, tested on old data) to achieve any high skill score values is
discouraging, and shows that there is still plenty of opportunity to improve
forecast methods.

In summary, numerous parameters and different statistical forecast methods are
shown here to provide improved prediction over climatological forecasts, but
none achieve large skill score values.  This result may improve when updated
methods and data are used.  For those who could separate the prediction process
into first characterizing an active region by way of one or more parameters,
and then separately using a statistical technique to arrive at a prediction,
both parameters and predictions were collected separately.  It may be possible
to increase the skill score values by combining the prediction algorithm from
one method with the parameterization from a different method.  Nonetheless,
this study presents the first systematic, focused head-to-head comparison of
many ways of characterizing solar magnetic regions and different statistical
forecasting approaches.

\acknowledgments

This work is the outcome of many collaborative and cooperative efforts.  The
2009 ``Forecasting the All-Clear'' Workshop in Boulder, CO was sponsored by
NASA/Johnson Space Flight Center's Space Radiation Analysis Group, the National
Center for Atmospheric Research, and the NOAA/Space Weather Prediction Center,
with additional travel support for participating scientists from NASA LWS TRT
NNH09CE72C to NWRA.  The authors thank the participants of that workshop, in
particular Drs.~Neal Zapp, Dan Fry, Doug Biesecker, for the informative
discussions during those three crazy days, and NCAR's Susan Baltuch and NWRA's
Janet Biggs for organizational prowess.  Workshop preparation and analysis
support was provided for GB, KDL by NASA LWS TRT NNH09CE72C, and NASA
Heliophysics GI NNH12CG10C.  PAH and DSB received funding from the European
Space Agency PRODEX Programme, while DSB and MKG also received funding from the
European Union's Horizon 2020 research and innovation programme under grant
agreement No.~640216 (FLARECAST project).  MKG also acknowledges research
performed under the A-EFFort project and subsequent service implementation,
supported under ESA Contract number  4000111994/14/D/MPR.  YY was supported by
the National Science Foundation under grants ATM 09-36665, ATM 07-16950,
ATM-0745744 and by NASA under grants NNX0-7AH78G, NNXO-8AQ90G.  YY owes his
deepest gratitude to his advisers Prof.~Frank Y.~Shih, Prof.~Haimin Wang and
Prof.~Ju Jing for long discussions, for reading previous drafts of his work and
providing many valuable comments that improved the presentation and contents of
this work.  JMA was supported by NSF Career Grant AGS-1255024 and by a NMSU
Vice President for Research Interdisciplinary Research Grant.

\bibliographystyle{apj}
\bibliography{biblio}

\begin{thebibliography}{68}
\expandafter\ifx\csname natexlab\endcsname\relax\def\natexlab#1{#1}\fi

\bibitem[{{Abramenko}(2005)}]{Abramenko2005b}
{Abramenko}, V.~I. 2005, ApJ, 629, 1141

\bibitem[{{Abramenko} {et~al.}(2003){Abramenko}, {Yurchyshyn}, {Wang},
  {Spirock}, \& {Goode}}]{Abramenko_etal_2003}
{Abramenko}, V.~I., {Yurchyshyn}, V.~B., {Wang}, H., {Spirock}, T.~J., \&
  {Goode}, P.~R. 2003, ApJ, 597, 1135

\bibitem[{{Ahmed} {et~al.}(2013){Ahmed}, {Qahwaji}, {Colak}, {Higgins},
  {Gallagher}, \& {Bloomfield}}]{Ahmedetal2013}
{Ahmed}, O.~W., {Qahwaji}, R., {Colak}, T., {Higgins}, P.~A., {Gallagher},
  P.~T., \& {Bloomfield}, D.~S. 2013, \solphys, 283, 157

\bibitem[{{Al-Ghraibah} {et~al.}(2015){Al-Ghraibah}, {Boucheron}, \&
  {McAteer}}]{AlGhraibahetal2015}
{Al-Ghraibah}, A., {Boucheron}, L.~E., \& {McAteer}, R.~T.~J. 2015, \aap, 579,
  A64

\bibitem[{{Alissandrakis}(1981)}]{Alissandrakis1981}
{Alissandrakis}, C.~E. 1981, \aap, 100, 197

\bibitem[{{Balch}(2008)}]{Balch2008}
{Balch}, C.~C. 2008, Space Weather, 6, 1001

\bibitem[{{Barnes}(2007)}]{Barnes2007}
{Barnes}, G. 2007, \apj, 670, L53

\bibitem[{{Barnes} \& {Leka}(2006)}]{dfa2}
{Barnes}, G. \& {Leka}, K.~D. 2006, \apj, 646, 1303

\bibitem[{{Barnes} \& {Leka}(2008)}]{BarnesLeka2008}
---. 2008, ApJL, 688, L107

\bibitem[{Barnes {et~al.}(2016)Barnes, Leka, Schrijver, Colak, Qahwaji, Yuan,
  Zhang, McAteer, Higgins, Conlon, Falconer, Georgoulis, Wheatland, \&
  Balch}]{ffc2}
Barnes, G., Leka, K.~D., Schrijver, C.~J., Colak, T., Qahwaji, R., Yuan, Y.,
  Zhang, J., McAteer, R.~T.~J., Higgins, P.~A., Conlon, P.~A., Falconer, D.~A.,
  Georgoulis, M.~K., Wheatland, M.~S., \& Balch, C. 2016, ApJ, in preparation

\bibitem[{{Barnes} {et~al.}(2007){Barnes}, {Leka}, {Schumer}, \&
  {Della-Rose}}]{SWJ}
{Barnes}, G., {Leka}, K.~D., {Schumer}, E.~A., \& {Della-Rose}, D.~J. 2007,
  Space Weather, 5, 9002

\bibitem[{{Barnes} {et~al.}(2005){Barnes}, {Longcope}, \& {Leka}}]{mct}
{Barnes}, G., {Longcope}, D., \& {Leka}, K.~D. 2005, ApJ, 629, 561

\bibitem[{{Berger} \& {Lites}(2003)}]{bergerlites2003}
{Berger}, T.~E. \& {Lites}, B.~W. 2003, Sol.~Phys., 213, 213

\bibitem[{{Bloomfield} {et~al.}(2012){Bloomfield}, {Higgins}, {McAteer}, \&
  {Gallagher}}]{Bloomfield_etal_2012}
{Bloomfield}, D.~S., {Higgins}, P.~A., {McAteer}, R.~T.~J., \& {Gallagher},
  P.~T. 2012, ApJL, 747, L41

\bibitem[{{Boser} {et~al.}(1992){Boser}, {Guyon}, \&
  {Vapnik}}]{BoserGuyonVapnik1992}
{Boser}, B.~E., {Guyon}, I.~M., \& {Vapnik}, V. 1992, in Proceedings of the
  Fifth Annual Workshop on Computational Learning Theory, COLT 92 (ACM)

\bibitem[{{Boucheron} {et~al.}(2015){Boucheron}, {Al-Ghraibah}, \&
  {McAteer}}]{Boucheronetal2015}
{Boucheron}, L.~E., {Al-Ghraibah}, A., \& {McAteer}, R.~T.~J. 2015, \apj, 812,
  51

\bibitem[{{Chang} \& {Lin}(2011)}]{ChangLin2011}
{Chang}, C.-C. \& {Lin}, C.-J. 2011, ACM Trans. Intell. Syst. Technol., 2, 27:1

\bibitem[{{Colak} \& {Qahwaji}(2008)}]{ColakQahwaji2008}
{Colak}, T. \& {Qahwaji}, R. 2008, \solphys, 248, 277

\bibitem[{{Colak} \& {Qahwaji}(2009)}]{ColakQahwaji2009}
---. 2009, Space Weather, 7, 6001

\bibitem[{{Conlon} {et~al.}(2008){Conlon}, {Gallagher}, {McAteer}, {Ireland},
  {Young}, {Kestener}, {Hewett}, \& {Maguire}}]{Conlon_etal_2008}
{Conlon}, P.~A., {Gallagher}, P.~T., {McAteer}, R.~T.~J., {Ireland}, J.,
  {Young}, C.~A., {Kestener}, P., {Hewett}, R.~J., \& {Maguire}, K. 2008,
  Sol.~Phys., 248, 297

\bibitem[{{Conlon} {et~al.}(2010){Conlon}, {McAteer}, {Gallagher}, \&
  {Fennell}}]{Conlon_etal_2010}
{Conlon}, P.~A., {McAteer}, R.~T.~J., {Gallagher}, P.~T., \& {Fennell}, L.
  2010, ApJ, 722, 577

\bibitem[{{Cortes} \& {Vapnik}(1995)}]{CortesVapnik1995}
{Cortes}, C. \& {Vapnik}, V. 1995, Mach. Learn., 20, 273

\bibitem[{{Crown}(2012)}]{Crown2012}
{Crown}, M.~D. 2012, Space Weather, 10, 6006

\bibitem[{{Efron} \& {Gong}(1983)}]{EfronGong1983}
{Efron}, B. \& {Gong}, G. 1983, Am.~Statist., 37, 36

\bibitem[{{Emslie} {et~al.}(2012){Emslie}, {Dennis}, {Shih}, {Chamberlin},
  {Mewaldt}, {Moore}, {Share}, {Vourlidas}, \& {Welsch}}]{Emslieetal2012}
{Emslie}, A.~G., {Dennis}, B.~R., {Shih}, A.~Y., {Chamberlin}, P.~C.,
  {Mewaldt}, R.~A., {Moore}, C.~S., {Share}, G.~H., {Vourlidas}, A., \&
  {Welsch}, B.~T. 2012, \apj, 759, 71

\bibitem[{{Falconer} {et~al.}(2011){Falconer}, {Barghouty}, {Khazanov}, \&
  {Moore}}]{Falconer_etal_2011}
{Falconer}, D., {Barghouty}, A.~F., {Khazanov}, I., \& {Moore}, R. 2011, Space
  Weather, 9, 4003

\bibitem[{{Falconer} {et~al.}(2012){Falconer}, {Moore}, {Barghouty}, \&
  {Khazanov}}]{Falconer_etal_2012}
{Falconer}, D.~A., {Moore}, R.~L., {Barghouty}, A.~F., \& {Khazanov}, I. 2012,
  ApJ, 757, 32

\bibitem[{{Falconer} {et~al.}(2014){Falconer}, {Moore}, {Barghouty}, \&
  {Khazanov}}]{Falconeretal2014}
---. 2014, Space Weather, 12, 306

\bibitem[{{Falconer} {et~al.}(2002){Falconer}, {Moore}, \&
  {Gary}}]{Falconer_etal_2002}
{Falconer}, D.~A., {Moore}, R.~L., \& {Gary}, G.~A. 2002, ApJ, 569, 1016

\bibitem[{{Falconer} {et~al.}(2003){Falconer}, {Moore}, \&
  {Gary}}]{Falconer_etal_2003}
---. 2003, Journal of Geophysical Research (Space Physics), 108, 1

\bibitem[{{Falconer} {et~al.}(2008){Falconer}, {Moore}, \&
  {Gary}}]{Falconer_etal_2008}
---. 2008, ApJ, 689, 1433

\bibitem[{{Gallagher} {et~al.}(2002){Gallagher}, {Moon}, \&
  {Wang}}]{Gallagheretal2002}
{Gallagher}, P., {Moon}, Y.-J., \& {Wang}, H. 2002, \solphys, 209, 171

\bibitem[{{Georgoulis}(2012)}]{Georgoulis2012}
{Georgoulis}, M.~K. 2012, \solphys, 276, 161

\bibitem[{{Georgoulis} \& {Rust}(2007)}]{GeorgoulisRust2007}
{Georgoulis}, M.~K. \& {Rust}, D.~M. 2007, ApJL, 661, L109

\bibitem[{{Georgoulis} {et~al.}(2012){Georgoulis}, {Tziotziou}, \&
  {Raouafi}}]{Georgoulis_etal_2012a}
{Georgoulis}, M.~K., {Tziotziou}, K., \& {Raouafi}, N.-E. 2012, ApJ, 759, 1

\bibitem[{{Hanssen} \& {Kuipers}(1965)}]{HanssenKuipers1965}
{Hanssen}, A.~W. \& {Kuipers}, W.~J.~A. 1965, Meded. Verhand, 81, 2

\bibitem[{{Higgins} {et~al.}(2011){Higgins}, {Gallagher}, {McAteer}, \&
  {Bloomfield}}]{Higgins_etal_2011}
{Higgins}, P.~A., {Gallagher}, P.~T., {McAteer}, R.~T.~J., \& {Bloomfield},
  D.~S. 2011, Advances in Space Research, 47, 2105

\bibitem[{{Jaynes} \& {Bretthorst}(2003)}]{Jaynes2003}
{Jaynes}, E.~T. \& {Bretthorst}, G.~L. 2003, {Probability Theory} (Cambridge
  University Press), 154--156

\bibitem[{{Jing} {et~al.}(2006){Jing}, {Song}, {Abramenko}, {Tan}, \&
  {Wang}}]{Jing_etal_2006}
{Jing}, J., {Song}, H., {Abramenko}, V., {Tan}, C., \& {Wang}, H. 2006, \apj,
  644, 1273

\bibitem[{{Jolliffe} \& {Stephenson}(2003)}]{JolliffeStephenson2003}
{Jolliffe}, I.~T. \& {Stephenson}, D. 2003, Forecast Verification: A
  Practioner's Guide in Atmospheric Science (Wiley)

\bibitem[{{Kendall} {et~al.}(1983){Kendall}, {Stuart}, \& {Ord}}]{ken83}
{Kendall}, M., {Stuart}, A., \& {Ord}, J.~K. 1983, {The Advanced Theory of
  Statistics}, 4th edn., Vol.~3 (New York: Macmillan Publishing Co., Inc)

\bibitem[{{Kildahl}(1980)}]{Kildahl1980}
{Kildahl}, K.~J.~N. 1980, in Solar-Terrestrial Predictions Proceedings, ed.
  R.~F. {Donnelly}, Vol.~3 (Boulder: U.S.~Dept.~of Commerce), 166

\bibitem[{{Leka} \& {Barnes}(2003{\natexlab{a}})}]{params}
{Leka}, K.~D. \& {Barnes}, G. 2003{\natexlab{a}}, ApJ, 595, 1277

\bibitem[{{Leka} \& {Barnes}(2003{\natexlab{b}})}]{dfa}
---. 2003{\natexlab{b}}, ApJ, 595, 1296

\bibitem[{{Leka} \& {Barnes}(2007)}]{dfa3}
---. 2007, ApJ, 656, 1173

\bibitem[{{Mason} \& {Hoeksema}(2010)}]{MasonHoeksema2010}
{Mason}, J.~P. \& {Hoeksema}, J.~T. 2010, ApJ, 723, 634

\bibitem[{{McAteer}(2015)}]{McAteer2015}
{McAteer}, R.~T.~J. 2015, \solphys, 290, 1897

\bibitem[{{McAteer} {et~al.}(2015){McAteer}, {Aschwanden}, {Dimitropoulou},
  {Georgoulis}, {Pruessner}, {Morales}, {Ireland}, \&
  {Abramenko}}]{McAteeretal2015}
{McAteer}, R.~T.~J., {Aschwanden}, M.~J., {Dimitropoulou}, M., {Georgoulis},
  M.~K., {Pruessner}, G., {Morales}, L., {Ireland}, J., \& {Abramenko}, V.
  2015, \ssr

\bibitem[{{McAteer} {et~al.}(2010){McAteer}, {Gallagher}, \&
  {Conlon}}]{McAteer_etal_2010}
{McAteer}, R.~T.~J., {Gallagher}, P.~T., \& {Conlon}, P.~A. 2010, Advances in
  Space Research, 45, 1067

\bibitem[{{McAteer} {et~al.}(2005{\natexlab{a}}){McAteer}, {Gallagher}, \&
  {Ireland}}]{McAteer_etal_2005}
{McAteer}, R.~T.~J., {Gallagher}, P.~T., \& {Ireland}, J. 2005{\natexlab{a}},
  \apj, 631, 628

\bibitem[{{McAteer} {et~al.}(2005{\natexlab{b}}){McAteer}, {Gallagher},
  {Ireland}, \& {Young}}]{McAteer_etal_2005b}
{McAteer}, R.~T.~J., {Gallagher}, P.~T., {Ireland}, J., \& {Young}, C.~A.
  2005{\natexlab{b}}, \solphys, 228, 55

\bibitem[{McIntosh(1990)}]{mcintosh90}
McIntosh, P.~S. 1990, \solphys, 125, 251

\bibitem[{{Moon} {et~al.}(2001){Moon}, {Choe}, {Yun}, \&
  {Park}}]{Moon_etal_2001}
{Moon}, Y.-J., {Choe}, G.~S., {Yun}, H.~S., \& {Park}, Y.~D. 2001, \jgr, 106,
  29951

\bibitem[{{Murphy}(1996)}]{Murphy1996}
{Murphy}, A.~H. 1996, Wea.~Forecasting, 11, 3

\bibitem[{{Qahwaji} {et~al.}(2008){Qahwaji}, {Colak}, {Al-Omari}, \&
  {Ipson}}]{Qahwaji_etal_2008}
{Qahwaji}, R., {Colak}, T., {Al-Omari}, M., \& {Ipson}, S. 2008, Sol.~Phys.,
  248, 471

\bibitem[{{Scherrer} {et~al.}(1995){Scherrer}, {Bogart}, {Bush}, {Hoeksema},
  {Kosovichev}, {Schou}, {Rosenberg}, {Springer}, {Tarbell}, {Title},
  {Wolfson}, {Zayer}, \& {MDI Engineering Team}}]{mdi}
{Scherrer}, P.~H., {Bogart}, R.~S., {Bush}, R.~I., {Hoeksema}, J.~T.,
  {Kosovichev}, A.~G., {Schou}, J., {Rosenberg}, W., {Springer}, L., {Tarbell},
  T.~D., {Title}, A., {Wolfson}, C.~J., {Zayer}, I., \& {MDI Engineering Team}.
  1995, Sol.~Phys., 162, 129

\bibitem[{{Schrijver}(2007)}]{Schrijver2007}
{Schrijver}, C.~J. 2007, ApJL, 655, L117

\bibitem[{{Silverman}(1986)}]{Silverman1986}
{Silverman}, B.~W. 1986, Density Estimation for Statistics and Data Analysis
  (London: Chapman and Hall)

\bibitem[{{Vapnik}(2000)}]{Vapnik2000}
{Vapnik}, V. 2000, The Nature of Statistical Learning Theory, 2nd edn.
  (Springer-Verlag)

\bibitem[{{Wheatland}(2001)}]{Wheatland2001}
{Wheatland}, M.~S. 2001, Sol.~Phys., 203, 87

\bibitem[{{Wheatland}(2004)}]{whe04a}
---. 2004, ApJ, 609, 1134

\bibitem[{{Wheatland}(2005)}]{Wheatland2005}
---. 2005, Space Weather, 3, 7003

\bibitem[{{Woodcock}(1976)}]{Woodcock1976}
{Woodcock}, F. 1976, Monthly Weather Review, 104, 1209

\bibitem[{{Yang} {et~al.}(2013){Yang}, {Lin}, {Zhang}, \&
  {Mao}}]{YangX_etal_2013}
{Yang}, X., {Lin}, G., {Zhang}, H., \& {Mao}, X. 2013, \apjl, 774, L27

\bibitem[{{Yu} {et~al.}(2010){Yu}, {Huang}, {Wang}, {Cui}, {Hu}, \&
  {Zhou}}]{Yu_etal_2010b}
{Yu}, D., {Huang}, X., {Wang}, H., {Cui}, Y., {Hu}, Q., \& {Zhou}, R. 2010,
  ApJ, 710, 869

\bibitem[{{Yuan} {et~al.}(2011){Yuan}, {Shih}, {Jing}, \&
  {Wang}}]{Yuan_etal_2011}
{Yuan}, Y., {Shih}, F.~Y., {Jing}, J., \& {Wang}, H. 2011, in {IAU Symposium},
  Vol. 273, {IAU Symposium}, 446--450

\bibitem[{{Yuan} {et~al.}(2010){Yuan}, {Shih}, {Jing}, \&
  {Wang}}]{Yuan_etal_2010}
{Yuan}, Y., {Shih}, F.~Y., {Jing}, J., \& {Wang}, H.-M. 2010, Research in
  Astronomy and Astrophysics, 10, 785

\bibitem[{{Zhang} {et~al.}(2010){Zhang}, {Wang}, \& {Liu}}]{ZhangWangLiu2010}
{Zhang}, J., {Wang}, Y., \& {Liu}, Y. 2010, \apj, 723, 1006

\end{thebibliography}

\appendix

\section{Prediction Method Descriptions}
\label{app:methods}

A more detailed description of each method referred to in the text is provided
here.  Since several methods implement similar-sounding techniques in slightly
different manners, a few salient points are included as appropriate, such as
specifics of data analysis and any free parameters available.  A summary of
each method's performance is presented for its optimal application.  That is,
skill scores are computed based only on the data for which the method provided
a forecast, and a different probability threshold is selected to maximize each
presented categorical skill score for each event definition.  Many methods have
a restriction on when a forecast is made, for example a restriction on the
observing angle of the active region.  Such restrictions reduce the sample
sizes from the original full dataset, and these reductions (if any) are
indicated.  As such, the summary metrics presented for the optimal performance
give an indication of \deleted{the} each method's performance, but should not
generally be used to directly compare methods.  A comparison of methods on
common data sets is presented in Section~\ref{sec:results}.

Even when the methods are evaluated on different data sets, there are some
common trends in the results.  As was the case for the common data sets, most
methods have higher skill scores for smaller event magnitudes, with the event
statistics methods again being the exception.  However, the ROC plots, the
\True, and the rate correct show the opposite trend for most methods, with
larger area under the ROC curve, a higher \True, and a higher rate correct for
larger event magnitudes.  This is an indication that the populations of events
and non-events are well separated in parameter space for large events, but the
prior probability of an event is sufficiently small that most measures of the
forecast performance have a small value.

\subsection{The Effective Connected Magnetic Field - M.~Georgoulis}
\label{sec:beff}

The analysis presented in \citet{GeorgoulisRust2007} describes the coronal
magnetic connectivity using the effective connected magnetic field strength
$B_{\rm eff}$. The $B_{\rm eff}$ parameter is calculated following inference
of a connectivity matrix in the magnetic-flux distribution of the target active
region (employing $\Bz \approx \Bl / \cos \theta$, where $\theta$ is the angle
from disk center).  An additional multiplicative correction factor of 1.56
(plage) and 1.45 (sunspots) was applied to correct for systematic insensitivity
\citep{bergerlites2003}.  

\begin{figure}[t]
\plottwo{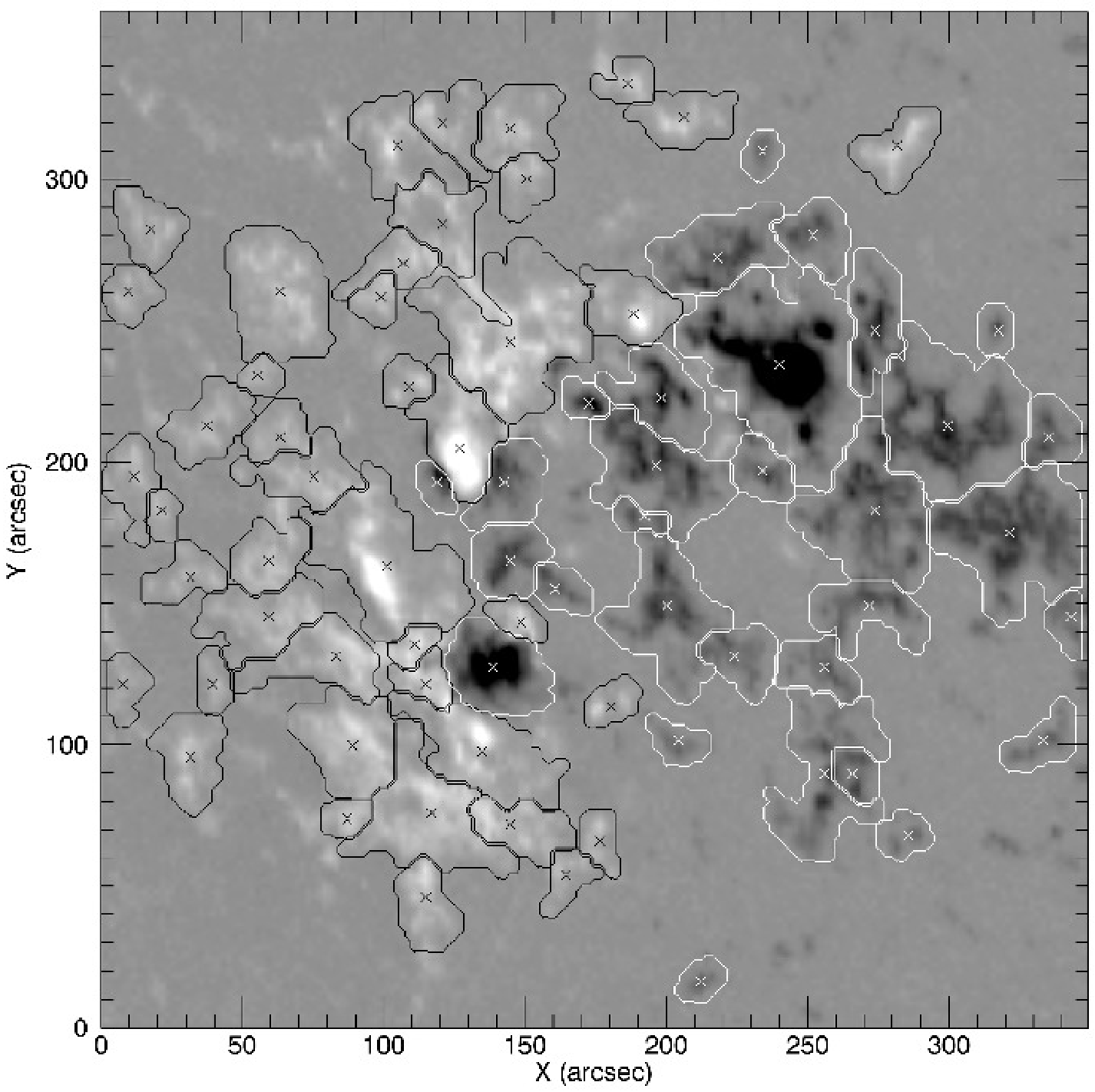}{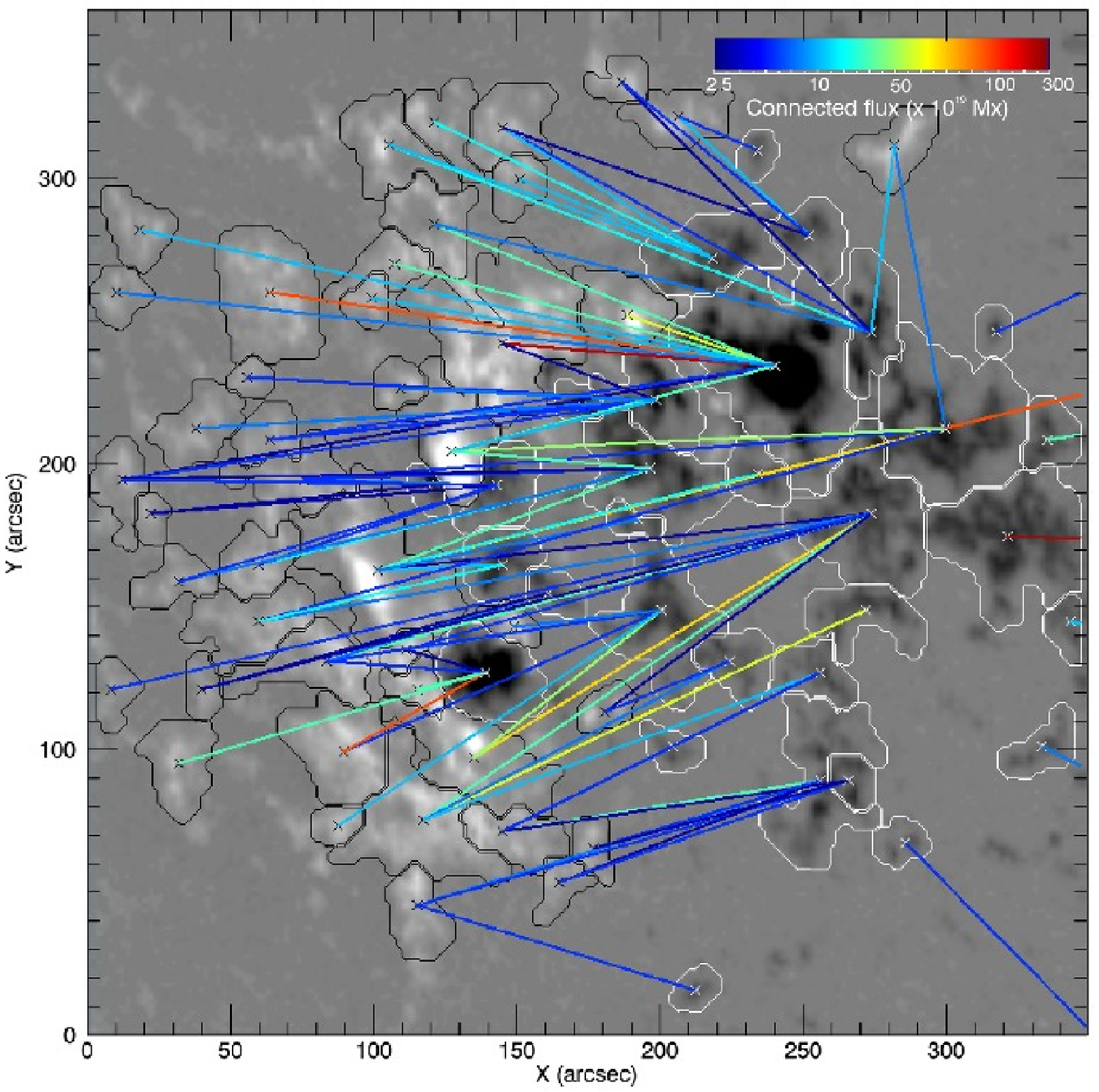}
\caption{Effective connected magnetic field (\S\ref{sec:beff}) for NOAA AR
09767 on the 2002 January 3 MDI magnetogram (shown in Figure~\ref{fig:MDI}).
Left: The line-of-sight magnetic field, saturated at $\pm$1500\,G, with the
outlines of the flux partitions (black/white contours for positive/negative
polarity). A ``$\times$'' symbol marks the flux-weighted center of each
partition.  
Right: The connectivity matrix, with color indicating the flux in the
identified connections (in units of $10^{19}\,{\rm Mx}$). Flux connecting to
the flux-balance ring is shown connecting outside the box.  In this example,
$B_{\rm eff} = 789.2$\,G.
}
\label{beff_fig}
\end{figure}

The resulting flare forecast relies on an event probability using Bayesian
inference and Laplace's rule of succession \citep{Jaynes2003}. In particular,
the steps used to produce a forecast are:
\begin{itemize} 
\item Partition the photospheric vertical magnetic field into ({\it e.g.}) $N_+$
positive-polarity and $N_-$ negative-polarity magnetic-flux concentrations
(Figure~\ref{beff_fig}, left), generally following \citet{mct}.  Particulars include: 
50\,G noise threshold,
minimum partition flux of $10^{20}$\,Mx, minimum partition size of 40 pixels,
and a slight smoothing (mean-neighborhood by a factor 2) applied 
prior to partitioning, only in order to draw smoother partition outlines.

\item Determine a connectivity matrix, $\psi_{ij}$ (Figure~\ref{beff_fig},
right), whose elements are the magnetic flux connecting the pairs of sources
$(i,j)$, by using simulated annealing to minimize the functional 
\begin{displaymath} F = \sum_{i,j} \bigg ({\vert \x_i -
\x_j \vert \over R_{\rm max}} + {\vert \Phi_i' +
\Phi_j' \vert \over \vert \Phi_i' \vert + \vert \Phi_j' \vert} \bigg ),
\end{displaymath}
where $\x_i$, $\x_j$ are the vector positions of the flux-weighted
centroids of two opposite-polarity partitions $i$ and $j$ ($i \equiv \{
1,...,N_+ \}$, $j \equiv \{ 1,...,N_- \}$) with respective flux contents
$\Phi_i'$ and $\Phi_j'$ and $R_{\rm max}$ is a constant, maximum distance
within the studied magnetogram, typically its diagonal length. The
implementation of $R_{\rm max}$ in the functional $F$ is a refinement over
the initial approach of \citet{GeorgoulisRust2007}. Another refinement
is the introduction of a mirror ``flux ring'' at a distance well outside 
of the studied magnetogram, that makes the flux distribution
exactly balanced prior to annealing \citep{Georgoulis_etal_2012a}.
This step consists of introducing a ring of mirror flux (as much
positive-/negative-polarity flux as the negative-/positive-polarity flux
of the active region) at large distances from the region, typically
three times larger than the largest dimension of the immediate region. The ring
of flux participates in the connectivity process via the simulated
annealing. Magnetic connections between active-region flux patches and
the flux ring are considered open, that is, closing beyond the
confines of the active region. These connections are not taken into
account in the calculation of $B_{\rm eff}$.

\item Define the effective connected magnetic field as
\begin{displaymath} 
B_{\rm eff} = \sum_{i=1}^{N_+} \sum_{j=1}^{N_-} {\psi_{ij} \over \vert \x_i - \x_j \vert^2}. 
\end{displaymath}
\item Construct the predictive conditional probability of an event above a
certain size according to Laplace's rule of succession: 
\begin{displaymath} 
P_{\rm flare}^{\rm th} = {N_{\rm mag(flare)}(B_{\rm eff} > B_{\rm eff}^{\rm th}) + 1 \over 
		 N_{\rm mag}(B_{\rm eff} > B_{\rm eff}^{\rm th}) + 2}, 
\end{displaymath}
where $N_{\rm mag(flare)}$ is the number of event-producing magnetic structures
(magnetograms) with $B_{\rm eff}$ greater than a given threshold $B_{\rm eff}^{\rm th}$
(for a particular event definition),
and $N_{\rm mag}$ is the total number of magnetograms with $B_{\rm eff}$ greater than
the same threshold.  Increasing $B_{\rm eff}$-thresholds are successively selected,
and the resulting curve of $P_{\rm flare}^{\rm th}$ as a function of $B_{\rm eff}^{\rm th}$
(for the targeted event definition) is then fitted by a sigmoidal curve; this
curve returns the flaring probability for an incoming $B_{\rm eff}$ measure.
The lowest threshold used for a particular event definition 
is the minimum $B_{\rm eff}$ found for all magnetograms for which an event
(as defined) was recorded.  Probabilities for magnetograms with $B_{\rm eff}$
less than this cutoff value are set to zero.

\end{itemize}
This approach does not use magnetic-field extrapolation.  Instead, the coronal
model relies on the minimum value of the functional $F$ which finds the
shortest connections between opposite-polarity flux concentrations.  The
$B_{\rm eff}$ values were calculated for all but one dataset.  However,
following \cite{GeorgoulisRust2007}, a limit of $\pm 41^\circ$ from disk center
is imposed to minimize projection-effect artifacts for the measures of this
method's optimal performance shown in Table~\ref{tbl:mkg_best}, thus the values
differ slightly from those presented in
Tables~\ref{tbl:bestAD}-\ref{tbl:bestMCD2} as a different subset of data is
used here.  All values of $B_{\rm eff}$ are used in the later AD comparisons
(\S~\ref{sec:best}).  

\begin{figure}[p!]
\centerline{
\includegraphics[width=0.33\textwidth, clip]{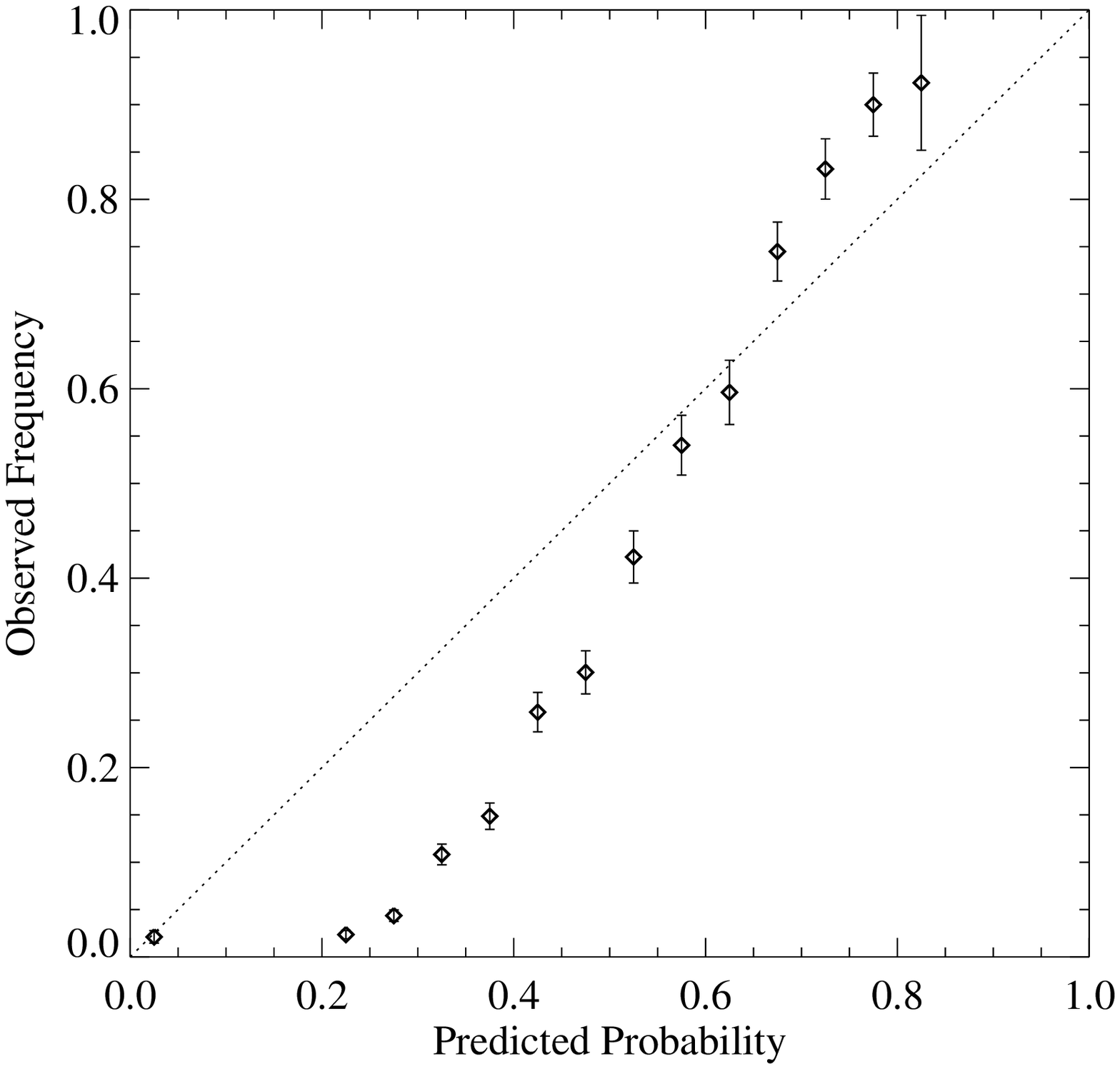}
\includegraphics[width=0.33\textwidth, clip]{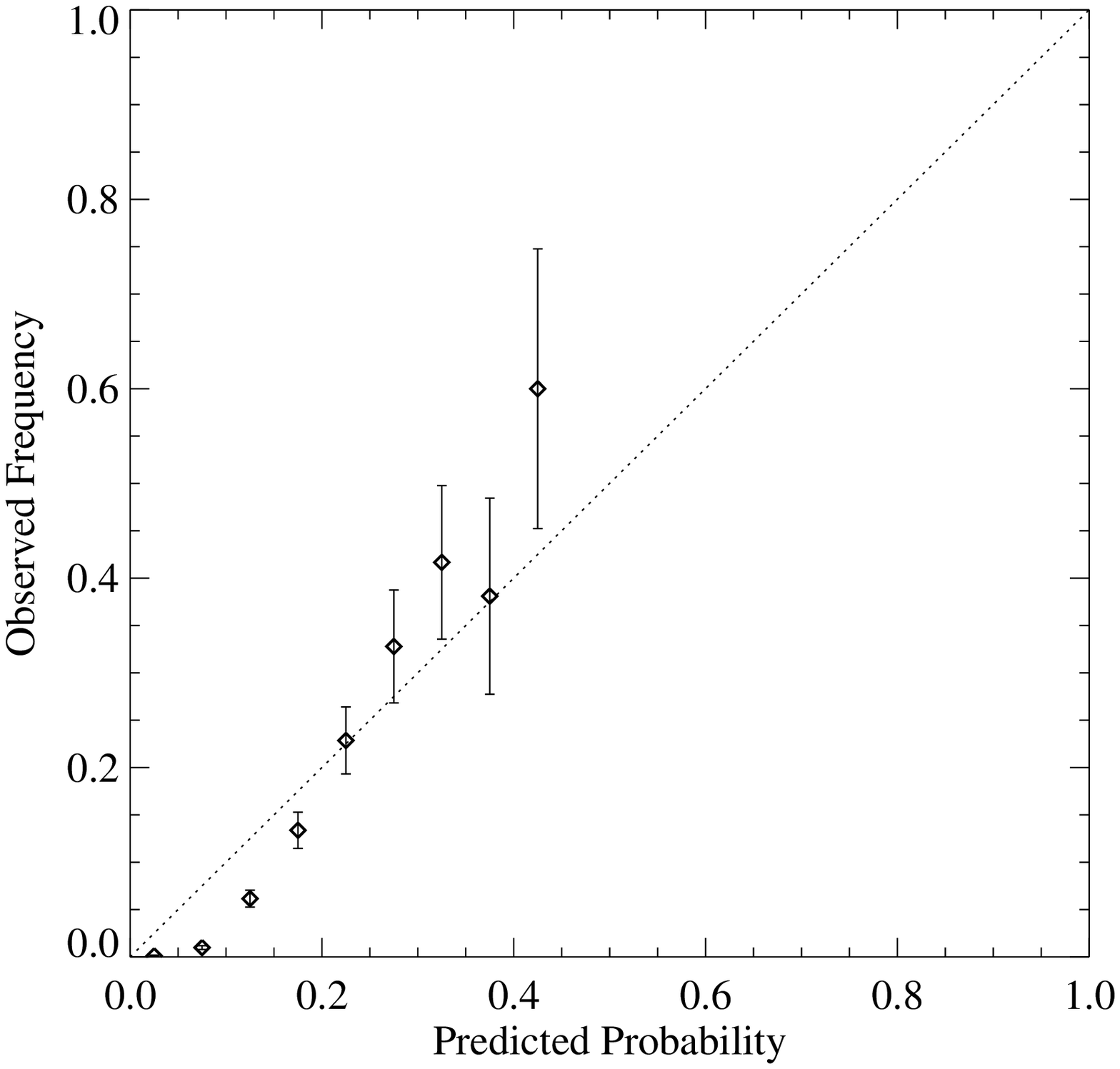}
\includegraphics[width=0.33\textwidth, clip]{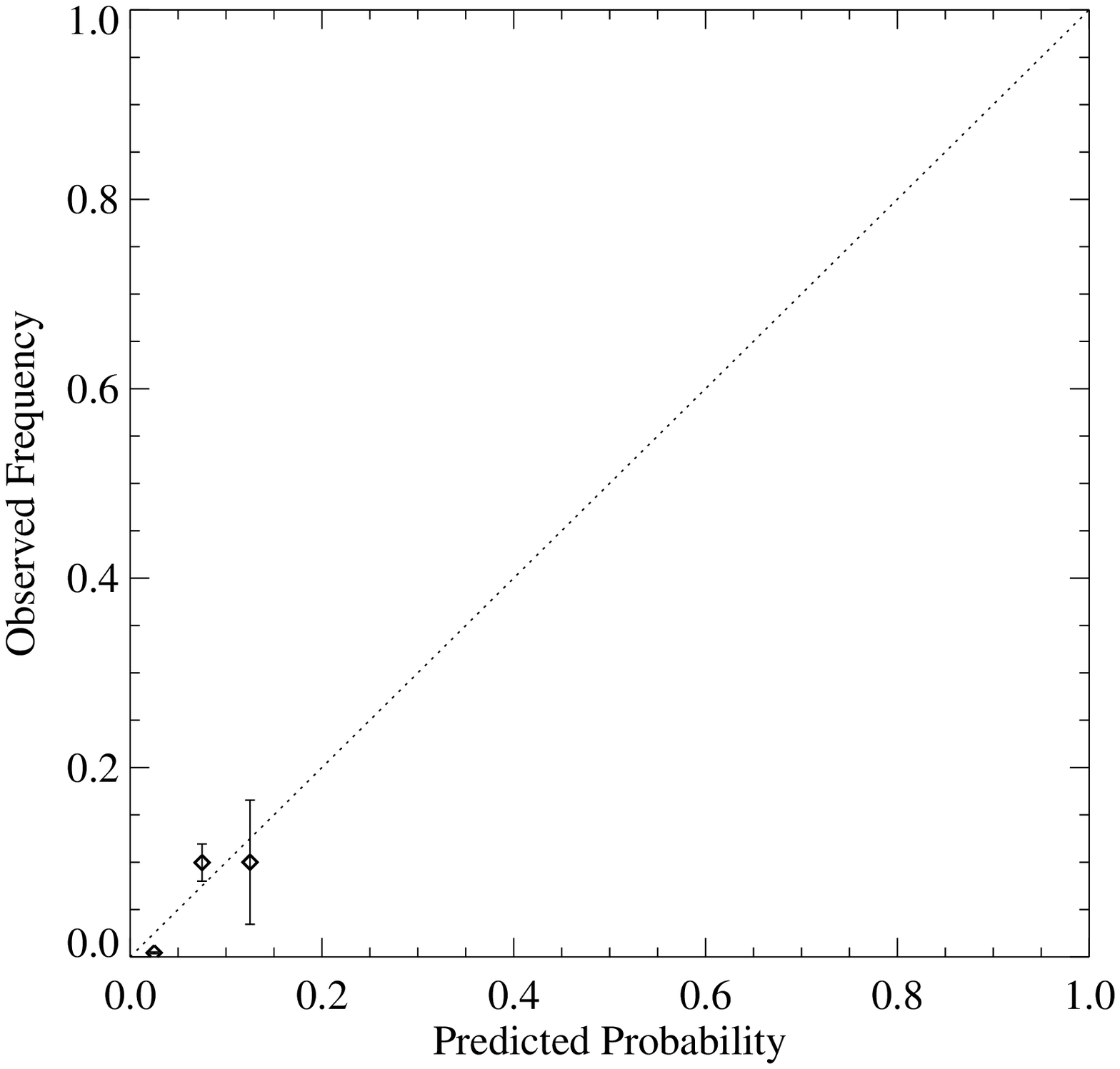}}
\centerline{
\includegraphics[width=0.33\textwidth, clip]{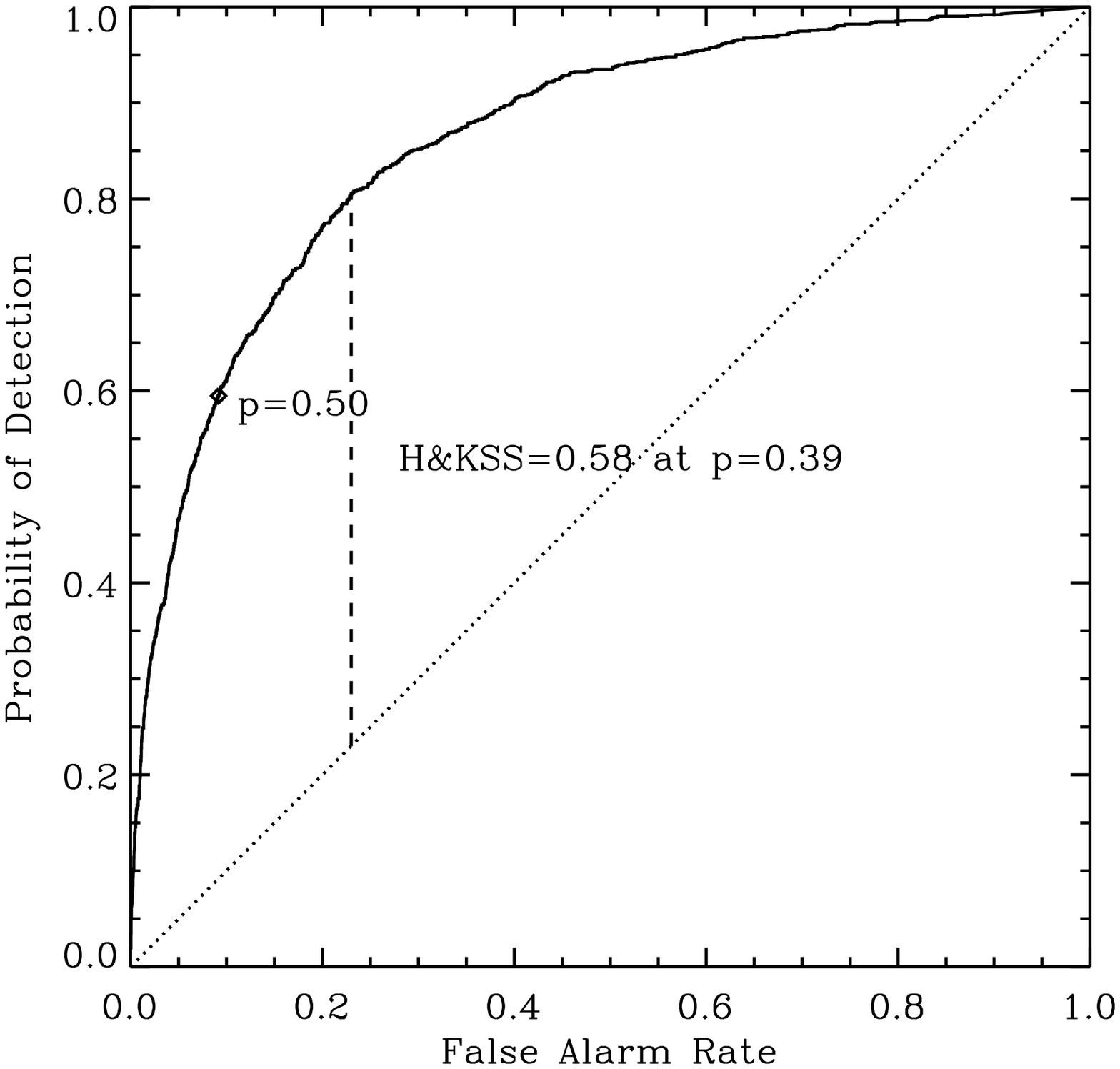}
\includegraphics[width=0.33\textwidth, clip]{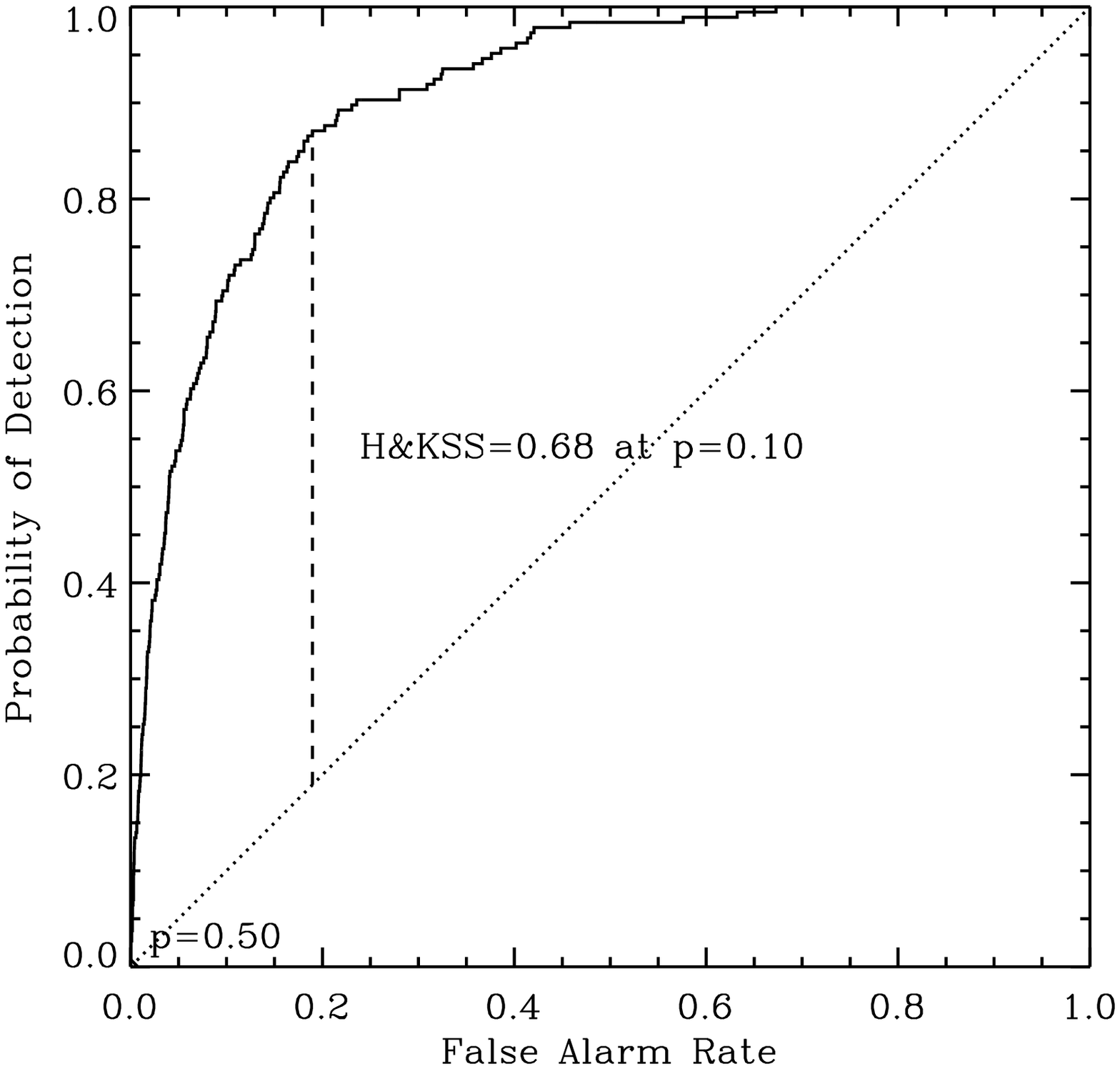}
\includegraphics[width=0.33\textwidth, clip]{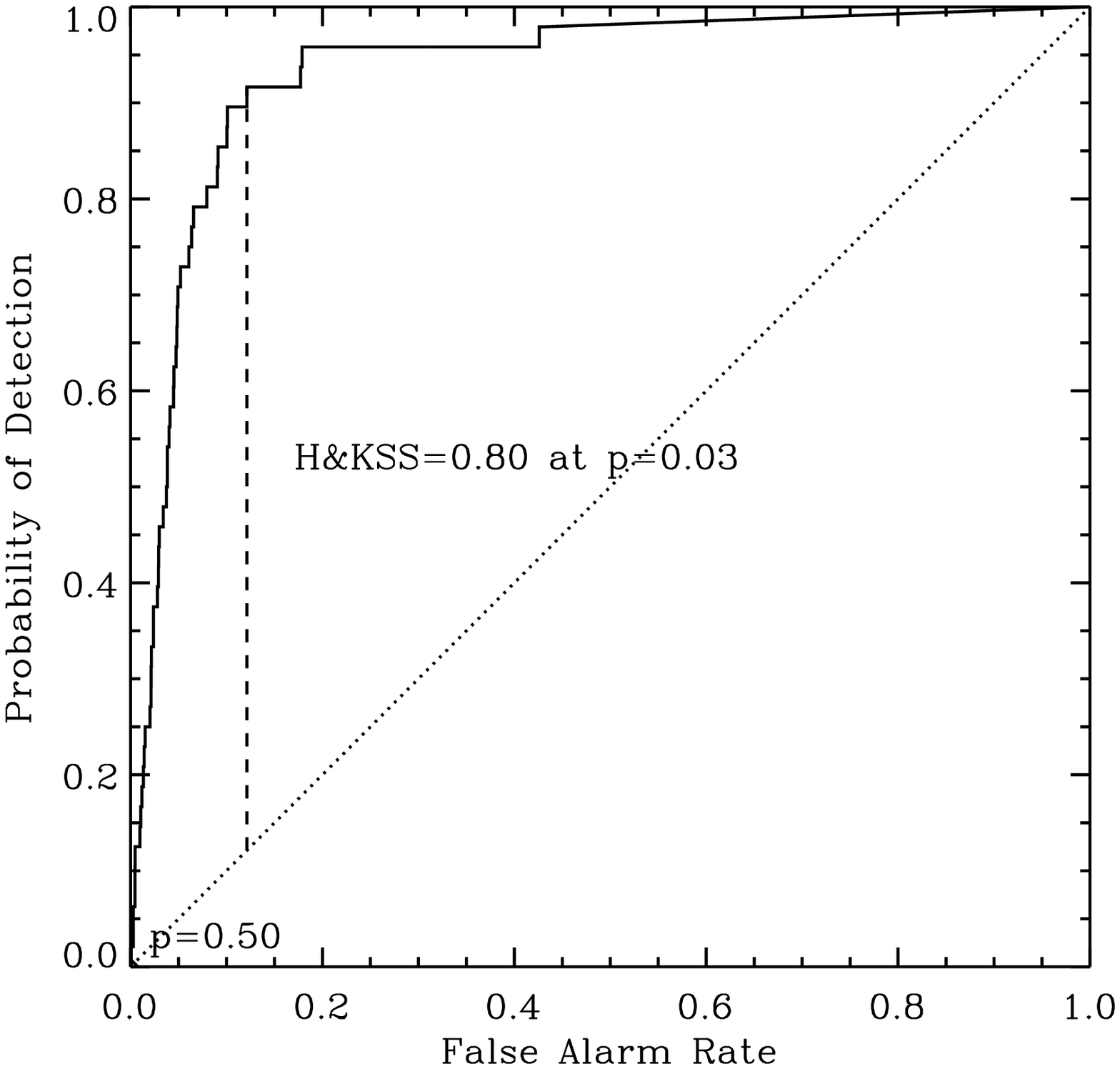}}
\caption{Summary plots (see \S\ref{sec:sss} and Figure~\ref{fig:rocrel}) of the
forecasting performance of $B_{\rm eff}$ for the three different event
definitions left:right \CC, \MS, \ML.  Top: Reliability plots, including
indications of sample-size within bin (error bars) and the $x=y$ line
($\cdots$), Bottom: Relative Operating Characteristic (ROC) curve, with
annotation indicating the location on the curve of the 50\% probability results
(quoted elsewhere), the location of the peak \True\ score on the curve (dashed
line) and at what probability threshold that occurs.  The $x=y$ line ($\cdots$)
is also included for reference.}
\label{fig:beff_plots}
\end{figure}

The skill-score results are unusual in that there is a large discrepancy in the
performance based on which skill score is used to evaluate the method, with the
\Heidke\ and \True\ skill score values being much larger than the \Appleman\ or
\Brier\ for \CC.  The reliability plots (Figure~\ref{fig:beff_plots}, top) show
a systematic over-prediction for lower probabilities and under-prediction for
higher probabilities. As is typical for most methods, the maximum probability
forecast decreases as the event threshold increases, while the maximum \True\
value and the ROC curve (Figure~\ref{fig:beff_plots}, bottom) improve with
increasing event-threshold magnitude. 

\begin{center}
\begin{deluxetable}{cccccccc}
\tablecolumns{8}
\tablewidth{0pc}
\tablecaption{Optimal Performance Results and Probability Thresholds: $B_{\rm eff}$}
\tablehead{ 
\colhead{Event} & \colhead{Sample} & \colhead{Event} & \colhead{\RC} & \colhead{\Heidke} & \colhead{\Appleman} & \colhead{\True} & \colhead{\Brier} \\
\colhead{Definition} & \colhead{Size} &  \colhead{Rate} & \colhead{(threshold)} & \colhead{(threshold)} & \colhead{(threshold)} & \colhead{(threshold)} & \colhead{} 
} 
\startdata 
\CC & 6234 & 0.197 & 0.85 (0.55) & 0.51 (0.50) & 0.26 (0.55) & 0.58 (0.39) & 0.12\\
\MS &  ''  & 0.030 & 0.97 (0.40) & 0.33 (0.22) & 0.01 (0.40) & 0.68 (0.10) & 0.07\\
\ML &  ''  & 0.008 & 0.99 (0.14) & 0.14 (0.05) & 0.00 (0.14) & 0.80 (0.03) & 0.03\\
\enddata 
\label{tbl:mkg_best} 
\end{deluxetable}
\end{center}

\subsection{Automated Solar Activity Prediction (``ASAP'') - T.~Colak, R.~Qahwaji}
\label{sec:asap}

A real-time automated flare prediction tool has been developed at the
University of Bradford/Centre for Visual Computing
\citep{ColakQahwaji2008,ColakQahwaji2009}.  The {\it Automated Solar Activity
Prediction} (``ASAP'')\footnote{See
\url{http://spaceweather.inf.brad.ac.uk/asap/}.} system uses the following
steps to predict the likelihood of a solar flare:
\begin{enumerate}
\item A feature-recognition system generates McIntosh classifications 
\citep{mcintosh90} for active regions from MDI white-light images.
\item A Machine Learning System is trained using these classifications and
flare event databases from NCEI.
\item New data are then used to generate real-time predictions.
\end{enumerate}

The system relies upon both MDI magnetic and white-light data, and hence is
unable to make a prediction for those data for which the white-light data are
unavailable.  Generally, ASAP generates a McIntosh classification, as
determined by step~\#1 above.  This is difficult for the active region clusters
made up of multiple active regions in the database we presented.  Thus, the
recorded McIntosh classifications included in the file headers were used for
the predictions ({\it i.e.} step~\#1 was essentially skipped), and forecasts
were made for the entire database.  However, the performance may have been less
than optimal due to this peculiarity.

\begin{figure}[p]
\centerline{
\includegraphics[width=0.33\textwidth, clip]{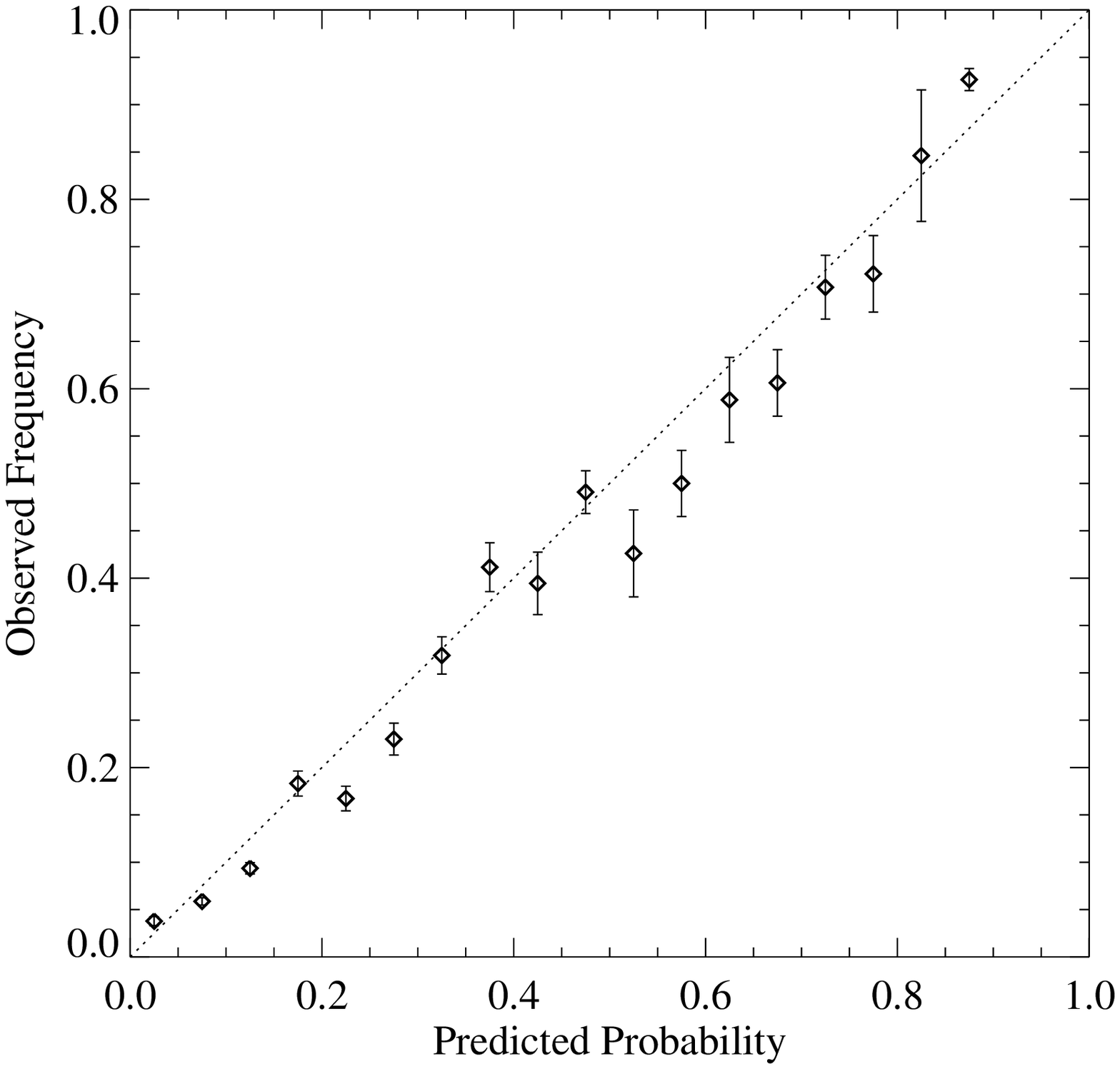}
\includegraphics[width=0.33\textwidth, clip]{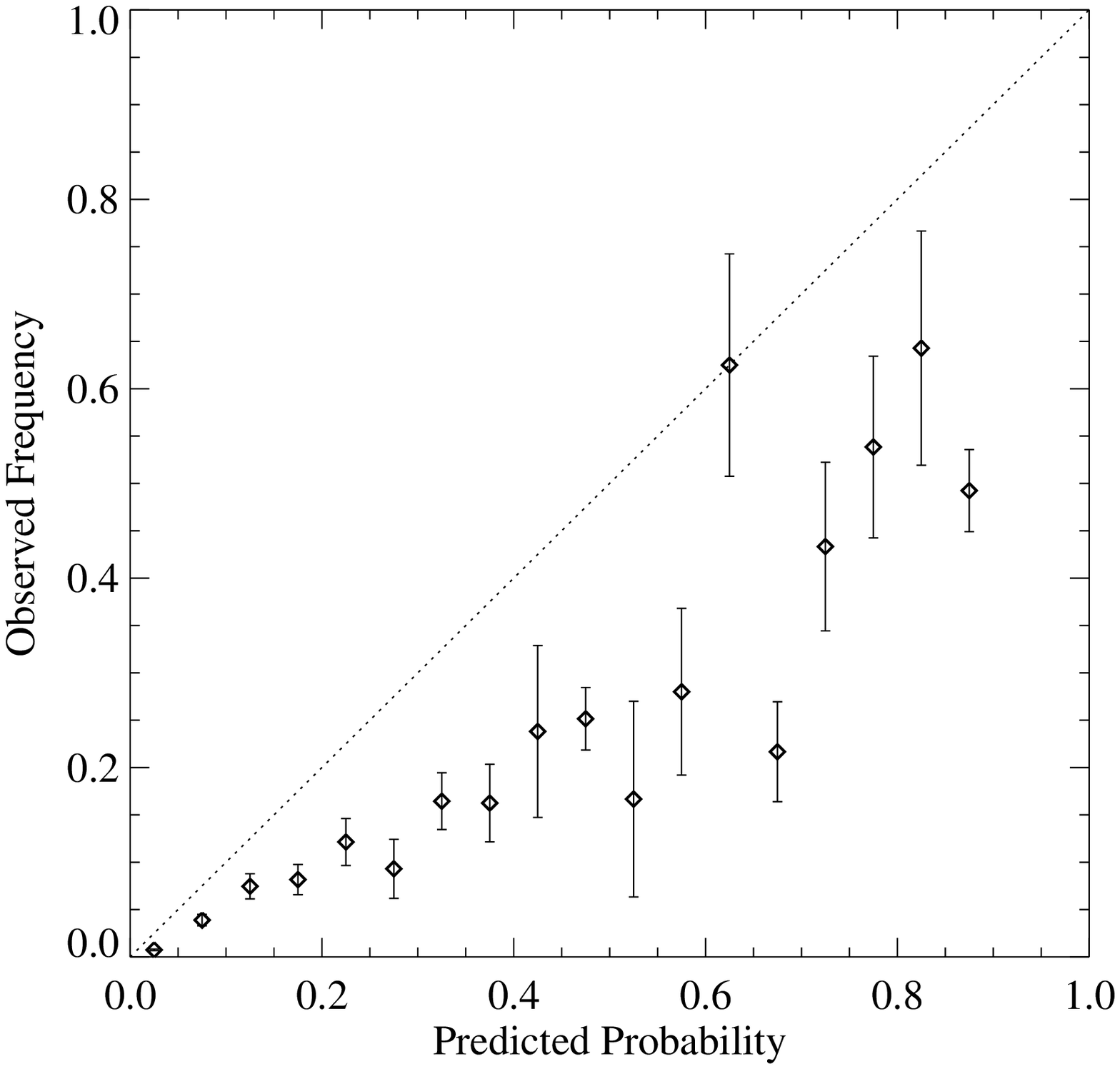}
\includegraphics[width=0.33\textwidth, clip]{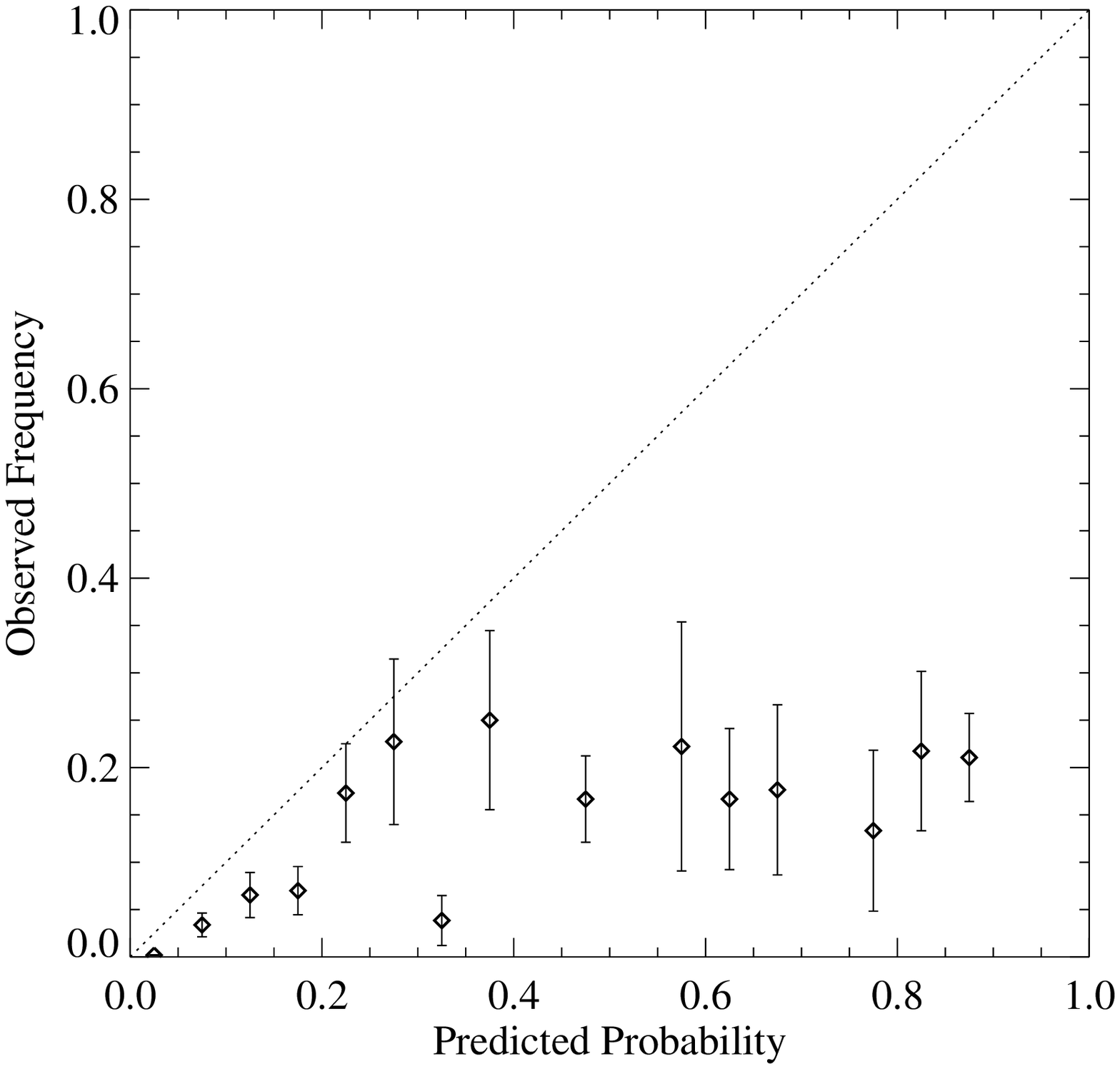}}
\centerline{
\includegraphics[width=0.33\textwidth, clip]{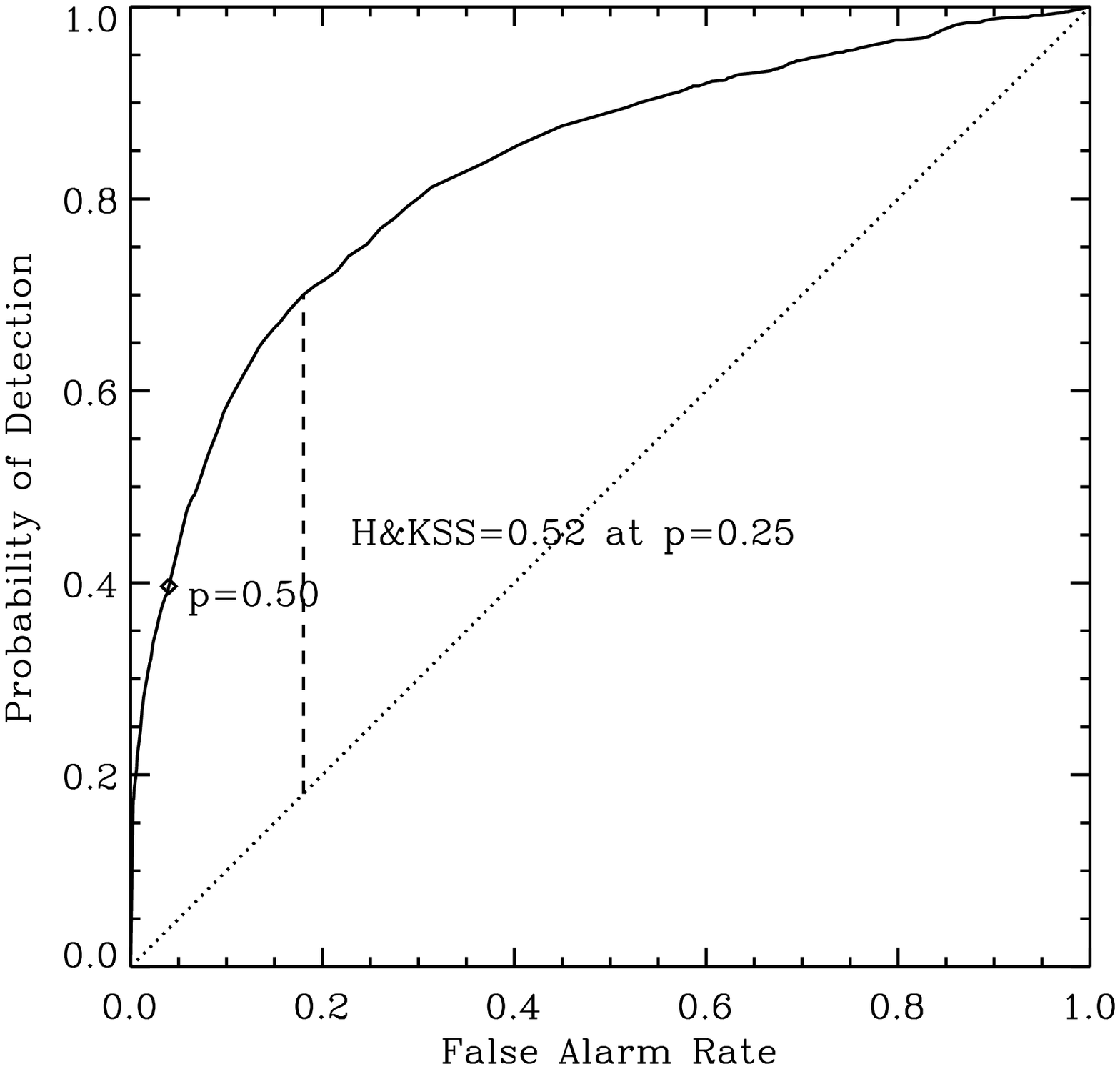}
\includegraphics[width=0.33\textwidth, clip]{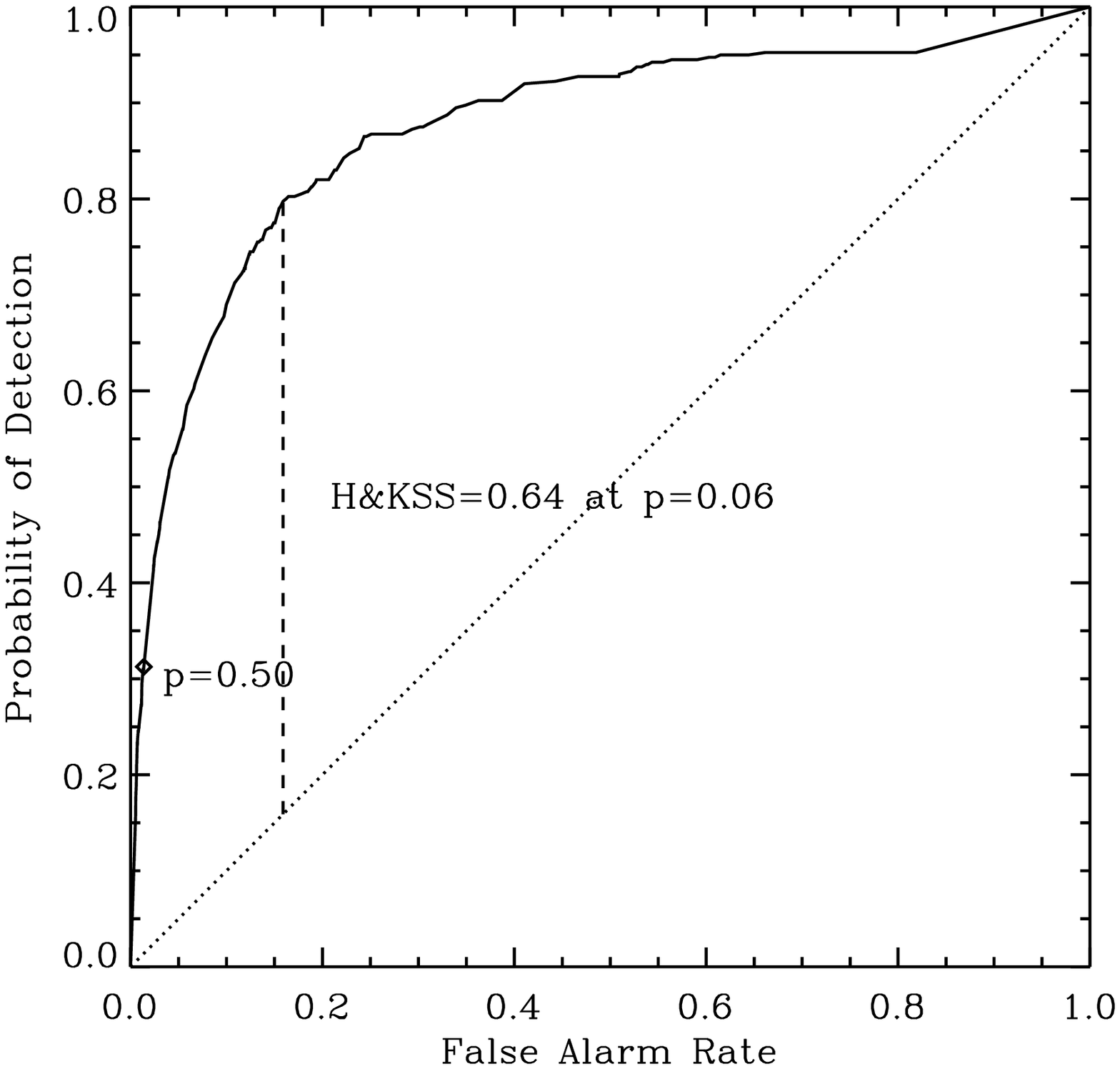}
\includegraphics[width=0.33\textwidth, clip]{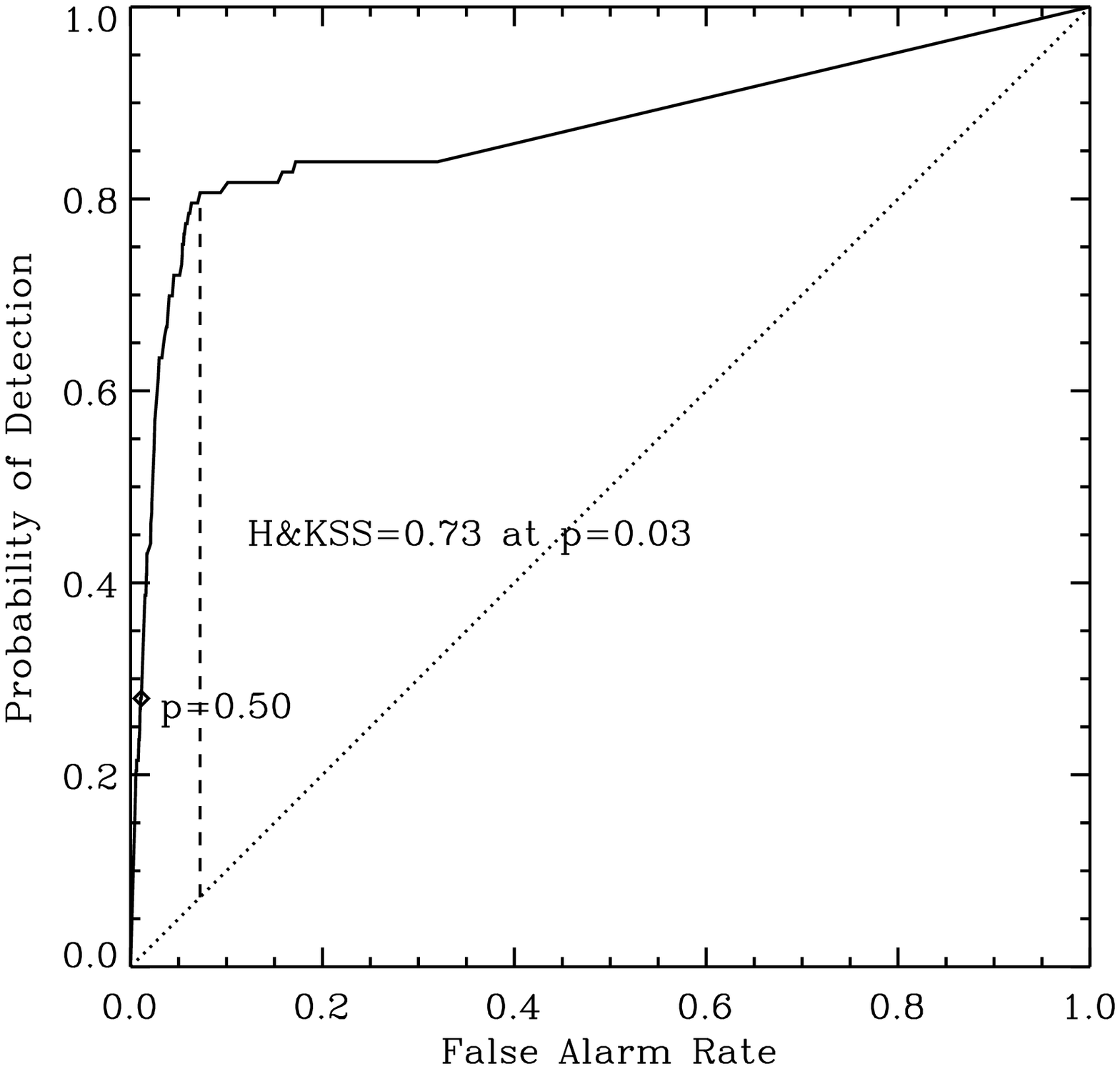}}
\caption{Same as Figure~\ref{fig:beff_plots} but for ASAP.}
\label{fig:asap_plots}
\end{figure}

The results for ASAP compared to the average show higher values for \Heidke\
and \True\ skill but lower for \Appleman\ and \Brier, a common result for
machine-learning based forecasting algorithms which are typically trained to
produce the largest \Heidke\ or \True.  The reliability plots
(Figure~\ref{fig:asap_plots}, top) show a systematic over-prediction for the
larger event thresholds, and ROC plots (Figure~\ref{fig:asap_plots}, bottom)
show lower probability of detection for high false alarm rates than most other
methods.

\begin{center}
\begin{deluxetable}{cccccccc}
\tablecolumns{8}
\tablewidth{0pc}
\tablecaption{Optimal Performance Results: ASAP} 
\tablehead{ 
\colhead{Event} & \colhead{Sample} & \colhead{Event} & \colhead{\RC} & \colhead{\Heidke} & \colhead{\Appleman} & \colhead{\True} & \colhead{\Brier} \\
\colhead{Definition} & \colhead{Size} &  \colhead{Rate} & \colhead{(threshold)} & \colhead{(threshold)} & \colhead{(threshold)} & \colhead{(threshold)} & \colhead{} 
} 
\startdata 
\CC & 12965 & 0.201 & 0.85 (0.58) & 0.49 (0.35) & 0.25 (0.58) & 0.52 (0.25) &\ 0.30\\
\MS &  ''   & 0.031 & 0.97 (0.74) & 0.36 (0.42) & 0.01 (0.74) & 0.64 (0.06) & -0.01\\
\ML &  ''   & 0.007 & 0.99 (0.90) & 0.22 (0.37) & 0.00 (0.90) & 0.73 (0.03) & -0.84\\
\enddata 
\label{tbl:asap_best} 
\end{deluxetable}
\end{center}

\subsection{Big Bear Solar Observatory/Machine Learning Techniques - Y.~Yuan}
\label{sec:bbso}

Another approach that uses a machine learning technique as the statistical
forecasting method has been developed at the New Jersey Institute of Technology
\citep{Yuan_etal_2010,Yuan_etal_2011}.  The steps in this method are:

\begin{itemize}
\item Compute three parameters describing an active region:
\begin{itemize}
\item total unsigned magnetic flux, computed using only the pixels for which
the absolute value of the field is greater than the mean value plus three times
the standard deviation of all field in the area under consideration;
\item the length of the strong-gradient neutral line (above
50\,G\,Mm$^{-1}$);
\item the total magnetic energy dissipation $E_{\rm diss} =
\int 4 [(\partial \Bz/\partial x)^2+(\partial \Bz/\partial y)^2] + 2(\partial
\Bz/\partial x + \partial \Bz/\partial y)^2 dA$, following
\citet{Abramenko_etal_2003}.
\end{itemize}
\item Use ordinal logistic regression and support vector machines (SVM) to make
predictions.
\end{itemize}

An SVM is a supervised learning method used for classification
\citep{BoserGuyonVapnik1992}, whose principle is to minimize the structural
risk \citep{Vapnik2000}. An SVM tries to find a plane in an n-dimensional space
that separates input data into two classes. The larger the distance from the
plane to the two different classes of data points in the n-dimensional space,
the smaller the classification error \citep{CortesVapnik1995}. The results
presented here used the open-source SVM implementation called LIBSVM
\citep{ChangLin2011}.

\begin{figure}
\centerline{
\includegraphics[width=0.33\textwidth, clip]{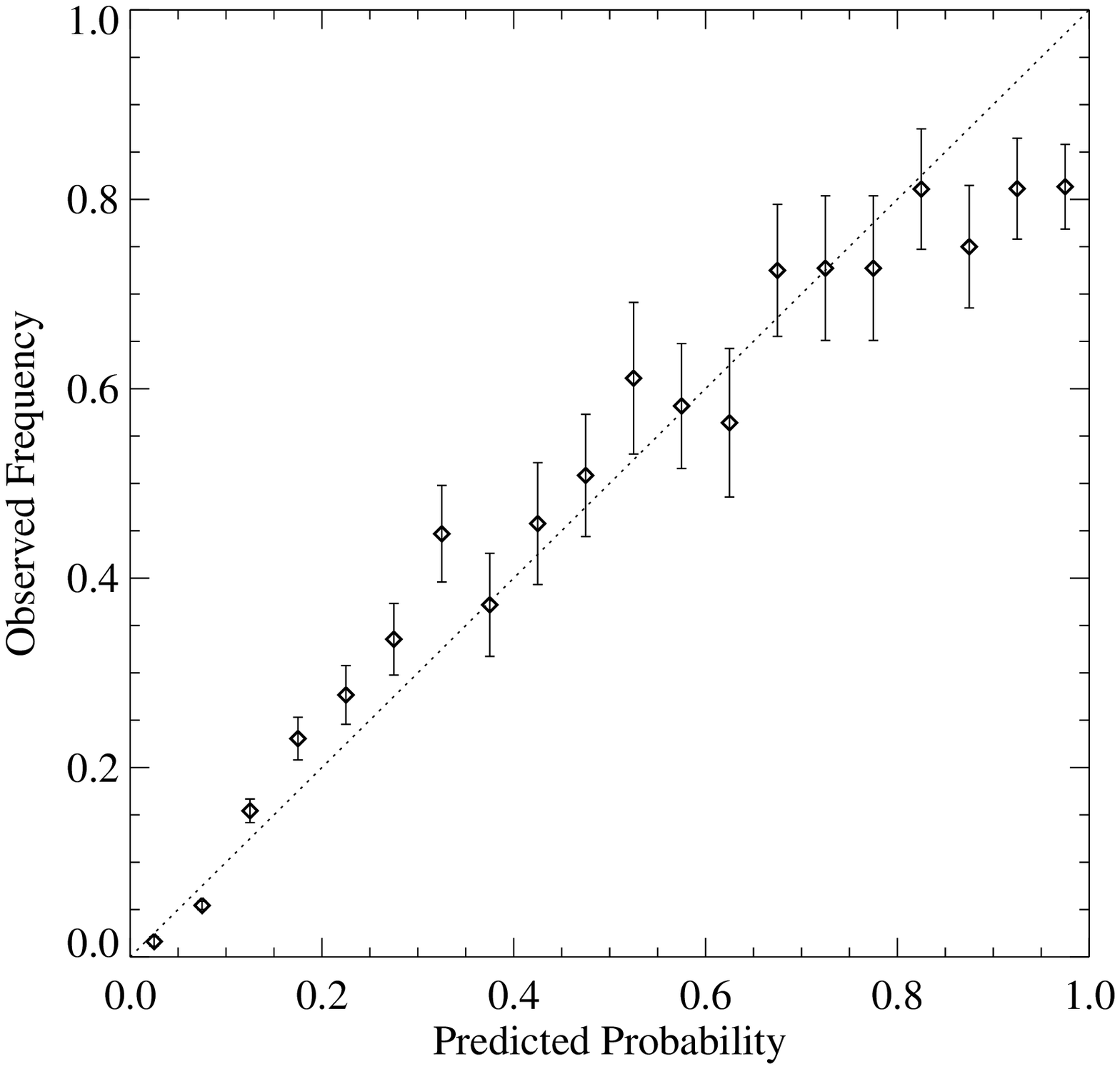}
\includegraphics[width=0.33\textwidth, clip]{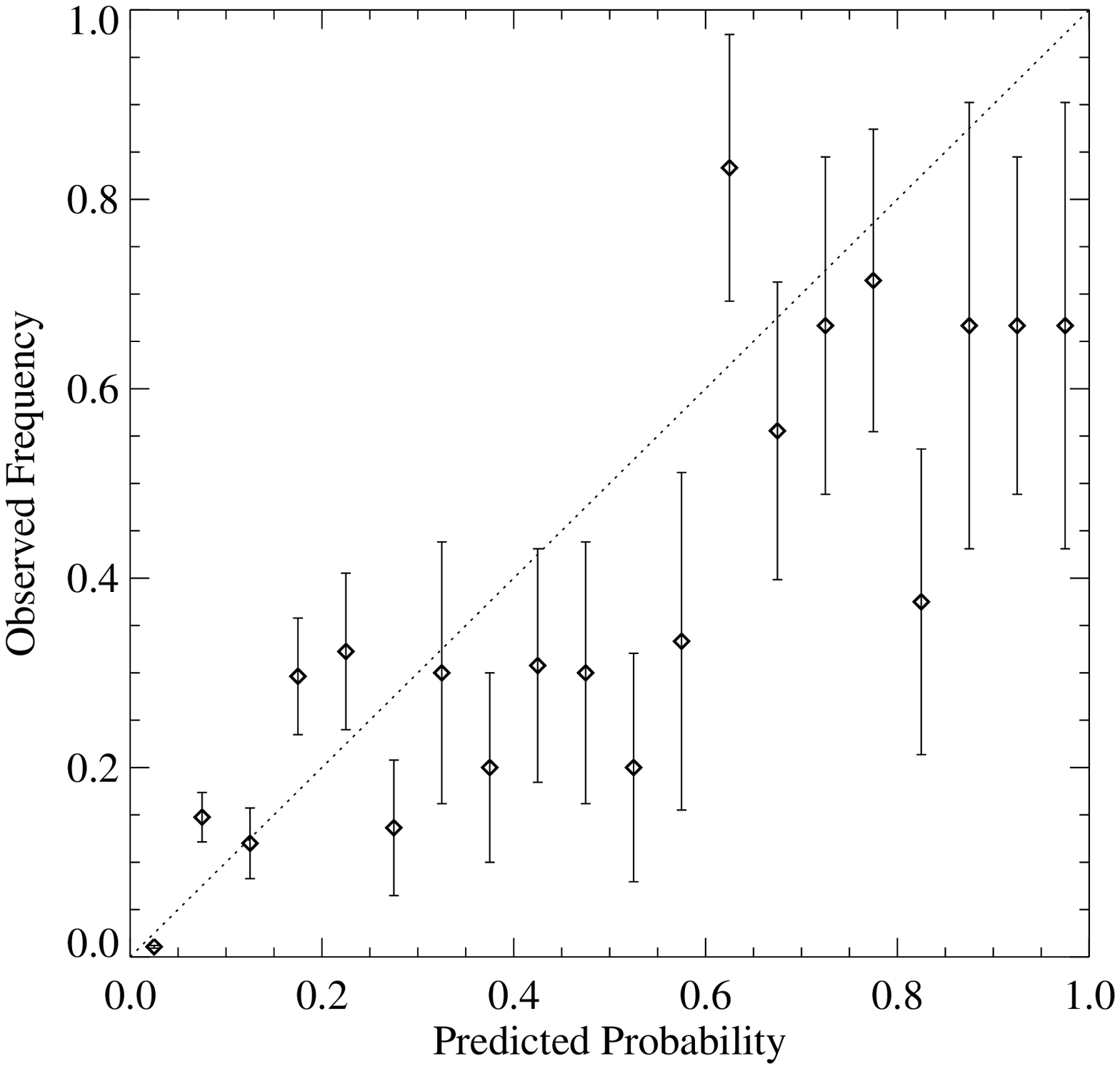}
\includegraphics[width=0.33\textwidth, clip]{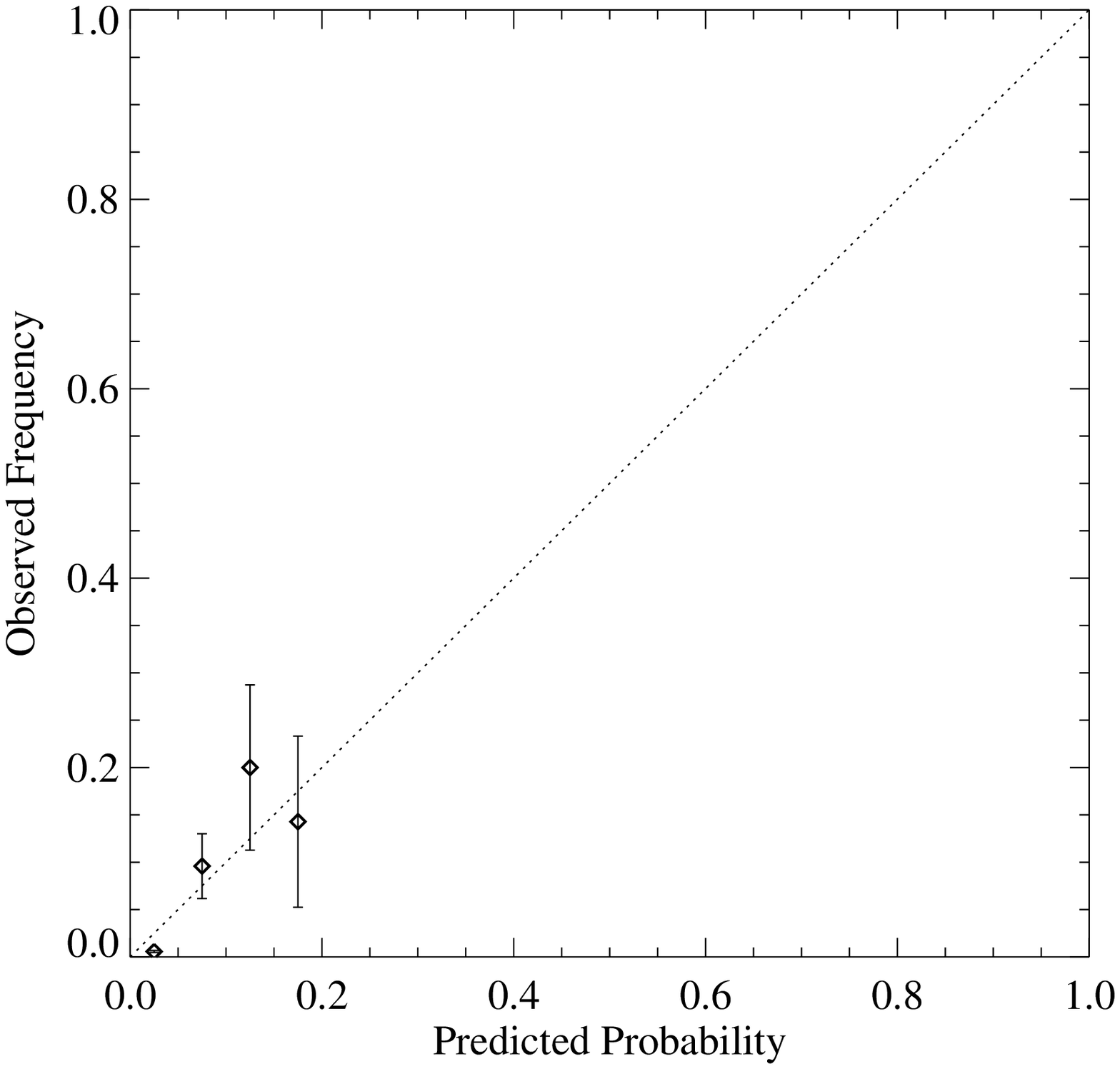}}
\centerline{
\includegraphics[width=0.33\textwidth, clip]{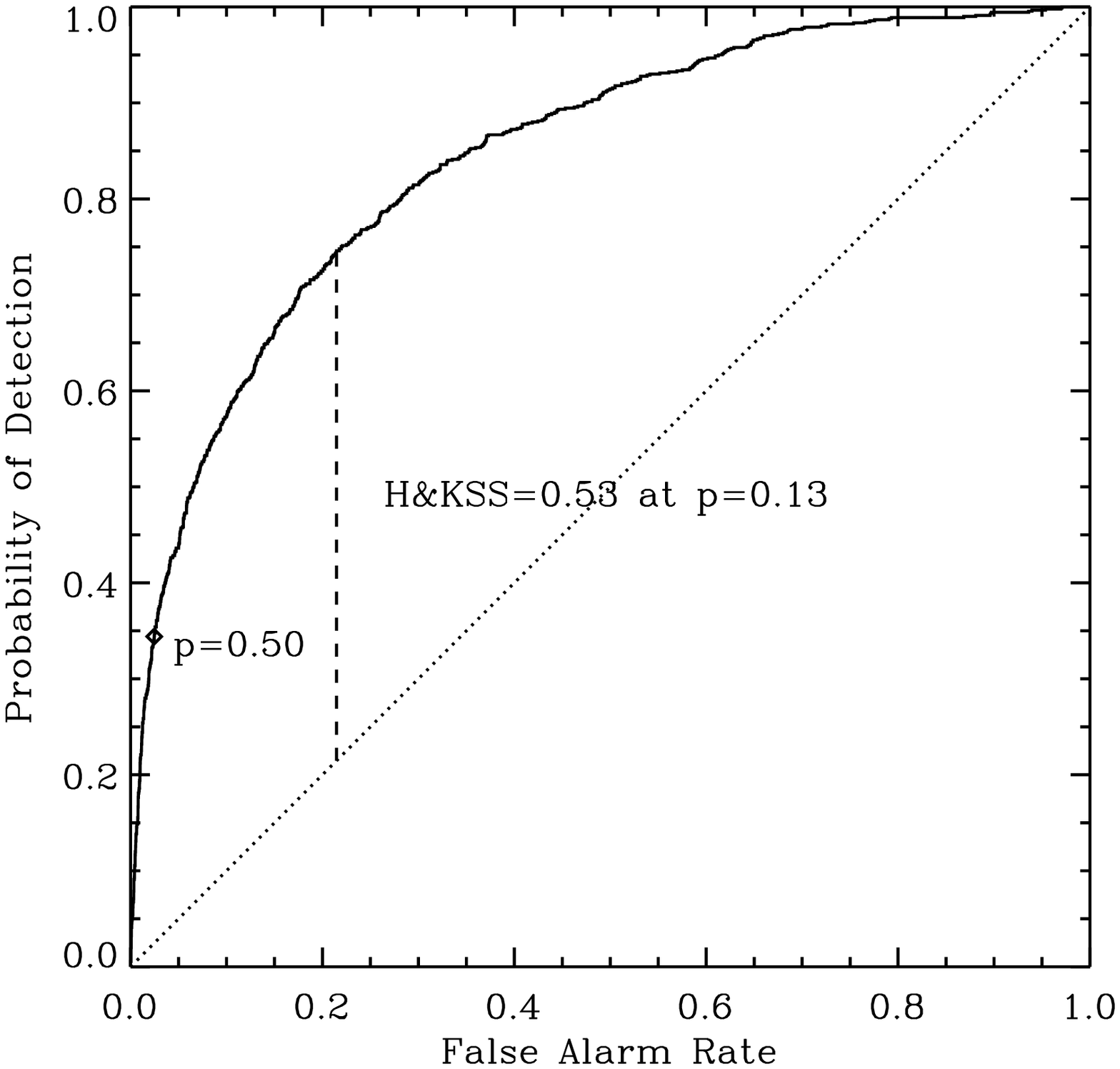}
\includegraphics[width=0.33\textwidth, clip]{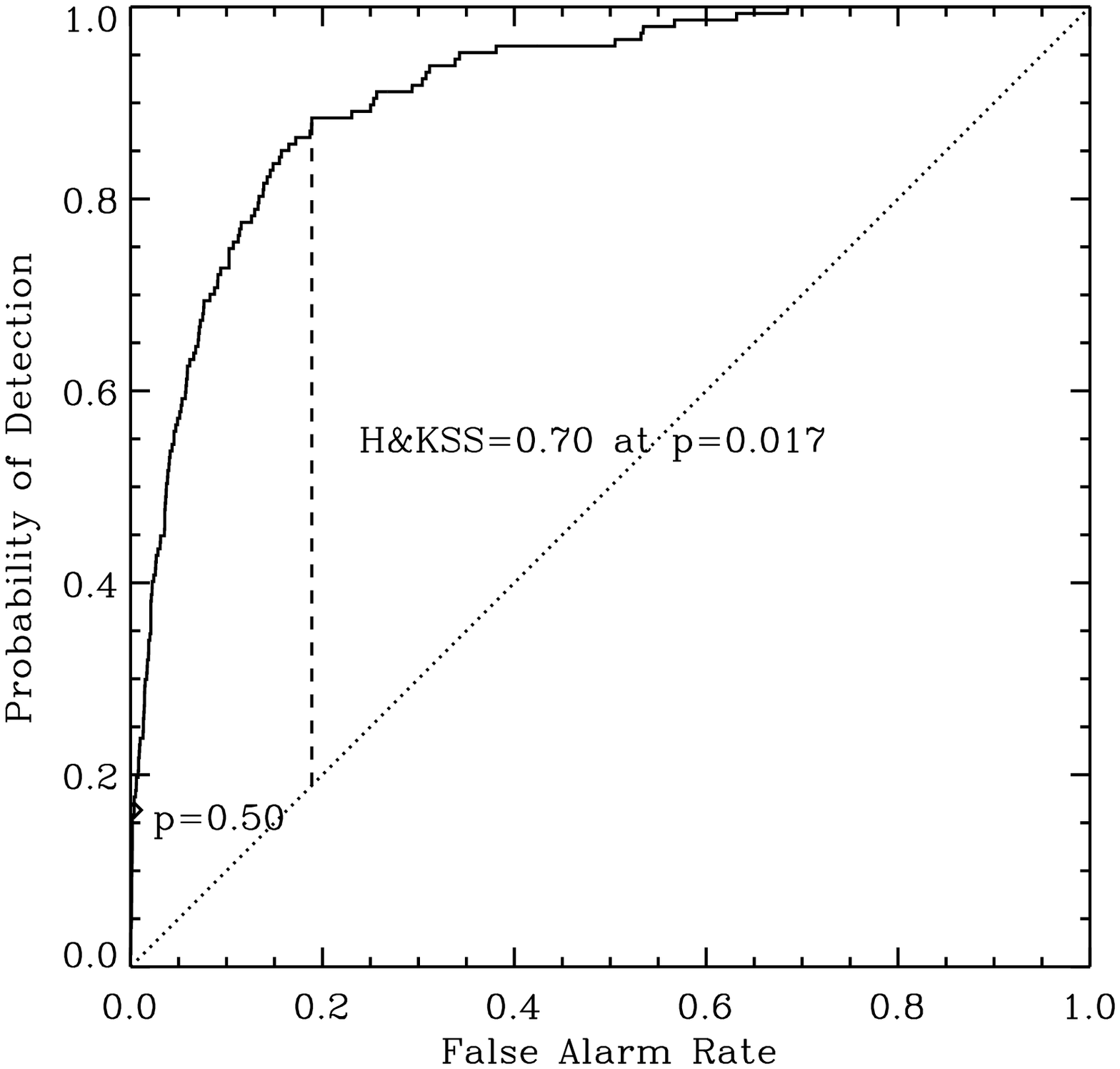}
\includegraphics[width=0.33\textwidth, clip]{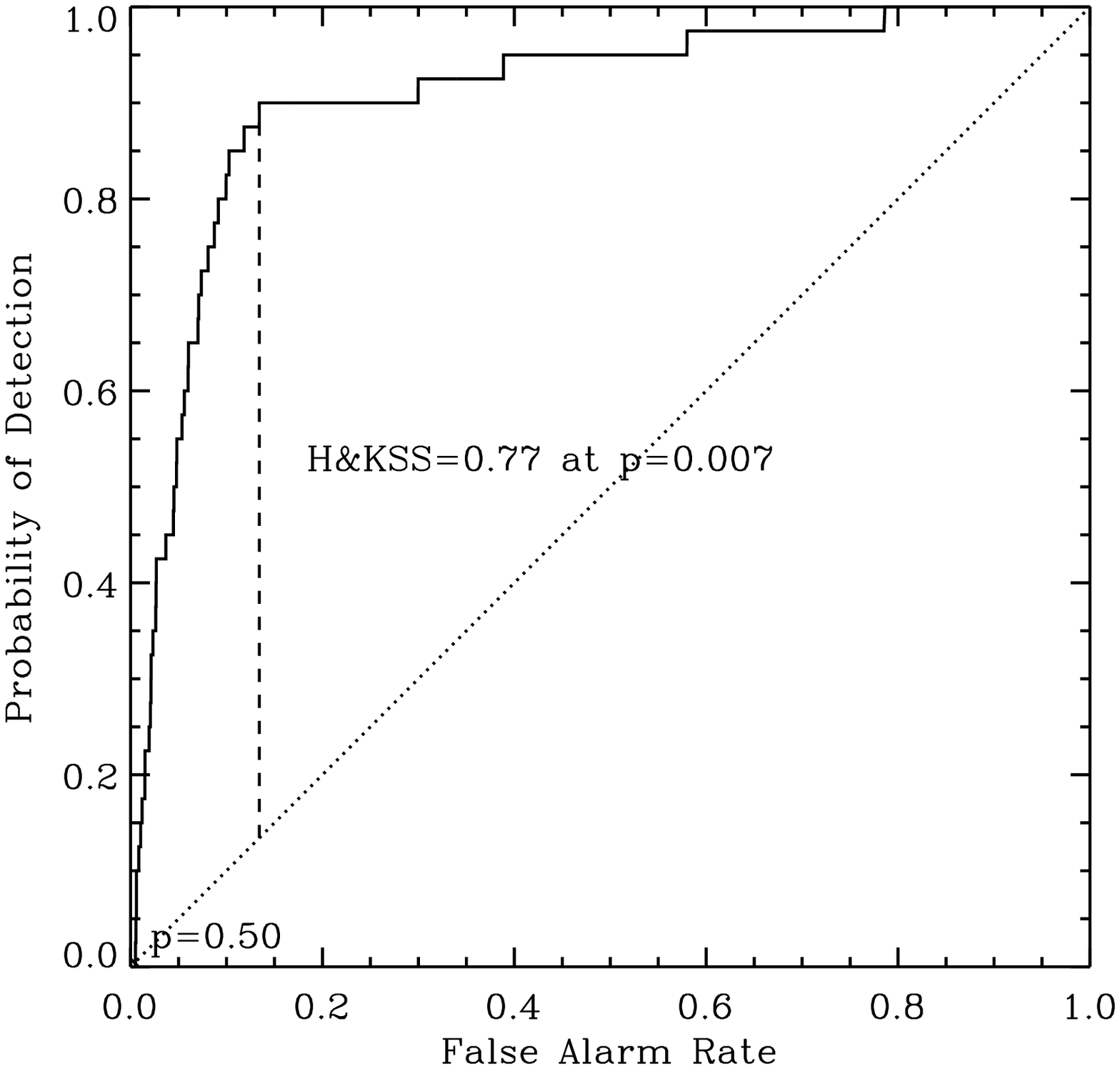}}
\caption{Same as Figure~\ref{fig:beff_plots} but for BBSO/Machine Learning.}
\label{fig:bbso_plots}
\end{figure}

The summary of results is given in Table~\ref{tbl:bbso_best}.  The method used
and made forecasts only on extracted data that had a single NOAA Active Region
region within $\pm40^\circ$ of disk center, reducing the sample to less than
half that provided.  This restriction also presents an example of a method
whose requirements are not well met by the data used for this workshop, and
thus whose skill scores may be penalized as a result.  When calculating the
flux, only pixels for which the absolute value of the field is greater than the
mean value plus three times the standard deviation of all field in the area
under consideration are included.  The summary plots
(Figure~\ref{fig:bbso_plots}) show a weak trend toward over-prediction at
higher probabilities, but with larger error bars due to the smaller sample
sizes.  The ROC plots are fairly typical. 

\begin{center}
\begin{deluxetable}{cccccccc}  
\tablecolumns{8}
\tablewidth{0pc}
\tablecaption{Optimal Performance Results: BBSO/Machine Learning} 
\tablehead{ 
\colhead{Event} & \colhead{Sample} & \colhead{Event} & \colhead{\RC} & \colhead{\Heidke} & \colhead{\Appleman} & \colhead{\True} & \colhead{\Brier} \\
\colhead{Definition} & \colhead{Size} &  \colhead{Rate} & \colhead{(threshold)} & \colhead{(threshold)} & \colhead{(threshold)} & \colhead{(threshold)} & \colhead{} 
} 
\startdata 
\CC & 5560 & 0.162 & 0.87 (0.46) & 0.47 (0.26) & 0.22 (0.46) & 0.53 (0.13) &\ 0.28\\ 
\MS &  ''  & 0.026 & 0.98 (0.59) & 0.34 (0.15) & 0.09 (0.59) & 0.70 (0.02) &\ 0.15\\
\ML &  ''  & 0.007 & 0.99 (0.71) & 0.14 (0.04) & 0.00 (0.71) & 0.76 (0.01) & -0.03\\
\enddata 
\label{tbl:bbso_best} 
\end{deluxetable}
\end{center}

\subsection{Total Nonpotentiality of Active Regions - D.~Falconer}
\label{sec:WLSG2}

\citet{Falconer_etal_2008} presented a prediction method based on
parameterizing active region magnetic morphology that was first applied to CME
prediction \citep{Falconer_etal_2002,Falconer_etal_2003} but more recently
extended to flares.  The modern version has been implemented in a code called
MAG-4 \citep{Falconer_etal_2011}, running at the Community Coordinated Modeling
Center and elsewhere.  The results presented here are from a predecessor of
MAG-4.

\begin{figure}
\plotone{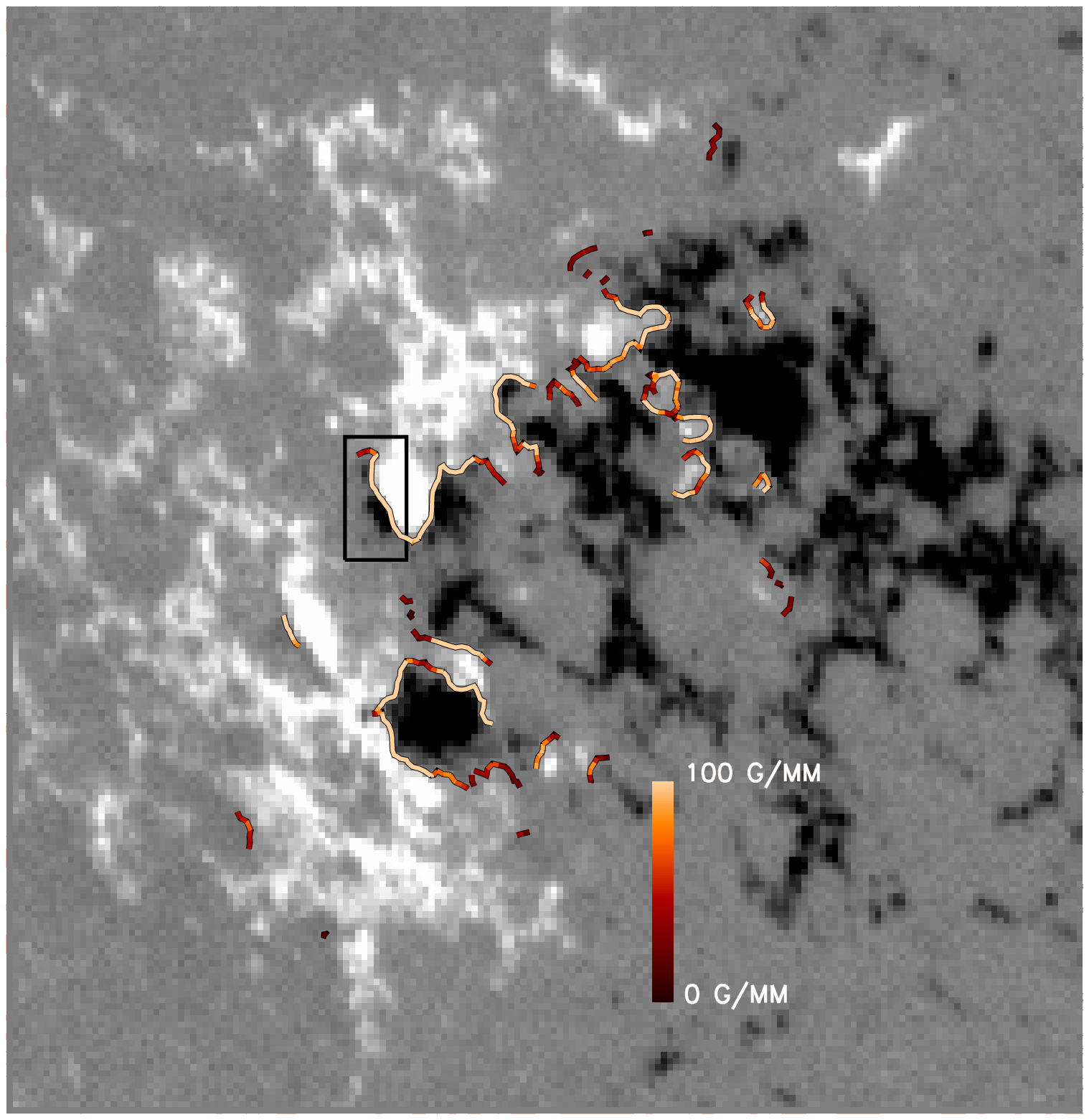}
\caption{Total Nonpotentiality of Active Regions (\S\ref{sec:WLSG2}) for NOAA
AR 09767, 2002 January 3, showing the line-of-sight field ($\Bl$), with
white/black indicating field aligned as positive/negative along the line of
sight.  The strong-field intervals of the neutral line used for calculating the
$WL_{SG2}$ parameter are shown by the color curves, with the color indicating
the strength of the gradient of the line-of-sight field ($\vert \grad \Bl
\vert$).   The black rectangle indicates a possibly fictitious neutral line
(one that occurs in the line-of-sight magnetic field, but not in the vertical
magnetic field).  For this active region, $WL_{SG2}=1.4\times 10^7\,{\rm
G}^2\,{\rm Mm}^{-1}$. Excluding the boxed area gives $WL_{SG2}=9.7\times
10^6\,{\rm G}^2\,{\rm Mm}^{-1}$.
}
\label{fig:WLSG2}
\end{figure}

Two parameters are calculated.  A measure of the free magnetic energy
is estimated based on the presence of strong gradient neutral lines:
\begin{equation} 
WL_{SG2} = \int_{NL} (\grad B_{\rm los})^2\,dl.
\end{equation}
An example of strong gradient neutral lines is shown in Figure~\ref{fig:WLSG2}.
The measure $WL_{SG2}$ is supplemented with the total unsigned magnetic flux
(for pixels above 100\,G) in the target active region.  No correction to the
$\Bl$ data is performed.  The utility of $WL_{SG2}$ was shown in
\citet{Falconer_etal_2008} where parameterizations using vector magnetogram
data were favorably compared to the $WL_{SG2}$ proxy, which can be calculated
from line-of-sight data.  In this study, the parameterizations for both
$WL_{SG2}$ and total magnetic flux $\Phi_{\rm tot}$ were calculated for all
extracted data sets, but the method is generally restricted to regions within
$30^\circ$ of disk center.  Table~\ref{tbl:fal_best}, summarizing the method's
performance, is restricted to this subset of predictions.  It is expected that
line-of-sight data beyond $30^\circ$ do not provide reliable estimates of
$WL_{SG2}$.

With these parameterizations, a least squares power-law fit to the event rates
as a function of $WL_{SG2}$ and total flux was constructed for each event
definition using only the data within the $30^\circ$ limit.  These event rates
are converted into probabilities as a function of the length of time $t$ of the
forecast interval, assuming that the event rate is constant over the larger
forecast interval (in this case 2000-2005).  The conversion from event rate to
probability assumes Poisson statistics \citep{Wheatland2001,Moon_etal_2001}:
the probability of an event is $P(t) = 100\times(1-e^{-\lambda t})$ where
$\lambda$ is the flaring rate for the particular event definition. 

\begin{figure}
\centerline{
\includegraphics[width=0.33\textwidth, clip]{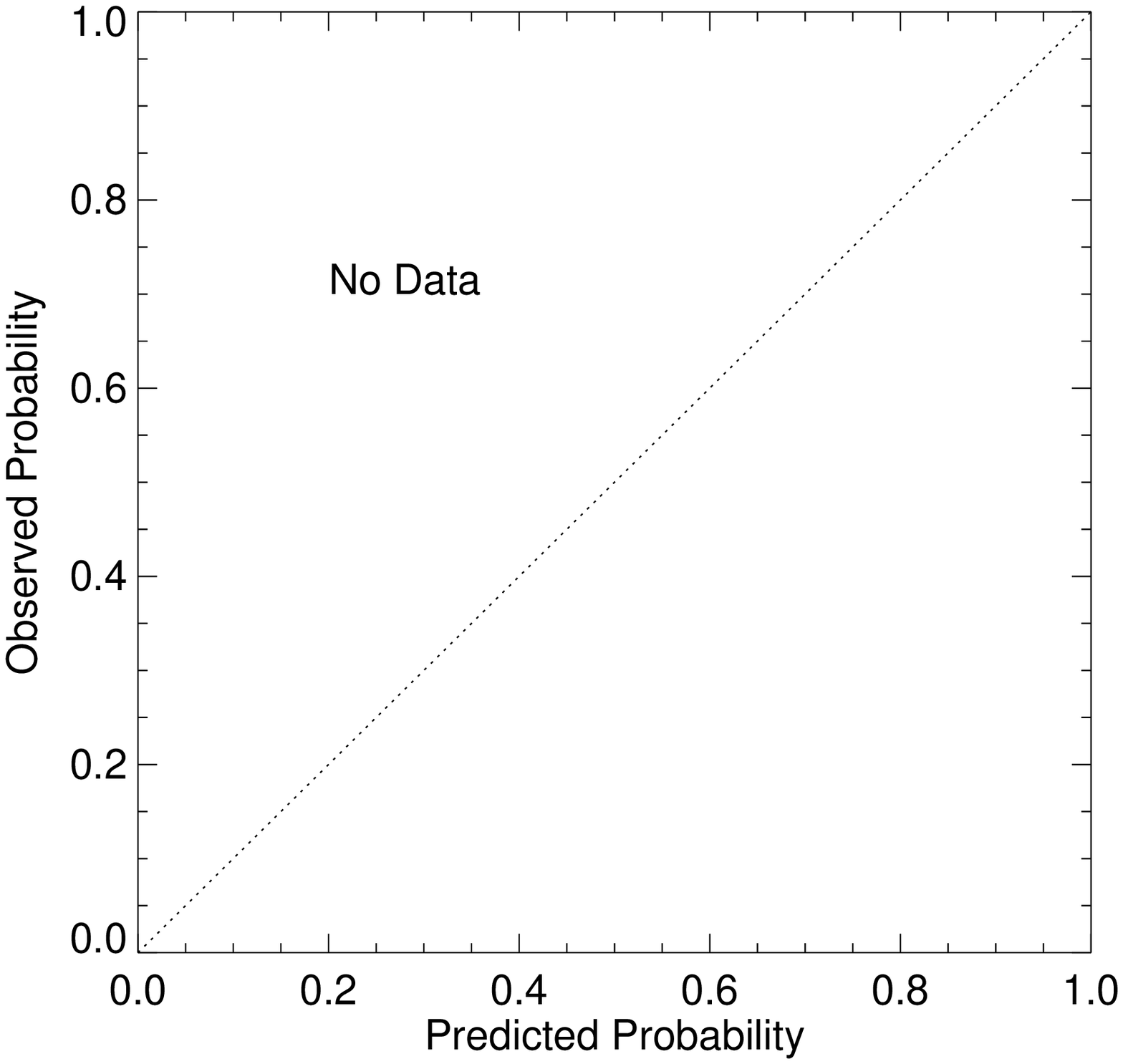}
\includegraphics[width=0.33\textwidth, clip]{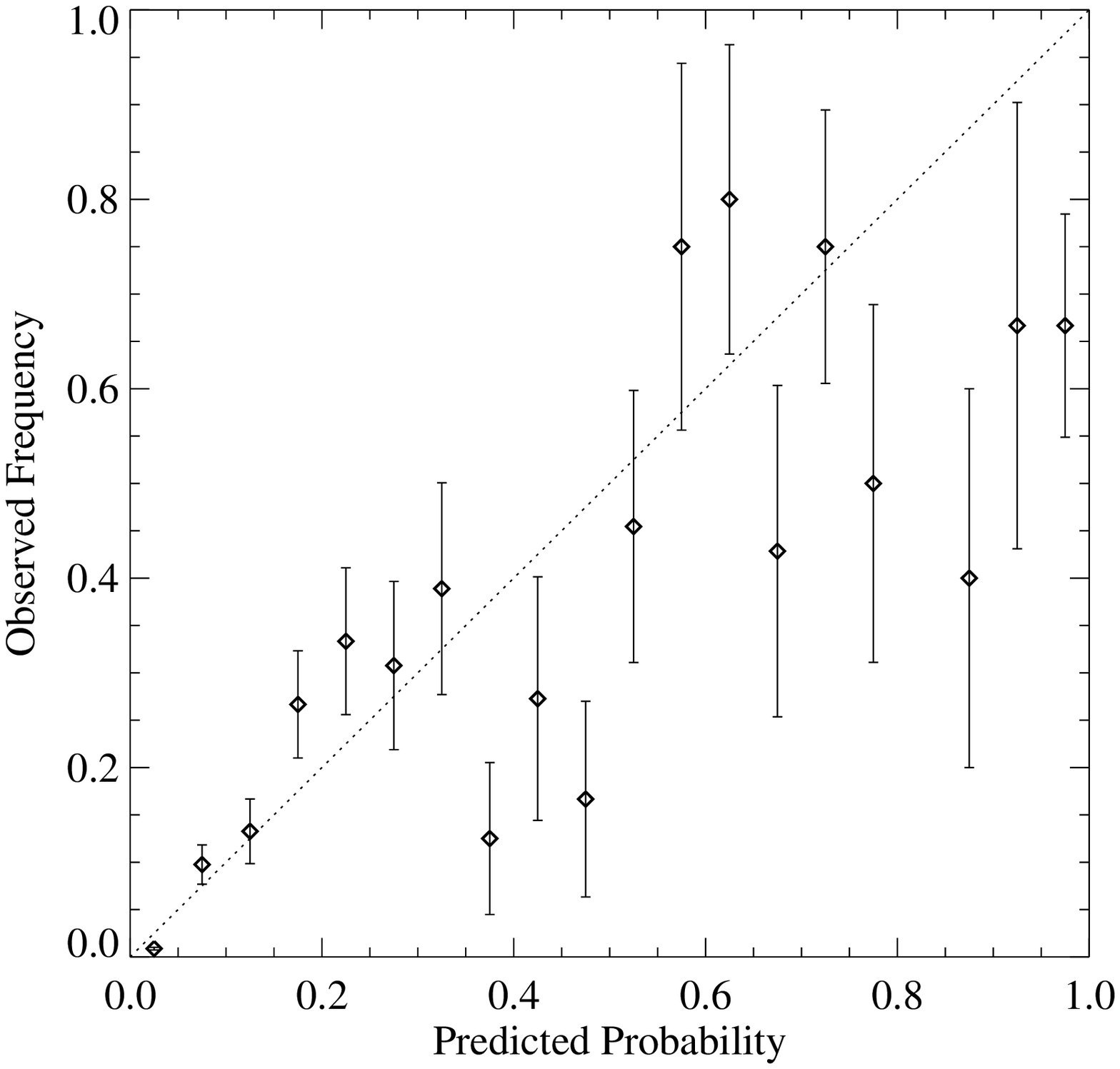}
\includegraphics[width=0.33\textwidth, clip]{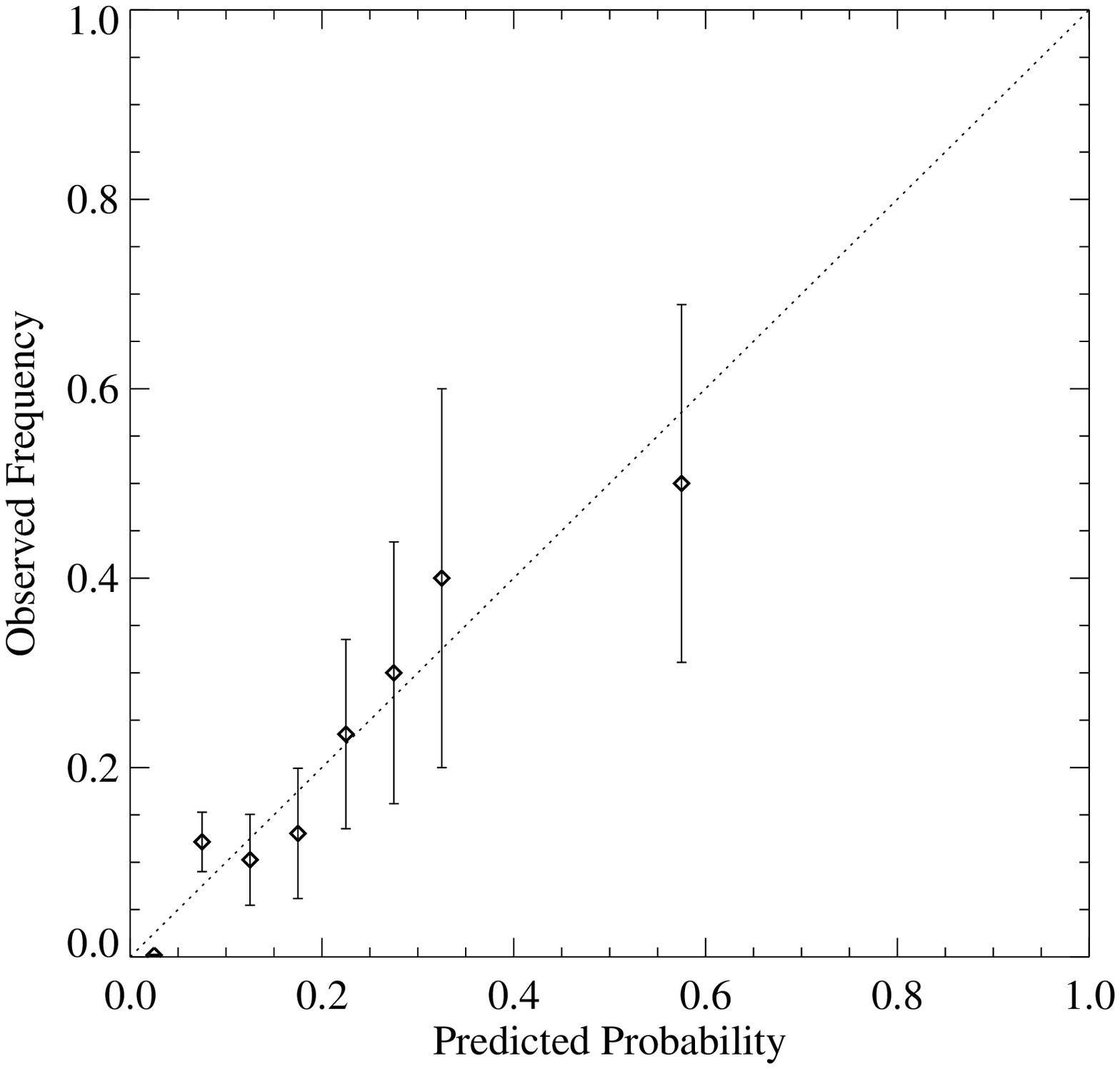}}
\centerline{
\includegraphics[width=0.33\textwidth, clip]{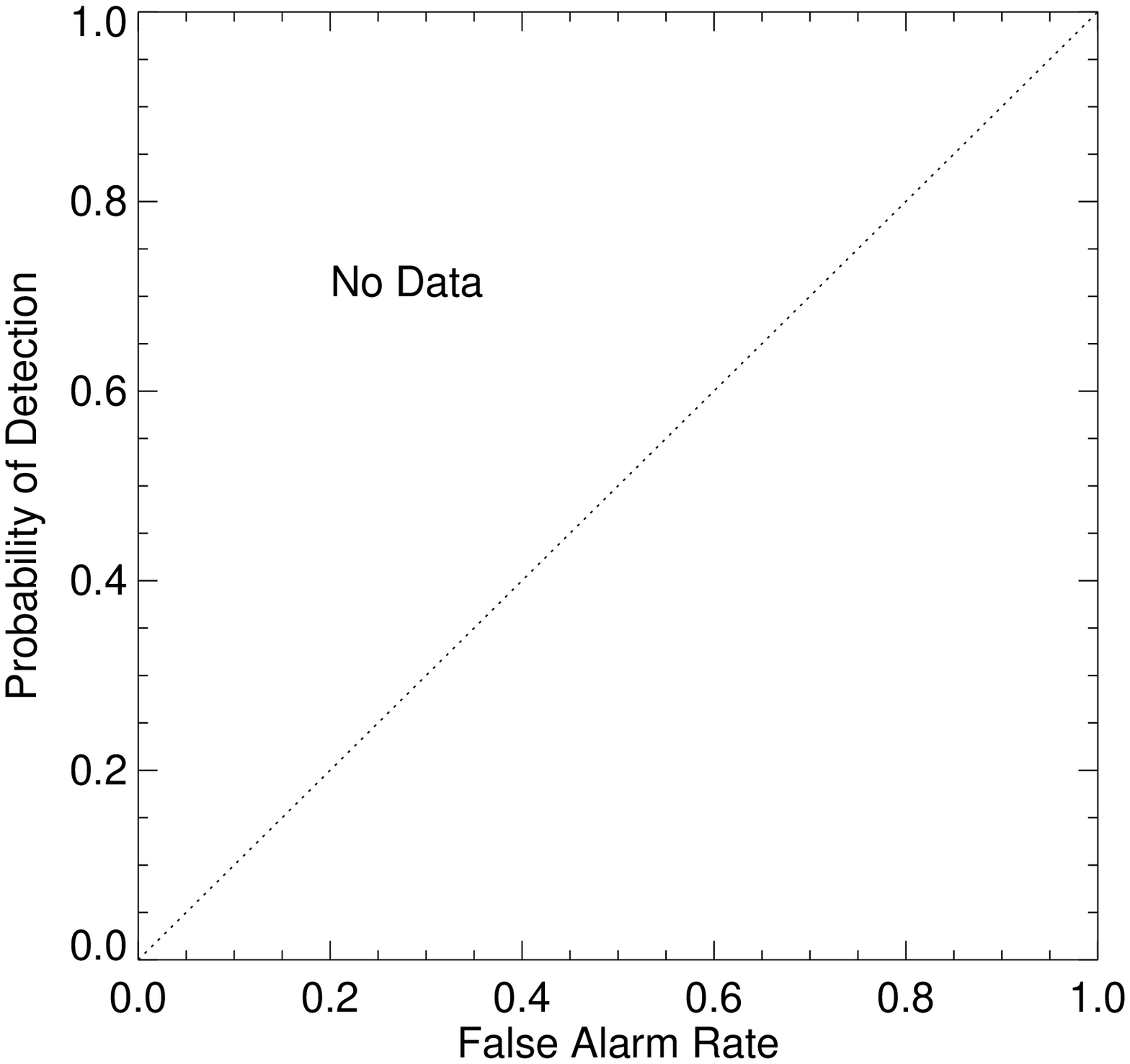}
\includegraphics[width=0.33\textwidth, clip]{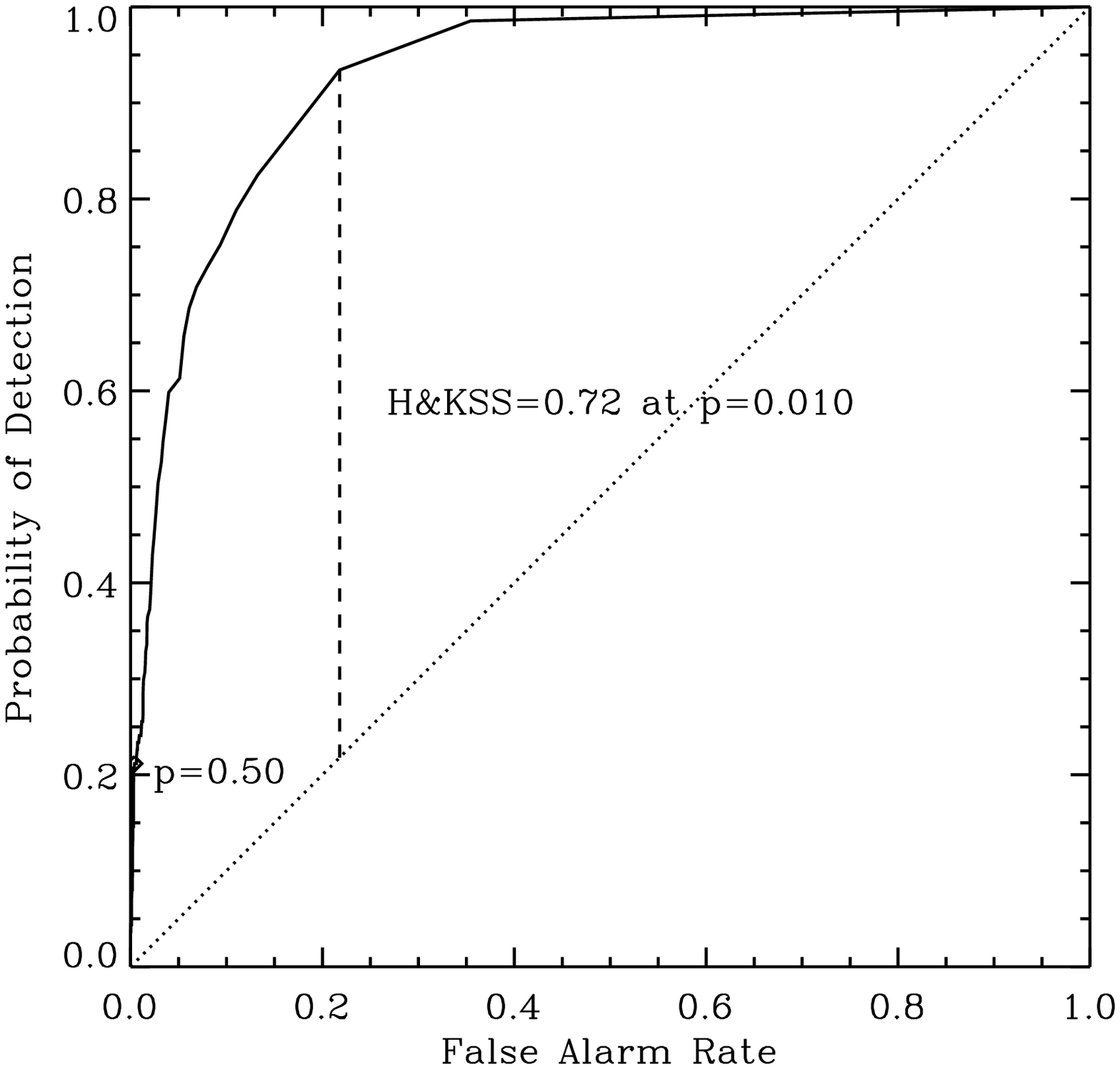}
\includegraphics[width=0.33\textwidth, clip]{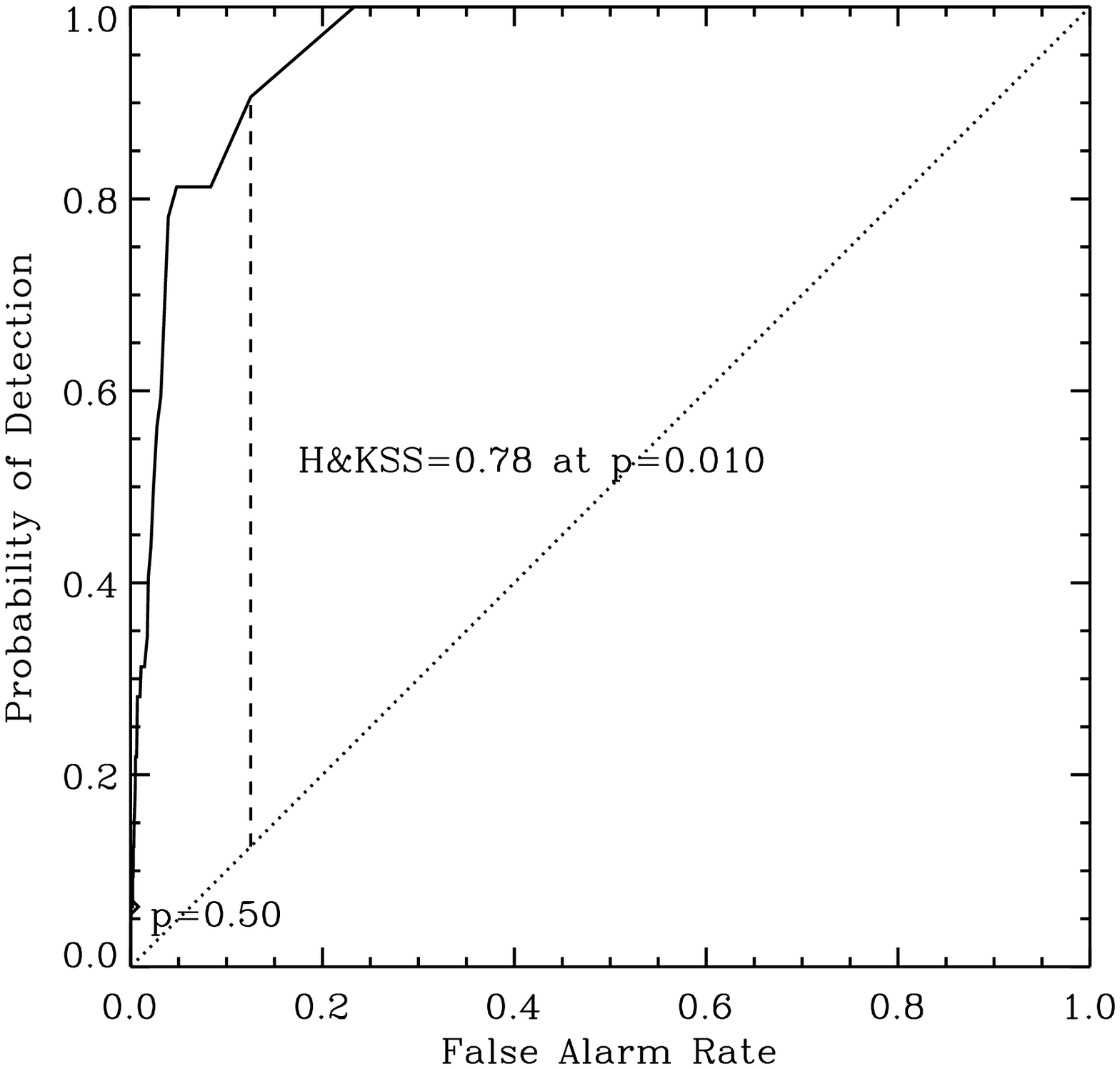}}
\caption{Same as Figure~\ref{fig:beff_plots} but for MSFC method. In this case,
predictions were not constructed for the \CC\ event definition.}
\label{fig:fal_plots}
\end{figure}

The rate-fitting algorithm is best for larger flares, and so no forecasts were
made for the \CC\ events.  It has been found that combining this strong
gradient neutral line with secondary measures \citep[e.g., total magnetic flux,
and more recently with prior flare history,][]{Falconer_etal_2012} is likely to
give more accurate forecasts, but these secondary measures are not included for
the forecasts presented here.  The results for the workshop dataset are shown
in Table~\ref{tbl:fal_best}, and only include the regions with higher
confidence.  The values of the \Brier\ are among the best, although the small
sample sizes are reflected in the larger error bars in the reliability plots
(Figure~\ref{fig:fal_plots}, top).  The ROC curves show that the probability of
detection remains near one to relatively small values of the false alarm rate,
hence this is one of the better methods for issuing all-clear forecasts.

\begin{center}
\begin{deluxetable}{cccccccc}
\tablecolumns{8}
\tablewidth{0pc}
\tablecaption{Optimal Performance Results: MSFC} 
\tablehead{ 
\colhead{Event} & \colhead{Sample} & \colhead{Event} & \colhead{\RC} & \colhead{\Heidke} & \colhead{\Appleman} & \colhead{\True} & \colhead{\Brier} \\
\colhead{Definition} & \colhead{Size} &  \colhead{Rate} & \colhead{(threshold)} & \colhead{(threshold)} & \colhead{(threshold)} & \colhead{(threshold)} & \colhead{} 
} 
\startdata 
\CC & N/A  &   N/A &  N/A        &         N/A &         N/A &         N/A &  N/A\\
\MS & 4510 & 0.030 & 0.97 (0.52) & 0.40 (0.16) & 0.10 (0.52) & 0.72 (0.01) & 0.19\\
\ML &  ''  & 0.007 & 0.99 (0.53) & 0.22 (0.17) & 0.00 (0.53) & 0.78 (0.01) & 0.05\\
\enddata 
\label{tbl:fal_best} 
\end{deluxetable}
\end{center}

\subsection{Magnetic Field Moment Analysis and Discriminant Analysis - K.D.~Leka, G.~Barnes}
\label{sec:nwra_mag}

The NWRA moment analysis parameterizes the observed magnetic field and its
spatial derivatives using the first four moments (mean, standard deviation,
skew and kurtosis), plus totals and net values when appropriate \citep{params}.
The moments of the observed field strength describe its distribution, while an
approximation of the total flux indicates how large the region is based on both
its area and field strength.  The higher-order moments can be sensitive to the
presence of small areas of complex field that are missed in the lower-order
moments and summations \citep{params}.

The default boundary is the observed line-of-sight data.  A correction is
applied to these data: $\Bz(x,y) \approx \Bl(x,y)/\cos \theta(x,y)$, meaning
that each observed point is divided by the cosine of its observing angle,
$\theta$, rather than an average observing angle or the observing angle of the
center of the field-of-view. This correction is especially important at large
observing angles.  Additionally, a magnetic-field boundary that more
closely approximates the radial field is prepared by computing the potential field that
matches the line-of-sight observed component \citep[e.g.,][]{Alissandrakis1981}
and using the resulting radial field as a boundary for some parameters.   This
approach has the effect of mitigating the appearance of the magnetic neutral
lines that are solely a manifestation of projection effects.  For the
potential-field approximation of the radial field, no additional correction
using the observing angle is needed.  For both boundaries, however, a limit of
$\cos \theta > 0.1$ is imposed: parameters are not computed for observing
angles $\ga 85^\circ$.

The parameters considered are the subset of those presented in
\cite{params} which can be computed using solely the line-of-sight
component of the field.  Over 50 parameters are considered, describing
the photospheric field distribution using three basic categories:
\begin{itemize}
\item The distribution of the approximated radial (vertical) component of the magnetic field.
\item The horizontal gradient of that distribution $|\mbox{\boldmath$\nabla$} \Bz|$.
\item The character of inferred magnetic neutral lines.
\end{itemize}
\noindent
The last category includes a variation on the $\mathcal{R}$ parameter
described in Section~\ref{sec:R}.

\begin{figure}
\epsscale{0.5}
\plotone{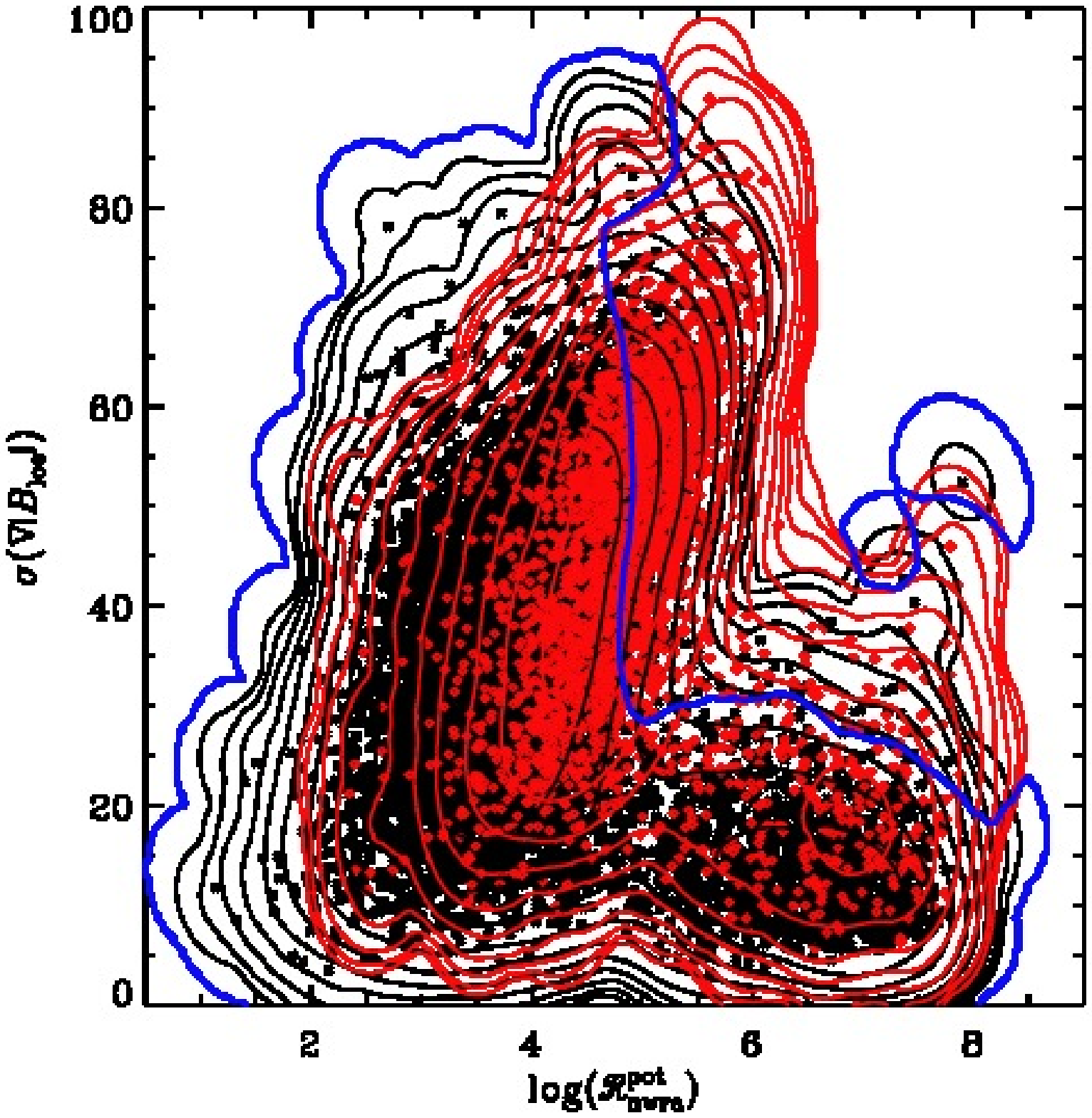}
\caption{Magnetic Field Moment Analysis and Discriminant Analysis
(\S\ref{sec:nwra_mag}) for two variable nonparametric Discriminant Analysis of
the log of C.~Schrijver's $\mathcal{R}$ parameter and the standard deviation of
the distribution of the magnitude of the horizontal gradient of the line of
sight field, $\sigma(|\grad B_{\rm los}|)$.  Red/black curves are contours of
the probability density estimate for the flaring/non-flaring samples
(specifically for a flaring threshold of C1.0 and prediction window of 24\,hr).
The probability of a flare occurring is estimated from the ratio of the density
estimates at any specified value of ($\mathcal{R}$, $\sigma(|\grad B_{\rm
los}|)$).  The blue line indicates where the probability density estimates for
the flaring and non-flaring regions are equal. Regions within the blue curves
are predicted to remain flare-quiet (i.e., have a greater non-flaring
probability than flaring probability). 
}
\label{fig:da}
\end{figure}

The forecast is produced with Discriminant Analysis \citep[DA,
e.g.,][]{ken83,Silverman1986}.  This statistical approach classifies new
measurements as belonging to one of two populations by dividing parameter space
into two regions based on where the probability density of one population ({\it
e.g.}, flaring regions) exceeds the other (flare-quiet regions).  A set of new
measurements, {\it i.e.} a new active region, that falls on the appropriate
side of the division is then predicted to flare (see Fig.~\ref{fig:da}).  Using
Bayes's theorem, the probability that a new measurement belongs to a given
population can be estimated from the probability density estimates \citep{SWJ}.
Given an accurate representation of the probability density of a parameter, DA
will maximize the overall accuracy of predictions \citep[see][for
examples]{dfa,dfa3}.  

The probability density is typically either assumed to be Gaussian, which
results in a linear discriminant function, or it is estimated
nonparametrically.  For the results presented here, a nonparametric density
estimate was made using the Epanechnikov kernel with the smoothing parameter
chosen optimal for a Gaussian distribution \citep{Silverman1986}.  One strength
of DA is that multiple variables can be considered simultaneously. Two-variable
combinations are employed here, as noted in the tables.  Cross-validation is
employed by default in order to remove bias in the skill scores.  

\begin{figure}
\centerline{
\includegraphics[width=0.33\textwidth, clip]{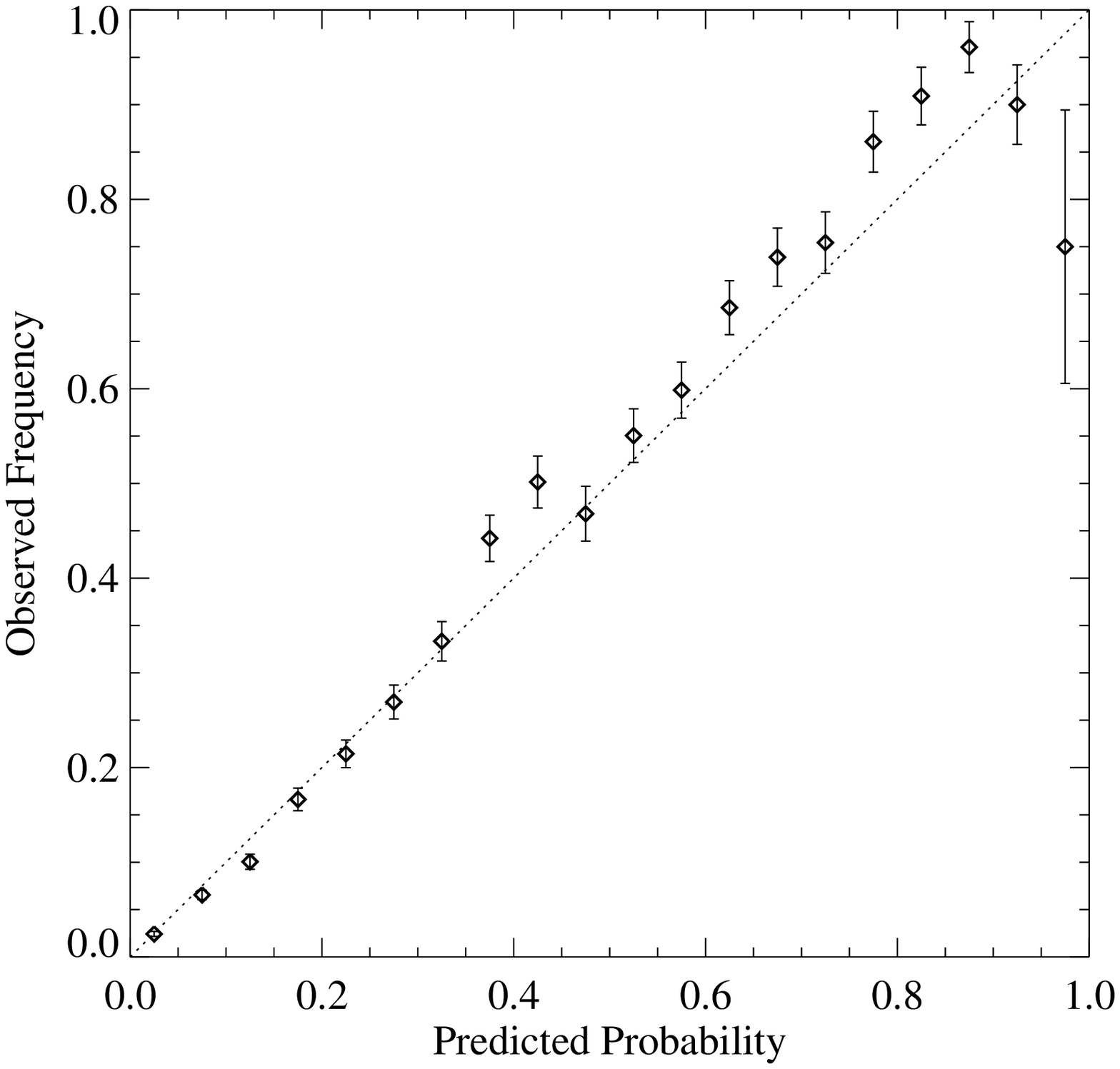}
\includegraphics[width=0.33\textwidth, clip]{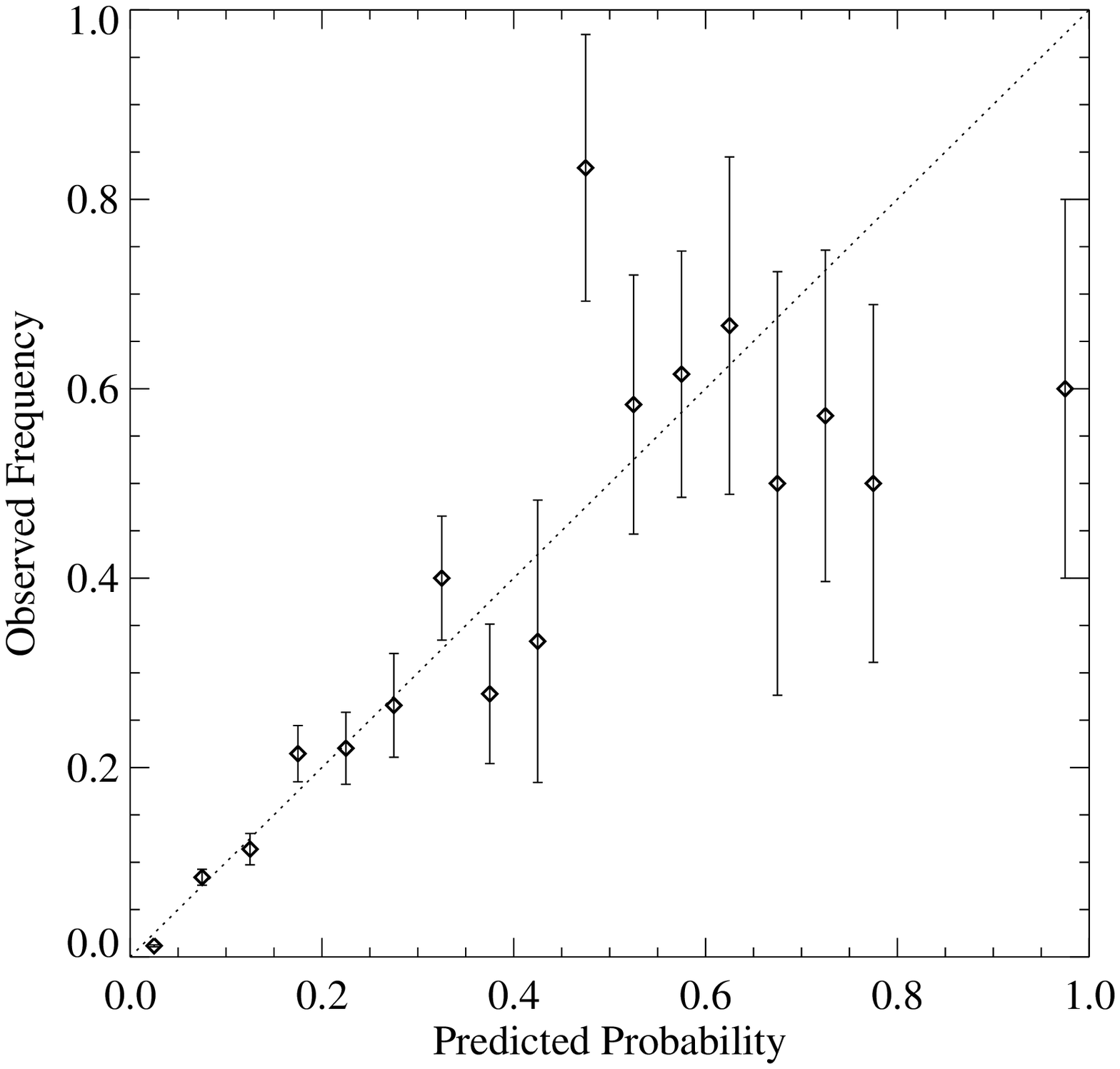}
\includegraphics[width=0.33\textwidth, clip]{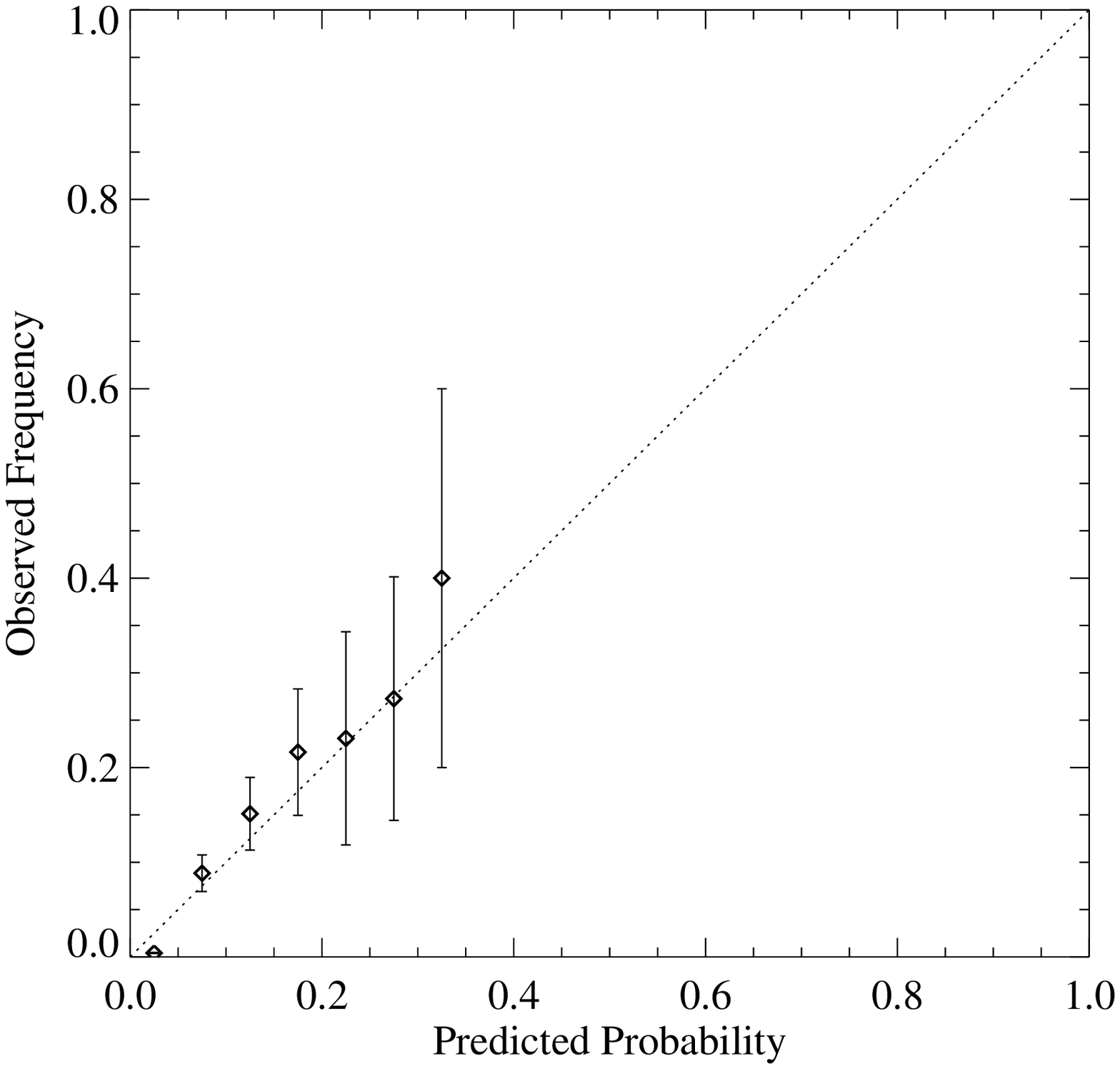}}
\centerline{
\includegraphics[width=0.33\textwidth, clip]{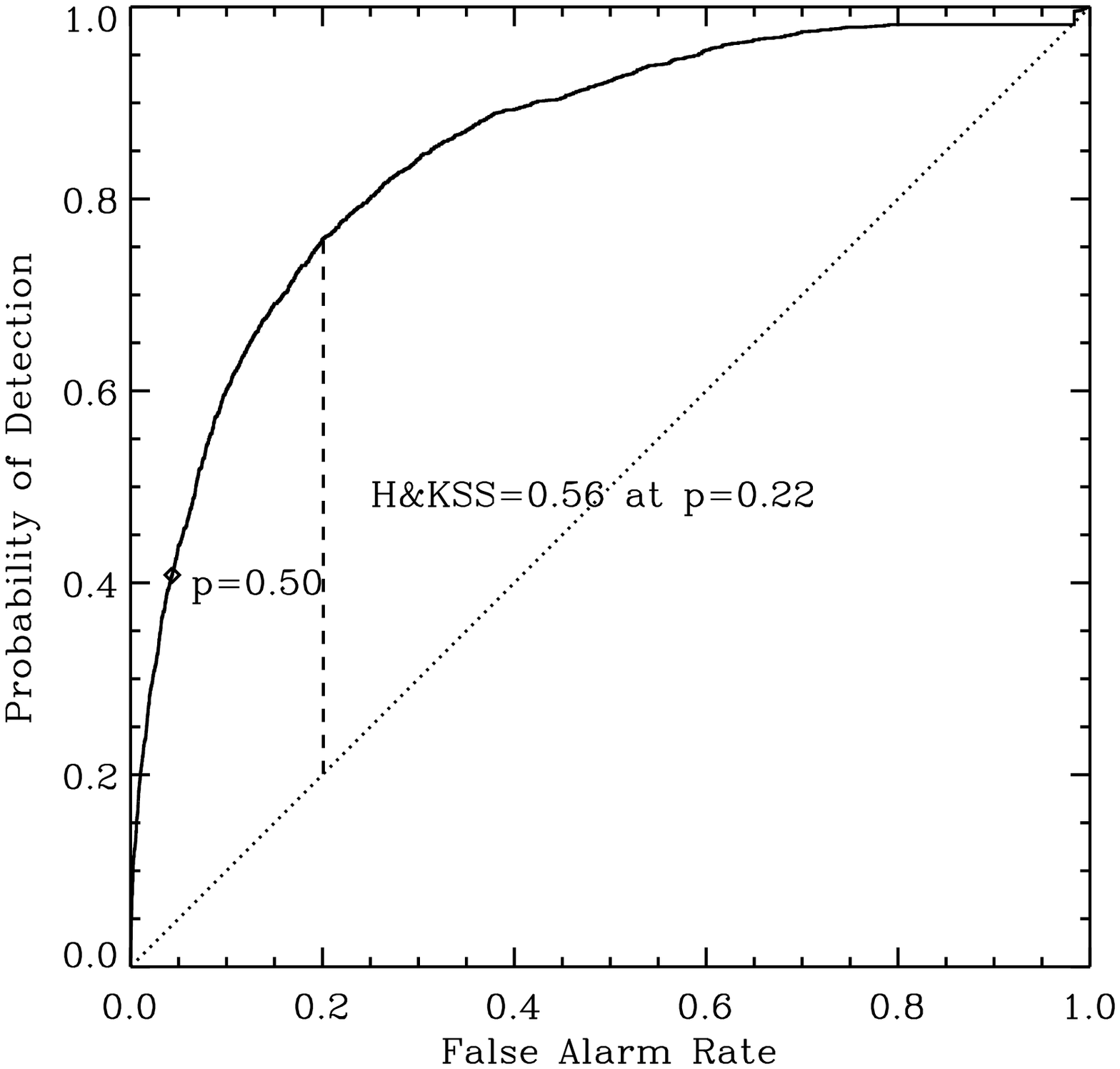}
\includegraphics[width=0.33\textwidth, clip]{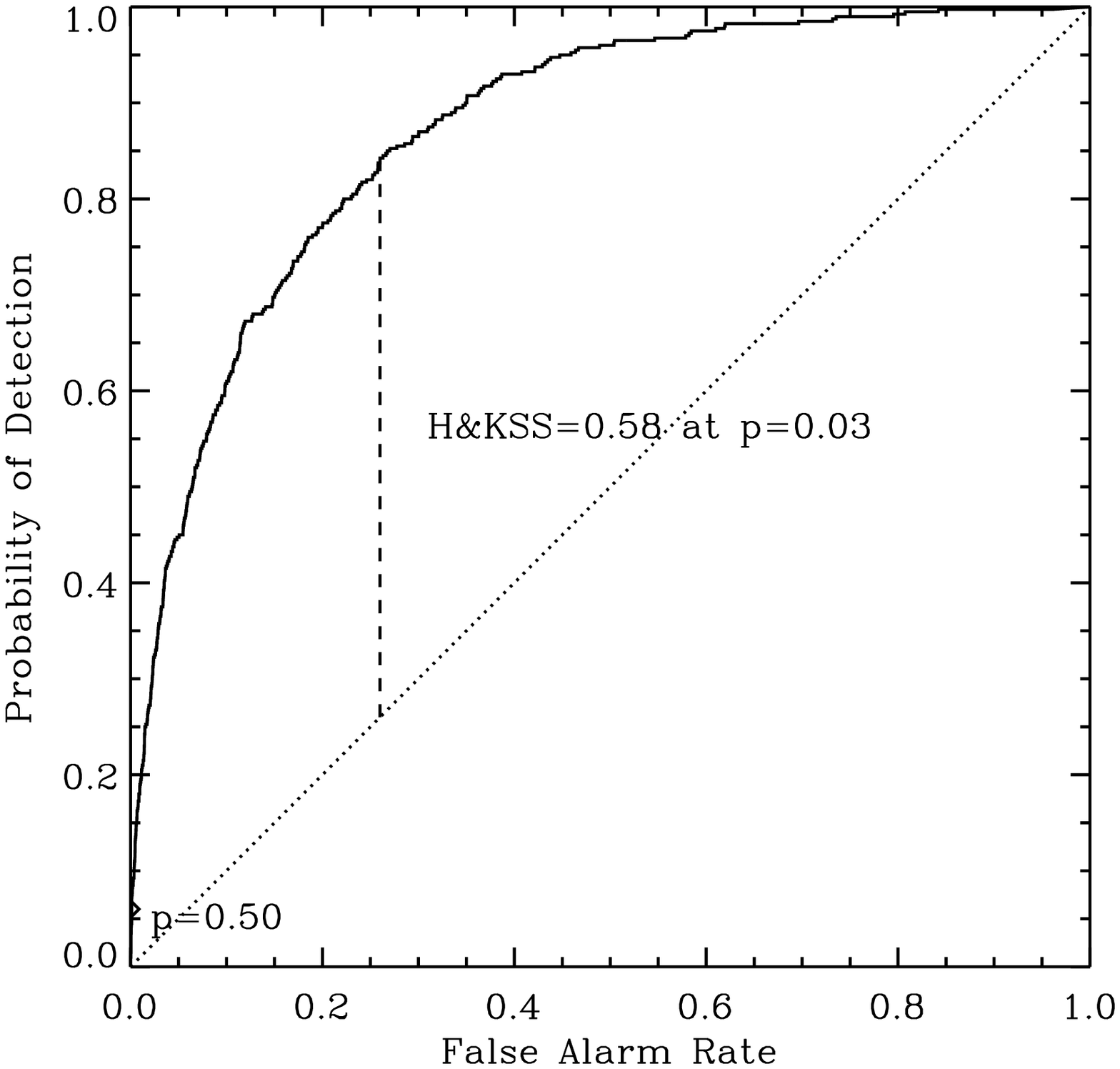}
\includegraphics[width=0.33\textwidth, clip]{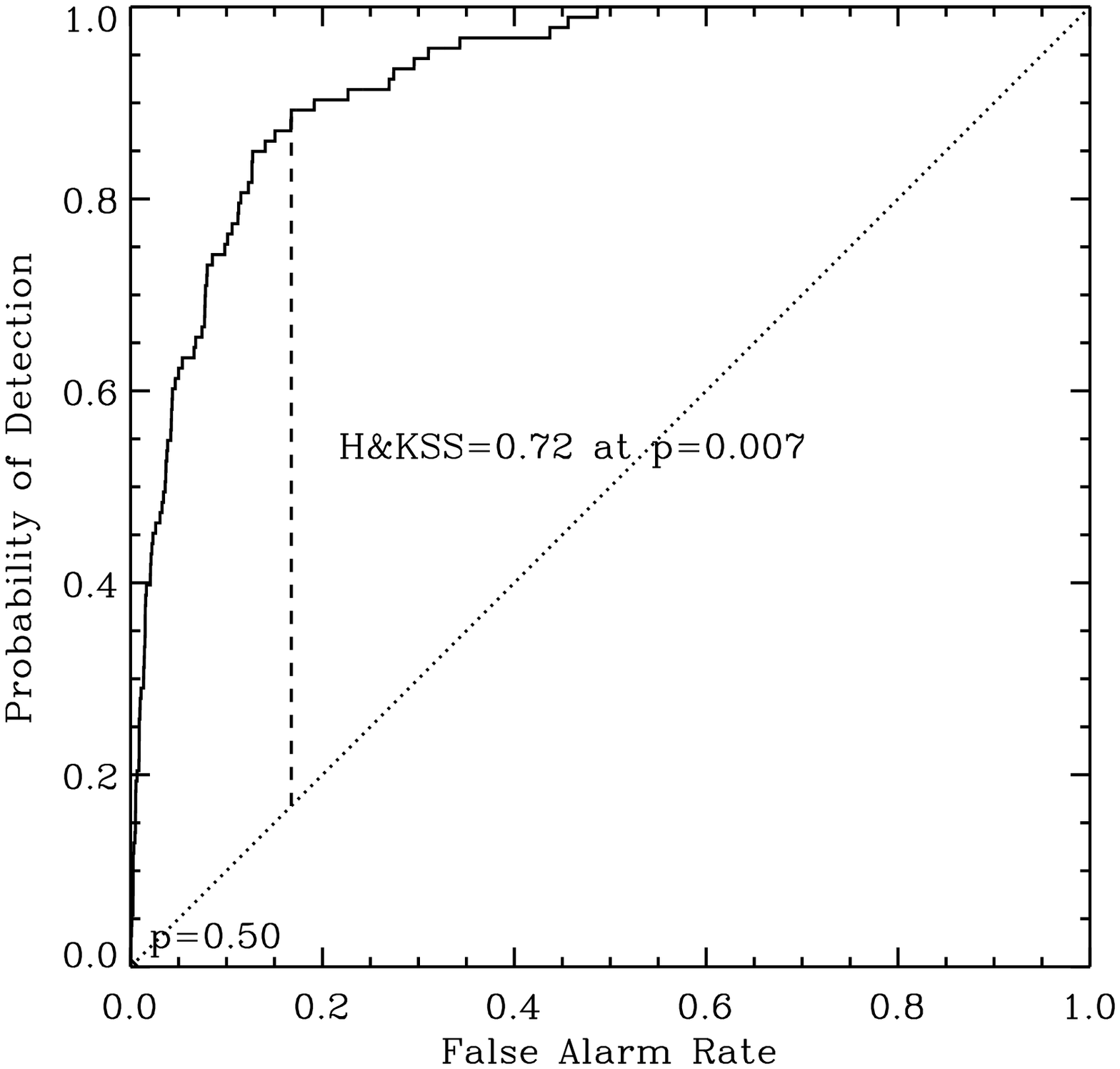}}
\caption{Same as Figure~\ref{fig:beff_plots} but for NWRA/Discriminant Analysis method.}
\label{fig:nwra_plots}
\end{figure}

The results for the variable combination with the highest Brier skill score for
each event definition are given in Table~\ref{tbl:nwra_best}.  Other variable
combinations may produce higher values of other skill scores, but only the
results for the variable combination with the highest Brier skill score are
listed here.  The reliability plots for Schrijver's implementation of
$\mathcal{R}$ and one-variable DA show a slight tendency for over-prediction
for the \MS\ threshold at higher probabilities, but less of such a trend for
\ML (Figure~\ref{fig:nwra_plots}, top).

\begin{center}
\begin{deluxetable}{cccccccc}  
\tablecolumns{8}
\tablewidth{0pc}
\tablecaption{Optimal Performance Results: NWRA Field Parameterizations, Non-Parametric Discriminant Analysis, Two-Variable Combinations} 
\tablehead{ 
\colhead{Event} & \colhead{Sample} & \colhead{Event} & \colhead{\RC} & \colhead{\Heidke} & \colhead{\Appleman} & \colhead{\True} & \colhead{\Brier} \\
\colhead{Definition} & \colhead{Size} &  \colhead{Rate} & \colhead{(threshold)} & \colhead{(threshold)} & \colhead{(threshold)} & \colhead{(threshold)} & \colhead{} 
} 
\startdata 
\CC\tablenotemark{a} 
 & 12965 & 0.201 & 0.85 (0.48) & 0.50 (0.35) & 0.24 (0.48) & 0.56 (0.22) & 0.32\\
\MS\tablenotemark{b} 
 &   ''  & 0.031 & 0.97 (0.42) & 0.29 (0.12) & 0.04 (0.42) & 0.58 (0.03) & 0.13\\
\ML\tablenotemark{c} 
 &   ''  & 0.007 & 0.99 (0.47) & 0.20 (0.07) & 0.00 (0.47) & 0.72 (0.01) & 0.06\\
\enddata 
\tablenotetext{a}{Variable combination: [$\sigma(\mbox{\boldmath$\nabla$} |\Bl|),\ \  \log(\mathcal{R}^{\rm pot}_{\rm nwra}) \ $].}
\tablenotetext{b}{Variable combination: [$\varsigma(\mbox{\boldmath$\nabla$} \Bl), \ \ \sigma(\Bz^{\rm pot})\ $].}
\tablenotetext{c}{Variable combination: [$\varsigma(\mbox{\boldmath$\nabla$} \Bl), \ \ \sigma(\Bz^{\rm pot})\ $].}
\label{tbl:nwra_best} 
\end{deluxetable}
\end{center}

\subsection{Magnetic Flux Close to High-Gradient Polarity Inversion Lines - C.~Schrijver}
\label{sec:R}

In \citet{Schrijver2007}, a parameter $\mathcal{R}$ was proposed as a proxy for
the emergence of current-carrying magnetic flux, {\it i.e.}, magnetic systems
with significant free magnetic energy which would be carried through the
photosphere and into the solar corona, enabling solar flares.  The
$\mathcal{R}$ (and, specifically, $\log(\mathcal{R}$)) parameter was proposed
and demonstrated as useful for forecasting solar flares in
\citet{Schrijver2007} by determining an empirical threshold above which flare
activity was significant.

\begin{figure}
\epsscale{1.0}
\plottwo{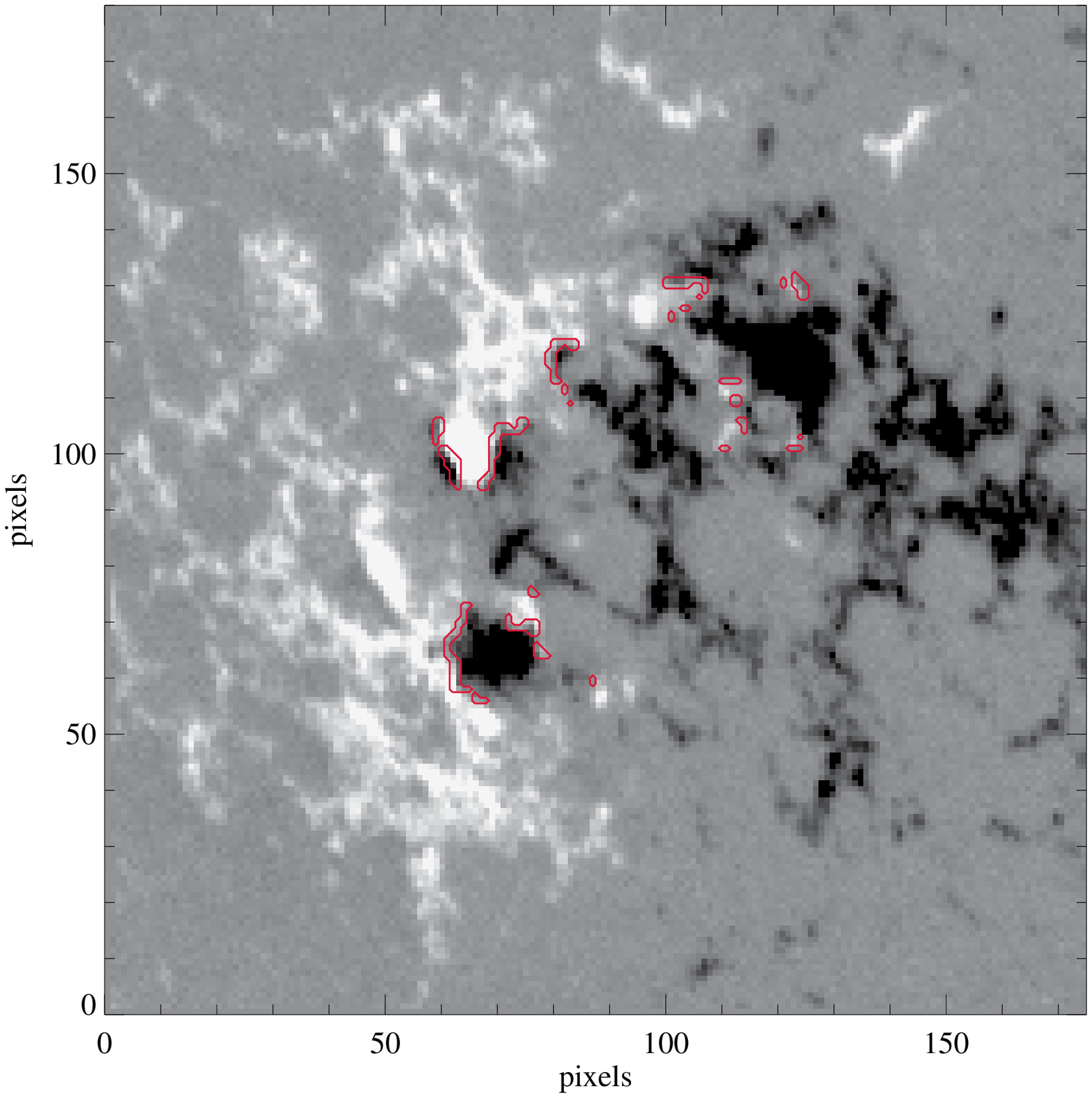}{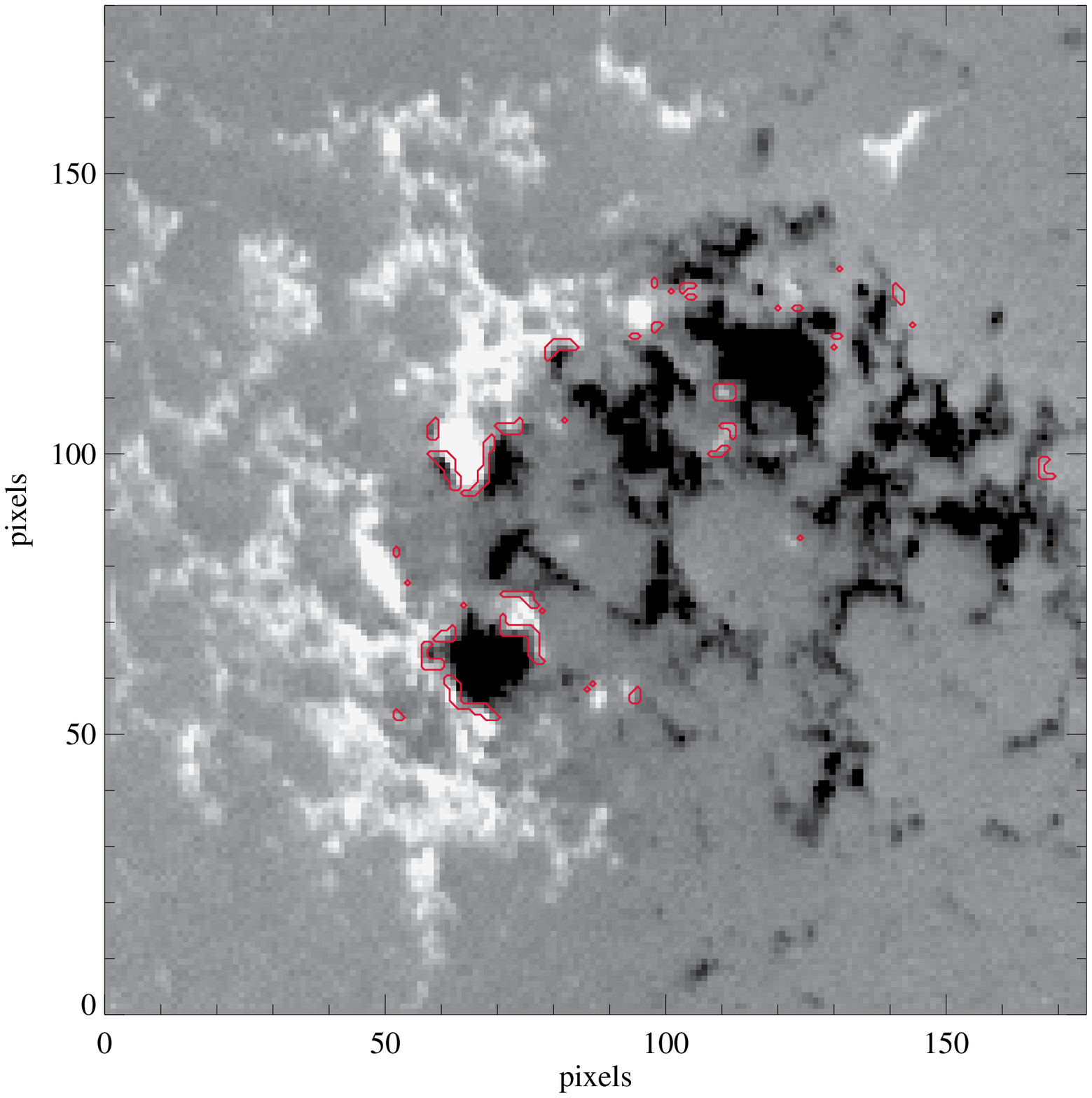}
\caption{Magnetic Flux Close to High-Gradient Polarity Inversion Lines
(\S\ref{sec:R}).  Red curves show the areas near high-gradient magnetic neutral
lines used for computing the $\mathcal{R}$ parameter \citep{Schrijver2007} for
NOAA AR 09767, 2002 January 3, following previous figures.  The results from
two methods are shown. The axis labels are in pixels, and the underlying
magnetic boundary images are saturated at $\pm 500$\,G.  
Left: the approach which replicates \citet{Schrijver2007} using the
line-of-sight magnetic field component and a Gaussian weighting function of
fixed 10-pixels width.  
Right: the NWRA implementation using the potential-field ${\rm B}_z$
boundary, and a Gaussian width that preserves the 15\,Mm distance on the Sun.
While the projection-effect polarity-inversion lines are not strong in this
example (since it is relatively close to disk center), there are some
differences between the two implementations, such as near the sunspot at $(x,y)
\approx (70,70)$.}
\label{fig:R}
\end{figure}

The parameter $\mathcal{R}$ was computed from the line-of-sight magnetic field
maps using the following steps:
\begin{itemize}
\item Dilate bitmaps of the magnetograms where the positive or negative flux
density exceeds a threshold (150\,Mx\,cm$^{-2}$).
\item Define high-gradient polarity-separation lines as areas where the bitmaps
overlap.
\item Convolve the resulting high-gradient polarity-separation line bitmap with a
Gaussian to obtain a ``weighting map''.
\item Obtain $\mathcal{R}$ by multiplying the weighting map by the unsigned
line-of-sight field and computing the total.
\end{itemize}
For the results here, the $\mathcal{R}$ parameter was calculated as part of the
NWRA magnetic field analysis (Section~\ref{sec:nwra_mag}), but is also included
in the parameterizations by other groups ({\it e.g.} SMART, see
Section~\ref{sec:smart_ccnn}), with slightly different implementations (see
Section~\ref{sec:implement}).  Within the NWRA magnetic parameterization,
$\mathcal{R}$ is calculated first so as to replicate the parameter in
\citet{Schrijver2007}: $\Bl$ is used directly, and a fixed width of 10 MDI
pixels is used for the Gaussian used in the convolution, to identify an area
within roughly 15\,Mm of a magnetic neutral line.  The targets are restricted
to those within $45^\circ$ of disk center, which limits the sample size.  An
image of the boundary of the inferred high-gradient neutral lines from this
replication of  \citet{Schrijver2007} is shown in Figure~\ref{fig:R}.

The second implementation within the NWRA magnetic parameterization uses the
$\Bz$ potential-field boundary as described in Section~\ref{sec:nwra_mag}, and
varies the width of the convolution function such that it effectively preserves
the target 15\,Mm physical distance on the Sun and performs all calculations in
a helioplanar coordinate system.  A comparison of this implementation with the
original is also shown in Figure~\ref{fig:R}.

\begin{figure}
\centerline{
\includegraphics[width=0.33\textwidth, clip]{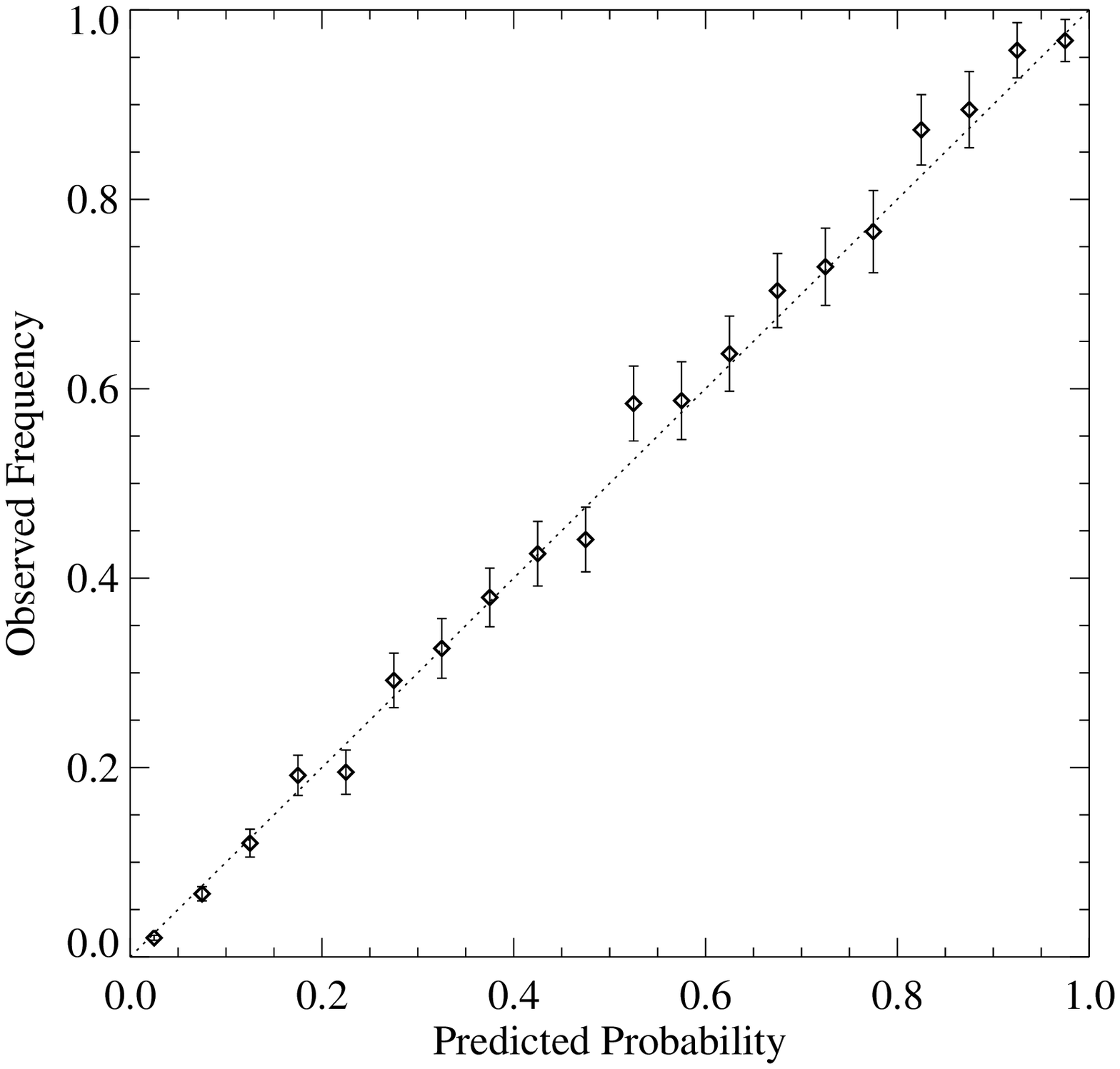}
\includegraphics[width=0.33\textwidth, clip]{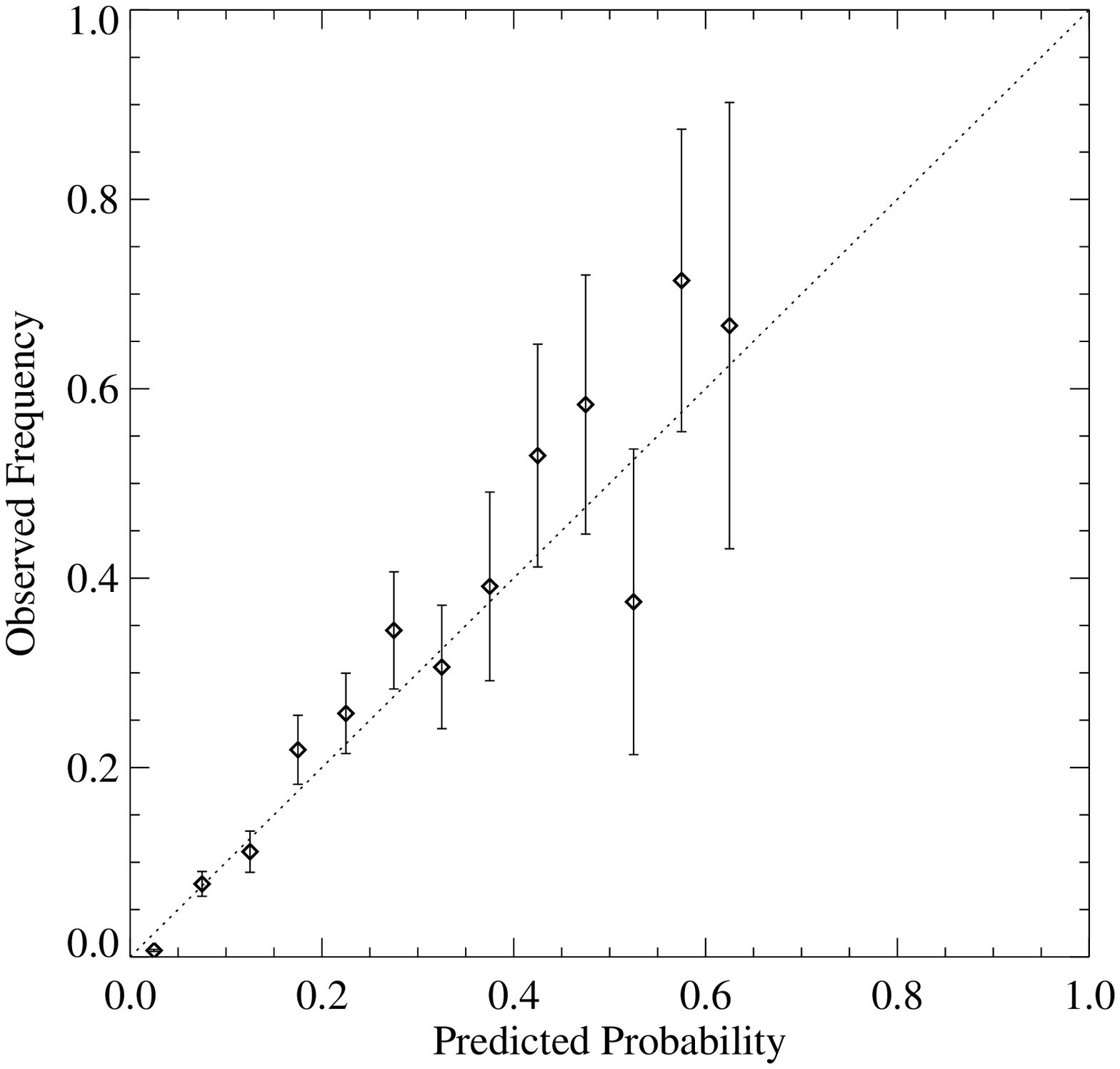}
\includegraphics[width=0.33\textwidth, clip]{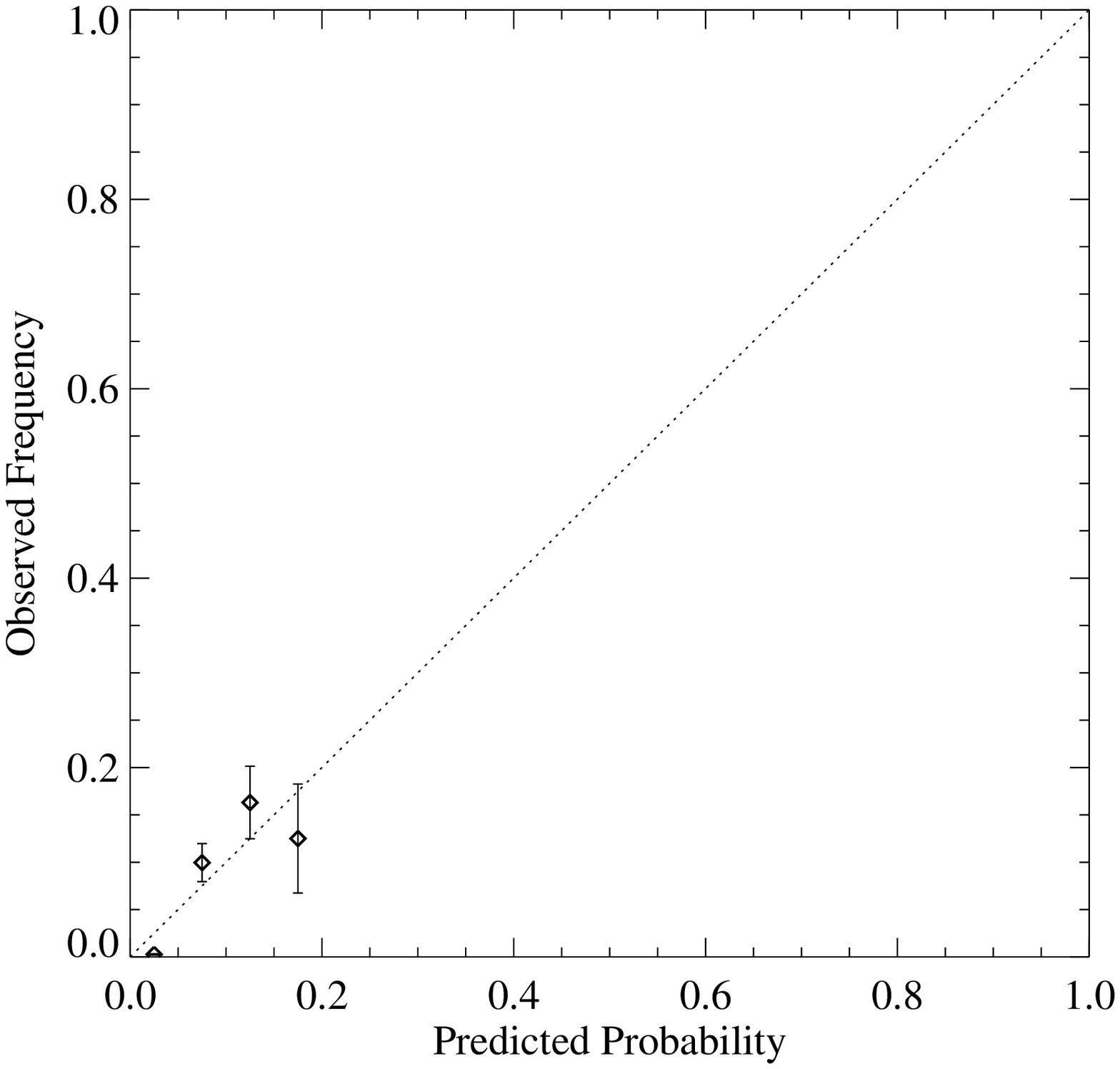}}
\centerline{
\includegraphics[width=0.33\textwidth, clip]{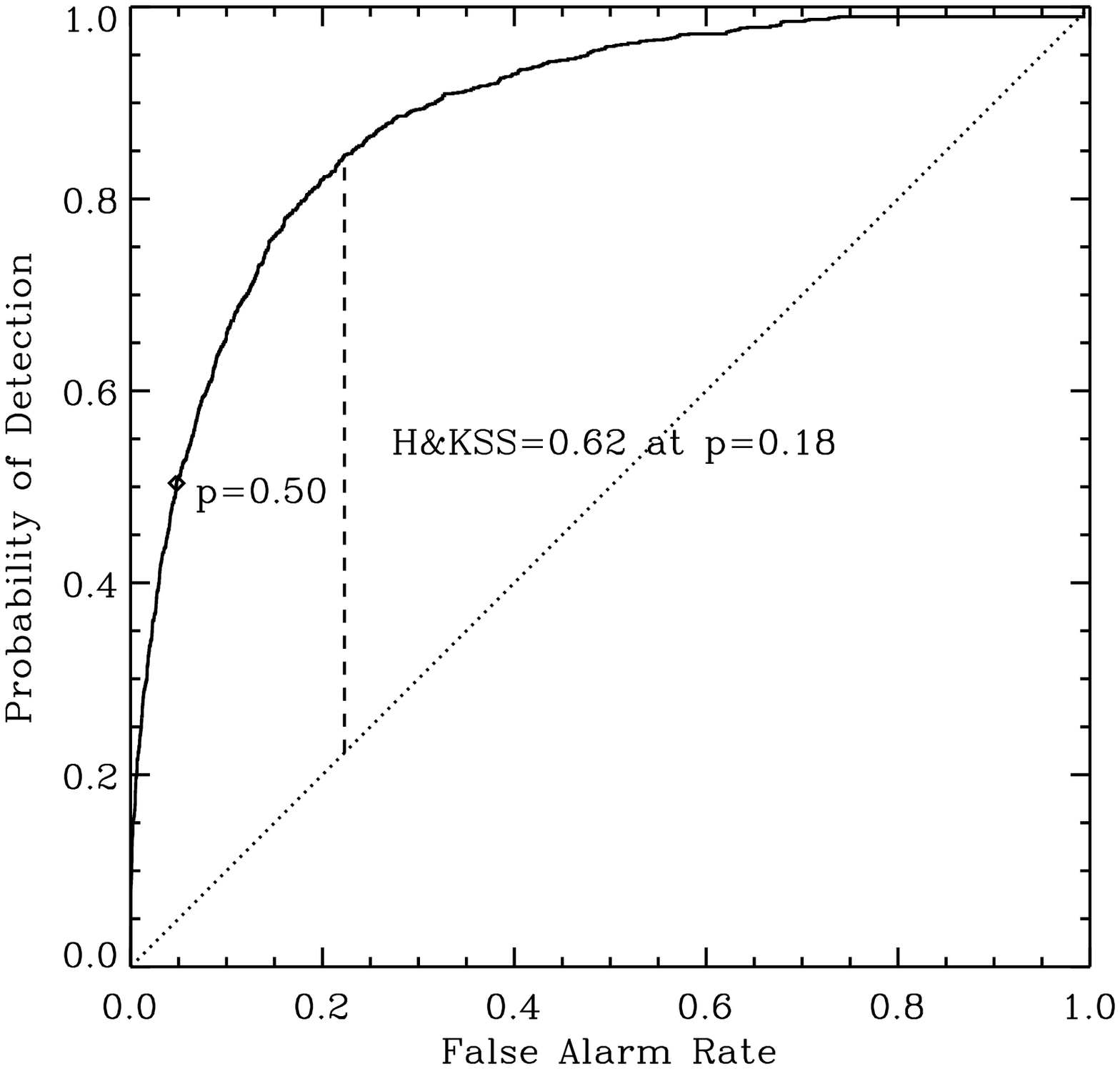}
\includegraphics[width=0.33\textwidth, clip]{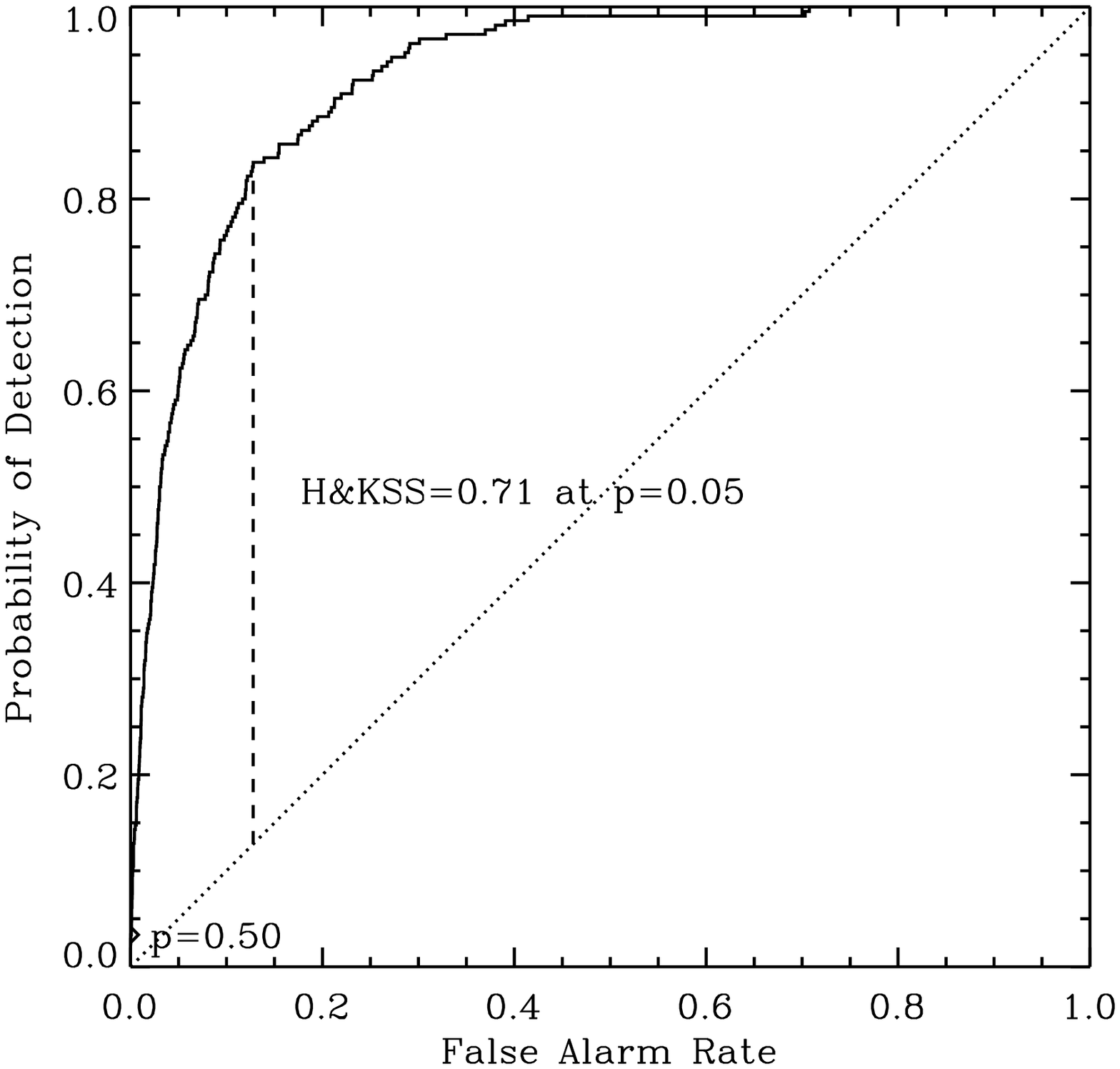}
\includegraphics[width=0.33\textwidth, clip]{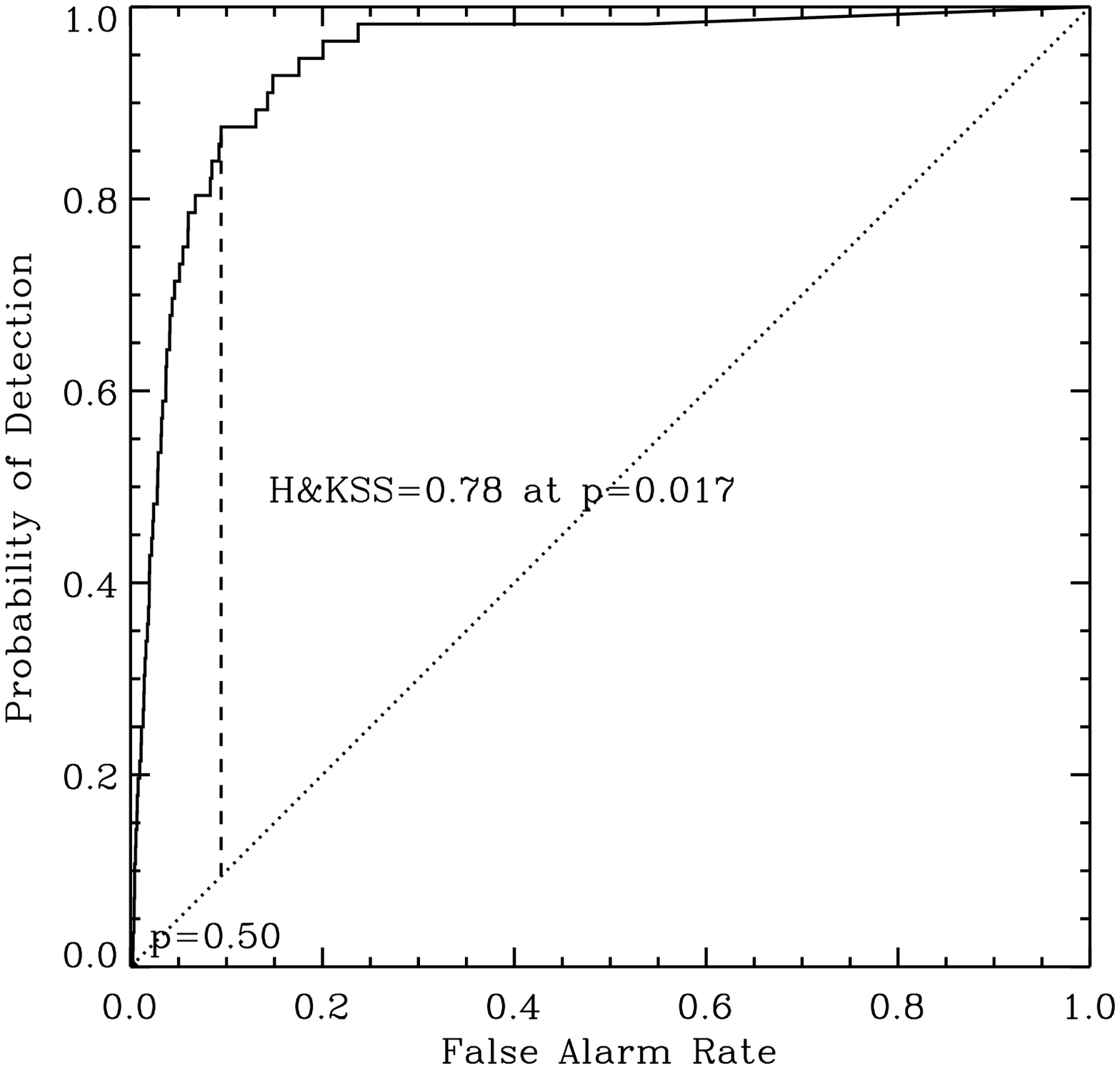}}
\caption{Same as Figure~\ref{fig:beff_plots} but for the Schrijver
$\ln(\mathcal{R})$ parameter with NPDA.}
\label{fig:ksr_plots}
\end{figure}

For the results here, predictions using $\mathcal{R}$ were made using
one-variable nonparametric Discriminant Analysis with cross-validation
(Section~\ref{sec:nwra_mag}), and the resulting skill scores are shown in
Table~\ref{tbl:ksr_best}.  The NWRA implementation is not included here, but it
is included in one high-performing two-variable combination in
Table~\ref{tbl:nwra_best}.  The reliability plots (Figure~\ref{fig:ksr_plots},
top) show a slight under prediction at the lower probabilities for \CC, and
little trending otherwise with the exception of points in the highest predicted
probability bins for both \MS\ and \ML.  The ROC curves
(Figure~\ref{fig:ksr_plots}, bottom) show that the probability of detection for
the \MS\ and \ML\ event definitions remain close to one for relatively small
values of the false alarm rate, thus this is another method that may be well
suited to issuing all-clear forecasts.

\begin{center}
\begin{deluxetable}{cccccccc}  
\tablecolumns{8}
\tablewidth{0pc}
\tablecaption{Optimal Performance Results: Schrijver $\log(\mathcal{R})$ (+ NPDA)} 
\tablehead{ 
\colhead{Event} & \colhead{Sample} & \colhead{Event} & \colhead{\RC} & \colhead{\Heidke} & \colhead{\Appleman} & \colhead{\True} & \colhead{\Brier} \\
\colhead{Definition} & \colhead{Size} &  \colhead{Rate} & \colhead{(threshold)} & \colhead{(threshold)} & \colhead{(threshold)} & \colhead{(threshold)} & \colhead{} 
} 
\startdata 
\CC & 7299 & 0.200 & 0.86 (0.50) & 0.55 (0.36) & 0.31 (0.50) & 0.62 (0.18) & 0.38\\
\MS &  ''  & 0.031 & 0.97 (0.42) & 0.38 (0.17) & 0.03 (0.42) & 0.71 (0.04) & 0.18\\
\ML &  ''  & 0.007 & 0.99 (0.20) & 0.19 (0.08) & 0.00 (0.20) & 0.78 (0.02) & 0.07\\
\enddata 
\label{tbl:ksr_best} 
\end{deluxetable}
\end{center}

\subsection{Generalized Correlation Dimension - R.T.J.~McAteer}
\label{sec:GCD}

The morphology of a flux concentration (active region) can be described
by the fractal dimension and related quantities.  This approach is
useful for understanding the underlying influence of turbulence on
solar structures \citep{McAteer_etal_2010}.

The Generalized Correlation (akin to a fractal) Dimension $D_{\rm BC}$ of an
active region is computed using images of the magnetic field distribution.
Images were processed in the same manner as \citet{McAteer_etal_2005b}, which
includes applying multiplicative corrections to stronger/weaker areas and an
observing-angle correction $\Bz \approx \Bl / \cos \theta$.  The images were
also subjected to a thresholding algorithm in the same manner as
\citet{McAteer_etal_2005b} and a box counting algorithm (with cancellation) was
applied to the resulting binary images.  Only regions within $\theta <
60^\circ$ were considered.  The generalized correlation dimensions were
calculated for $q$ values of $0.1,~0.5,~1.5,~2.0,~8.0$  \citep[$q$ referring to
the ``$q$-moment'', which governs the influence of strong {\it vs.} weak areas
in the $D_{\rm BC}$ measure, see][]{McAteer_etal_2010}.

The fractal dimension is a parameterization of the active region, and does not
include a statistical flare prediction method {\it per se}.  In
\citet{McAteer_etal_2005}, the fractal dimension was shown to be related to a
region's flare productivity: lower limits of $D_{\rm BC} = 1.2 (1.25) $ were
found to be required for M-class (X-class) flares (with $q = 8$).  Extensions
of this work involving the multifractal spectrum \citep{Conlon_etal_2008,
Conlon_etal_2010} were not computed for this study.  \citet{Georgoulis2012}
show that one multifractal algorithm is resolution-dependent, However
\citet{McAteer2015} and \citet{McAteeretal2015} show that a multifractal
analysis that uses the wavelet transform modulus maxima approach, performed
over a series of images, may produce a useful measure of the energy stored in
the coronal magnetic field.

For the results here, predictions were made using one-variable nonparametric
Discriminant Analysis with cross-validation (Section~\ref{sec:nwra_mag}) with
these fractal-related parameterizations.  The best-performing parameterization
is shown in Table~\ref{tbl:gcd_best}, and is the $q=8.0$ variable as expected
from prior work, and thus is subsequently the only variable presented here.
The performance as measured by the \Brier\ is comparatively higher than for the
\True\, as expected from the NPDA.  The ROC curves show the usual improvement
with increasing event magnitude, but overall show worse performance than most
other methods. 

\begin{center}
\begin{deluxetable}{cccccccc}
\tablecolumns{8}
\tablewidth{0pc}
\tablecaption{Optimal Performance Results: Generalized Correlation Dimensions (+ NPDA)}
\tablehead{
\colhead{Event} & \colhead{Sample} & \colhead{Event} & \colhead{\RC} & \colhead{\Heidke} & \colhead{\Appleman} & \colhead{\True} & \colhead{\Brier} \\
\colhead{Definition} & \colhead{Size} &  \colhead{Rate} & \colhead{(threshold)} & \colhead{(threshold)} & \colhead{(threshold)} & \colhead{(threshold)} & \colhead{} 
}
\startdata
\CC & 9339 & 0.195 & 0.81 (0.49) & 0.27 (0.30) & 0.04 (0.49) & 0.31 (0.18) & 0.11\\ 
\MS &  ''  & 0.027 & 0.97 (0.30) & 0.15 (0.07) & 0.00 (0.30) & 0.48 (0.02) & 0.05\\
\ML &  ''  & 0.007 & 0.99 (0.13) & 0.09 (0.05) & 0.00 (0.13) & 0.62 (0.01) & 0.02\\
\enddata
\label{tbl:gcd_best}
\end{deluxetable}
\end{center}

\begin{figure}
\centerline{
\includegraphics[width=0.33\textwidth, clip]{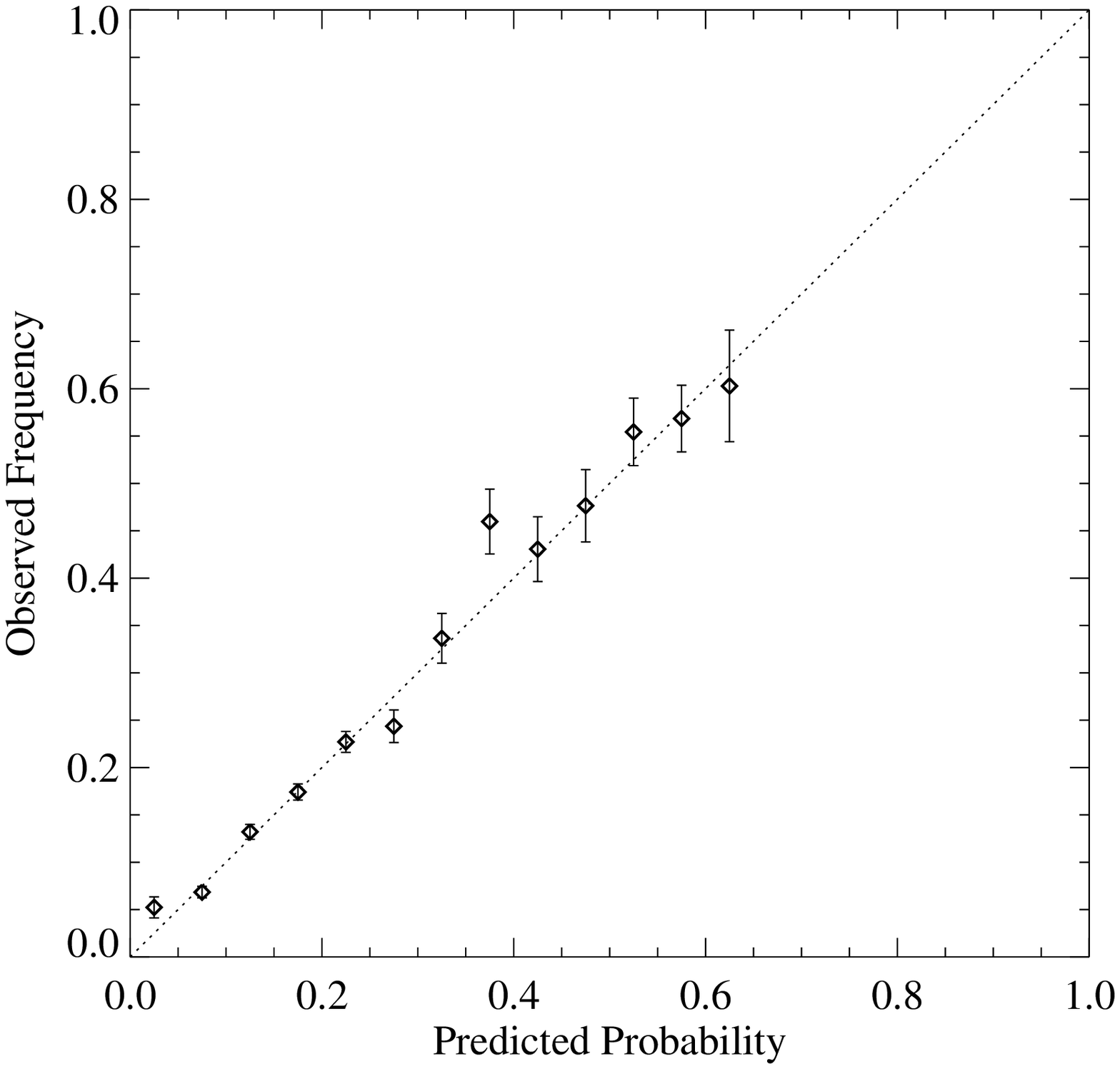}
\includegraphics[width=0.33\textwidth, clip]{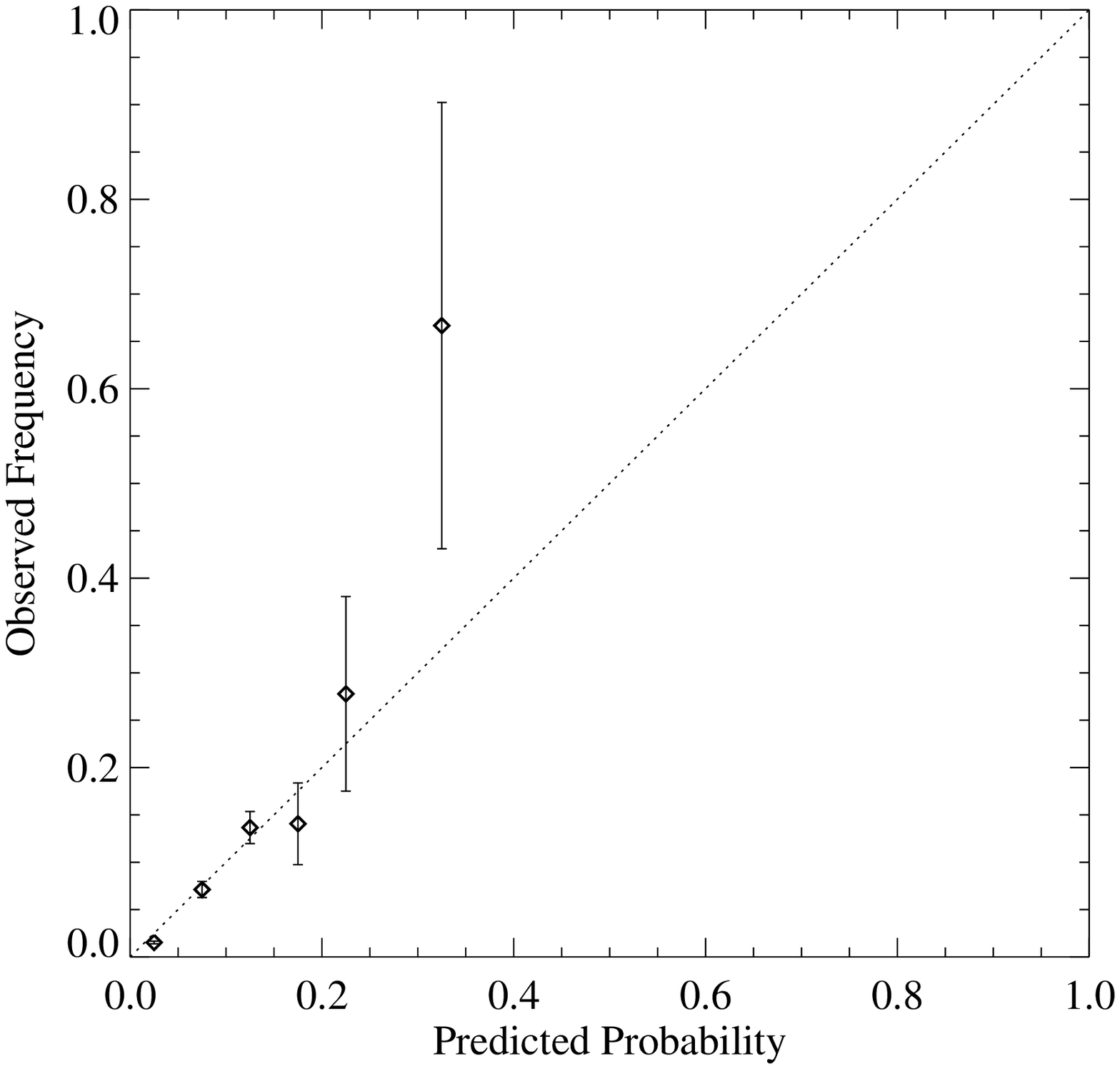}
\includegraphics[width=0.33\textwidth, clip]{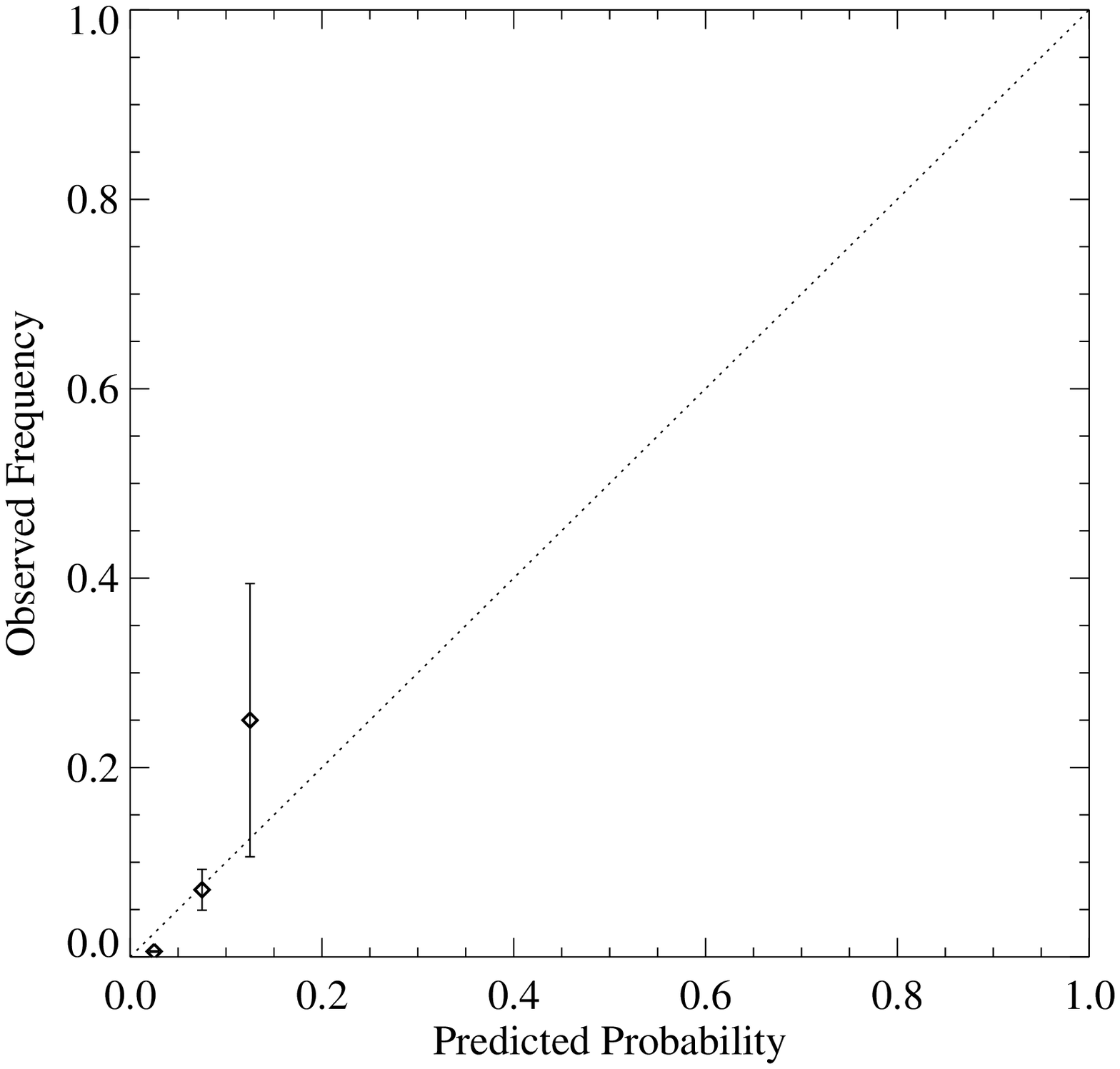}}
\centerline{
\includegraphics[width=0.33\textwidth, clip]{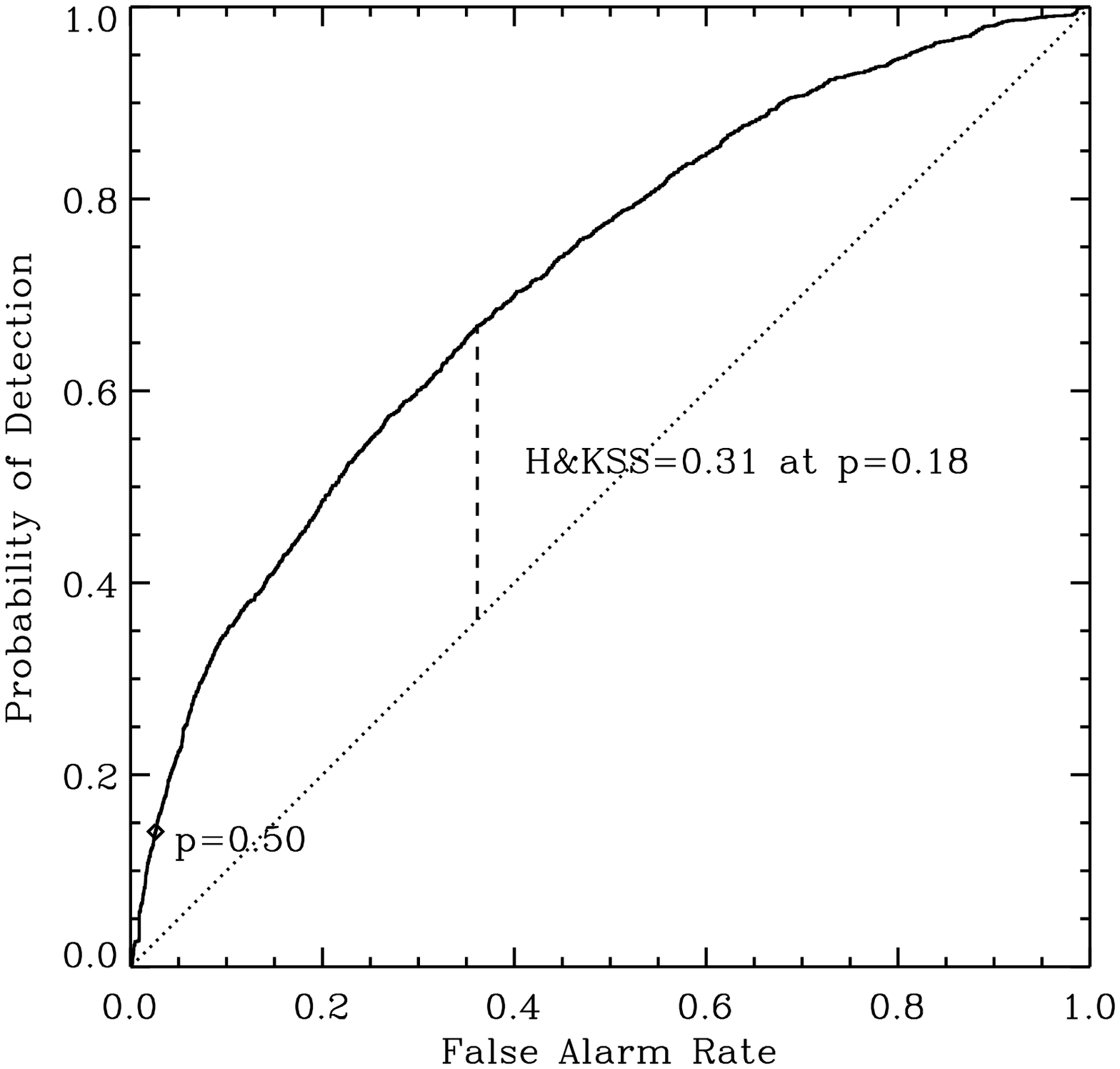}
\includegraphics[width=0.33\textwidth, clip]{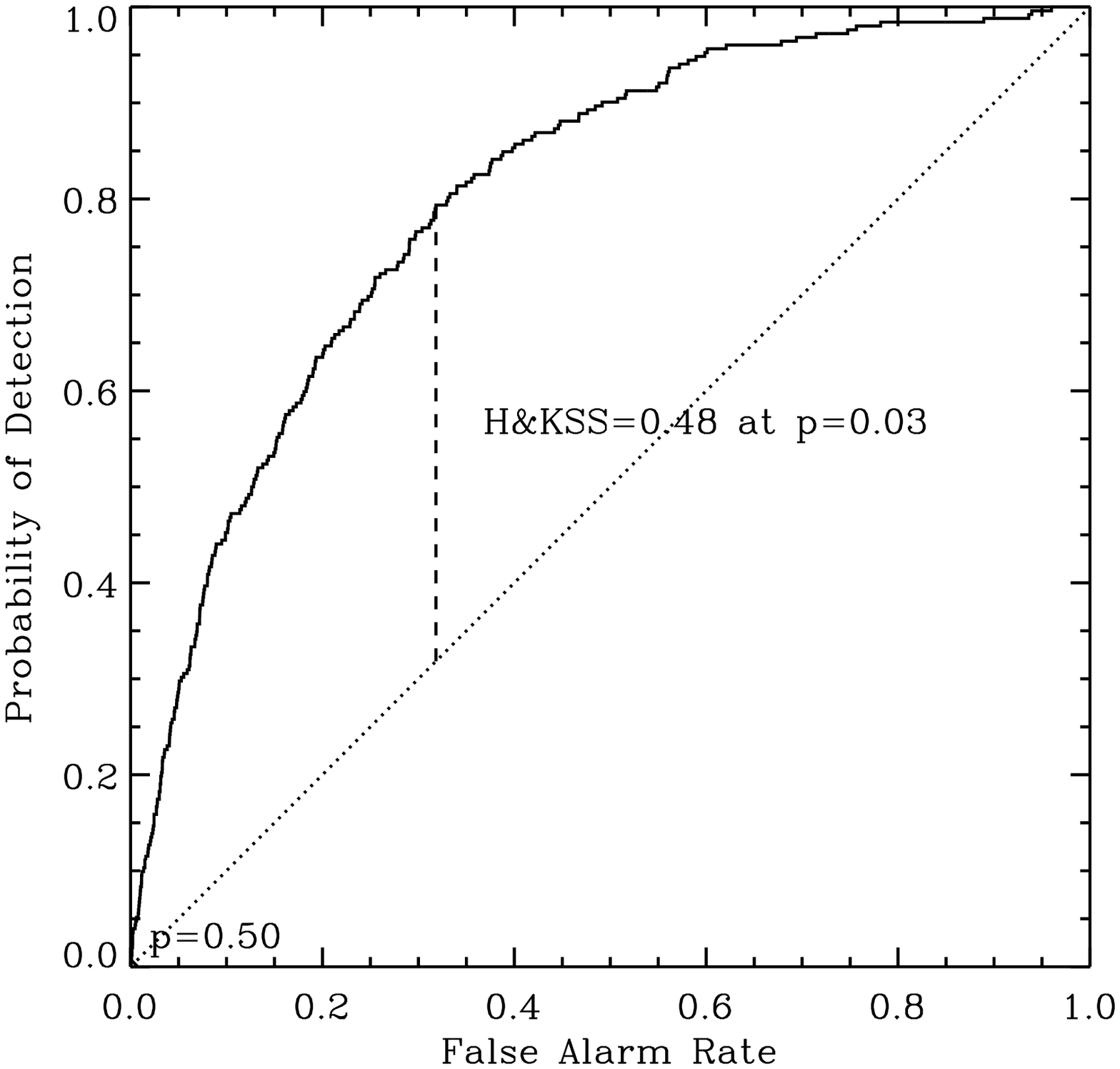}
\includegraphics[width=0.33\textwidth, clip]{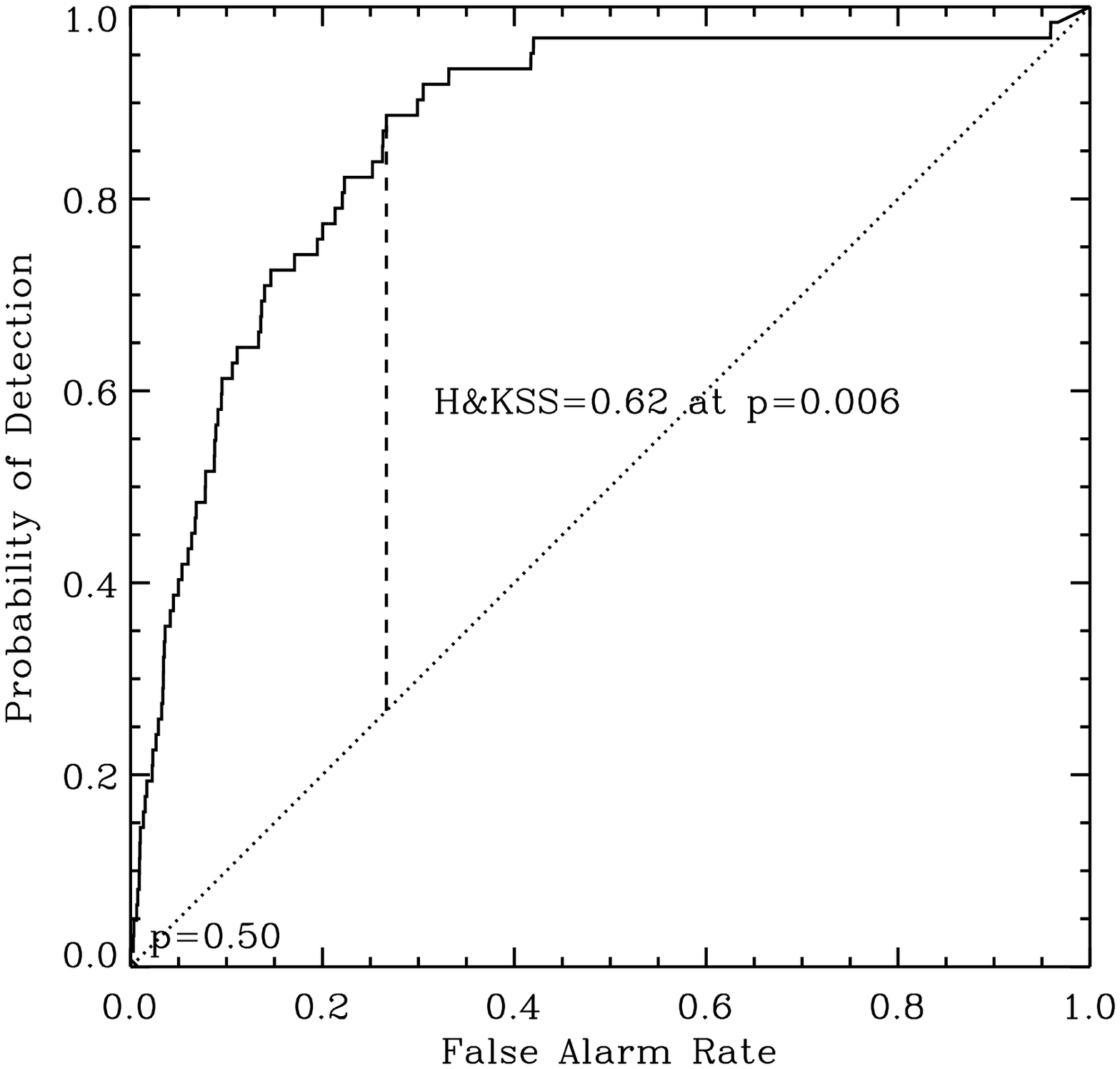}}
\caption{Same as Figure~\ref{fig:beff_plots} but for the Generalized Correlation Dimensions
parameter with NPDA.}
\label{fig:gcd_plots}
\end{figure}

\subsection{Magnetic Charge Topology and Discriminant Analysis - G.~Barnes,
K.D.~Leka}\label{sec:MCT}

\begin{figure}
\plottwo{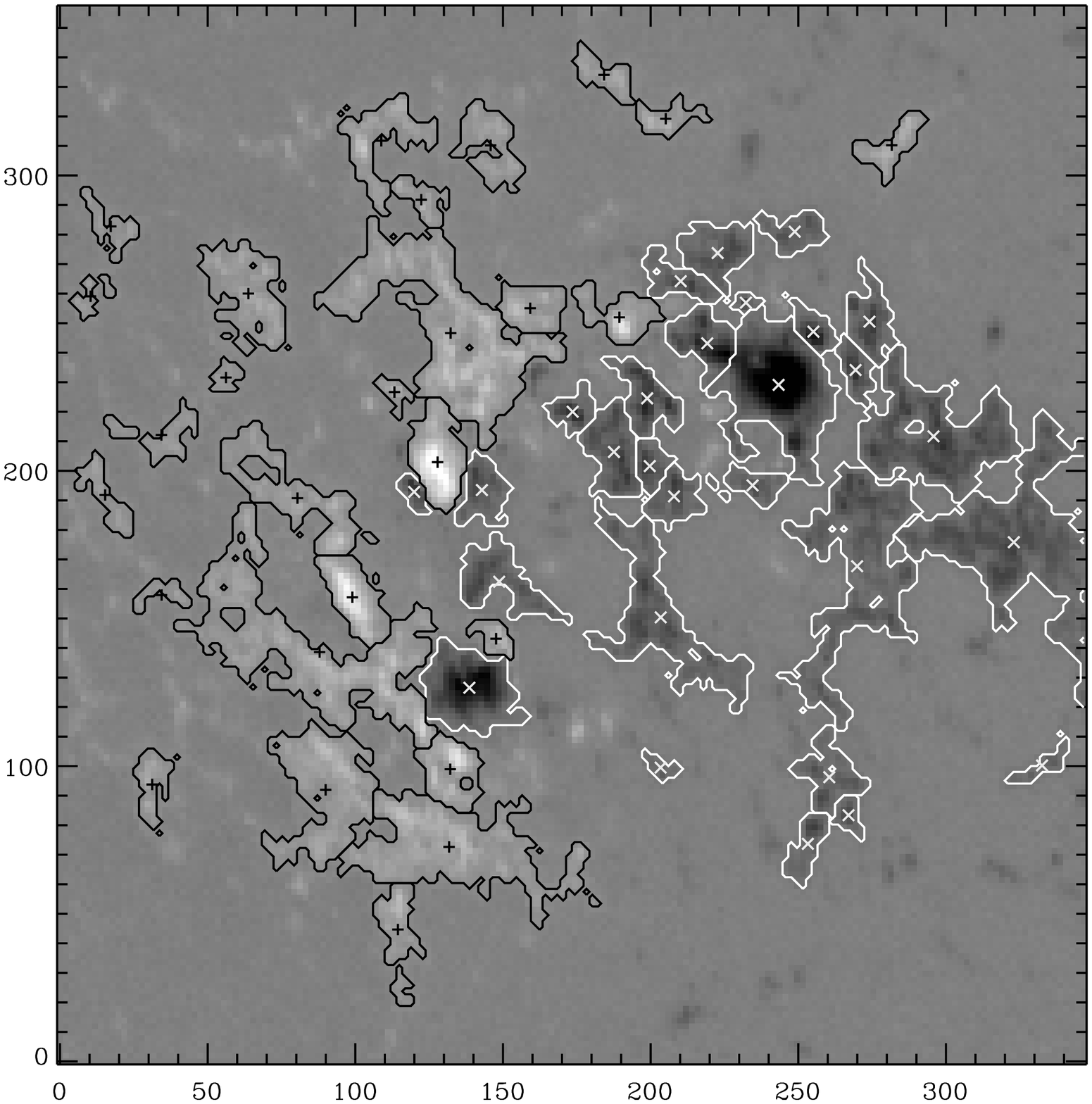}{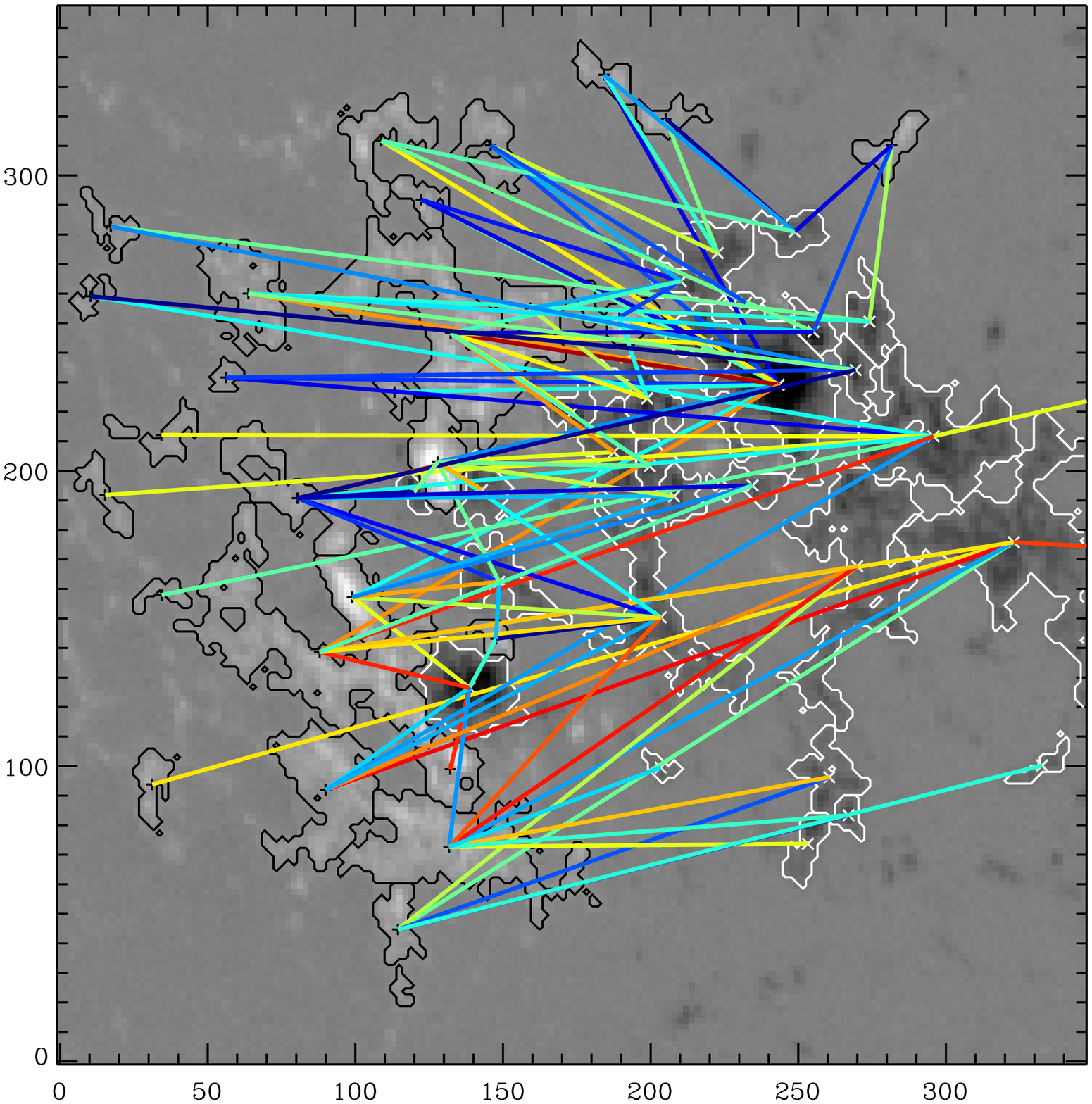}
\caption{Magnetic Charge Topology method. Example of the partitioning and
connectivity for NOAA AR\,9767 on 2002 January 3, for the MCT model described
in Section~\ref{sec:MCT}, in the same format as Fig.~\ref{beff_fig}.  Note that
there are substantial differences in the partitions (left panel), with this
implementation typically representing the plage with fewer sources than that
used to compute $B_{\rm eff}$ (\S\ref{sec:beff}). Likewise, there are
noticeable differences in the connectivity matrix, with this implementation
typically having more connections per source, despite fewer overall sources.
}
\label{fig:MCT}
\end{figure}

A corona with a very complex magnetic topology is one which will more readily
allow initiation of fast magnetic reconnection, and hence an energetic event.
Indeed, \citet{dfa2} demonstrated that parameters derived using a Magnetic
Charge Topology coronal model \citep[MCT;][and reference therein]{mct}
out-performed those describing the photospheric field in distinguishing
flare-imminent from flare-quiet times for seven active regions \citep[{\it
c.f.}, Table 5 of][]{dfa2}.  To implement the MCT model, the following steps
are followed:
\begin{itemize}
\item Partition the photospheric field into magnetic field concentrations and 
represent each partition as a single point source, located at the
flux-weighted center of the partition, $\x_i$, with magnitude equal to the flux in the
partition, $\Phi_i$.   
\item Represent the coronal magnetic field as the potential field associated
with this collection of point sources. 
\item Define the magnetic connectivity matrix as the amount of magnetic flux 
in each coronal connection, $\psi_{ij}$, and estimate it by tracing field lines.
\item Calculate the location of magnetic null points (places where the magnetic
field vanishes) using the method described in \cite{Barnes2007}.
\end{itemize}
An example of the results is shown in Figure~\ref{fig:MCT}.  Compared to the
partitioning used to compute $B_{\rm eff}$ (\S\ref{sec:beff}), this
implementation represents the plage areas with fewer sources.  Despite this,
the resulting connectivity matrix typically has more connections per source.

\begin{figure}
\centerline{
\includegraphics[width=0.33\textwidth, clip]{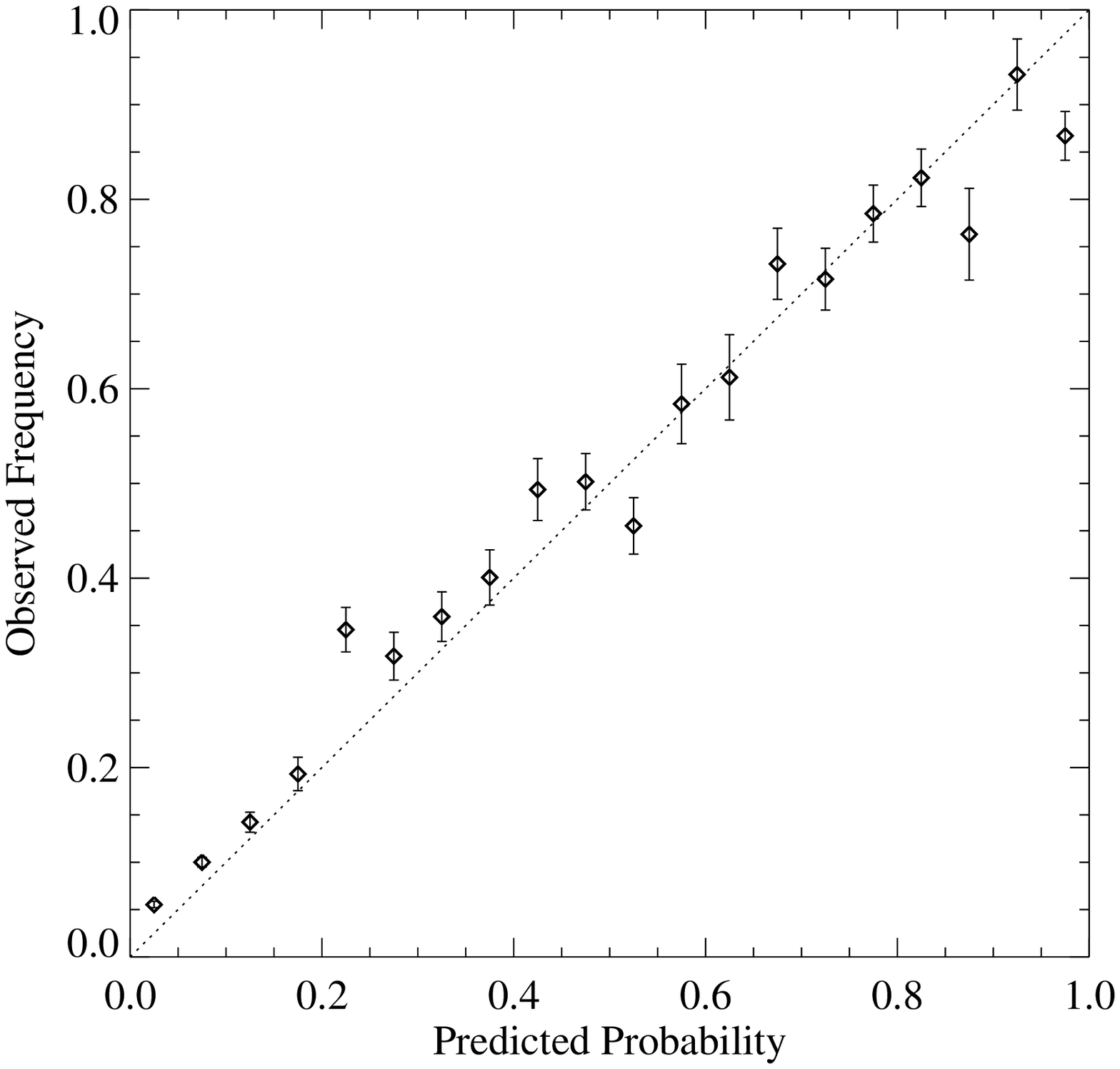}
\includegraphics[width=0.33\textwidth, clip]{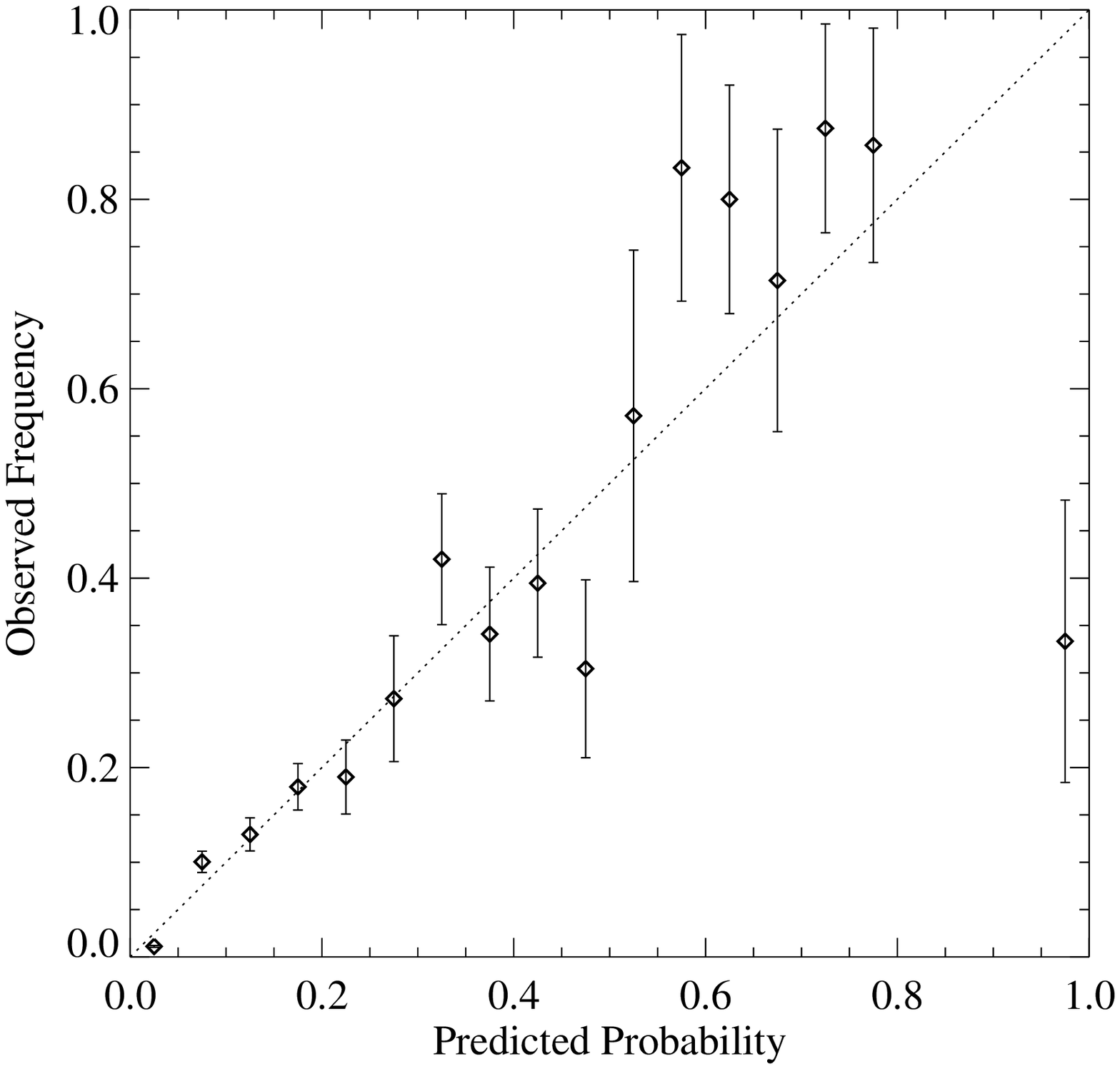}
\includegraphics[width=0.33\textwidth, clip]{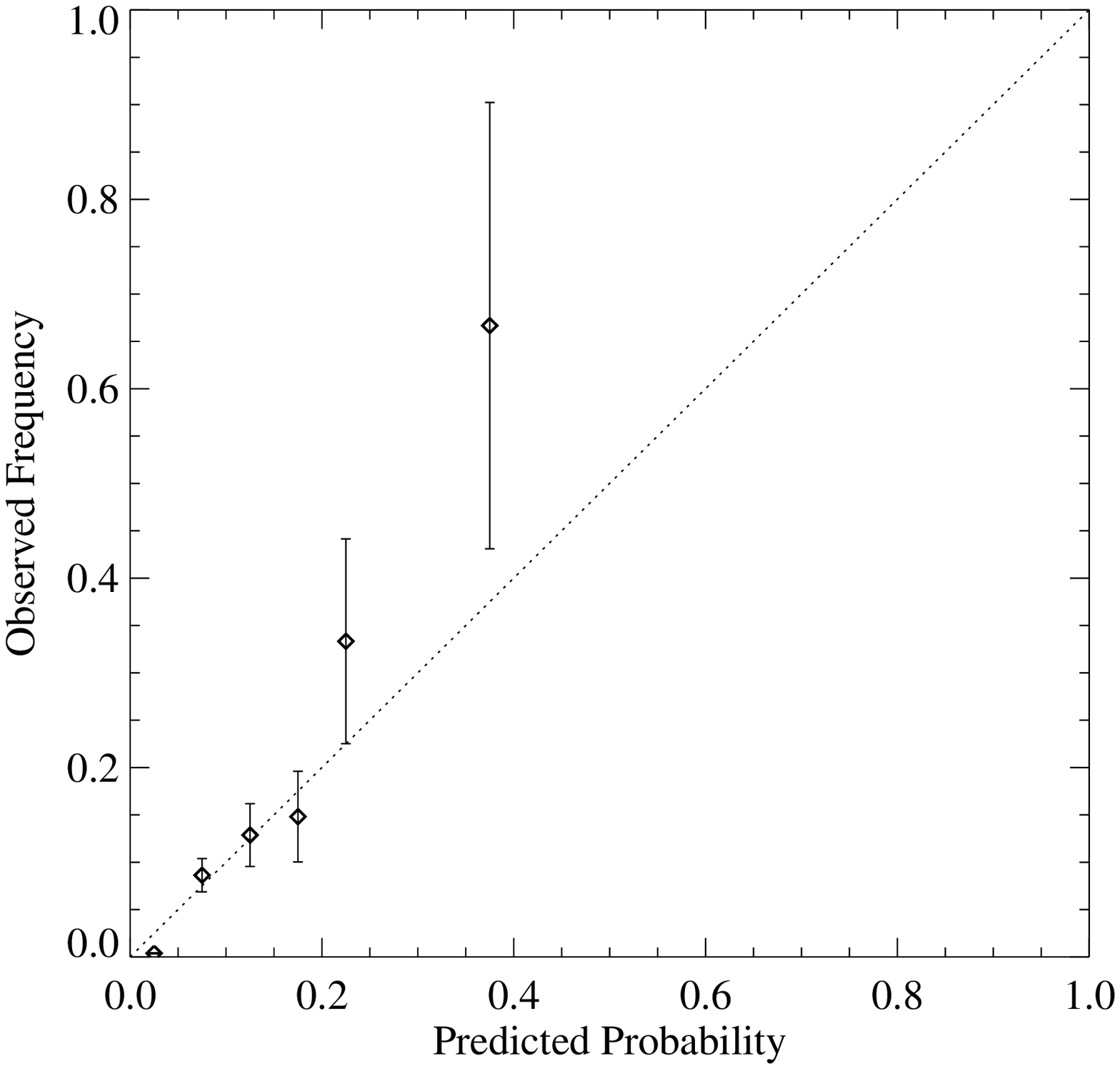}}
\centerline{
\includegraphics[width=0.33\textwidth, clip]{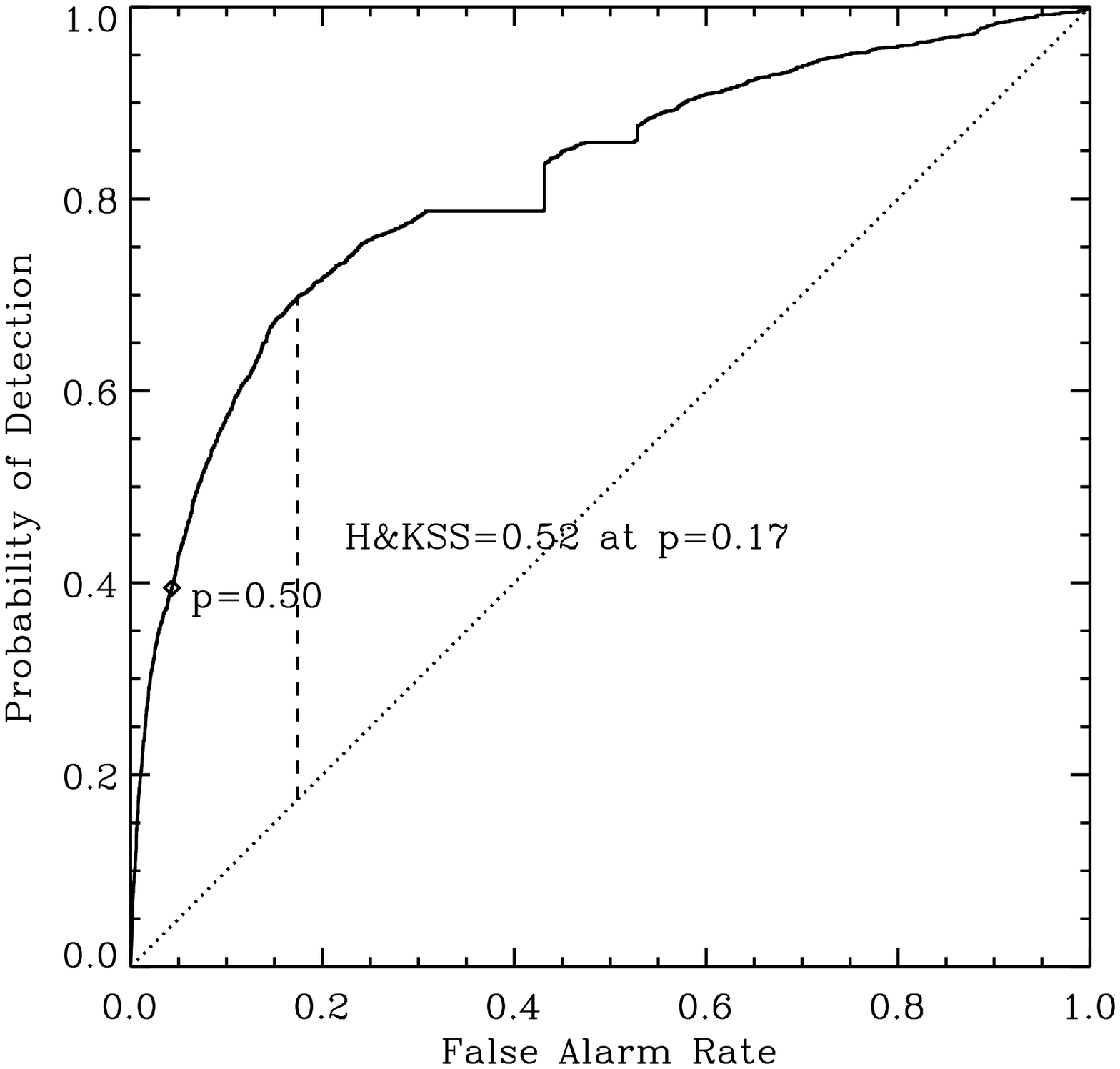}
\includegraphics[width=0.33\textwidth, clip]{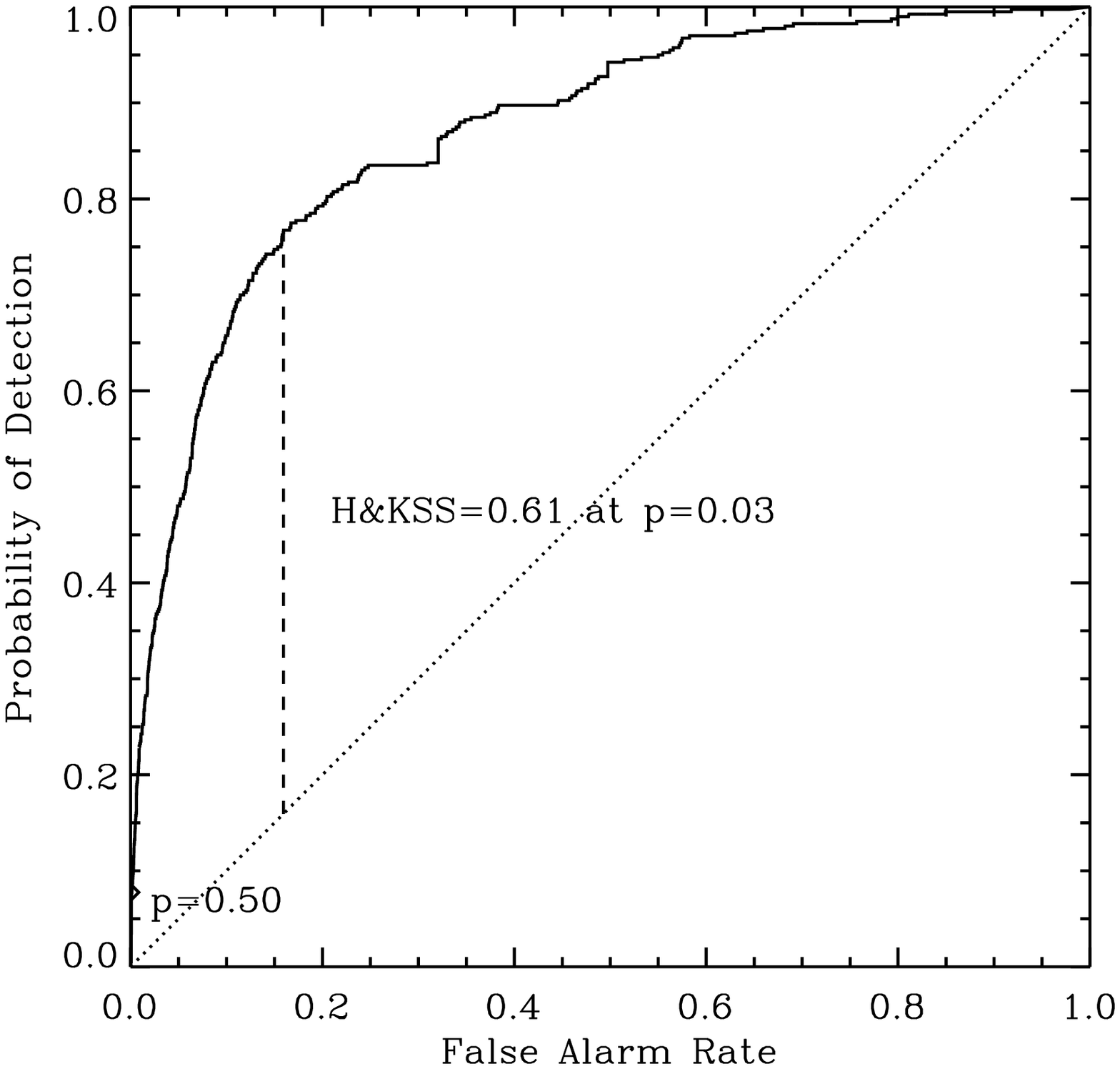}
\includegraphics[width=0.33\textwidth, clip]{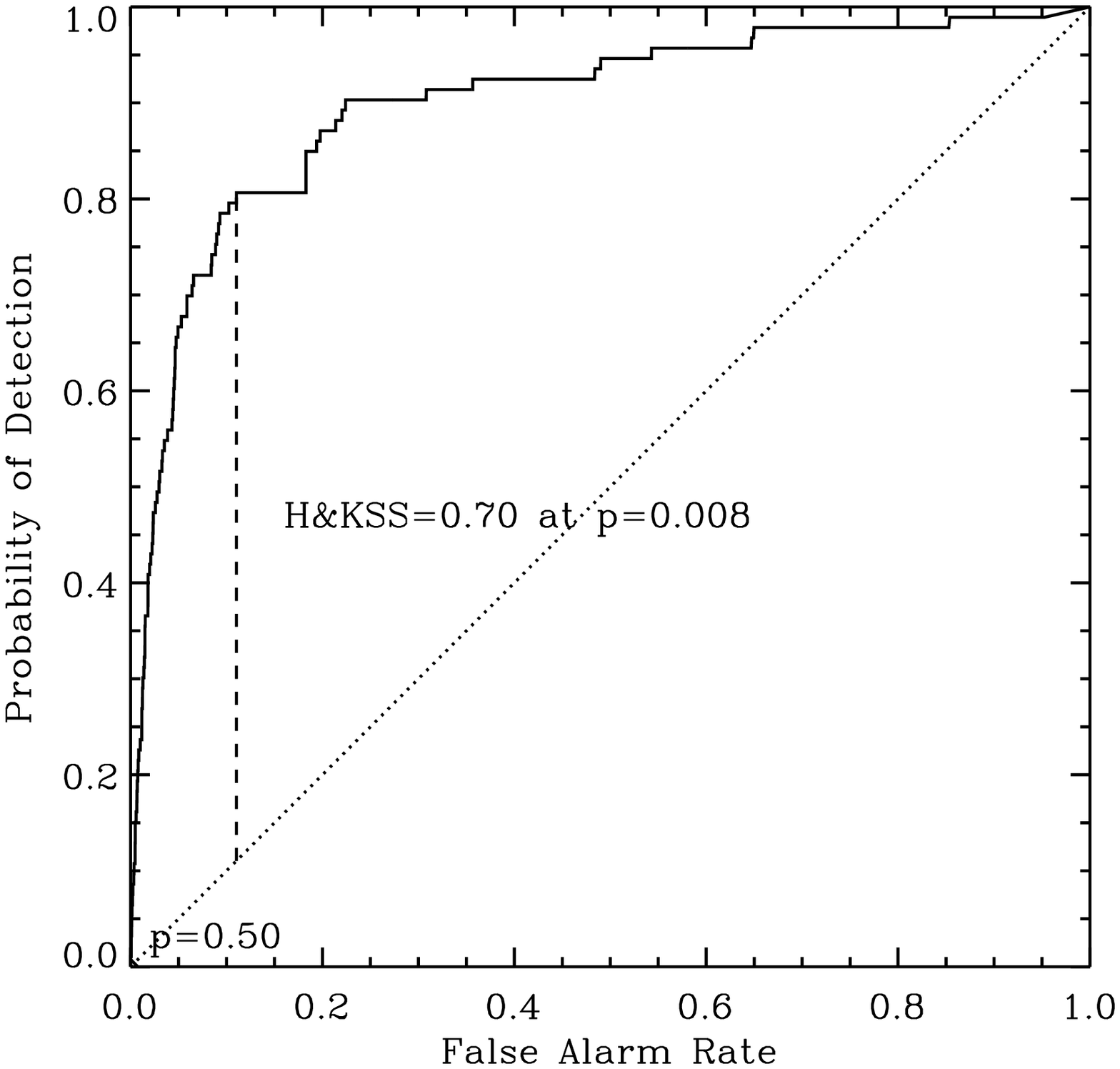}}
\caption{Same as Figure~\ref{fig:beff_plots} but for Magnetic Charge Topology 
parameters and forecasts using 2-variable NPDA.}
\label{fig:mct_plots}
\end{figure}

Almost 50 parameters were calculated based on moments and totals of the
distributions of properties of the sources and the connectivity matrix.  These
properties characterize quantities such as the number, orientation, and flux in
the connections \citep{dfa2,mct,BarnesLeka2008}, and include $\phi_{\rm tot} =
\sum \psi_{ij}/ \vert {\bf x}_i - {\bf x}_j \vert$, the total of each
connection's distance-weighted connection flux, as well as moments of the
distribution of $\psi_{ij}/ \vert {\bf x}_i - {\bf x}_j \vert$, and the moments
of the distribution of $\psi_{ij}/ \vert {\bf x}_i - {\bf x}_j \vert^2 $.  The
total of this quantity, denoted $\phi_{2, \rm tot}$, is essentially the same as
$B_{\rm eff}$ \citep[Section~\ref{sec:beff};][but see the discussion in
\citet{BarnesLeka2008}]{GeorgoulisRust2007} except in how the connectivity matrix is inferred.  Nonparametric Discriminant
Analysis with cross-validation is used to make a prediction
\citep[Section~\ref{sec:nwra_mag},][]{dfa2}, with the resulting skill scores
for the variable combinations that resulted in the best Brier skill score
values shown in Table~\ref{tbl:mct_best}; the \ML\ \Brier\ is one of the better
quoted in this study, but still not overly impressive.    The reliability plots
(Figure~\ref{fig:mct_plots}, top) show a slight tendency for underprediction at
high probabilities for the \MS\ and \ML\ categories. The ROC curves
(Figure~\ref{fig:mct_plots}, bottom) are fairly typical.

\begin{center}
\begin{deluxetable}{cccccccc}  
\tablecolumns{8}
\tablewidth{0pc}
\tablecaption{Optimal Performance Results: Magnetic Charge Topology, 2-Variable Combinations, Non-Parametric Discriminant Analysis} 
\tablehead{ 
\colhead{Event} & \colhead{Sample} & \colhead{Event} & \colhead{\RC} & \colhead{\Heidke} & \colhead{\Appleman} & \colhead{\True} & \colhead{\Brier} \\
\colhead{Definition} & \colhead{Size} &  \colhead{Rate} & \colhead{(threshold)} & \colhead{(threshold)} & \colhead{(threshold)} & \colhead{(threshold)} & \colhead{} 
} 
\startdata 
\CC\tablenotemark{a} 
 & 12965 & 0.201 & 0.85 (0.55) & 0.48 (0.29) & 0.23 (0.55) & 0.52 (0.17) & 0.28\\
\MS\tablenotemark{b} 
 &   ''  & 0.031 & 0.97 (0.50) & 0.31 (0.19) & 0.05 (0.50) & 0.61 (0.03) & 0.14\\
\ML\tablenotemark{b} 
 &   ''  & 0.007 & 0.99 (0.33) & 0.20 (0.08) & 0.00 (0.33) & 0.70 (0.01) & 0.06\\
\enddata 
\tablenotetext{a}{Variable combination: [$\phi_{2,\rm tot},\ \sigma(\phi_{2})$].}
\tablenotetext{b}{Variable combination: [$\phi_{\rm tot},\ \overline{|\x^{\rm null}_i - \x^{\rm null}_j|}\ $].}
\tablenotetext{c}{Variable combination: [$\phi_{\rm tot},\ \overline{|\x^{\rm null}_i - \x^{\rm null}_j|}\ $].}
\label{tbl:mct_best} 
\end{deluxetable}
\end{center}

\subsection{Solar Monitor Active Region Tracker with Cascade Correlation Neural
Networks - P.A.~Higgins, O.W.~Ahmed}
\label{sec:smart_ccnn}

The Solar Monitor Active Region Tracker method is incorporated into the
well-known {\tt SolarMonitor.org} resource, as the SMART2 code package, which
performs a combination of detecting, tracking, and characterizing active
regions \citep{Higgins_etal_2011}.   For the present comparison, the provided
patches are used rather than the full-disk MDI data, but all subsequent
analysis proceeds as described in \citet{Higgins_etal_2011}.  This includes
smoothing and thresholding to differentiate plage from spot regions, and an
observing angle correction similar to other methods, but no additional
multiplicative factor correction.  

Since only one magnetogram per day was provided, the parameters that
characterize temporal variations were not calculated.  Twenty parameters were
calculated including:
\begin{itemize}
\item Area of the final dilated mask, and relevant totals, Net, min/max and moments of 
the flux within the final dilated mask, and the area of the mask.
\item Maximum, mean, and median of the field spatial gradients.
\item Length of the line-of-sight polarity separation lines and 
strong-gradient neutral lines.
\item Amount of Flux near strong-gradient polarity separation lines and
non-potentiality gauges following \citet{Schrijver2007} (Section~\ref{sec:R})
and \citet{Falconer_etal_2008} (Section~\ref{sec:WLSG2}), with two different
thresholds applied.
\end{itemize}

These parameters are used to make forecasts using the Cascade Correlation
Neural Networks method \citep[CCNN;][]{Qahwaji_etal_2008} implemented in
\cite{Ahmedetal2013}.  For the machine-learning task, all active regions with a
finite value of the parameters were considered, providing 8137 forecasts, and
the six years are alternately rotated to enable training on five years of data
at a time, in a jackknife manner \citep[e.g.,][]{EfronGong1983}.  This training
is performed separately for each flare event definition, using all parameters
simultaneously.  The CCNN is trained to optimize \True.  The skill results
shown in Table~\ref{tbl:ccnn_best} confirm this, with the highest \True\ and
\Heidke\ scores of any method, especially for the larger thresholds -- but with
the worst \Brier\ skill of any method, as well.  

\begin{figure}
\centerline{
\includegraphics[width=0.33\textwidth, clip]{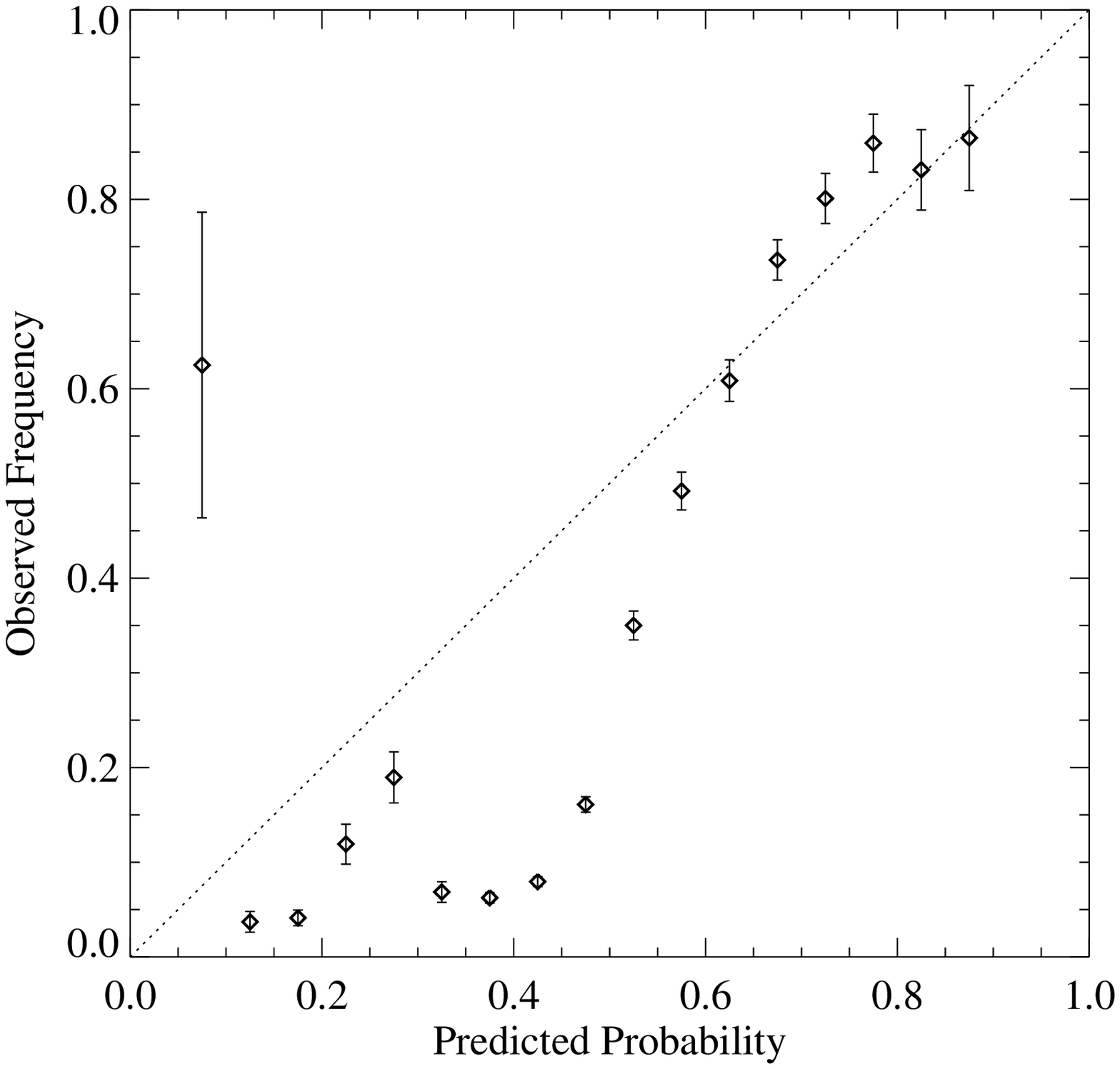}
\includegraphics[width=0.33\textwidth, clip]{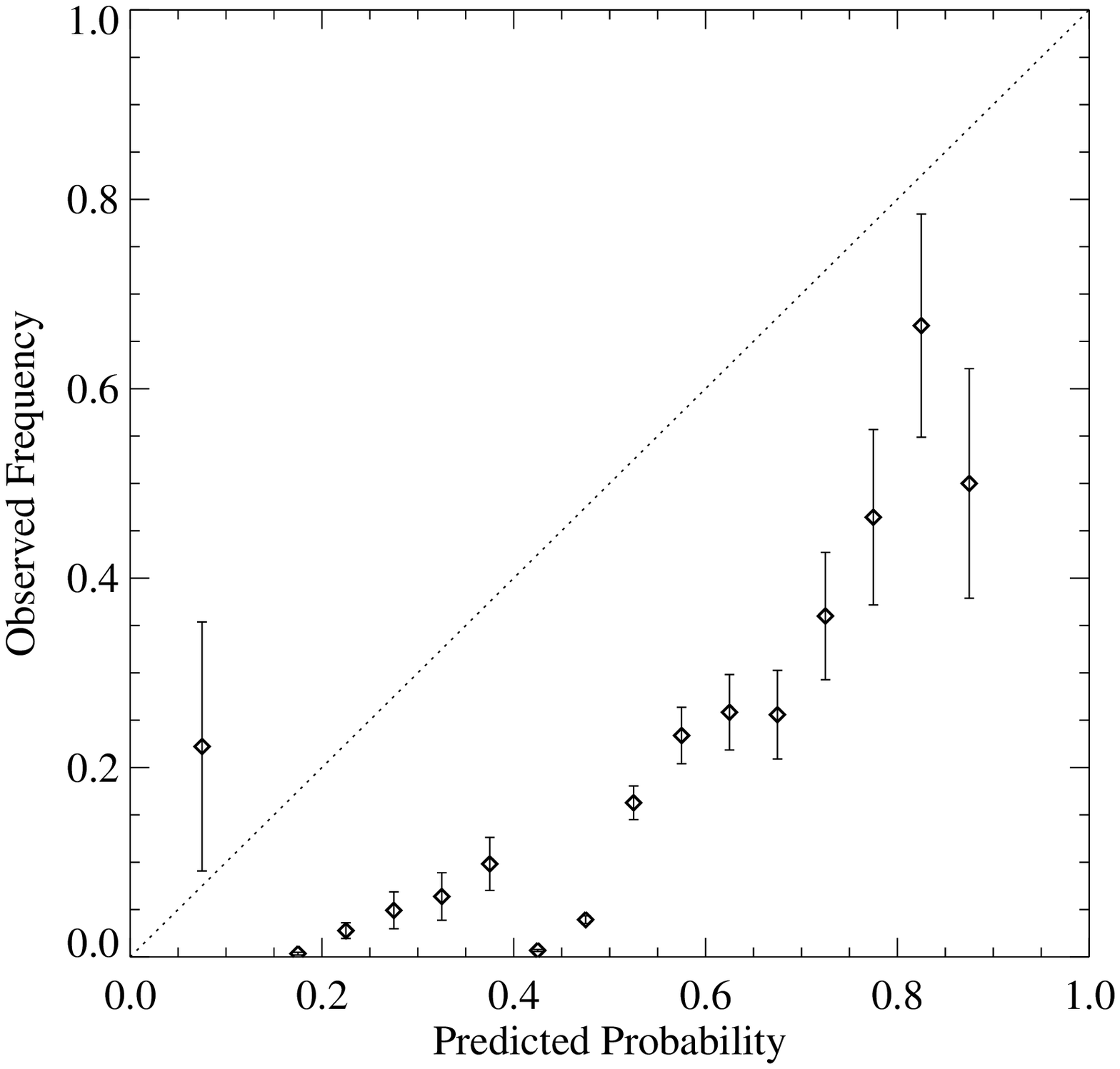}
\includegraphics[width=0.33\textwidth, clip]{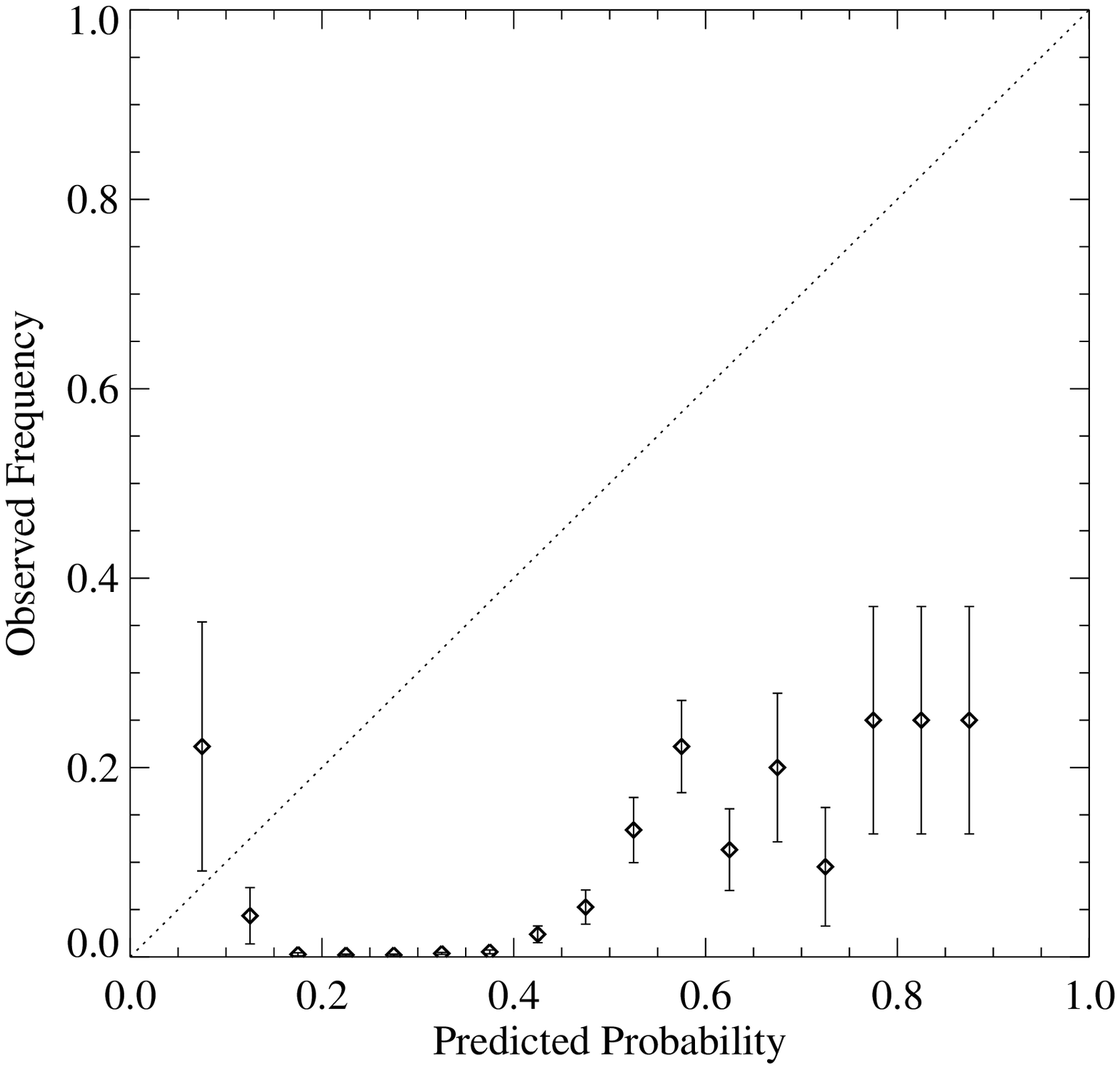}}
\centerline{
\includegraphics[width=0.33\textwidth, clip]{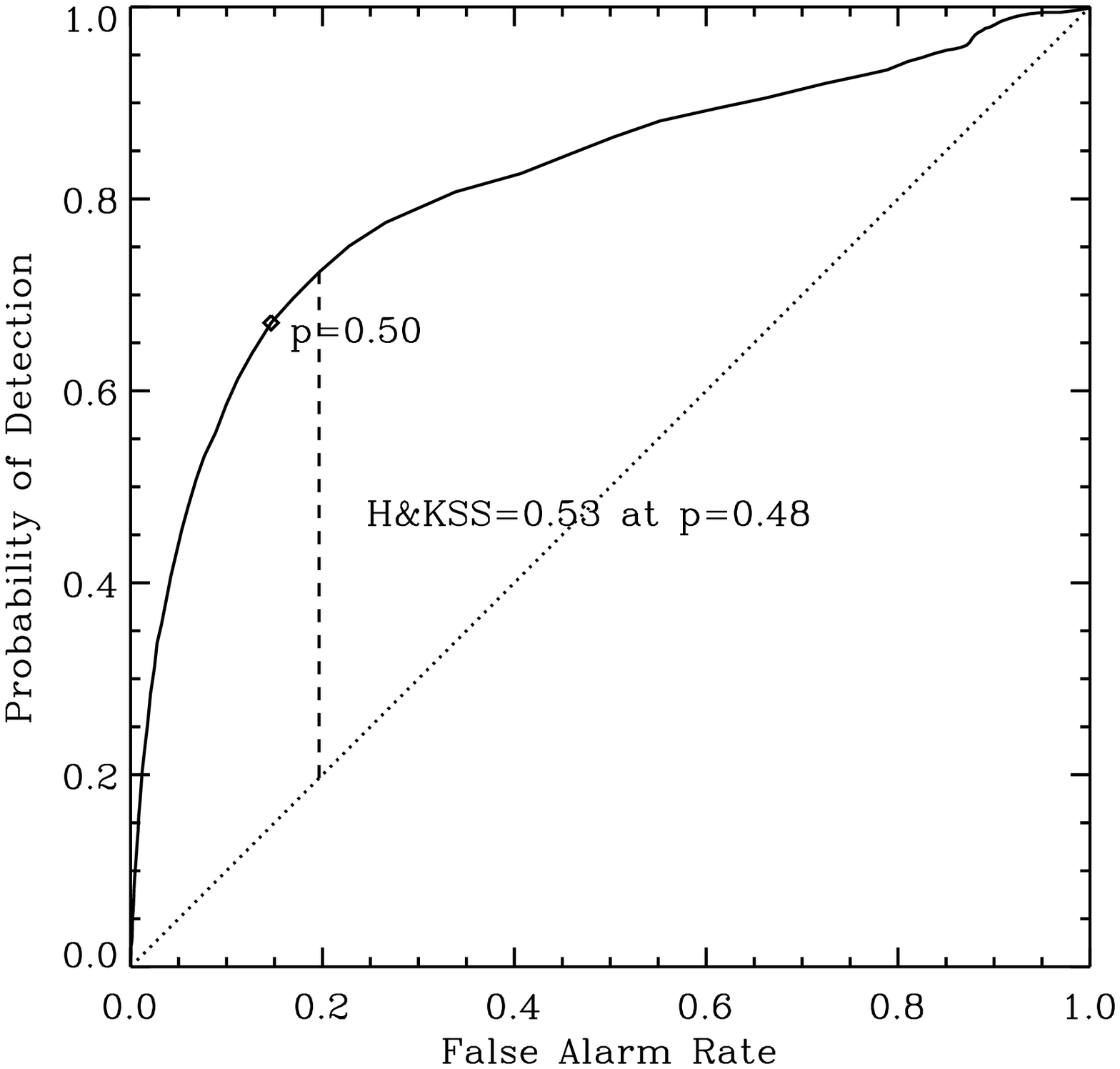}
\includegraphics[width=0.33\textwidth, clip]{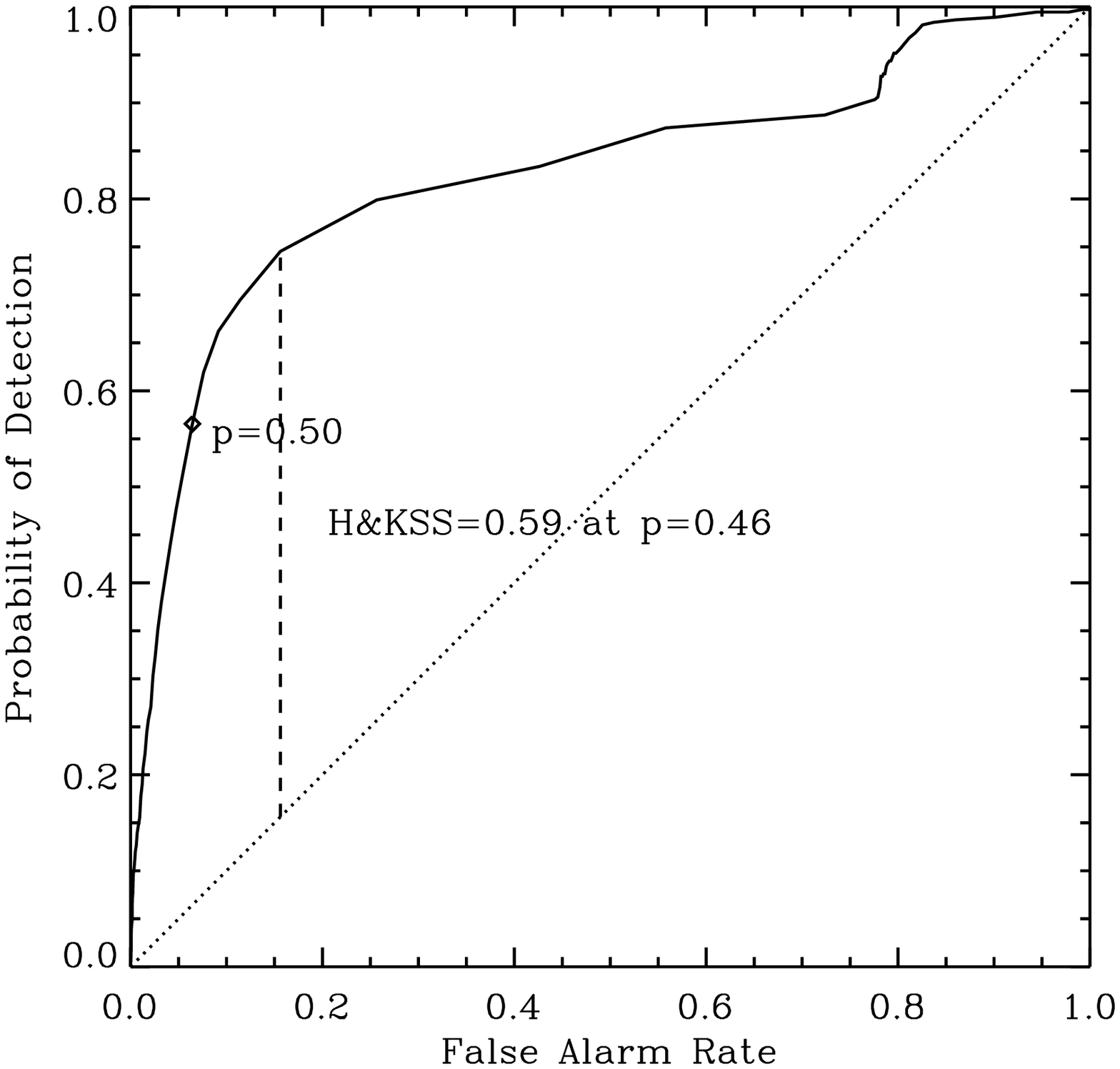}
\includegraphics[width=0.33\textwidth, clip]{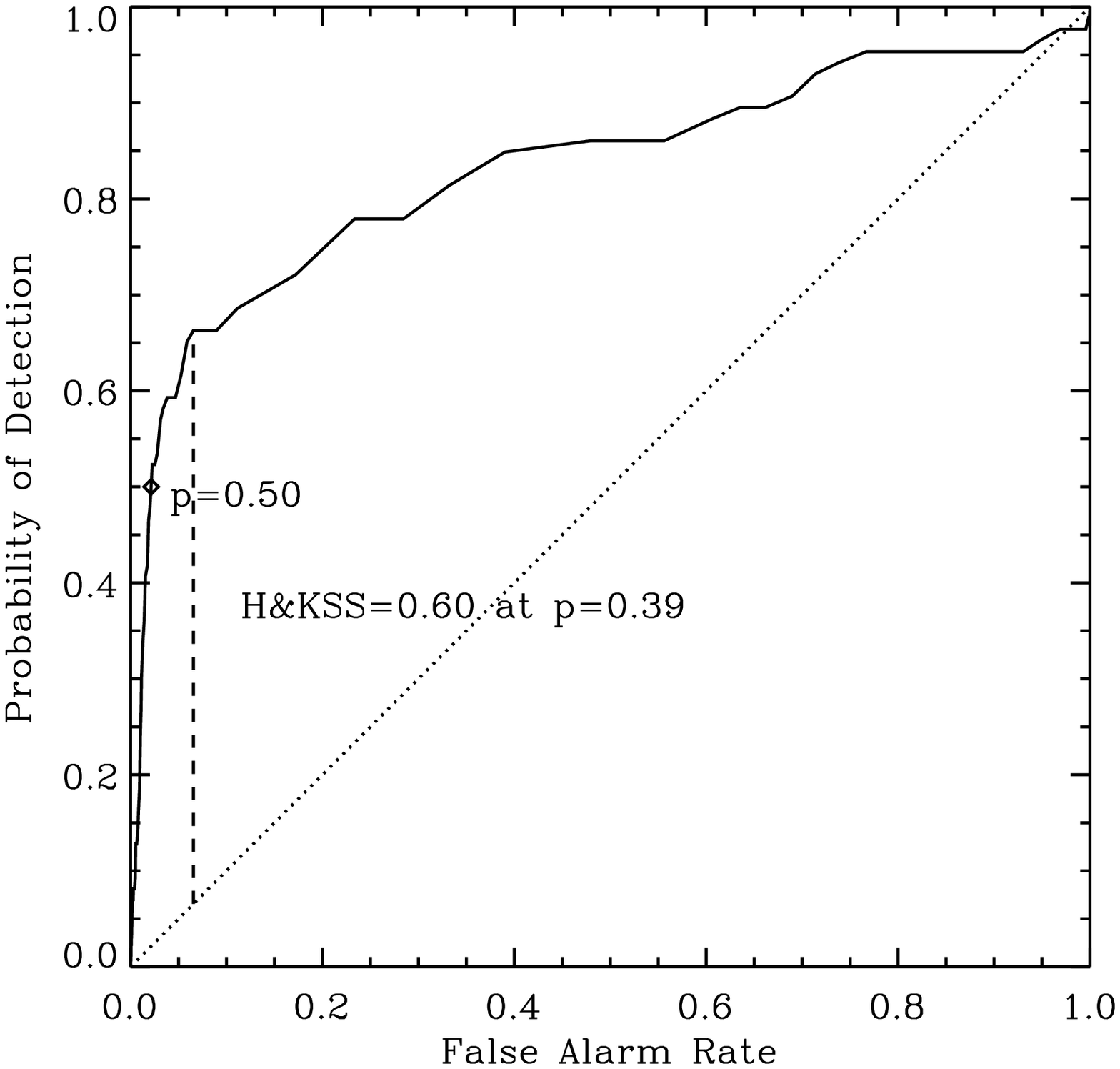}}
\caption{Same as Figure~\ref{fig:beff_plots} but for the SMART2 with Cascade Correlation 
Neural Networks.}
\label{fig:ccnn_plots}
\end{figure}

The reliability plots (Figure~\ref{fig:ccnn_plots}, top) show that the method
is substantially overpredicting events in almost all cases.  This is likely a 
consequence of optimizing on the \True: by overpredicting and using a threshold
of 0.5 to convert to a categorical forecast, a similar classification table 
is achieved to making accurate probabilistic forecasts but using a lower 
threshold for converting to categorical forecasts.  This appears to be supported
by the ROC plots (Figure~\ref{fig:ccnn_plots}, bottom), in which the largest 
\True\ is obtained for a threshold value close to 0.5, as compared to most 
methods for which the threshold for the maximum \True\ is much lower.  The ROC
curves also show a relatively rapid decrease in the probability of detection
as the false alarm rate decreases, indicating that this method may not be 
well suited to issuing all-clear forecasts.

\begin{center}
\begin{deluxetable}{cccccccc}  
\tablecolumns{8}
\tablewidth{0pc}
\tablecaption{Optimal Performance Results: SMART2 with Cascade Correlation Neural Networks} 
\tablehead{ 
\colhead{Event} & \colhead{Sample} & \colhead{Event} & \colhead{\RC} & \colhead{\Heidke} & \colhead{\Appleman} & \colhead{\True} & \colhead{\Brier} \\
\colhead{Definition} & \colhead{Size} &  \colhead{Rate} & \colhead{(threshold)} & \colhead{(threshold)} & \colhead{(threshold)} & \colhead{(threshold)} & \colhead{} 
} 
\startdata 
\CC & 11536 & 0.212 & 0.84 (0.58) & 0.50 (0.52) & 0.26 (0.58) & 0.53 (0.48) &\ -0.13\\
\MS &   ''  & 0.032 & 0.97 (0.76) & 0.30 (0.53) & 0.02 (0.76) & 0.59 (0.46) &\ -4.63\\
\ML &   ''  & 0.007 & 0.99 (0.90) & 0.23 (0.52) & 0.00 (0.90) & 0.60 (0.39) & -12.10\\
\enddata
\label{tbl:ccnn_best}
\end{deluxetable}
\end{center}

\subsection{Event Statistics - M.~Wheatland}
\label{sec:wheatland}

\citet{whe04a} presented a Bayesian method to predict flaring
probability for different flare sizes using only the flaring history of
observed active regions. Since it uses only flare history as input, the results
from this approach serve as a baseline for comparison with methods using
magnetogram and/or white-light image data. 

The event statistics method assumes that solar flares (the events) 
obey a power-law frequency-size distribution: 
\begin{equation} 
N(S) = \lambda_1 (\gamma - 1) S_1^{\gamma - 1} S^{-\gamma},
\end{equation}
where $N(S)$ denotes the number of events per unit time and per unit
size $S$, where $\lambda_1$ is the mean rate of events above size 
$S_1$, and where $\gamma$ is the power-law index.
The method also assumes that events occur randomly in time, on short 
timescales following a Poisson process with a constant mean rate, so
that the event waiting-time distribution above size $S$ is
\begin{equation} 
P(\tau) = \lambda \exp (- \lambda \tau),
\end{equation}
where $\tau$ denotes the waiting time for events above size $S$,
and where $\lambda$ is the corresponding mean rate of events.
Given a past history of events above a small size $S_1$, the method
infers the current mean rate $\lambda_1$ of events subject to the 
Poisson assumption, and then uses the power-law distribution to infer 
probabilities for occurrence of larger events within a given time.

\begin{figure}
\plotone{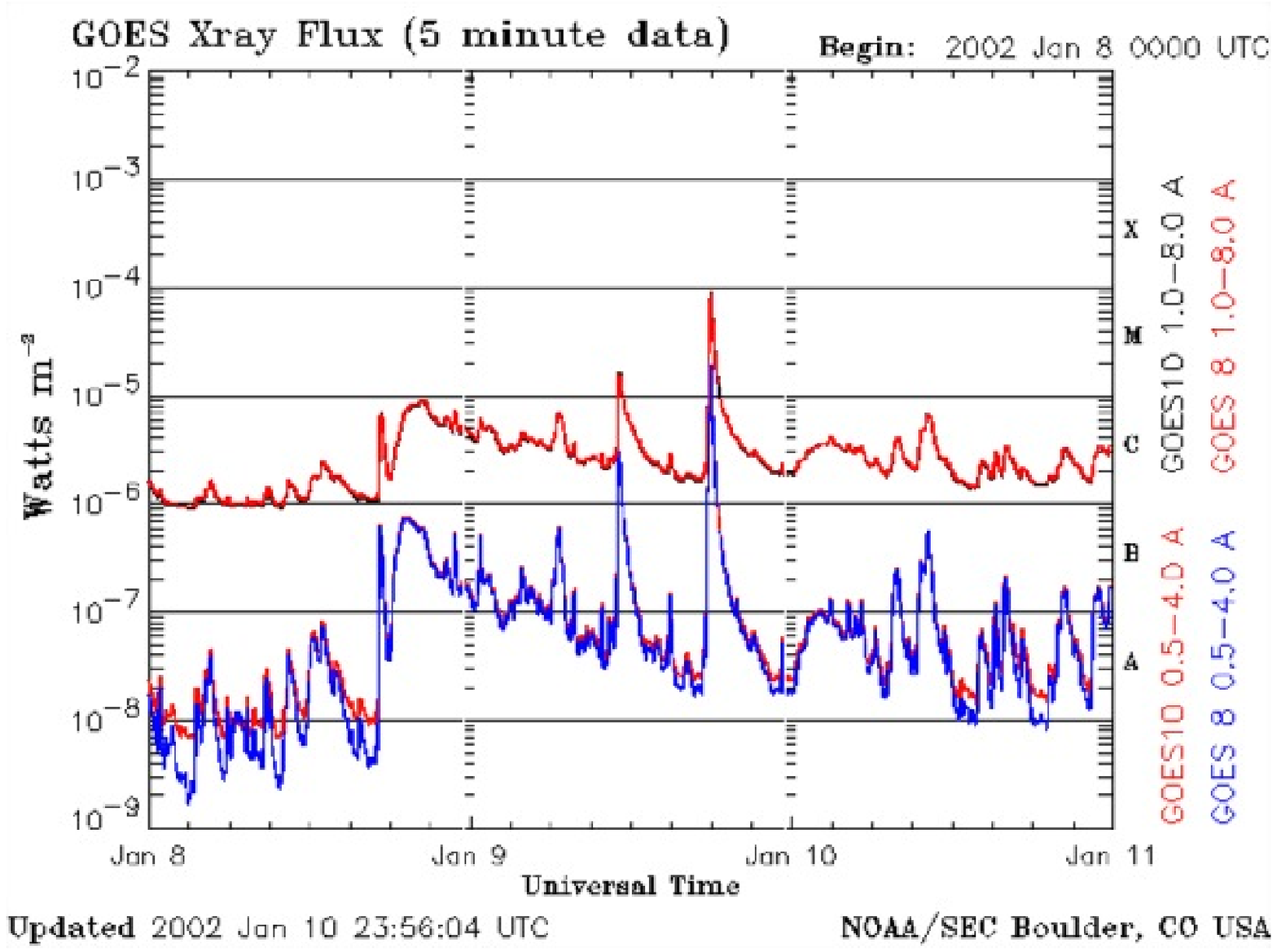}
\caption{Event Statistics method.  The number, timing, and size of flares as
tabulated using GOES data for each active region are the sole data required by
the event statistics prediction method. This figure shows a GOES time history
including the M-class flare on 2002 January 09.
}
\label{fig:GOES}
\end{figure}

\citet{Wheatland2005} presented an
implementation of the event statistics method for whole-Sun prediction 
of GOES soft X-ray events. Figure~\ref{fig:GOES} shows a GOES light 
curve, from which GOES event lists are routinely produced.
In the application to GOES data the event size $S$ is 
taken to be the peak 1-8\,\AA ~GOES flux in the GOES X-ray event 
lists, and the times of events are identified with the corresponding 
tabulated peak times. The whole-Sun method presented in Wheatland 
(2005) is used in this study, and the method is also applied (for the
first time) to events in individual active regions.

Three applications of the method were run for these tests: active region
forecasts for which a minimum of five prior events was required for a
prediction, active region forecasts for which ten prior events were required,
and a full-disk prediction.  Since none of the other methods produced a
full-disk flare probability, those results are included for this study just for
completeness.  This method has no explicit restriction on the position of an
active region on the disk, although the minimum number of events means that
predictions are not made for regions that have just rotated into view.  The
sample sizes for the region forecasts with ten prior events are the smallest of
any method, indicating that a substantial fraction of regions do not produce
even ten events. 

A summary of skill results is given in Table~\ref{tbl:estat_best}.  The results
for \CC\ are arguably the worst of any method, but the approach is biased
against regions which produce only a few small events.  The values of the skill
scores generally increase when more prior events are included, so the five
prior event case has most skill score values lower than the ten prior event
case, which in turn has most skill score values lower than the full disk case.
It is likely that the larger number of prior events simply allows for a better
estimate of the power-law index and the mean rate of events above a given size.
The accuracy of the forecast is expected to scale as $N_1^{-1/2}$, where $N_1$
is the number of small events observed and used to infer the rate $\lambda_1$
\citep{whe04a}.

\begin{figure}
\centerline{
\includegraphics[width=0.33\textwidth, clip]{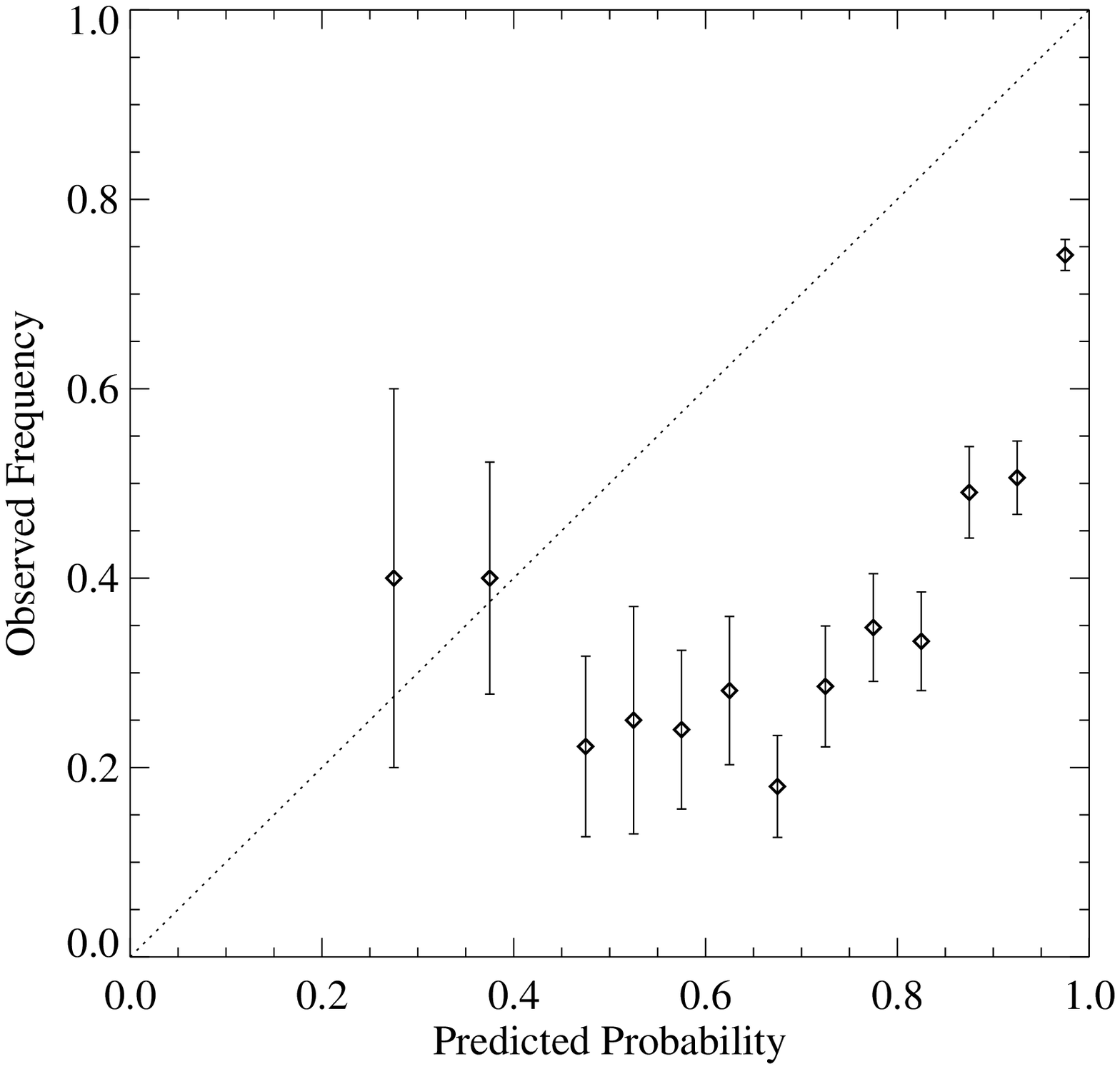}
\includegraphics[width=0.33\textwidth, clip]{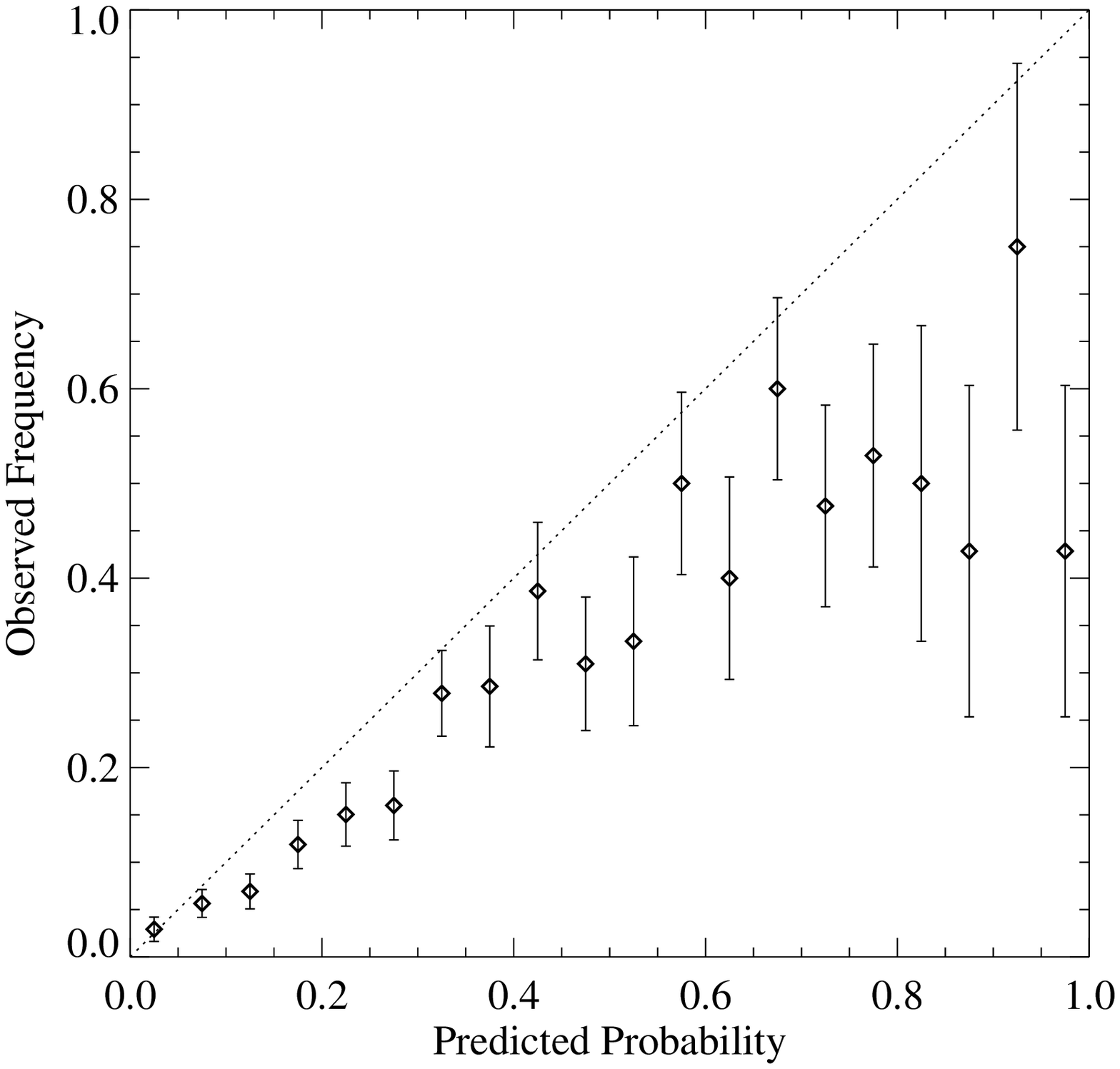}
\includegraphics[width=0.33\textwidth, clip]{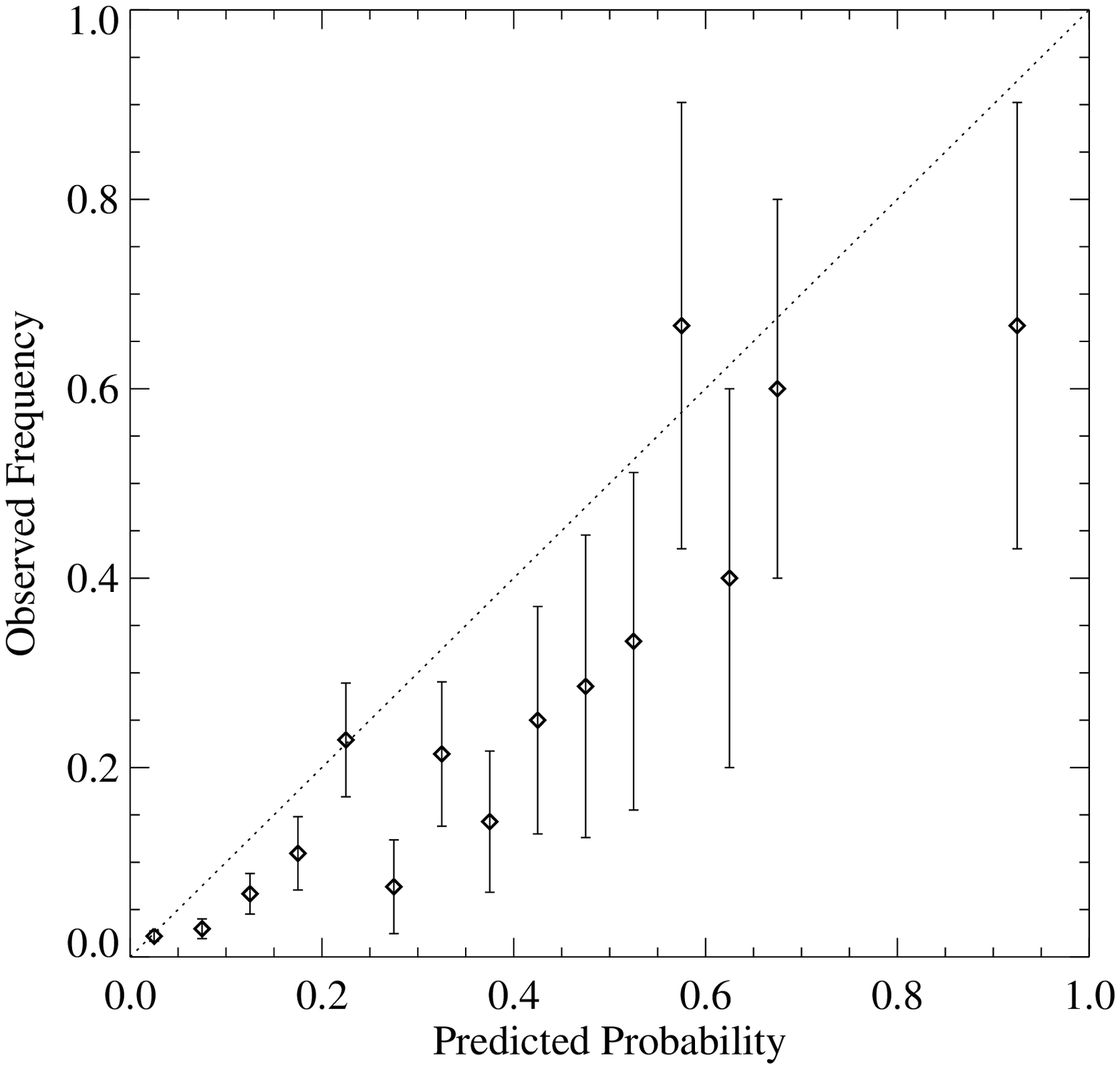}}
\centerline{
\includegraphics[width=0.33\textwidth, clip]{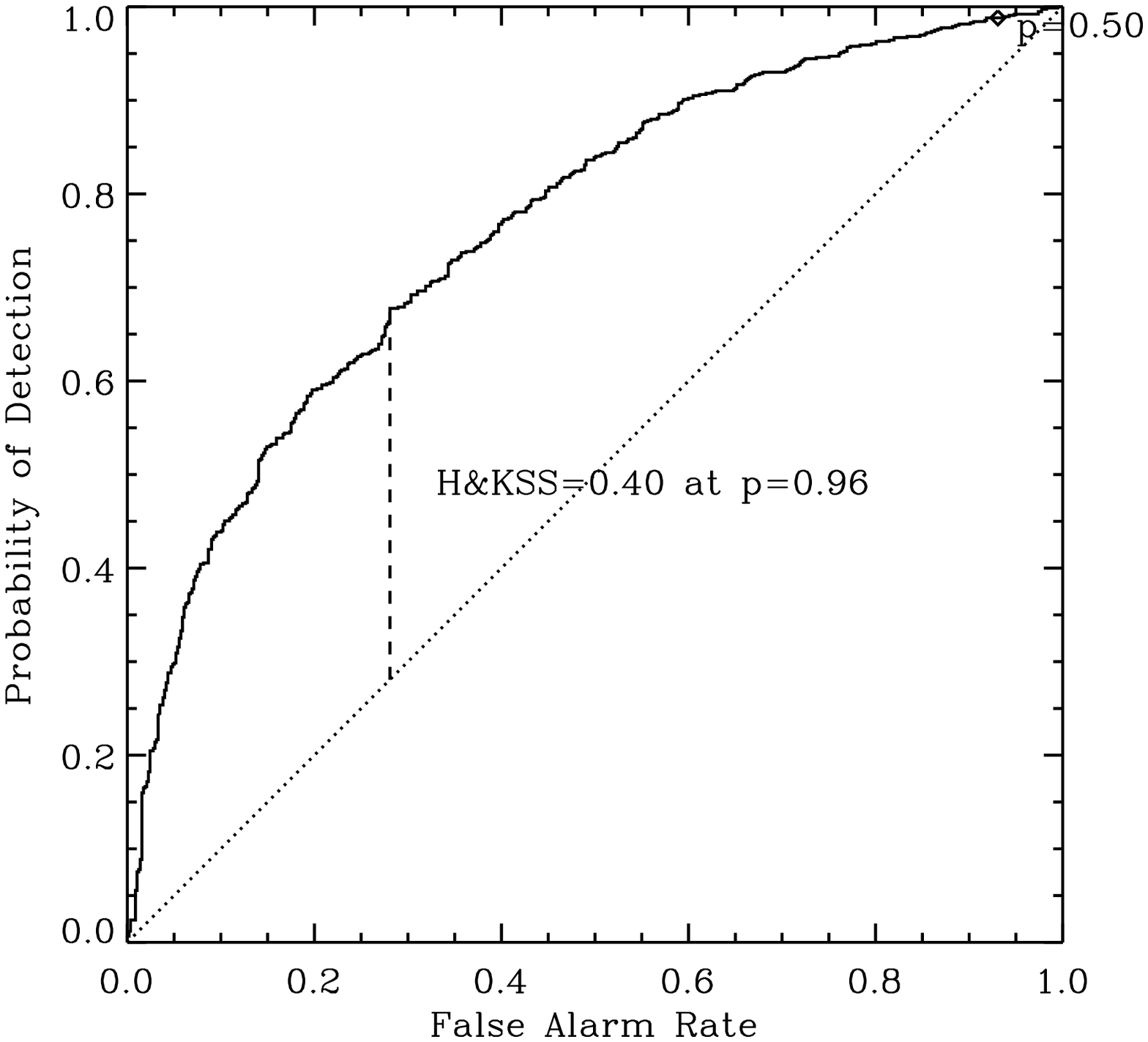}
\includegraphics[width=0.33\textwidth, clip]{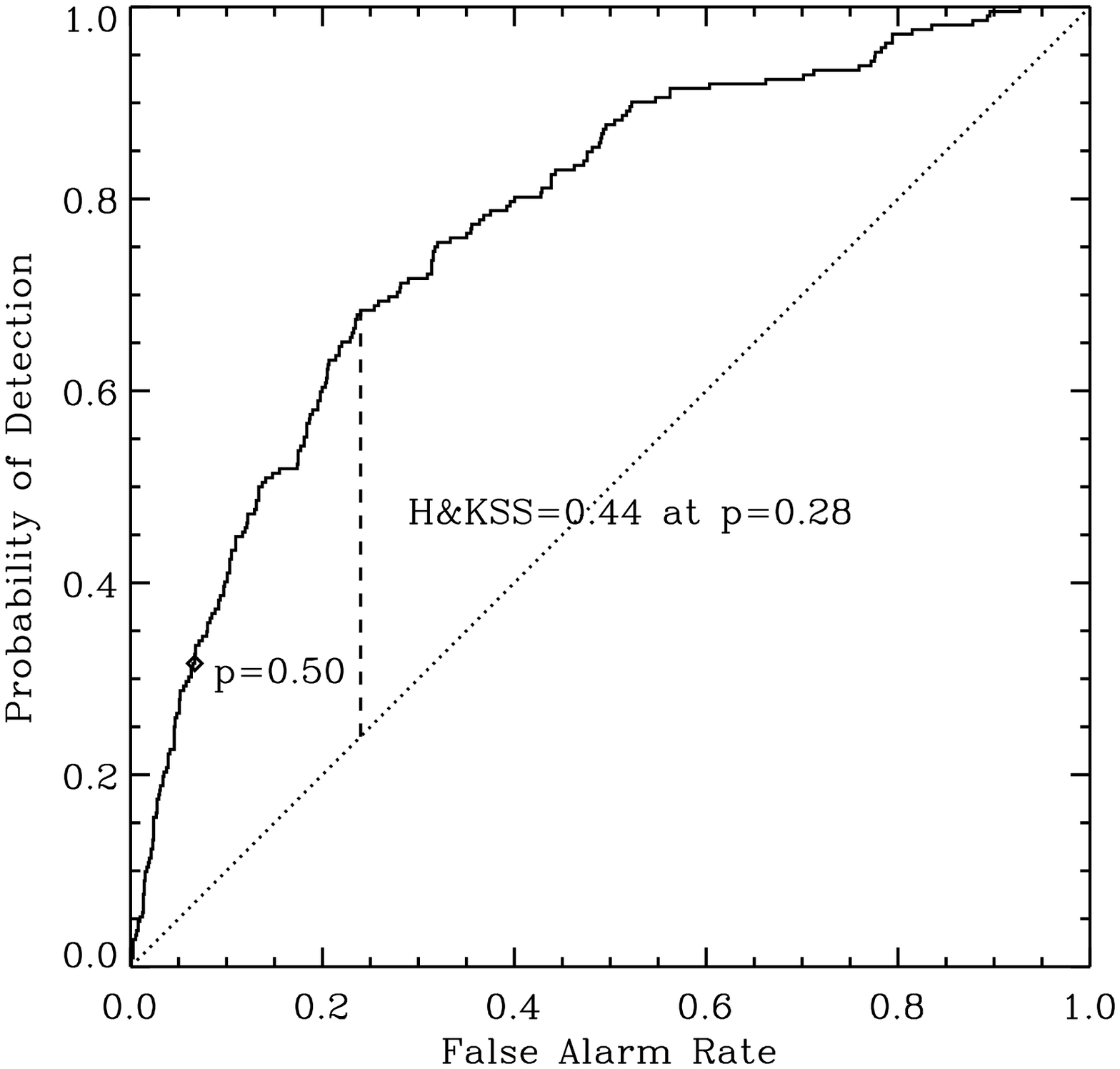}
\includegraphics[width=0.33\textwidth, clip]{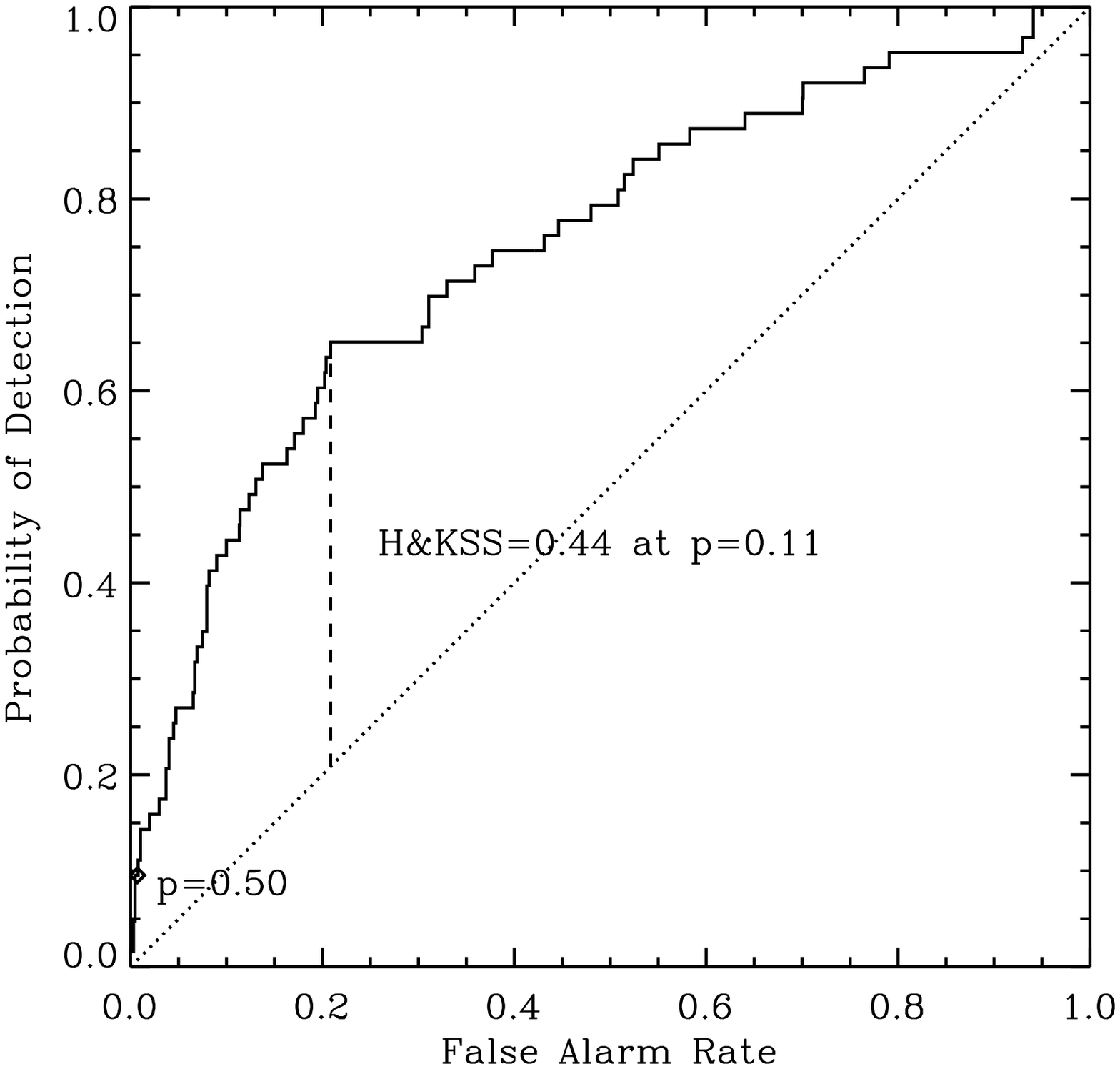}}
\caption{Same as Figure~\ref{fig:beff_plots}, but for the Event Statistics method for 
region forecasts, requiring 10 prior events.}
\label{fig:estat_plots}
\end{figure}

The summary plots (Figure~\ref{fig:estat_plots}) are shown only for the region
forecasts requiring ten prior events.  The reliability plots show a systematic
overprediction for all event definitions, and the ROC plots do not noticeably
improve with increasing threshold, in contrast to most other methods.  The peak
\True\ score is somewhat low and remarkably consistent across event
definitions.

\begin{center}
\begin{deluxetable}{cccccccc}
\tablecolumns{8}
\tablewidth{0pc}
\tablecaption{Optimal Performance Results: Event Statistics} 
\tablehead{ 
\colhead{Event} & \colhead{Sample} & \colhead{Event} & \colhead{\RC} & \colhead{\Heidke} & \colhead{\Appleman} & \colhead{\True} & \colhead{\Brier} \\
\colhead{Definition} & \colhead{Size} &  \colhead{Rate} & \colhead{(threshold)} & \colhead{(threshold)} & \colhead{(threshold)} & \colhead{(threshold)} & \colhead{} 
} 
\startdata 
\cutinhead{5 Prior Events:}
\CC & 2367 & 0.497 & 0.68 (0.92) & 0.35 (0.92) & 0.35 (0.92) & 0.35 (0.92) & -0.41\\
\MS &  ''  & 0.116 & 0.89 (0.65) & 0.34 (0.36) & 0.03 (0.65) & 0.46 (0.22) &\ 0.09\\
\ML &  ''  & 0.033 & 0.97 (0.88) & 0.22 (0.21) & 0.01 (0.88) & 0.51 (0.11) & -0.01\\
\cutinhead{10 Prior Events:}
\CC & 1334 & 0.567 & 0.70 (0.91) & 0.38 (0.94) & 0.30 (0.91) & 0.40 (0.96) & -0.28\\
\MS &  ''  & 0.159 & 0.84 (0.65) & 0.34 (0.36) & 0.02 (0.65) & 0.44 (0.28) &\ 0.09\\
\ML &  ''  & 0.047 & 0.95 (0.88) & 0.21 (0.21) & 0.02 (0.88) & 0.44 (0.11) & -0.03\\
\cutinhead{Full Disk:}
\CC & 12965 & 0.809 & 0.82 (0.66) & 0.36 (0.96) & 0.08 (0.66) & 0.41 (0.99) & -0.00\\
\MS &  ''   & 0.199 & 0.82 (0.62) & 0.34 (0.54) & 0.09 (0.62) & 0.39 (0.36) &\ 0.11\\
\ML &  ''   & 0.047 & 0.95 (0.41) & 0.19 (0.14) & 0.00 (0.41) & 0.39 (0.06) &\ 0.06\\
\enddata 
\label{tbl:estat_best} 
\end{deluxetable}
\end{center}

\subsection{Active Region McIntosh Class Poisson Probabilities -
D.S.~Bloomfield, P.A.~Higgins, P.T.~Gallagher}
\label{sec:poisson}

Another method that was applied to the active region patch dataset which did
not use the magnetogram data was one based on historical flare rates from
McIntosh active region classifications. This method is the same as that
presented in \cite{Bloomfield_etal_2012}, where the occurrence of GOES X-ray
flares from individual McIntosh classifications were collated over 1969-1976
and 1988-1996. These average 24-hr flaring rates, $\mu_{24}$, lead to a Poisson
probability of one or more flares occurring in any 24-hr interval from
\cite{Gallagheretal2002},
\begin{equation} 
P_{\mu_{24}} (N \geqslant 1) = 1 - \exp( -\mu_{24} ) \ ,
\end{equation}
or the probability of one or more flares in any 12-hr interval from,
\begin{equation}\label{eqn:P12} 
P_{\mu_{12}} (N \geqslant 1) = 1 - \exp( -\mu_{24}/2 )\ .
\end{equation}
The nature of the statistics collated in \cite{Bloomfield_etal_2012} places a
limitation on the forecasts outlined in Section~\ref{sec:eventdata} that are
able to be studied by this method. Forecasts of at least one C1.0 or greater
flare within 24 hr (\CC) are directly achieved by the Poisson flare
probabilities for ``Above GOES C1.0'' published in Table 2 of
\cite{Bloomfield_etal_2012}, while forecasts of at least one M1.0 or greater
flare within 12 hr (\MS) were calculated by combining the 24-hr M- and X- class
flare rates published in the same Table ($\mu_{24}$) and applying
Equation~\ref{eqn:P12} above. Forecasts of at least one M5.0 flare or greater
within 12 hr (\ML) were not capable of being achieved by this method because
the flares that were collated from 1969-1976 \citep{Kildahl1980} were
identified only by their GOES class and not the complete class and magnitude.

For each magnetogram patch the observation date and NOAA region number(s)
contained within it were used to cross-reference the McIntosh class of that
active region in that day's NOAA Solar Region Summary file. It should be noted
that, even for the \CC\ and \MS\ forecasts, flare probabilities were not able
to be issued for some patches because the magnetogram was recorded before the
active region had received a NOAA number designation. For the case of
magnetogram patches containing multiple active regions, the reported flare
probability was the largest probability from any of the regions within that
patch.

\begin{figure}[h!]
\centerline{
\includegraphics[width=0.33\textwidth, clip]{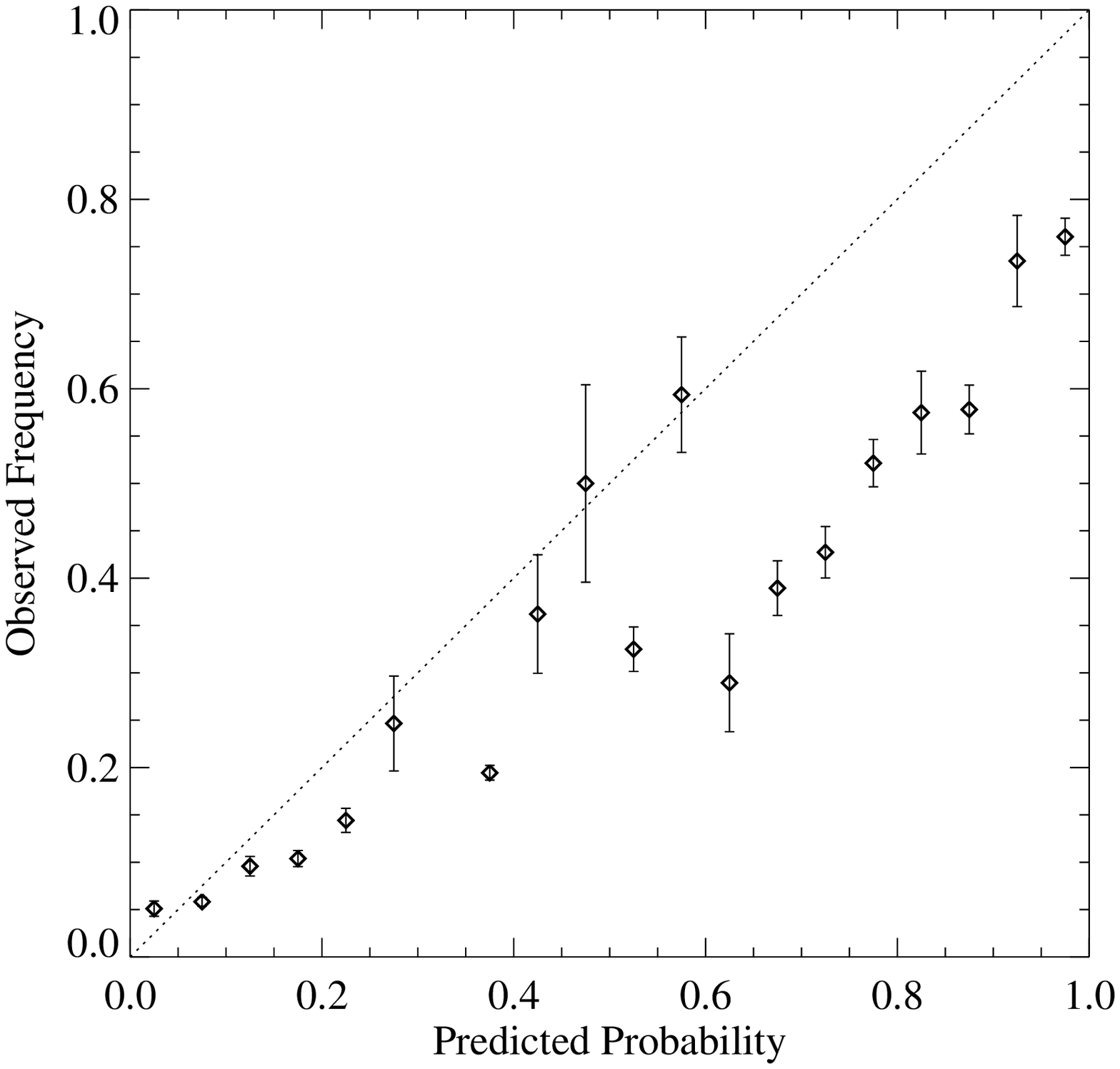}
\includegraphics[width=0.33\textwidth, clip]{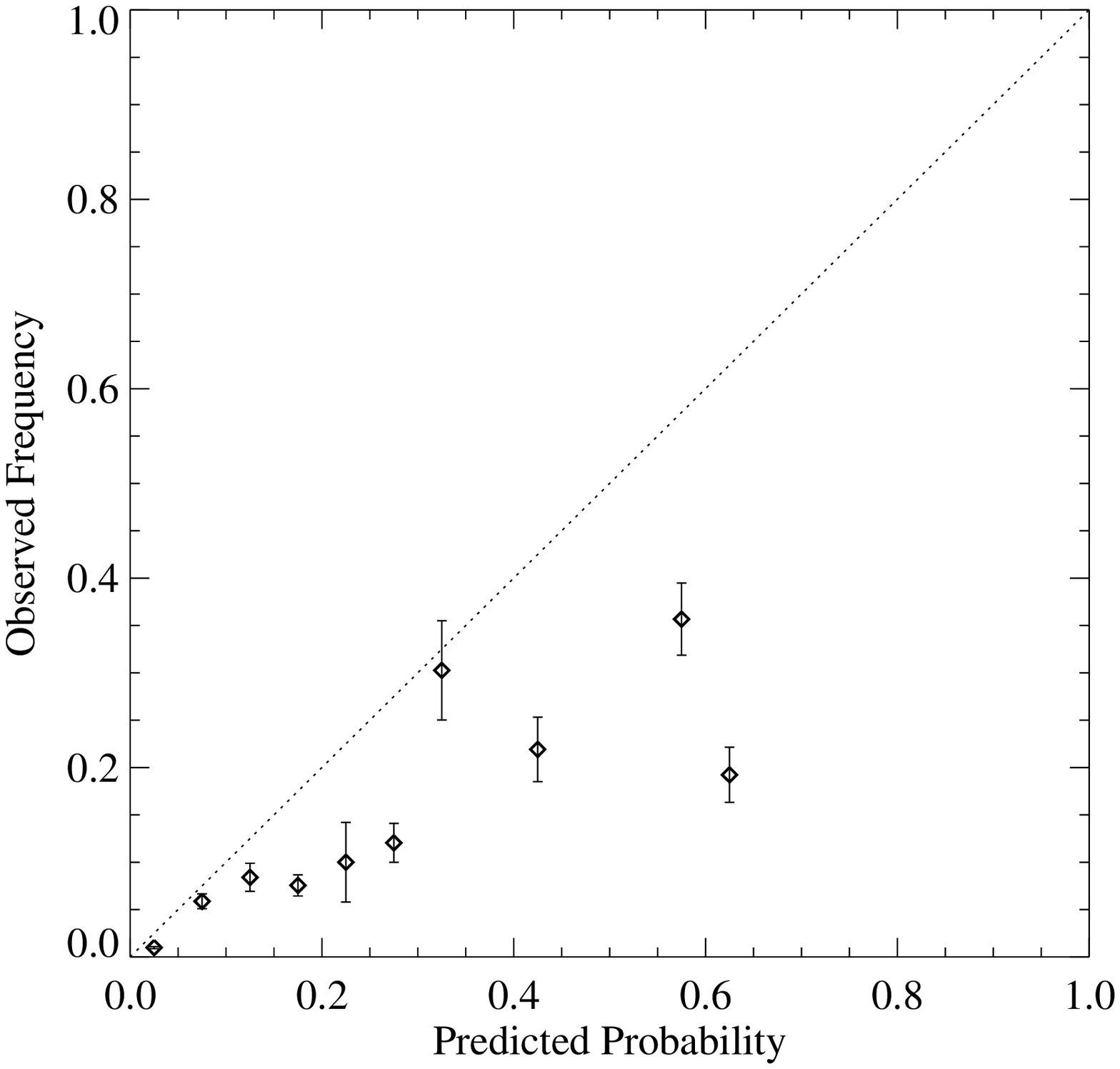}
\includegraphics[width=0.33\textwidth, clip]{Reliability_dummy.eps}}
\centerline{
\includegraphics[width=0.33\textwidth, clip]{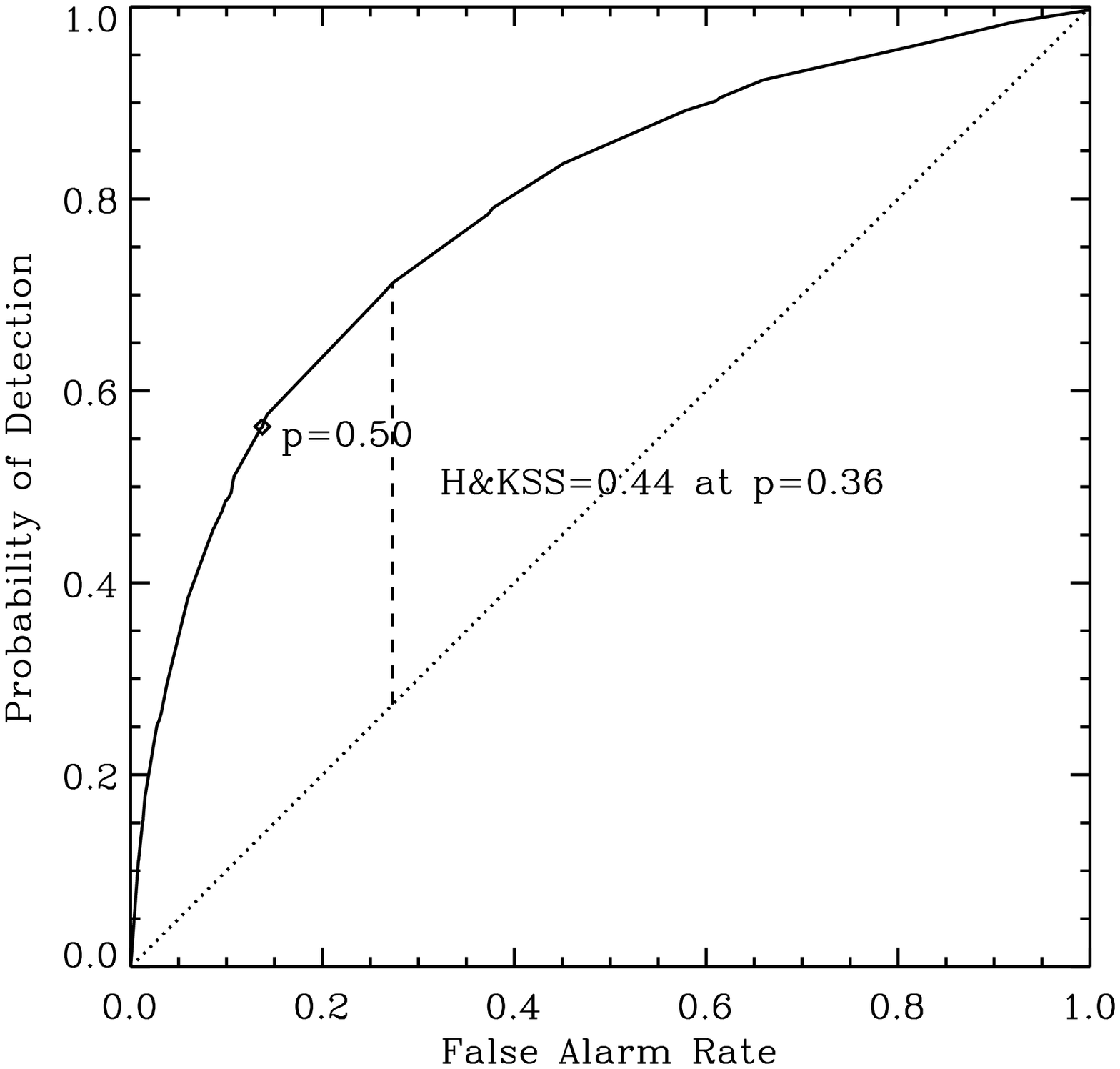}
\includegraphics[width=0.33\textwidth, clip]{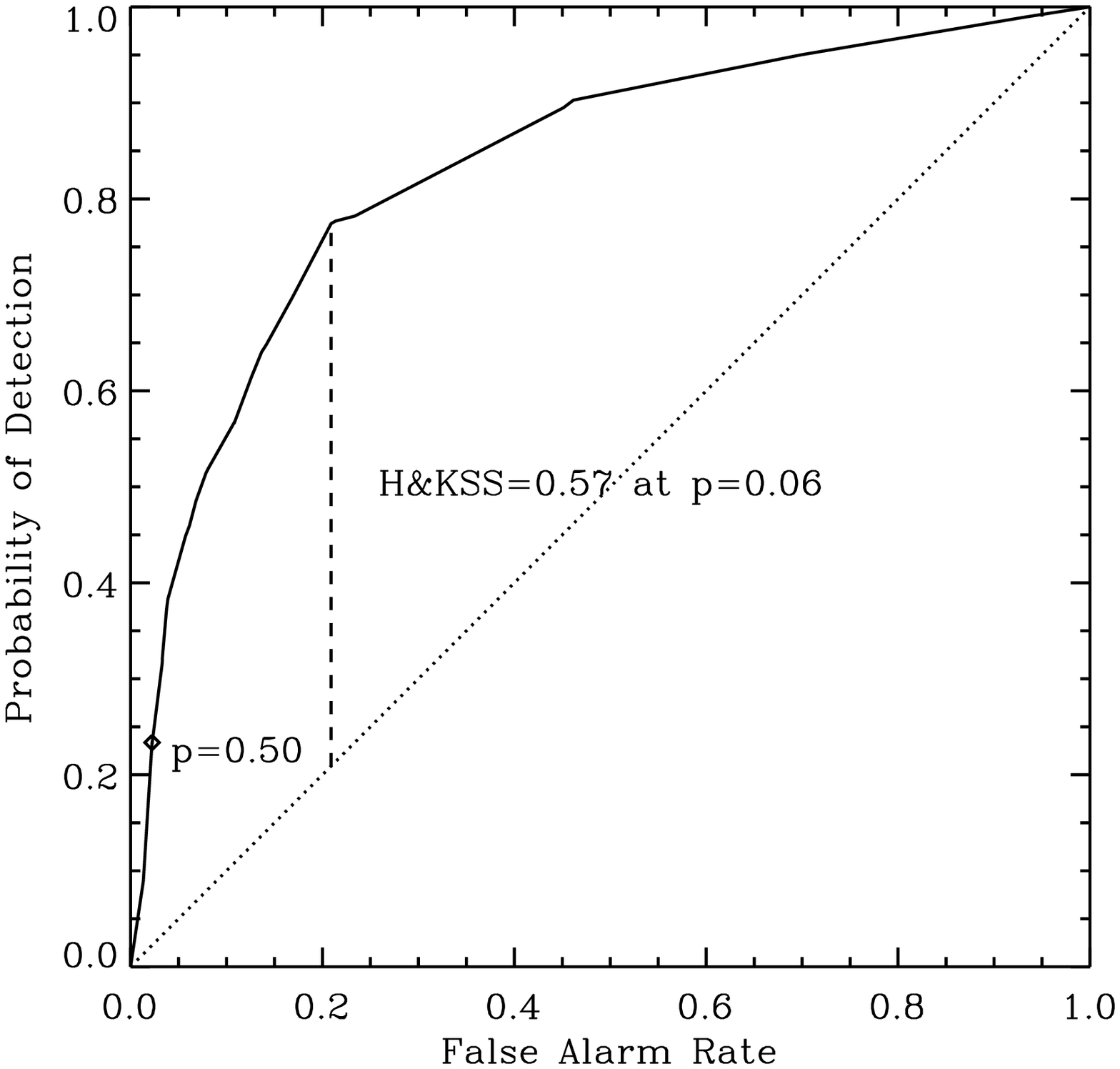}
\includegraphics[width=0.33\textwidth, clip]{ROC_dummy.eps}}
\caption{Same as Figure~\ref{fig:beff_plots} but for the Poisson Statistics method.  In this
case, no prediction could be made for the \ML\ event definition.}
\label{fig:poisson_plots}
\end{figure}

The summary plots (Figure~\ref{fig:poisson_plots}) for the \CC\ and \MS\
thresholds show an overprediction tendency in the reliability plots, with
marginal improvement with increased threshold in the ROC plots.  The Poisson
method as implemented here is an example of a method whose requirements are not
well met by the data used for this workshop, and the method is likely penalized
as a result.

\begin{center}
\begin{deluxetable}{cccccccc}
\tablecolumns{8}
\tablewidth{0pc}
\tablecaption{Optimal Performance Results: Poisson Statistics}
\tablehead{
\colhead{Event} & \colhead{Sample} & \colhead{Event} & \colhead{\RC} & \colhead{\Heidke} & \colhead{\Appleman} & \colhead{\True} & \colhead{\Brier} \\
\colhead{Definition} & \colhead{Size} &  \colhead{Rate} & \colhead{(threshold)} & \colhead{(threshold)} & \colhead{(threshold)} & \colhead{(threshold)} & \colhead{} 
}
\startdata
\CC & 11385 & 0.210 & 0.82 (0.73) & 0.42 (0.41) & 0.16 (0.73) & 0.44 (0.36) &\ 0.07\\
\MS &  ''   & 0.033 & 0.97 (0.62) & 0.28 (0.26) & 0.00 (0.62) & 0.56 (0.06) & -0.06\\
\ML &  ''   &   N/A &         N/A &         N/A &         N/A &         N/A &   N/A\\
\enddata
\label{tbl:poisson_best}
\end{deluxetable}
\end{center}

\section{Accessing the database}
\label{app:website}

The website for all data is: \url{http://www.cora.nwra.com/AllClear/}.
Users are required to register but otherwise access is completely open.
The MDI data sets are
provided, as well as event lists (boolean results of events for each data set
according to event definition).  Also provided are all of the parameters
calculated by each group participating, summaries of forecasts and resulting
skill scores.  There are multiple {\tt ``README''} files on format and content.
We request acknowledgement for use of the data, and that if you use the data to
make predictions, you agree to allow your results to be added to the database
(at an appropriate time with respect to relevant publications), as all
participants herein have agreed.

We invite groups doing research on active regions and flares, as well as
on statistical analysis, to become involved.  Instructions are posted on how 
to submit forecasts and/or new parameters, in order to (for example) benchmark
new techniques against those highlighted here.

Any questions regarding the website should be addressed to the NWRA authors.

\end{document}